\documentclass[12pt,prd, aps, showpacs, superscriptaddress,floatfix]{revtex4-1}
\usepackage{graphicx}
\usepackage{bm}
\usepackage{multirow}
\usepackage{axodraw}
\usepackage{epsfig}
\usepackage{psfrag}
\usepackage{soul}
\usepackage{multirow}
\usepackage{mathptmx}
\usepackage{mathrsfs}
\usepackage{amsmath, amssymb}
\usepackage{dcolumn}
\usepackage{epstopdf}
\usepackage{color}
\usepackage{hhline}
\usepackage{changebar}
\usepackage{subfigure}
\usepackage{placeins}

\usepackage{slashed}
\usepackage{longtable}
\usepackage{CJK}
\newcommand\adelaide{CSSM, School of Chemistry and Physics, University of Adelaide, Adelaide SA 5005, Australia}
\newcommand\riken{RIKEN-BNL Research Center, Brookhaven National
  Laboratory, Upton, NY 11973, USA}
\newcommand\bnl{Brookhaven National Laboratory, Upton, NY 11973, USA}
\newcommand\edinb{SUPA, School of Physics, The University of
  Edinburgh, Edinburgh EH9 3JZ, UK}

\newcommand\cu{Physics Department, Columbia University, New York,
  NY 10027, USA}

\newcommand\uconn{Physics Department, University of Connecticut,
  Storrs, CT 06269-3046, USA}
\newcommand\soton{School of Physics and Astronomy, University of
  Southampton,  Southampton SO17 1BJ, UK}
\newcommand\kek{Institute of Particle and Nuclear Studies,
  KEK, Tsukuba, Ibaraki, 305-0801, Japan}
\newcommand\sokendai{Department of Particle and Nuclear Physics, Sokendai
Graduate University of Advanced Studies, Hayama, Kanagawa 240-0193, Japan}

\newcounter{Outline}
\newcounter{Introduction}
\newcounter{ChPT}
\newcounter{SimDetail}
\newcounter{LatticePS}
\newcounter{Baryons}
\newcounter{SuTwo}
\newcounter{Bk}
\newcounter{Vector}
\newcounter{CombinedChiralFits}
\newcounter{Conclusions}
\newcounter{Acknowledgments}
\newcounter{Appendix}
\newcounter{Tables}
\newcounter{Figures}
\newcommand{\ba}{\begin{eqnarray}}
\newcommand{\ea}{\end{eqnarray}}
\newcommand{\bas}{\begin{eqnarray*}}
\newcommand{\eas}{\end{eqnarray*}}
\newcommand{\be}{\begin{equation}}
\newcommand{\ee}{\end{equation}}
\newcommand{\bes}{\begin{equation*}}
\newcommand{\ees}{\end{equation*}}
\newcommand{\bi}{\begin{itemize}}
\newcommand{\ei}{\end{itemize}}

\newcommand{\bcentre}{\begin{center}}
\newcommand{\ecentre}{\end{center}}

\font\tenmsb=msbm10 scaled\magstep1
\font\sevenmsb=msbm7 scaled\magstep1
\font\fivemsb=msbm5 scaled\magstep1
\newfam\msbfam
\textfont\msbfam=\tenmsb
\scriptfont\msbfam=\sevenmsb
\scriptscriptfont\msbfam=\fivemsb

\newcommand{\mres}{m_{\rm res}}

\def\msbar{\overline{\mbox{\scriptsize MS}}}
\def\rational#1#2{{\mathchoice{\textstyle{#1\over#2}}%
  {\scriptstyle{#1\over#2}}{\scriptscriptstyle{#1\over#2}}{#1/#2}}}
\def\half{\rational12}			    % One half
\def\rmsub#1#2{#1_{\mbox{\tiny #2}}}	    % Small roman subscript
\def\ln{\mathop{\rm ln}}		    % ln
\def\det{\mathop{\rm det}}		    % Determinant
\def\su3{SU(3)}

\def\tD{\mbox{D}\kern-0.65em\raise0.15ex\hbox{/}\kern0.15em} % D slash
\def\sD{\mbox{\scriptsize D}\kern-0.5em\raise0.15ex\hbox{\scriptsize/}}
\def\ssD{\mbox{\tiny D}\kern-0.42em\raise0.15ex\hbox{\tiny/}}

\def\dslash{\hbox{\(\partial\)}\kern-0.5em\raise0.15ex\hbox{/}} % d slash

\def\su3{SU(3)}

\def\mres{\rmsub{m}{res}}

\def\nicefrac#1#2{\leavevmode\kern.1em\raise.5ex\hbox{\the\scriptfont0 #1}\kern-
.1em/\kern-.15em\lower.25ex\hbox{\the\scriptfont0 #2}}

\newcommand{\nconfIDlight}{180 }
\newcommand{\nconfIDheavy}{148 }
\newcommand{\msIDbare}{0.0467(6)}
\newcommand{\mresIDsim}{0.001842(7)}
\newcommand{\mpicut}{350\ }
\newcommand{\mpicutanalytic}{260\ }

\newcommand{\ainvzero}{2.31(4)}
\newcommand{\ainvone}{1.75(4)}
\newcommand{\ainvtwo}{1.37(1)}

\newcommand{\bea}{\begin{eqnarray}}
\newcommand{\eea}{\end{eqnarray}}

\newcommand{\gmuL}{\gamma_\mu^{\,L}}

\newcommand{\GeV}{\;\rm GeV}

\newcommand{\schemeS}{{\rm S}}
\newcommand{\qslashs}{\not{\hspace{1pt}{q}}}
\newcommand{\qslash}{\not{\! q}}

\newcommand{\qslashL}{\not{\! \hat q}^{\,L}}

\begin{document}
\bibliographystyle{apsrev}

%%%%%%%%%%%%%%%%%%%%%%%%%%%%%% TITLEPAGE %%%%%%%%%%%%%%%%%%%%%%%%%%

%CJK
\begin{CJK*}{UTF8}{}
%CJK
\title{Domain wall QCD with near-physical pions}

\author{R.~Arthur}\affiliation{\edinb}
\author{T.~Blum}\affiliation{\uconn}\affiliation{\riken}
\author{P.A.~Boyle}\affiliation{\edinb}
\author{N.H.~Christ}\affiliation{\cu}
\author{N.~Garron}\affiliation{\edinb}
\author{R.J.~Hudspith}\affiliation{\edinb}
\author{T.~Izubuchi}\affiliation{\riken}\affiliation{\bnl}
\author{C.~Jung}\affiliation{\bnl}
\author{C.~Kelly}\affiliation{\cu}
\author{A.T.~Lytle}\affiliation{\soton}
\author{R.D.~Mawhinney}\affiliation{\cu}
\author{D.~Murphy}\affiliation{\cu}
%\author{S.~Ohta}\affiliation{\kek}\affiliation{\sokendai}\affiliation{\riken}
%CJK
\CJKfamily{min}
\author{S.~Ohta (太田滋生)}\affiliation{\kek}\affiliation{\sokendai}\affiliation{\riken}
 %\email{shigemi.ohta@kek.jp}
%CJK
\author{C.T.~Sachrajda}\affiliation{\soton}
\author{A.~Soni}\affiliation{\bnl}
\author{J.~Yu}\affiliation{\cu}
\author{J.M.~Zanotti}\affiliation{\adelaide}

\collaboration{RBC and UKQCD Collaborations}
%
% preprint numbers
%
\mbox{}\hfill\noaffiliation{ADP-12-15/T782, CU-TP-1203, Edinburgh 2012/05, KEK-TH-1540, RBRC 949, SHEP-1213}


\maketitle
%CJK
\end{CJK*}
%CJK
%%%%%%%%%%%%%%%%%%%%%%%%%%%%%% ABSTRACT %%%%%%%%%%%%%%%%%%%%%%%%%%%%%%%
\centerline{ABSTRACT}
We present physical results for a variety of light hadronic quantities obtained via a combined analysis of three 2+1 flavor domain wall fermion ensemble sets. For two of our ensemble sets we used the Iwasaki gauge action with $\beta=2.13$ ($a^{-1}=\ainvone$ GeV) and $\beta=2.25$ ($a^{-1}=\ainvzero$ GeV) and lattice sizes of $24^3\times 64$ and $32^3\times 64$ respectively, with unitary pion masses in the range 293(5)--417(10) MeV. The extent $L_s$ for the $5^{\textrm{th}}$ dimension of the domain wall fermion formulation is $L_s=16$ in these ensembles. In this analysis we include a third ensemble set that makes use of the novel Iwasaki+DSDR (dislocation suppressing determinant ratio) gauge action at $\beta = 1.75$ ($a^{-1}=\ainvtwo$ GeV) with a lattice size of $32^3\times 64$ and $L_s=32$ to reach down to partially-quenched pion masses as low as $143(1)$ MeV and a unitary pion mass of $171(1)$ MeV, while retaining good chiral symmetry and topological tunneling. We demonstrate a significant improvement in 
our control over the chiral extrapolation, resulting in much improved continuum predictions for the above quantities. The main results of this analysis include the pion and kaon decay constants, $f_\pi=127(3)_{\rm stat}(3)_{\rm sys}$ MeV and $f_K = 152(3)_{\rm stat}(2)_{\rm sys}$ MeV respectively ($f_K/f_\pi = 1.199(12)_{\rm stat}(14)_{\rm sys}$); the average up/down quark mass and the strange-quark mass in the $\msbar$-scheme at 3 GeV, $m_{ud}(\msbar, 3\ {\rm GeV}) = 3.05(8)_{\rm stat}(6)_{\rm sys}$ MeV and $m_s(\msbar, 3\ {\rm GeV}) = 83.5(1.7)_{\rm stat}(1.1)_{\rm sys}$; the neutral kaon mixing parameter in the $\msbar$-scheme at 3 GeV, $B_K(\msbar,3\;\GeV) = 0.535(8)_{\rm stat}(13)_{\rm sys}$, and in the RGI scheme, $\hat B_K = 0.758(11)_{\rm stat}(19)_{\rm sys}$; and the Sommer scales $r_1 = 0.323(8)_{\rm stat}(4)_{\rm sys}$ fm and $r_0 = 0.480(10)_{\rm stat}(4)_{\rm sys}$ ($r_1/r_0 = 0.673(11)_{\rm stat}(3)_{\rm sys}$). We also obtain values for the SU(2) chiral perturbation theory effective couplings, 
$\bar{l_3} = 2.91(23)_{\rm stat}(7)_{\rm sys}$ and $\bar{l_4} = 3.99(16)_{\rm stat}(9)_{\rm sys}$.
\refstepcounter{section}
\setcounter{section}{0}

%%%%%%%%%%%%%%%%%%%%%%%%%%%%  Section %%%%%%%%%%%%%%%%%%%%%%%%%%%%%%%%

%%%%%%%%%%%%%%%%%%%%%%%%%%%% Introduction %%%%%%%%%%%%%%%%%%%%%%%%%%%%
\newpage
\section{Introduction}
\label{sec:Introduction}

% \ifnum\theIntroduction=1
The RBC and UKQCD collaborations have recently published continuum limit results~\cite{Aoki:2010dy,Aoki:2010pe} for a variety of light hadronic quantities, including the pion and kaon decay constants, quark masses and the neutral kaon mixing parameter $B_K$, determined using two ensemble sets of $2+1$-flavor domain wall fermions (DWF) with the Iwasaki gauge action at $\beta = 2.25$ (corresponding to a lattice spacing of $a\approx 0.086$ fm) and $\beta=2.13$ ($a\approx 0.114$ fm), with lattice sizes of $32^3\times64$ and $24^3\times 64$ respectively and fifth-dimensional extents of $L_s=16$. We refer to this as the ``2010 analysis''. With precise nonperturbative renormalization methods made possible by the good chiral symmetry of the action, and a combined chiral/continuum fit analysis to maximise the use of the available data, our predictions were limited mainly by the $\mathcal{O}(5\%)$ systematic error on the extrapolation from the simulated $293(5)\ \mathrm{MeV}\leq m_\pi \leq 417(10)$ MeV pion mass-range 
to the physical point. In order to address this issue we must simulate with lighter quark masses, which necessitates an increase in the physical lattice volume in order to maintain small finite-volume corrections. As increasing the number of lattice sites is very costly we must use coarser lattices in order to perform the calculation with the currently available resources. Aside from the larger discretization errors, the only significant impact of simulating with a coarser lattice is an increase in the size of the residual mass $m_\mathrm{res}$, which parametrizes the explicit chiral symmetry breaking occurring due to the finite length of the fifth dimension. $\mres$ gets larger due to the increased number of low-modes of the Wilson Dirac operator in the infrared regime, that are likely caused by so-called ``dislocations'' -- localized instanton-like artifacts -- in the gauge fields. Configurations containing these low modes may be suppressed in the path integral via the introduction to the gauge action of an 
additional weighting factor known as the dislocation suppressing determinant ratio (DSDR)~\cite{Vranas:1999rz,Vranas:2006zk,Fukaya:2006vs,Renfrew:2009wu}.

In this paper we present the ``2012 analysis'' of the RBC and UKQCD collaboration's $\beta = 1.75$ $32^3\times 64\times 32$ DWF ensembles that make use of the Iwasaki+DSDR gauge action to reach unitary pion masses as low as $171(1)$ MeV and partially-quenched pion masses at a near-physical value of $143(1)$ MeV. The results for the physical quark masses and lattice spacings presented in this document were used in our recent calculation of the $\Delta I=3/2$ $K\rightarrow \pi\pi$ amplitudes with physical kinematics~\cite{KtopipiPRD}. 

Note that the pion masses in physical units quoted above and in the abstract, as well as those given in the remainder of this paper, were obtained by combining the data at the simulated strange quark mass with the final lattice spacings obtained in this analysis, and the error represents the combined systematic and statistical uncertainty.

Throughout this document we make use of the shorthand 32ID to refer to the $32^3\times 64\times 32\ $ Iwasaki+DSDR ensemble set, and 32I and 24I for the $32^3\times64\times 16$ and $24^3\times 64\times 16$ Iwasaki ensemble sets respectively. This notation differs slightly from ref.~\cite{KtopipiPRD}, where the Iwasaki+DSDR ensemble set was labelled 32IDSDR.

In this paper all dimensionful quantities are expressed in lattice units unless other units are explicitly specified or clarity is served by introducing explicit factors of the lattice spacing $a$.

In section~\ref{sec:SimulationDetails} we provide further details on the Iwasaki+DSDR gauge action and present the simulation parameters of our 32ID ensembles. In section~\ref{sec:DSDRresults} we present our results for the pseudoscalar masses and decay constants, the Omega baryon mass (used to set the scale), the Sommer scales $r_0$ and $r_1$ and also $B_K$, measured on these ensembles.

In the Symanzik effective action (up to and including dimension-5 terms), explicit chiral symmetry breaking effects manifest as a dimension-3 term closely related to the residual mass, and a dimension-5 clover term. The latter introduces ${\cal O}(a)$ discretization errors that make it difficult to perform continuum extrapolations with traditional Wilson fermions. In the domain wall formulation however, the clover term has a magnitude of $\mathcal O(a^2 \mres)$, and can therefore be discounted in our simulations, where $a\mres$ is always on the order of $10^{-3}$ or smaller. (In Appendix~\ref{appendix-cloverterm} we perform additional checks to ensure that this assumption remains true for our Iwasaki+DSDR ensembles.) Due to the excellent chiral symmetry, lattice artefacts involving odd powers of the lattice spacing are heavily suppressed and we gain automatic offshell $\mathcal{O}(a)$ improvement. As a result, the leading discretization effects appear at $\mathcal{O}(a^2)$, and the next-to-leading effects at 
$\mathcal O(a^4)$. Note that higher order corrections to the Symanzik expansion can lead to terms logarithmic in the lattice spacing that can, in extreme circumstances, spoil the neat power-law behaviour we have described; in Appendix~\ref{appendix-alphascorrections} we discuss this possibility further, and conclude that, providing the range of lattice spacings under consideration is not too large, such corrections introduce systematic errors into our continuum extrapolation similar to those that result from the neglected $O(a^4)$ terms and, for the 0.086--0.11 fm range of lattice spacings considered here, can be expected to be of a similar size. (At nonzero quark mass there can also arise terms $\mathcal{O}(a^2 m_q)$, which can also be expected to be of a similar size.) In our analysis of the present DSDR ensemble, we find the typical size of the $\mathcal{O}(a^2)$ terms to be $\lesssim 5\%$, hence we can expect the next-to-leading discretization errors to be roughly $\mathcal O(0.05^2) \sim 0.25\%$. These 
are an order of magnitude smaller than the errors arising from the chiral extrapolation and the nonperturbative renormalization (where appropriate), and can therefore be safely ignored. The only surviving dependence on the lattice spacing is therefore a single $\mathcal{O}(a^2)$ term for each measured quantity. Of course this term depends on the lattice action, but as all other parameters (the slopes with respect to the quark masses) describing the quantity are common between the Iwasaki and Iwasaki+DSDR actions, we can easily obtain the $a^2$ coefficients for the Iwasaki+DSDR action by comparing any single measured value on the 32ID ensemble set with the continuum limit obtained from the Iwasaki ensembles. In practice we include the Iwasaki+DSDR ensembles in our simultaneous chiral/continuum fitting framework, allowing these data to constrain the mass dependences close to the physical point, substantially reducing the chiral extrapolation systematic error on our continuum predictions, as well as allowing 
us to obtain the $a^2$ coefficients for the Iwasaki+DSDR data. In this framework, any remaining errors associated with the leading, $\mathcal{O}(a^2)$ effects are included in the statistical error. Since we have only two ensembles with different values for the lattice spacing that use the same lattice action, we can only make a simple $a^2\to0$ extrapolation to remove the ${\cal O}(a^2)$ artifacts.  Remaining lattice artifacts of order $a^4$ or higher, or possible $a^2 \ln(a^2)$ effects, can only be estimated from the size of the observed $a^2$ effect and contribute small systematic errors.

The chiral/continuum fitting framework is discussed in more detail in section~\ref{sec:CombinedChiralFits}. We use this procedure in sections~\ref{sec:FitResults} through~\ref{sec:r0r1} to simultaneously fit the aforementioned quantities over all three ensemble sets, from which we obtain the lattice spacings and physical quark masses as well as improved continuum predictions for the decay constants, Sommer scales and $B_K$.

%In practice we can include the Iwasaki+DSDR ensembles in our simultaneous chiral/continuum fitting framework to obtain these $a^2$ coefficients under a fit to all of the 32ID data, as well as allowing this data to further constrain the shared parameters. In this framework, any remaining errors associated with the leading, $\mathcal{O}(a^2)$ effects are included in the statistical error. This is discussed further in section~\ref{sec:CombinedChiralFits}. We use this procedure in sections~\ref{sec:FitResults} through~\ref{sec:r0r1} to simultaneously fit the aforementioned quantities over all three ensemble sets, from which we obtain the lattice spacings and physical quark masses as well as improved continuum predictions for the decay constants, Sommer scales and $B_K$.

In closing this section we would like to emphasize the importance of the discussion in the above paragraphs. Aside from the $\mathcal{O}(a^2)$ errors that are explicitly included in our fit, the next largest discretization effects arise at $\mathcal O(a^4)$. This level of control over the discretization effects can be achieved, as we demonstrated in our 2010 analysis and also in this document, using only two lattice spacings. To resolve the $\mathcal{O}(0.25\%)$ next-to-leading effects would require another lattice spacing (and likely a substantial increase in statistics), which we do not deem a sensible use of our resources in light of their expected size in comparison to our other systematic errors. This is in contrast to other lattice formulations which do not have automatic $\mathcal O(a)$ improvement, such as the Wilson approach, for which not three but five lattice spacings are required for an effect of this size to be measured.

% \fi

%%%%%%%%%%%%%%%%%%%%%%%% Simulation Details and Ensemble Properties %%%%%%%%%%%%%%%%%%%%%%%%%%%%%%
\section{Simulation Details and Ensemble Properties}
\label{sec:SimulationDetails}

We generated a set of domain wall fermion ensembles using the Iwasaki+DSDR gauge action, which allows for simulations to be performed on coarser lattices while retaining good chiral symmetry and topological tunneling. In this section we provide background on the DSDR term followed by a list of simulation parameters and an analysis of the integrated autocorrelation length and topological charge evolution.

\subsection{The DSDR term}

The explicit breaking of chiral symmetry in the domain wall fermion framework can be described by an additive mass renormalization parameter referred to as $\mres$, whose magnitude is related to the eigenvalue density $\rho(\lambda)$ of the logarithm of the transfer matrix in the fifth-dimension,
\begin{equation}
H_{\rm transfer} = 2\tanh^{-1}\left( \frac{H_W}{2+D_W} \right)\,,\label{eqn-HTdef}                                                                                                                                                                                                                                                                                                                                                                                                                                              \end{equation}
that describes the propagation of quarks through the fifth dimension, via the following relation~\cite{Antonio:2007tr}:
\begin{equation}
\mres = R^4\int_0^\infty {\rm d}\lambda\ \rho(\lambda)e^{-L_s \lambda}\,.\label{eqn-mresintlambda}
\end{equation}
Here $R$ is a (possibly eigenvalue-dependent) radius factor, $D_W$ is the Wilson Dirac operator and $H_W = \gamma^5 D_W$ is the hermitian Wilson Dirac operator. 

In the low-eigenvalue region the eigenmodes of $H_{\rm transfer}$ and those of $H_W$ are necessarily identical. It has been demonstrated~\cite{Aoki:2001su,Aoki:2001dea,Golterman:2003qe,Golterman:2004cy,Golterman:2005fe,Svetitsky:2005qa,Antonio:2007tr} that the modes of the latter can be divided into two regions, one containing only localized eigenmodes with small eigenvalues and one containing extended eigenmodes with large eigenvalues, separated by a mobility edge $\lambda_c$. Picking out the dominant contributions above and below the mobility edge from eqn.~\ref{eqn-mresintlambda}, we expect the following dependence of $\mres$ upon $L_s$:
\begin{equation}
\mres = R_e^4  \rho(\lambda_c)\frac{e^{-\lambda_c L_s}}{L_s} + R_l^4\rho(0)\frac{1}{L_s}\,,
\end{equation}
where $R_e$ and $R_l$ are the radius parameters for the extended and local modes respectively. The exponentially-decreasing contribution from the extended modes above the mobility edge can be controlled by increasing $L_s$, with a cost that rises at worst linearly. In our previous Iwasaki simulations the magnitude of $\mres$ was dominated by the term in $\rho(0)$, the density of near-zero eigenmodes. These modes are thought to be associated with localized and short-lived dislocations or ``tears'' in the gauge fields, which can cause changes in the field topology. As the strong coupling limit is approached, the gauge fields become more disordered and the density of near-zero modes increases sharply. In order to maintain good chiral symmetry properties at stronger coupling we must therefore seek to suppress the near-zero modes. On the other hand we must take care not to also remove the very-near-zero eigenmodes that are required for topological tunneling to occur during the gauge evolution.

The DSDR, or ``auxiliary determinant'' is applied to the gauge action as a multiplicative weight of the form~\cite{Vranas:1999rz,Vranas:2006zk,Fukaya:2006vs,Renfrew:2009wu}
\begin{equation}
\mathcal{W}(M;\epsilon_f;\epsilon_b) = \frac{\mathrm{det}\left[D_W(-M+i\epsilon_f\gamma^5)^\dagger D_W(-M+i\epsilon_f\gamma^5)\right]  }{\mathrm{det}\left[D_W(-M+i\epsilon_b\gamma^5)^\dagger D_W(-M+i\epsilon_b\gamma^5)\right]} = \prod_i \frac{\lambda_i^2+\epsilon_f^2}{\lambda_i^2+\epsilon_b^2}  \,,
\end{equation}
where $\epsilon_f$ and $\epsilon_b$ are tunable parameters with typical sizes $0<\epsilon_f^2 \ll \epsilon_b^2<1$. With this weighting, the contribution of a single eigenmode to the molecular dynamics force becomes a function of $\epsilon_f$ and $\epsilon_b$ of the form
\begin{equation}
\mathcal{F}_i(\epsilon_f,\epsilon_b) = \frac{d}{d\lambda_i}\left(-\log \frac{\lambda_i^2+\epsilon_f^2}{\lambda_i^2+\epsilon_b^2}\right)\,,
\end{equation}
which when plotted against the eigenvalue has a peak and tail which are independently tunable by varying the two parameters. It is therefore possible to tune the force to suppress near-zero eigenmodes while not completely suppressing the essential very-near-zero modes. 

Numerical studies~\cite{Renfrew:2009wu} have demonstrated a reduction in chiral symmetry breaking while retaining adequate topological tunneling through the use of this term. In Appendix~\ref{appendix-cloverterm} we demonstrate the lack of observable explicit chiral symmetry breaking effects on our Iwasaki+DSDR ensembles.

\subsection{Simulation parameters}

We generated DWF ensembles with the Shamir kernel and the Iwasaki+DSDR gauge action on a $32^3\times 64$ lattice volume with $L_s = 32$. We used a ``domain wall height'' of $M_5=1.8$ and a gauge coupling of $\beta = 1.75$, which as determined in section~\ref{sec:FitResults}, corresponds to an inverse lattice spacing of $\ainvtwo$ GeV. The parameters of the DSDR factor, $\epsilon_b = 0.5$ and $\epsilon_f = 0.02$, were chosen to minimize the residual mass while still allowing a reasonable rate of topological tunneling. We generated two ensembles with bare light-quark masses of $m_l=0.001$ and $m_l = 0.0042$, for which the corresponding unitary pion masses are $171(1)$ and $246(2)$ MeV. In this document we analyze $\sim 1400$ and $\sim 1200$ MD time units on these ensembles respectively (discarding 500 and 600 MD time units respectively for thermalization). On each of the ensembles we simulated with a single strange-quark mass close to the physical value and use reweighting to correct to the true physical value 
in our fits \textit{a posteriori}. Further details of the number of reweighting steps and stochastic samples are given in the following subsection.

\subsection{Ensemble generation}

In this section we provide a summary of the Monte Carlo algorithms that were employed for the gauge evolution. Further discussion of our algorithms, along with the full set of parameters, can be found in Appendix~\ref{appendix-integrators}.

For the fermionic contribution to the evolution of the $m_l=0.0042$ ensemble we employed the ``RHMC II'' algorithm~\cite{Allton:2008pn}, in which the calculation of the strange-quark determinant is broken into three factors and evaluated using the rational approximation with equal molecular dynamics time steps, and the determinant of the two degenerate light-quarks was preconditioned by the strange-quark determinant. With the notation $\mathcal{D}(m) = D^\dagger_{\rm DWF}(M_5,m)D_{\rm DWF}(M_5,m)$ for the Hermitian domain wall operator and using $\mathcal{R}_a(m)$ to represent the rational approximation to the $a^{\mathrm{th}}$ power of $\mathcal{D}$ for mass $m$, the algorithm can be written as
{\small
\begin{equation}
{\rm det}\left[\frac{\mathcal{D}(m_s)^{\frac{1}{2}}\mathcal{D}(m_l)}{\mathcal{D}(1)^{\frac{3}{2}}}\right] = {\rm det}\left[\mathcal{R}_{\frac{1}{2}}\left(\frac{\mathcal{D}(m_s)}{\mathcal{D}(1)}\right)\right]\cdot{\rm det}\left[\mathcal{R}_{\frac{1}{2}}\left(\frac{\mathcal{D}(m_s)}{\mathcal{D}(1)}\right)\right]\cdot{\rm det}\left[\mathcal{R}_{\frac{1}{2}}\left(\frac{\mathcal{D}(m_s)}{\mathcal{D}(1)}\right)\right]\cdot{\rm det}\left[\frac{\mathcal{D}(m_l)}{\mathcal{D}(m_s)}\right]\,,
\end{equation}}where each determinant is estimated using independent pseudofermion fields. We made use of an Omelyan integrator with parameter $\lambda=0.22$ during the evolution of this ensemble. 

For the lighter $m_l=0.001$ ensemble, we were able to achieve a significant speed-up~\cite{Yin:2011sz} in evaluating the light-quark contribution to the gauge field update using multiple Hasenbusch mass splittings~\cite{Hasenbusch:2001ne,Yin:2011sz}. Here the determinant is split into $k$ steps (with $k=6$ in our case), each evaluated using a shifted mass:
\begin{equation}
 {\rm det}\left[\frac{\mathcal{D}(m_l)}{\mathcal{D}(1)}\right] = \prod_{i=1}^{k+1}{\rm det}\left[\frac{\mathcal{D}(m_l+\mu_{i-1})}{\mathcal{D}(m_l+\mu_i)}\right]\,,
\end{equation}
where $0=\mu_0 < \mu_1 \ldots \mu_{k+1}=1-m_l$. The intermediate masses $\mu_i(i=1..k)$ can be continuously tuned, enabling us to evaluate the individual determinants at a reduced precision -- $10^{-6}$ residual as opposed to $10^{-8}$ -- considerably reducing the computational cost. The strange-quark determinants were again evaluated using the rational approximation. We obtained a further increase in speed by utilizing a force gradient integrator~\cite{Yin:2011sz,Kennedy:2009fe} in place of the Omelyan integrator.

In table~\ref{tab-simparams} we give details of the molecular dynamics time steps and the update ratios for each component of the force, alongside the total MD time, the Metropolis acceptance and the values of the average plaquette and chiral condensate on each ensemble.

\begin{table}[t]
\begin{tabular}{c|c|c|c|c|c|c|c|c}
\hline
$m_s$ & $m_l$ & $\tilde{m_s}/\tilde{m_l}$ & $\Delta t \times N_{\rm steps}$ & $N_G:N_{\rm DSDR}:N_{\rm ferm}$  &  $\tau(MD)$ & Acceptance & $\langle P \rangle$ & $\langle \bar\psi\psi(m_l) \rangle$\\
\hline
\multirow{2}{*}{0.045}& 0.0042 & 7.8 & $1/8\times 8$ & 64:8:(2:1)  & 1176 & 70\% & 0.512198(3) & 0.001579(5)\\
                      & 0.001  & 16.5 & $1/9\times 9$ & 12:6:1  & 1432 & 73\% & 0.512230(3) & 0.001202(3)\\
\end{tabular}
\caption{Simulation parameters for the 32ID ensembles. Here the fifth column contains a gross summary of the algorithm, giving the ratio of gauge field updates ($N_G$) to the number of DSDR updates ($N_{\rm DSDR}$) to the number of updates of the fermion force ($N_{\rm ferm}$). For the heavier ensemble, the fermion component is divided into the rational approximation for the strange-quark determinant and the light quark determinant; the former is updated twice as often as the latter. On the lighter ensemble the strange-quark determinant and the Hasenbusch-preconditioned light-quark determinant are not nested but instead are evaluated independently and their force contributions combined linearly. The Molecular Dynamics time step for the top-level integrator and the number of steps per trajectory ($N_{\rm steps}$) is given in the fourth column. The quantity $\tau(MD)$ is the length of the ensemble used for the analyzes in this document, measured in molecular dynamics time units.}
\label{tab-simparams}
\end{table}

\subsection{Ensemble properties}

\begin{figure}[tp]
\centering
\includegraphics[width=0.8\textwidth]{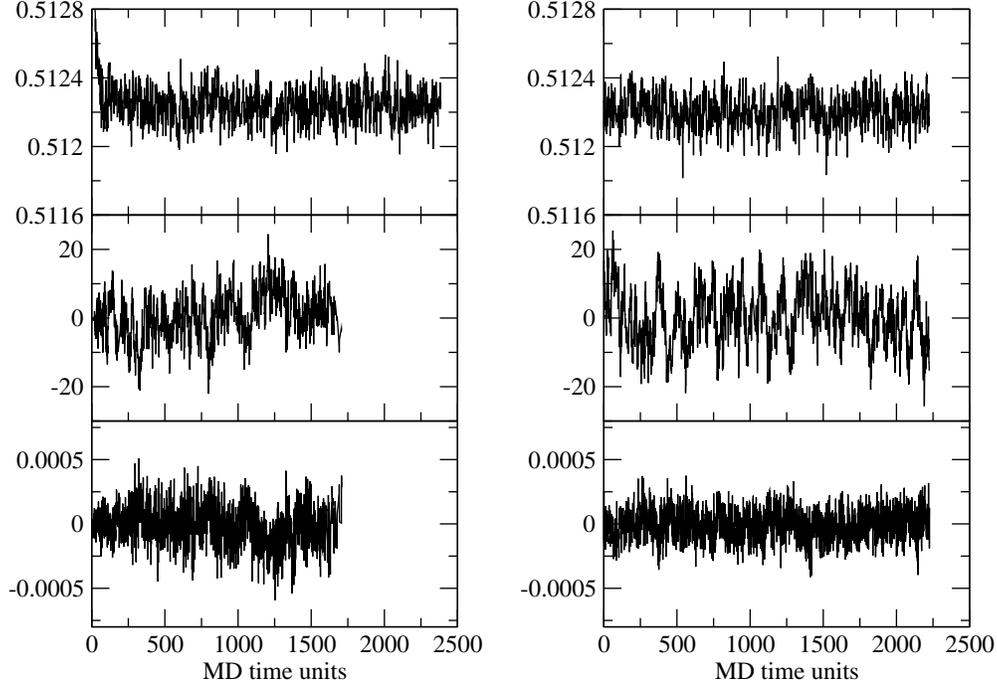}
\caption{ Monte Carlo evolution of the average plaquette (top), topological charge (middle), and light-quark pseudoscalar density (bottom) on the $m_l = 0.001$ (left) and $m_l = 0.0042$ (right) ensembles.\label{fig:signal}}
\end{figure}
\begin{figure}[tp]
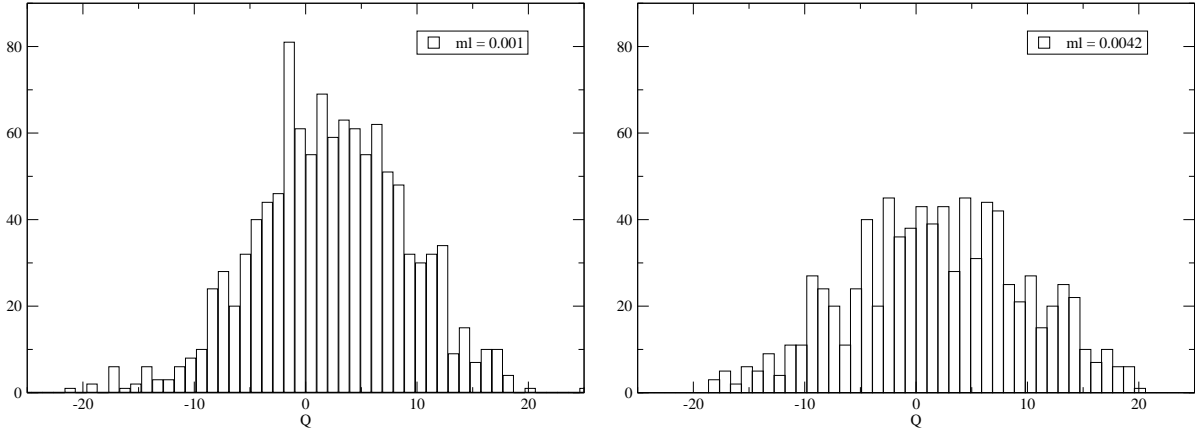

\centering
\includegraphics*[width=0.47\textwidth]{fig/ensemble_gen/mu0_001_q_hist.eps}\quad\includegraphics*[width=0.47\textwidth]{fig/ensemble_gen/mu0_0042_q_hist.eps}
\caption{Topological charge distributions for the $m_l = 0.001$ (left) and $m_l = 0.0042$ (right) ensembles.\label{fig:qhist}}
\end{figure}

In figure~\ref{fig:signal} we plot the Monte Carlo evolution of the plaquette, topological charge and the light-quark pseudoscalar density. We measured the topological charge directly using ``cloverleaf'' estimates of the field strength tensor, with 1x1, 1x2, 2x2, 1x3 and 3x3 Wilson loops calculated on APE-smeared gauge fields (with 60 smearing steps) and combined using the `5li' (five-loop improved) combination~\cite{deForcrand:1997sq} which eliminates the $\mathcal{O}(a^2)$ and $\mathcal{O}(a^4)$ terms at tree-level. We show histograms of the topological charge distribution in figure~\ref{fig:qhist}.

\begin{figure}[tbp]
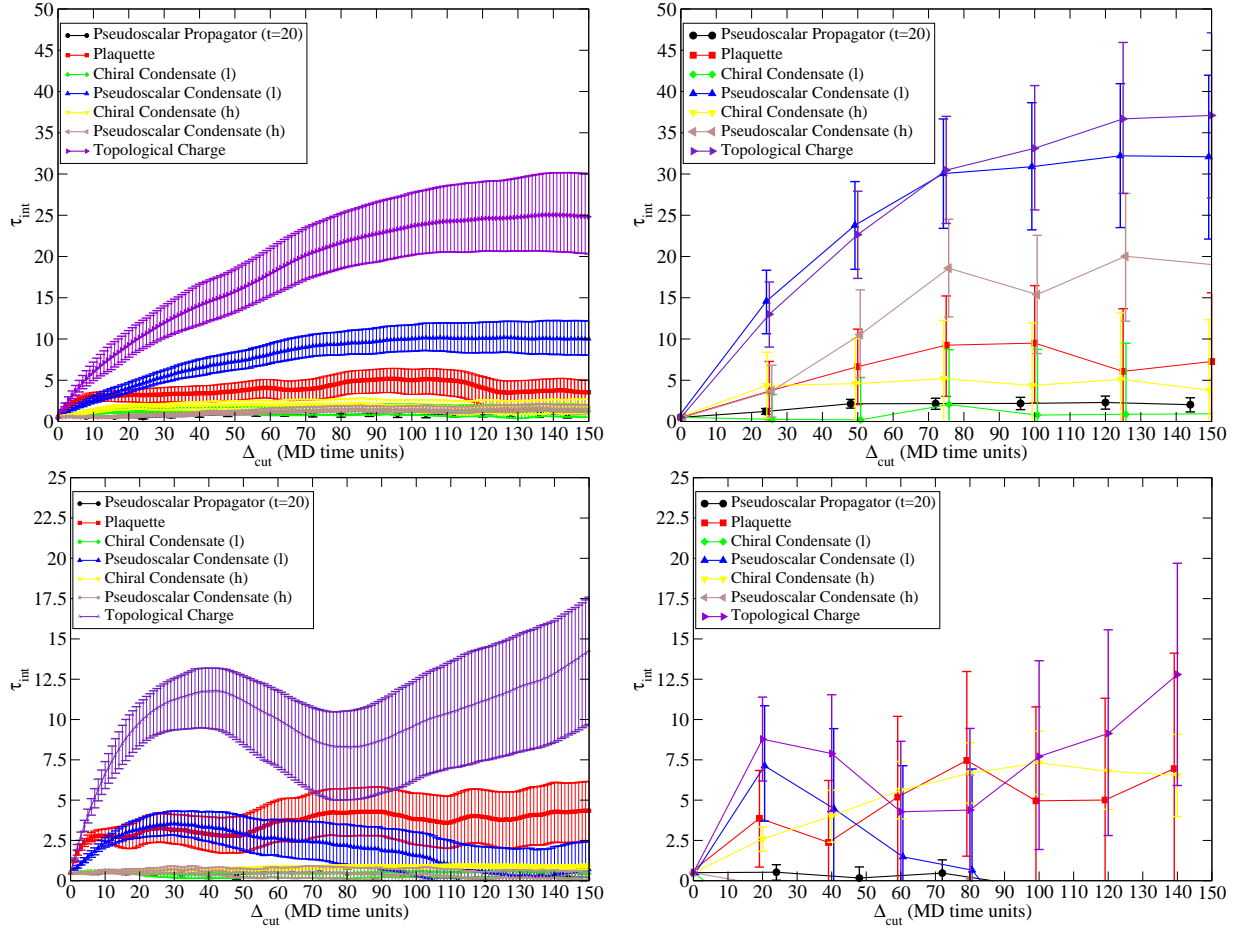

\centering
\includegraphics*[width=0.48\textwidth]{fig/ensemble_gen/mu0.001_bincorr_binsize25.eps}\quad
\includegraphics*[width=0.48\textwidth]{fig/ensemble_gen/mu0.001_bint_binsize25.eps}\\
\includegraphics*[width=0.48\textwidth]{fig/ensemble_gen/mu0.0042_bincorr_binsize20.eps}\quad
\includegraphics*[width=0.48\textwidth]{fig/ensemble_gen/mu0.0042_bint_binsize20.eps}\\
\caption{The integrated autocorrelation time is shown for the average plaquette, topological charge, the chiral condensate and pseudoscalar density for the light and heavy quark species (labelled `l' and `h' respectively), and the pseudoscalar two-point function at $t=20$, as a function of the upper bound on the integral $\Delta_{\rm cut}$, using data from the $m_l = 0.001$ (top) and $m_l = 0.0042$ (bottom) ensembles. For those plots on the left we estimated the errors by binning the set of correlations between measurements at fixed MD time separation (the first strategy discussed in the text), and in those on the right we block over the data and measure the correlation between blocks (the second strategy). We chose bin sizes of 25 and 20 on the lighter and heavier ensembles respectively. The pseudoscalar two-point function was only measured every 8 MD time units, hence for both methods we bin these data with a bin size of 24 MD time units. In the right-hand plots the data have been shifted slightly for 
clarity.\label{fig:autocorr_time}}
\end{figure}

Figure~\ref{fig:autocorr_time} contains plots of the integrated autocorrelation time for various quantities on the $m_l = 0.001$ and $m_l = 0.0042$ ensembles as a function of the cut on the upper bound of the integral, $\Delta_{\rm cut}$:
\begin{equation}
\tau_{int}(\Delta_{\rm cut}) = \half + \sum_{\Delta=1}^{\Delta_{\rm cut}}C(\Delta)\,,
\label{eqn-autocorr-tauint}
\end{equation}
where
\begin{equation}
C(\Delta) = \Big\langle\frac{ \left(Y(t)-\bar Y\right)\left(Y(t+\Delta)-\bar Y\right)}{\sigma^2}\Big\rangle_t
\label{eqn-autocorr-c}
\end{equation}
for a quantity $Y$, where $\bar Y$ is the expectation value over the ensemble, $\sigma^2$ its variance, and $\Delta$ is the molecular dynamics time separation between measurements. The average in the second equation is performed over the set of pairs of configurations separated by $\Delta$ MD time units. In order to correctly estimate the errors on the integrated autocorrelation time, we investigated two strategies:
\begin{enumerate}
 \item At each fixed $\Delta$ we formed a bootstrap distribution to estimate the error on the mean $\langle...\rangle_t$ in eqn.~\ref{eqn-autocorr-c}. Prior to performing the bootstrap resampling, we binned the set of measurements $\left(Y(t)-\bar Y\right)\left(Y(t+\Delta)-\bar Y\right)$ over neighboring configurations (indexed here by $t$). The bin size was successively increased until the errors stopped growing, which we found to be at bin sizes of $25$ and $20$ on the $m_l=0.001$ and $m_l=0.0042$ ensembles respectively. The error on $\tau_{\rm int}$ was obtained from the bootstrap sum over $C(\Delta)$ according to eqn.~\ref{eqn-autocorr-tauint}. This method closely resembles the standard strategy for binning equivalent quantities over a set of correlated measurements under a bootstrap.

\item We took the full set of measurements $Y(t)$ over the ensemble and formed blocks by averaging over neighboring configurations. We then measured the correlations between these blocks, taking the center-point of each block as the associated MD time. This has the effect of averaging over short-range correlations, exposing those with longer range, but also results in changes to the central value of $\tau_{\rm int}$ at fixed $\Delta_{\rm cut}$ as the bin size is increased, as at each bin size we are measuring a different quantity. We chose the optimal bin size to be the point where further increases resulted in statistically consistent central values. This strategy was used in our 2010 analysis for estimating the autocorrelation length of the two Iwasaki ensemble sets.

\end{enumerate}
The aforementioned figure contains plots for both of these strategies. We see that they give consistent results. The integrated autocorrelation time for the majority of the quantities we looked at appears to lie between 5 and 10 MD time units. However, as is typically the case, the topological charge (and of course the pseudoscalar condensate) display considerably larger autocorrelation lengths, around 25 MD time units on the ligher ensemble and 15 on the heavier ensemble, reflecting their sensitivity to the underlying global gauge field topology. The larger autocorrelation length suggests a lower topological tunneling rate for our lighter ensemble. However we emphasize that these autocorrelation times are considerably shorter than those of the Iwasaki lattices, which were estimated to be $\mathcal{O}(80)$ MD time units~\cite{Aoki:2010dy} from the topological charge measurements.

For the simulation parameters and properties of the 32I and 24I Iwasaki ensemble sets we refer to reader to ref.~\cite{Aoki:2010dy}.

\subsection{Reweighting the strange quark}

We make use of reweighting in the strange sea-quark mass to obtain the mass dependence of our data, and hence interpolate to the physical value, without incurring the expense of simulating with additional masses. The reweighting factor $w_i$ for a particular reweighted mass $m_h^{rw}$ and configuration $i$ is determined by measuring the degree to which that configuration, as sampled from the un-reweighted path integral, contributes to the path integral with the reweighted mass; in practice this involves the calculation of the ratio of Dirac-matrix determinants with the reweighted and simulated masses respectively. The expectation value of an observable $\mathcal{O}$ with the shifted strange-quark mass is then obtained by first measuring on the original, unreweighted configurations, then applying the reweighting factors:
\begin{equation}
\langle \mathcal{O} \rangle _{m_h^{rw}} = \frac{\langle w\mathcal{O} \rangle_{m_h^{sim}} }{\langle w \rangle_{m_h^{sim}}}\,.
\end{equation}
The determinants are stochastically evaluated using several Gaussian sampled vectors and the weight factor obtained from the average over these samples. This procedure was used in the 2010 analysis, and more details can be found in ref~\cite{Aoki:2010dy}.

We performed measurements over incremental steps from the simulated mass of 0.045 up to 0.052. We previously found that the number of stochastic samples required for a reliable estimate of the weighting factor is dependent upon the size of the mass increments, with smaller increments requiring less samples. As a result, we use two stochastic samples and small increments of $\Delta m_h = 0.00025$ -- the same parameters as were used for the 24I ensembles.

The reweighting procedure naturally reduces the effective number of configurations $N_{\rm eff}$ in each ensemble set. In ref~\cite{Aoki:2010dy} we showed that a reliable estimate of this quantity can be determined via the following expression:
\begin{equation}
N_{\rm eff} = \frac{(\sum_i w_i)^2}{\sum w_i^2}\,.
\end{equation}
A value of unity indicates that the measurement is entirely dominated by a single configuration, whereas $N_{\rm eff}$ is equal to the original number of configurations $N_{\rm conf}$ when there are no fluctuation in the weighting factors. In section~\ref{sec:FitResults} we measure the physical strange quark mass to be $m_h^{\rm phys} = \msIDbare$, which is close to the simulated value. At the nearest reweighted mass-step to the physical mass, that with $m_h=0.0465$, we find $N_{\rm eff} =133$ ($N_{\rm conf} = 180$) and $N_{\rm eff} =119$ ($N_{\rm conf} = 148$) on the $m_l=0.001$ and $m_l=0.0042$ ensembles respectively, suggesting that reweighting to the physical strange-quark mass will result in only a 10\%-15\% increase in the statistical errors on these ensembles. This is of a similar magnitude to the increase suggested by the values of $N_{\rm eff}$ on the 32I ensembles, which are given in ref~\cite{Aoki:2010dy}. On the 24I ensembles we require a slightly larger extrapolation to reach the physical value, 
hence the reweighting introduces larger increases of 25\%-35\%.

%%%%%%%%%%%%%%%%%%%%%%%% Updated Results for 32^3 Simulations %%%%%%%%%%%%%%%%%%%%%%%%%%%%%%
\section{Results From The $32^3$ DWF+ID Ensembles}
\label{sec:DSDRresults}
\FloatBarrier

In this section we present the results of fitting to a number of observables on the 32ID ensembles. We performed measurements on \nconfIDlight configurations on the $m_l = 0.001$ ensemble and \nconfIDheavy on the $m_l = 0.0042$ ensemble, with each configuration separated by 8 MD time units. The analysis in the previous section suggests an autocorrelation length of $\sim 25$ on the $m_l=0.001$ ensemble and $\sim 7$ on the $m_l=0.0042$ ensemble, which can be overcome by binning the data before performing the fits. We shifted the gauge fields in the time-direction by 16 lattice spacings relative to the previous configuration prior to measuring the quark propagators. This has the effect of reducing the correlation between successive measurements, suggesting that binning the data may not be necessary. However, this does not apply to the measurements of the Sommer scales $r_0$ and $r_1$, which are formed using Wilson loops with origins on all lattice sites. In order to remain consistent, we decided to bin the data 
for all of our quantities over 4 successive measurements (32 MD time units) on both ensembles; although this is larger than the measured autocorrelation length, it matches the periodicity of the quark propagator measurements, and is therefore a more natural choice. We found no statistically significant dependence on the bin size in any of our measured error values, hence the choice of bin size has little effect on the final results of this analysis.

The pseudoscalar meson two-point correlation functions were calculated in the same manner as those on the 32I ensembles, namely using Coulomb gauge-fixed wall source propagators originating at the lattice time boundary $t=0$ with both periodic ($p$) and antiperiodic ($a$) boundary conditions in the temporal direction. Taking the $p+a$ combination of propagators to form each leg of the correlation function projects out the component travelling forwards in time. Likewise, the $p-a$ combination projects out the degenerate backwards-propagating state. The correlation functions formed using these combinations of propagators have a temporal periodicity of double the usual length, which results in a significant reduction in round-the-world propagation. The Omega baryon correlation functions were calculated separately using box-sources  with a spatial volume of $15^3$ lattice sites and with one corner at the spatial origin. These were placed on time-slices $t=0$ and $32$, and antiperiodic boundary conditions were 
used for the propagators. As mentioned above, the gauge fields were shifted in time by 16 units with respect to the previous configuration prior to performing all of these measurements.

For each quantity we tabulate the results of fitting to the time-dependence of the corresponding correlation functions measured at the simulated strange-quark mass, and we present example effective mass plots demonstrating the quality of our data. We also provide tables of data corrected to the physical strange-quark mass of $m_s = \msIDbare$ determined in section~\ref{sec:FitResults}, using the the NLO ChPT with finite-volume corrections parametrization for the mass dependence.

\subsubsection{The residual mass}

\begin{table}[tp]
\centering
\begin{tabular}{c|cc}
\hline
$m_x$ & \multicolumn{2}{c}{$m_l$}\\
\hline
 &0.001&0.0042\\
\hline
0.0001 & 0.0018447(60) & 0.0018888(48) \\
0.001 & 0.0018510(43) & 0.0018889(47) \\
0.0042 & 0.0018269(58) & 0.0018735(48) \\
0.008 & 0.0018025(57) & 0.0018500(48) \\
0.035 & 0.0016939(44) & 0.0017356(39) \\
0.045 & 0.0016739(39) & 0.0017141(37) \\
0.055 & 0.0016619(36) & 0.0017014(35) \\
\end{tabular}
\caption{$m_\mathrm{res}^\prime$ on the 32ID ensemble set at the simulated strange-quark mass.}
\label{tab:mresprime32ID}
\end{table}

\begin{figure}[hbt]
\centering
\includegraphics*[width=0.48\textwidth]{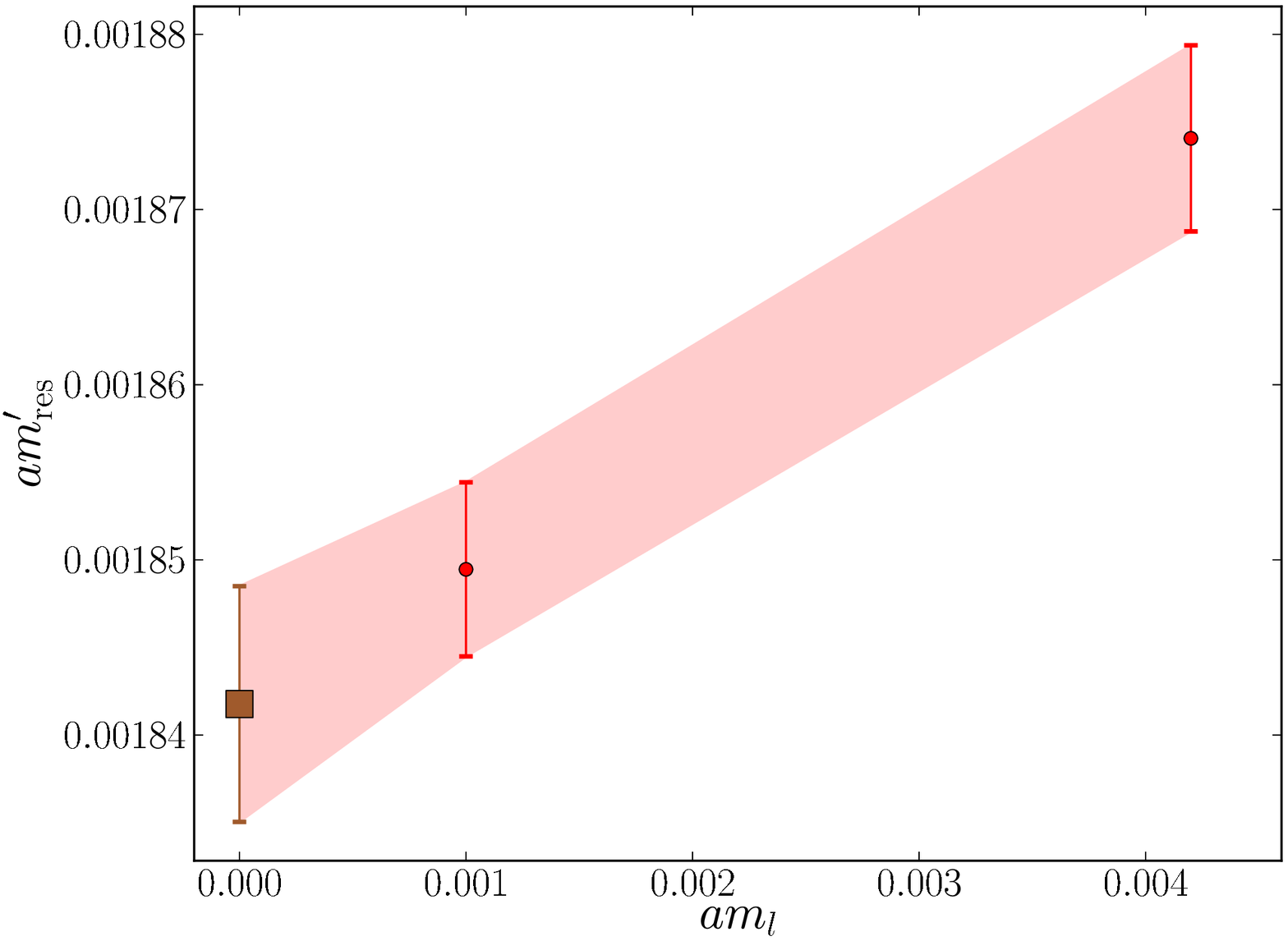}\quad
\includegraphics*[width=0.48\textwidth]{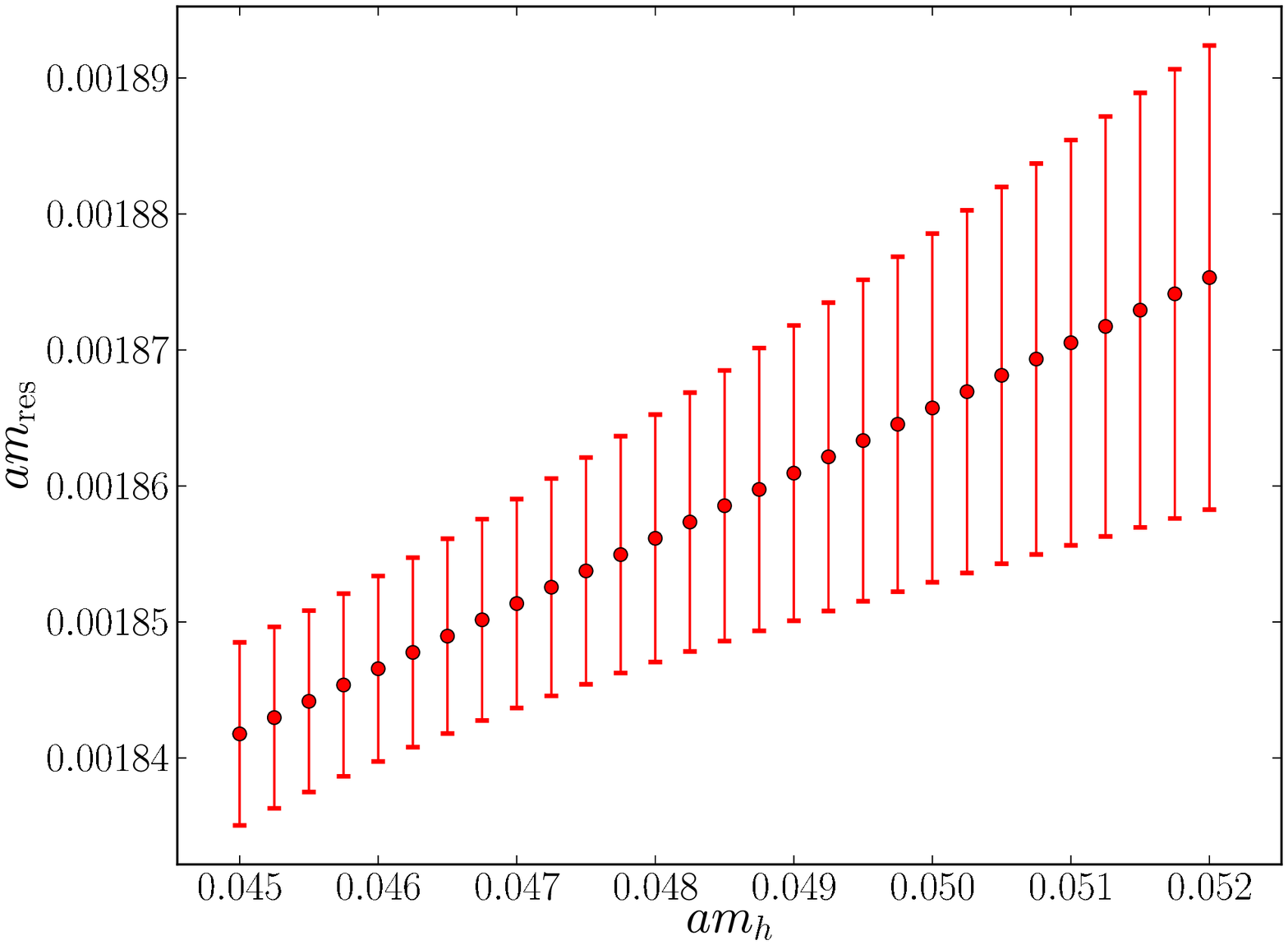}
\caption{
\label{fig:mres_chiral_rw_32ID}
The chiral extrapolation of $\mres^\prime$ over the unitary data points at the simulated strange-quark mass, with $\mres$ in the chiral limit denoted by the brown square point (left); and the strange-quark mass dependence of $\mres$ in the chiral limit (right).
}
\end{figure}

The residual mass at a nonzero (partially-quenched) quark mass $m_x$ may be determined via the following ratio:
\begin{equation}
\mres^\prime(m_x) = \frac{\langle 0|J^a_{5q}|\pi\rangle}{\langle 0|J^a_{5}|\pi\rangle}\,,
\end{equation}
where $J^a_{5q}$ is the pseudoscalar density at the midpoint of the fifth dimension, and $J^a_{5}$ is the physical pseudoscalar density constructed from the surface fields. The prime superscript is used to differentiate this quantity from the residual mass in the two-flavor chiral limit, $\mres = \mres^\prime (m_x=m_l=0)$. We averaged the data at $t$ and $T-t$ (we refer to this as folding the data), where $T$ is the lattice temporal extent, and fit over the time range 6--32 on both ensembles, obtaining the values given in table~\ref{tab:mresprime32ID}. Note that on the lighter ensemble, the nonunitary values were determined on a reduced data set of 92 configurations; these data were not used in the later analysis but are presented here for completeness. We obtained $\mres$ by extrapolating the unitary light-quark mass to the chiral limit at each reweighted strange quark mass. As discussed in refs.~\cite{Sharpe:2007yd} and \cite{Allton:2008pn}, defining the residual mass as this limit guarantees that the 
pion mass will vanish in the limit $m_f+\mres \rightarrow 0$ up to subpercent corrections of order $d m'_\mathrm{res}(m_f)/d m_f \cdot \mres$. A plot of the chiral extrapolation at the simulated strange-quark mass is shown in figure~\ref{fig:mres_chiral_rw_32ID}. Owing to the minor strange-quark mass dependence of this quantity evident in the right panel of figure~\ref{fig:mres_chiral_rw_32ID}, and the small separation between the simulated and physical strange quark masses, the value of $\mres$ at the physical strange quark mass is not measurably different from that at the simulated value of $\mresIDsim$.

\subsubsection{Pseudoscalar masses}

\begin{figure}[hbt]
\centering
\includegraphics*[width=0.45\textwidth]{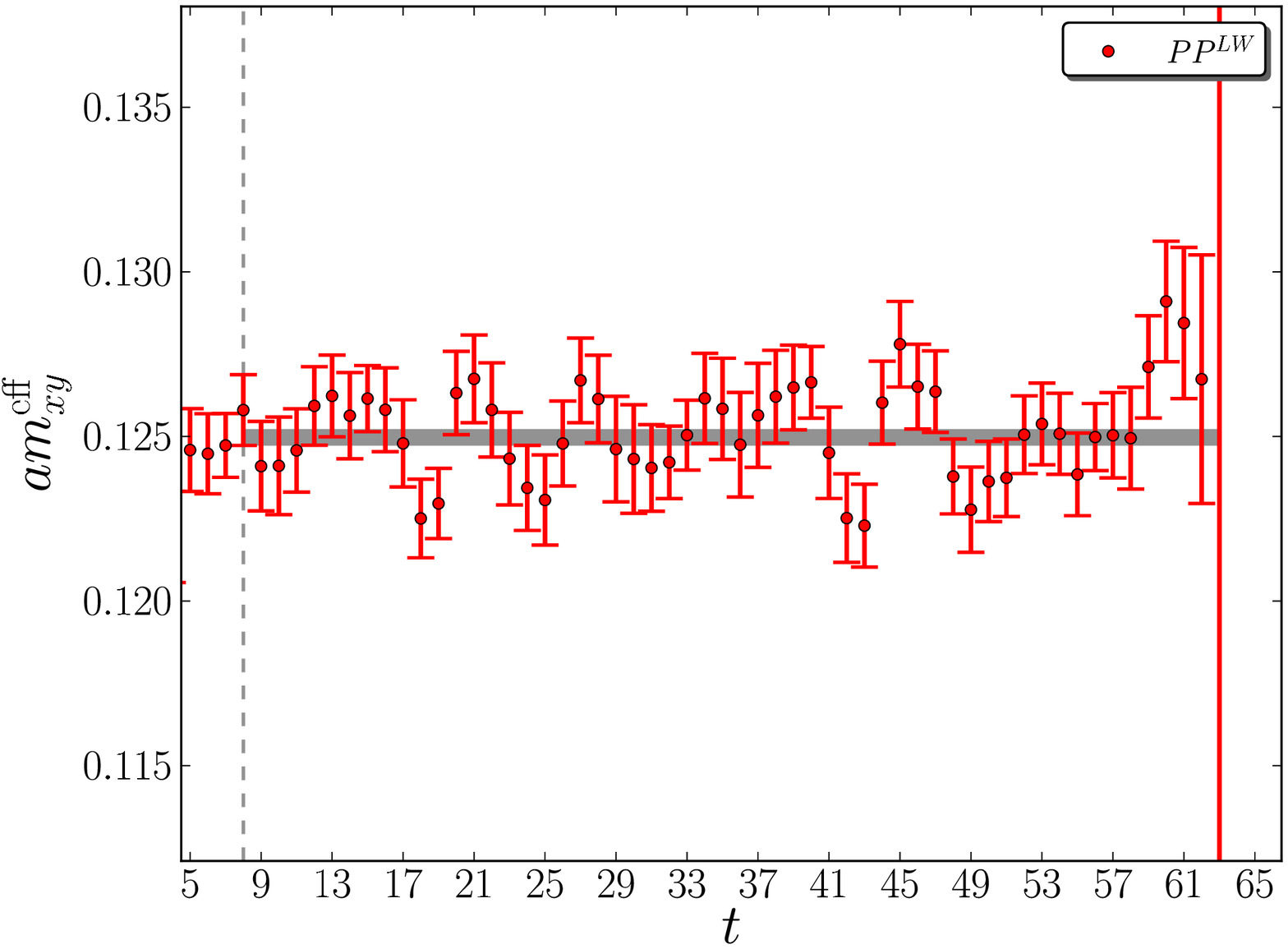}
\includegraphics*[width=0.45\textwidth]{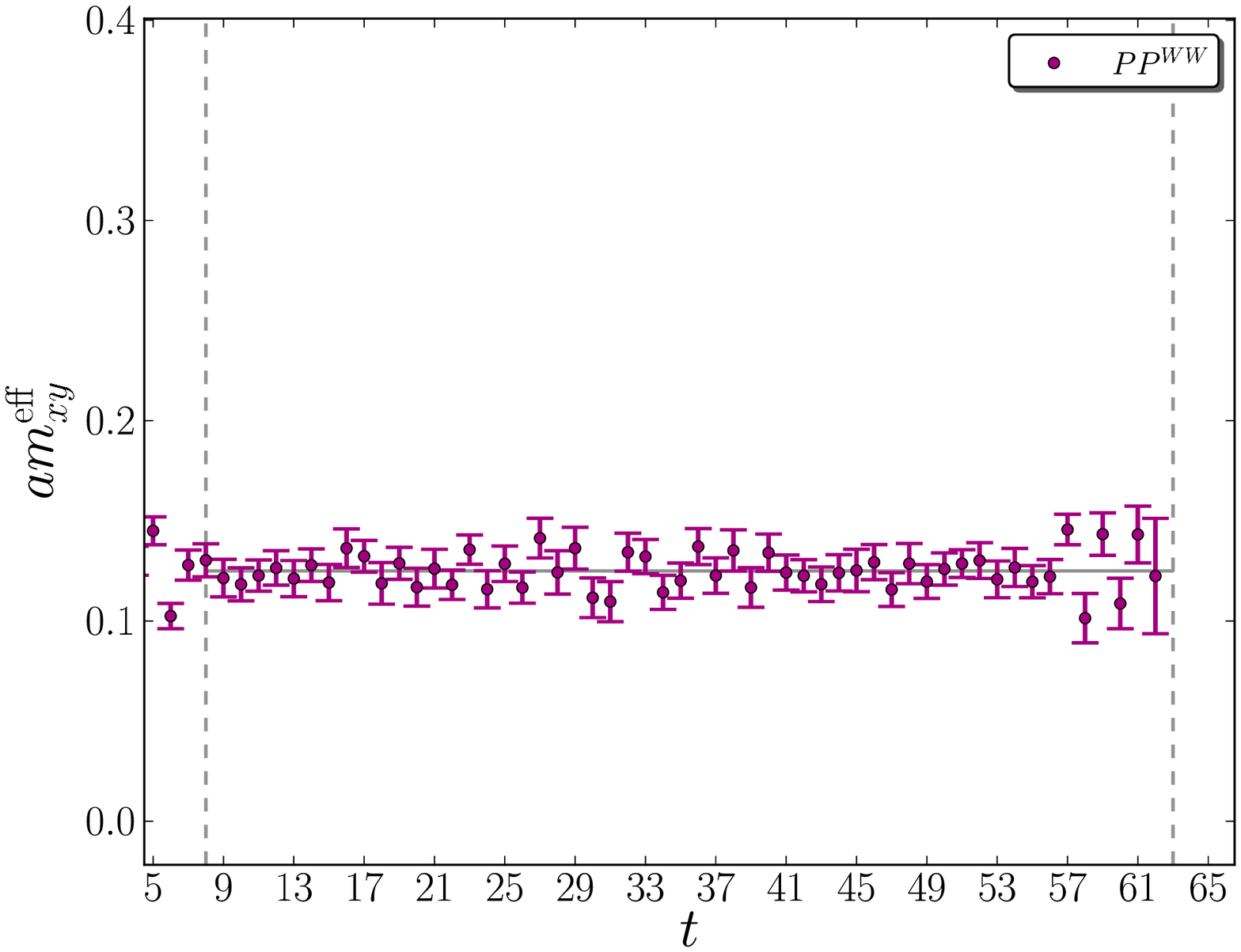}\\[0.1in]
\includegraphics*[width=0.45\textwidth]{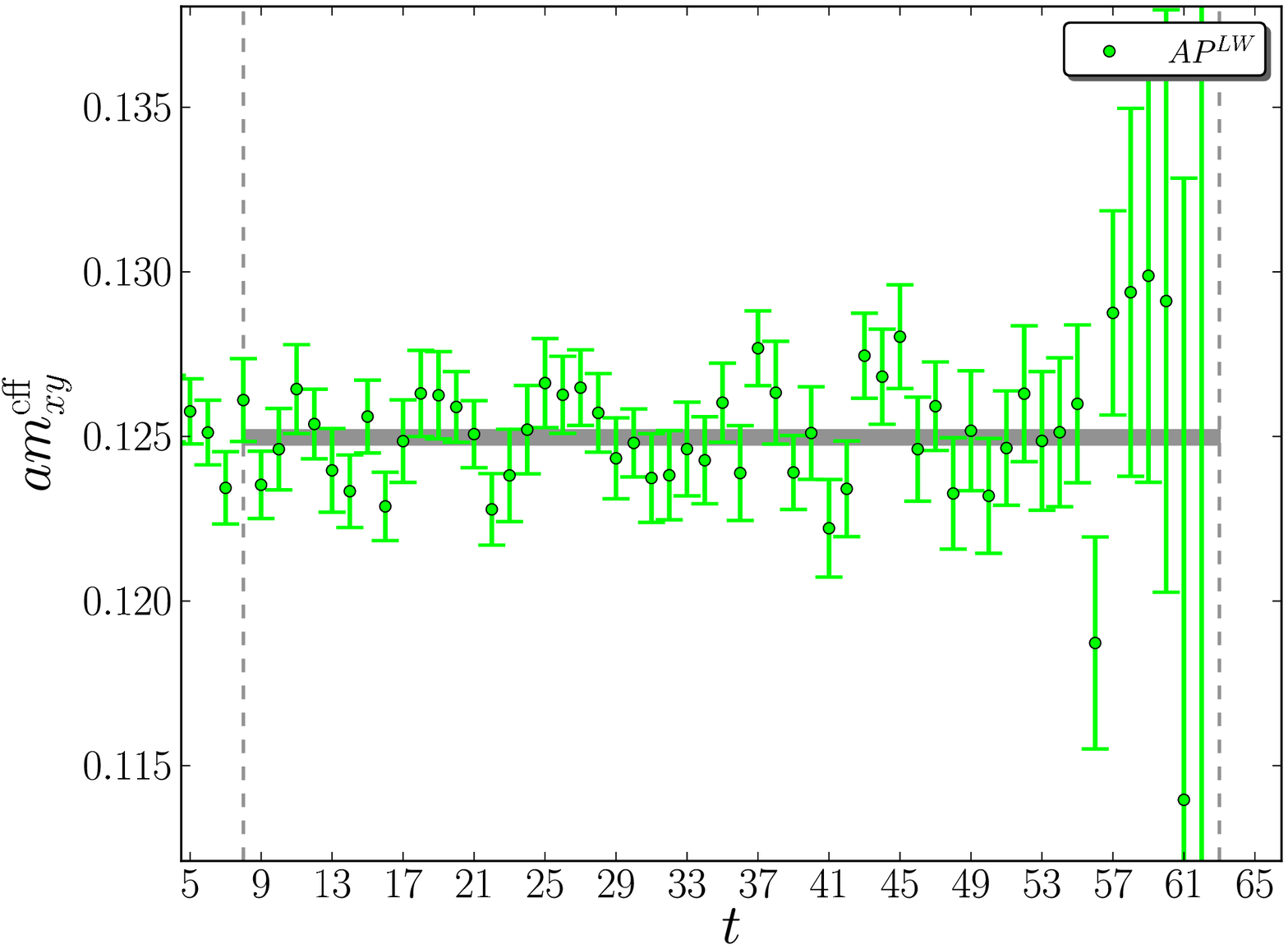}
\includegraphics*[width=0.45\textwidth]{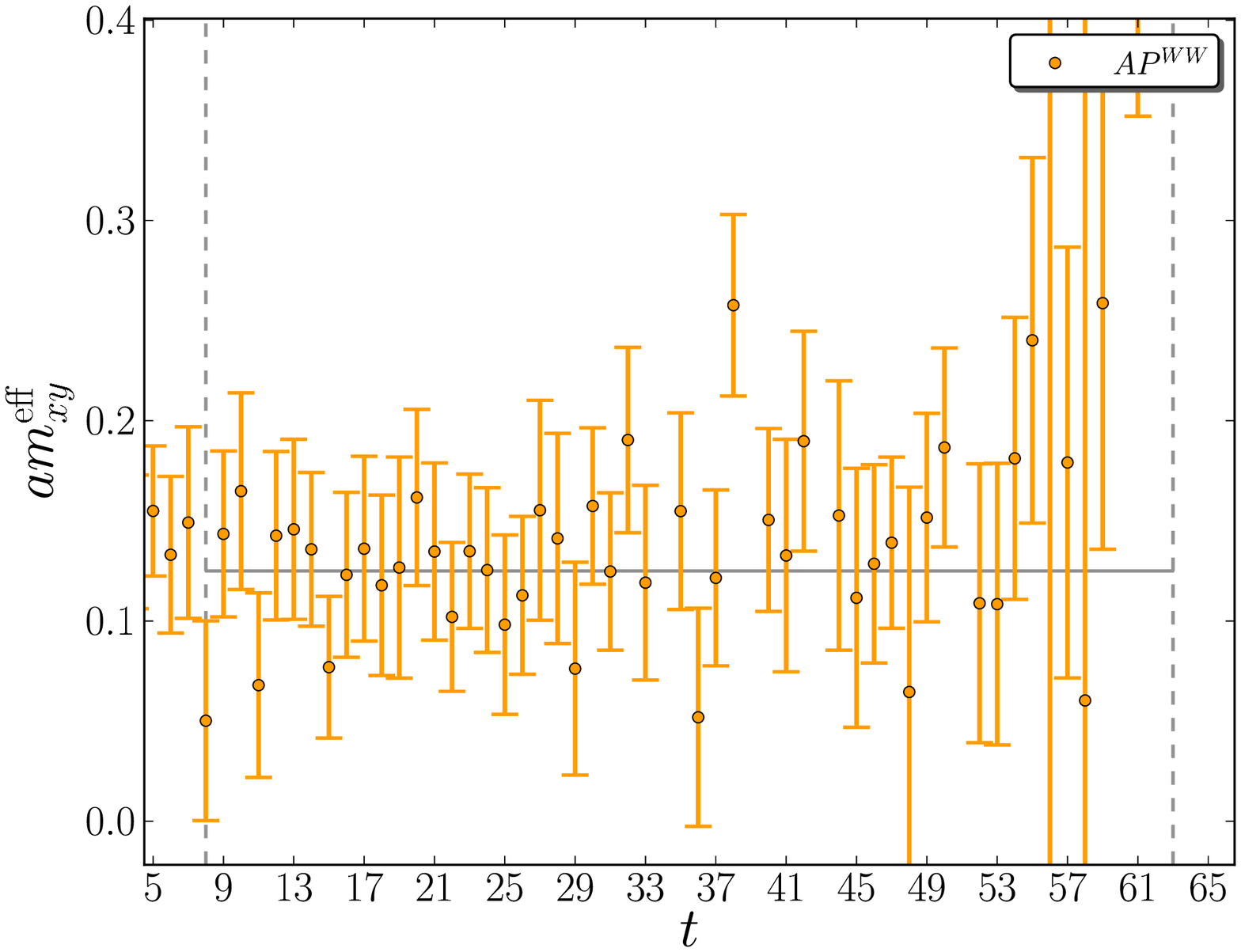}\\[0.1in]
\includegraphics*[width=0.45\textwidth]{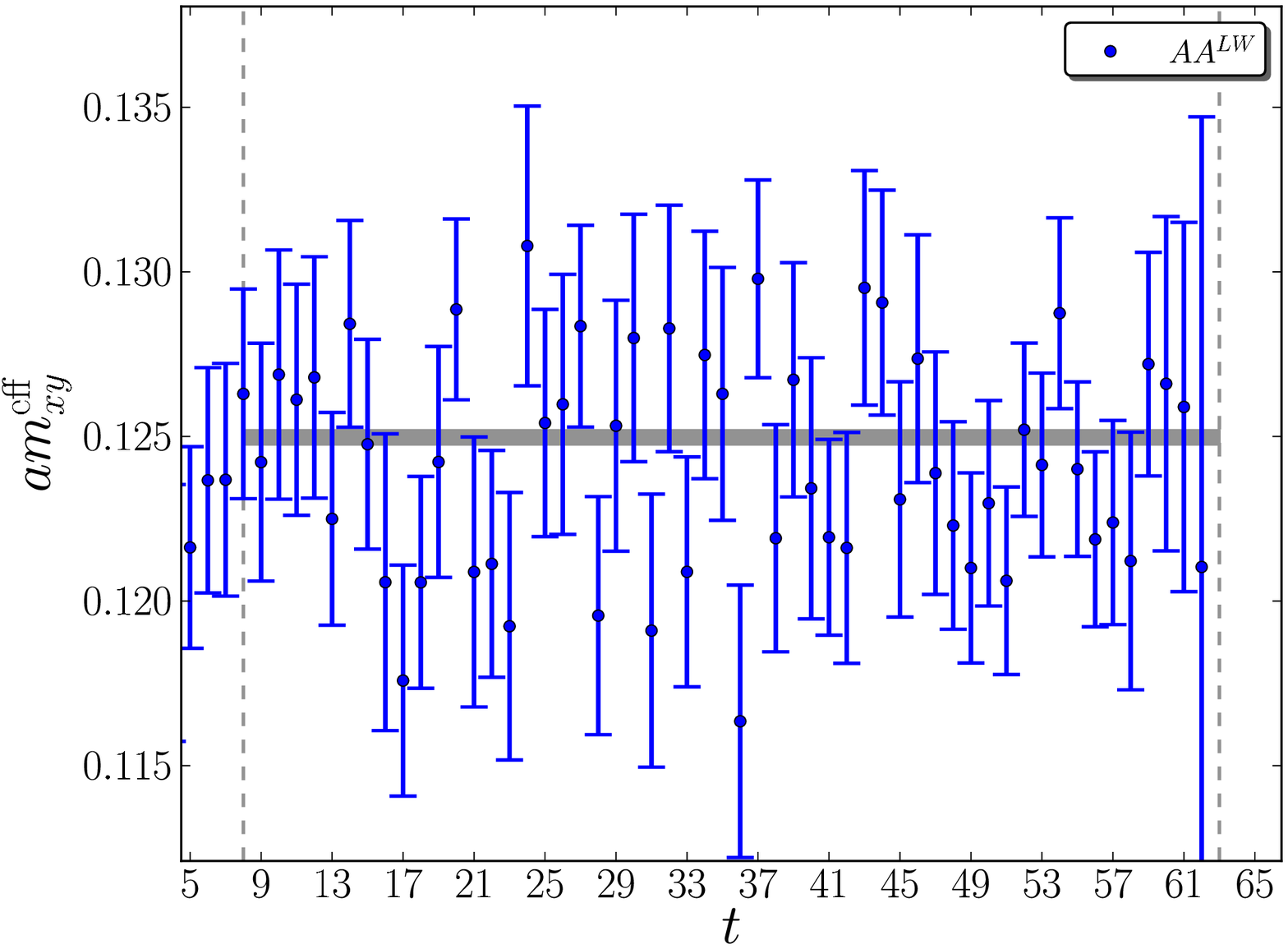}
\caption{
\label{fig:pionmeff32ID}
Effective unitary pion masses on the $m_l=0.001$ ensemble from the
PP LW correlator (top left), PP WW correlator (top right),
AP LW correlator (center left), AP WW (center right) and
AA LW correlator (bottom). Note the different vertical scale for the WW correlators. The horizontal bands represent the result for the mass from a simultaneous fit.
}
\end{figure}

\begin{figure}[hbt]
\centering
\includegraphics*[width=0.45\textwidth]{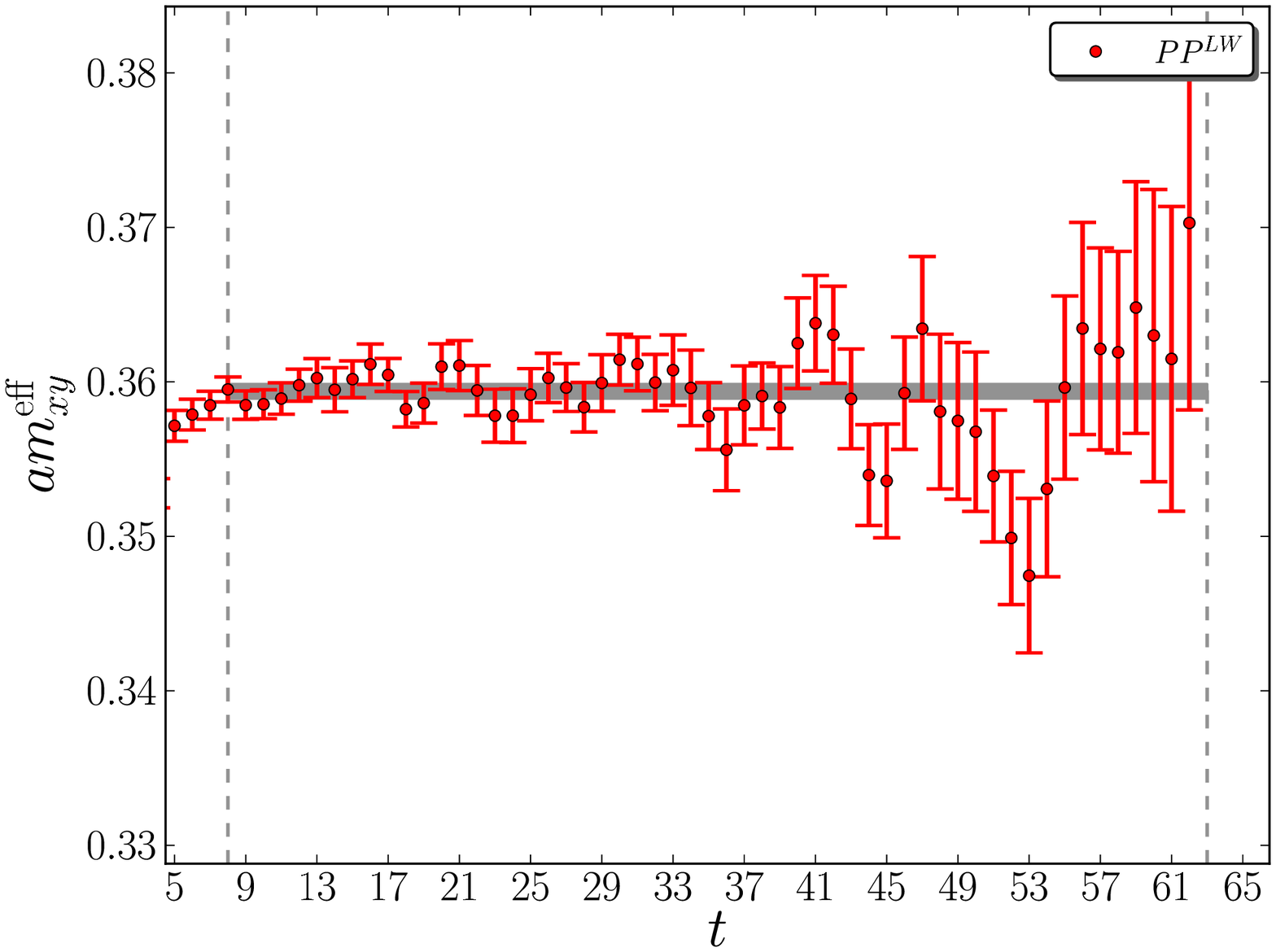}
\includegraphics*[width=0.45\textwidth]{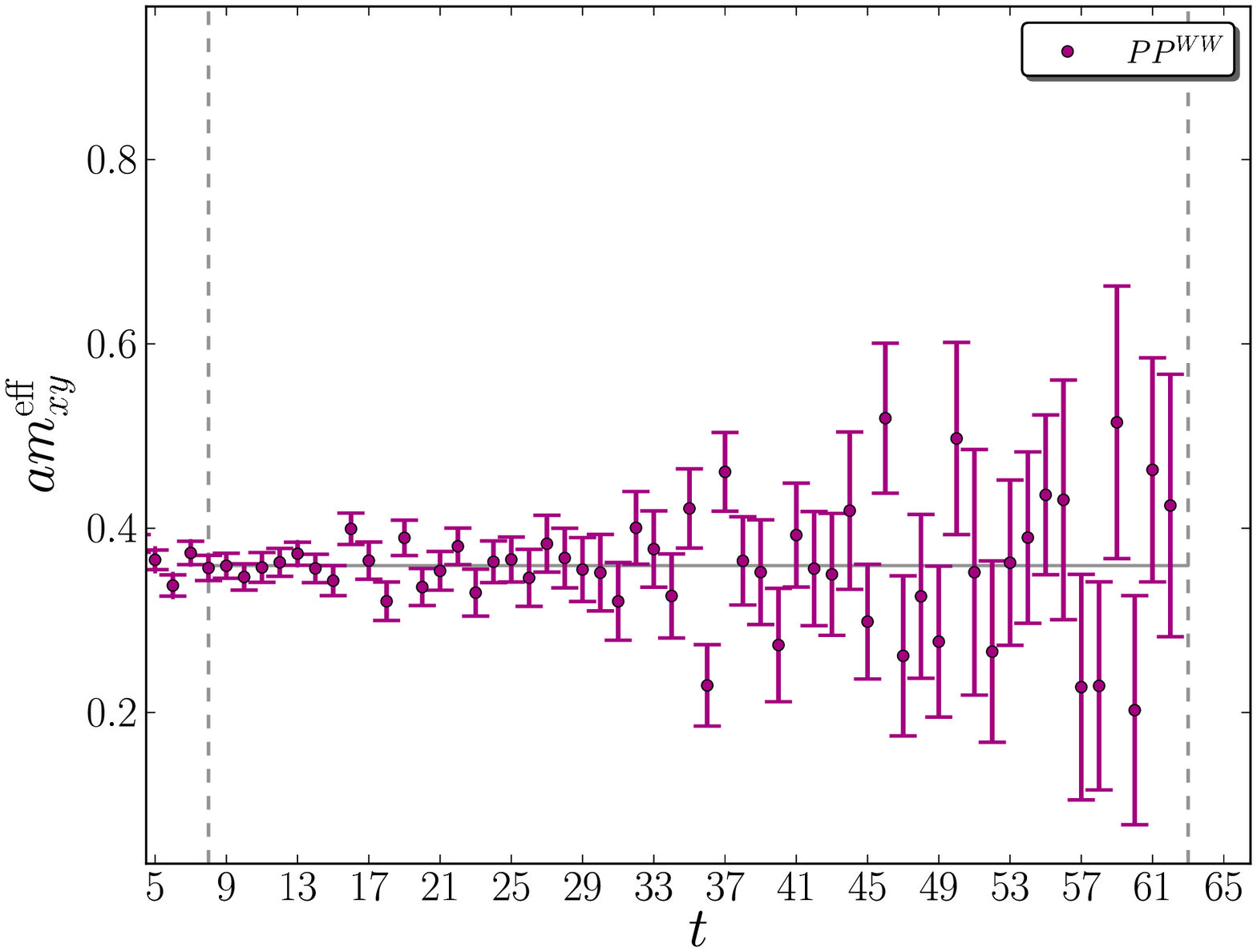}\\[0.1in]
\includegraphics*[width=0.45\textwidth]{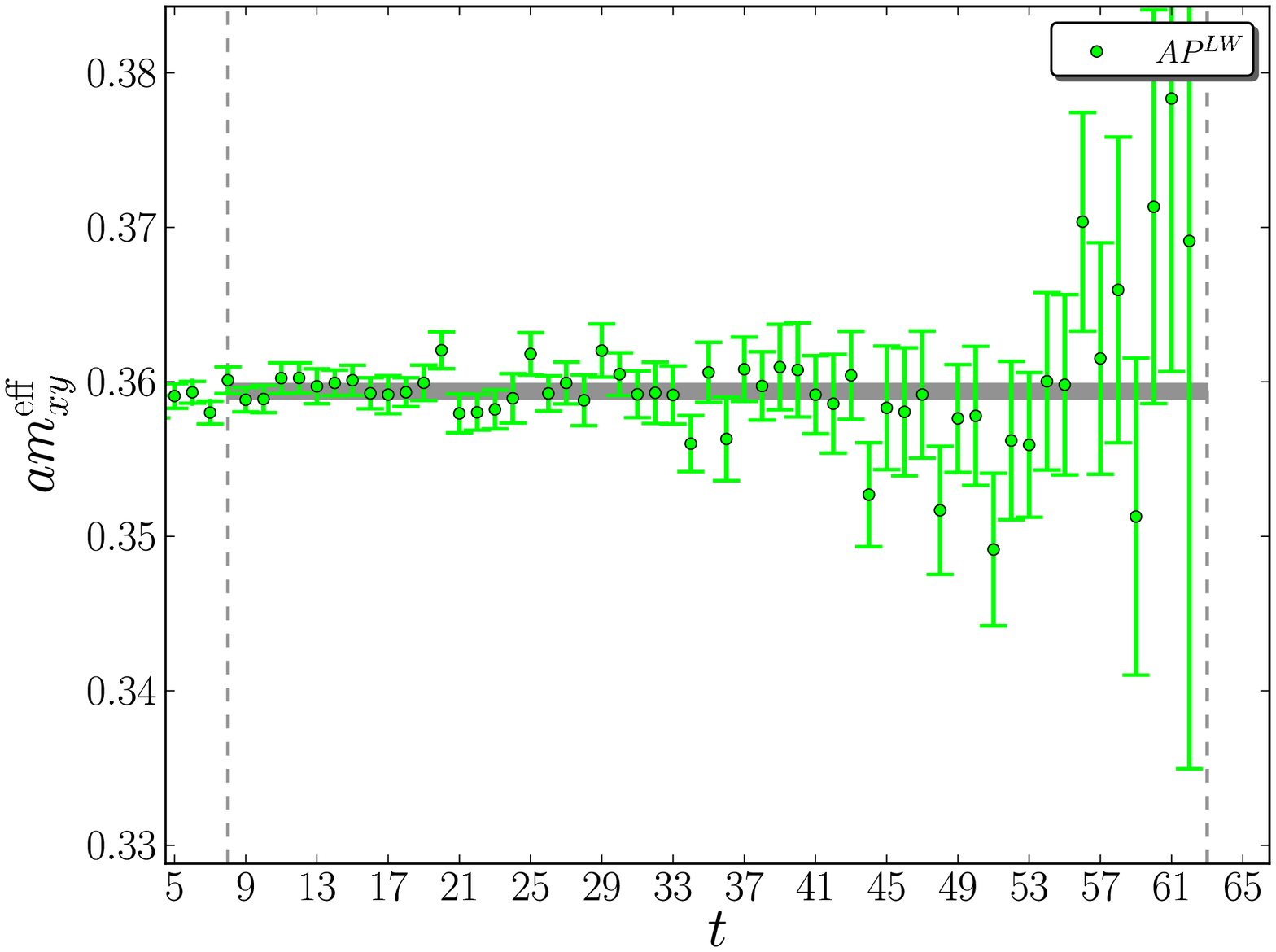}
\includegraphics*[width=0.45\textwidth]{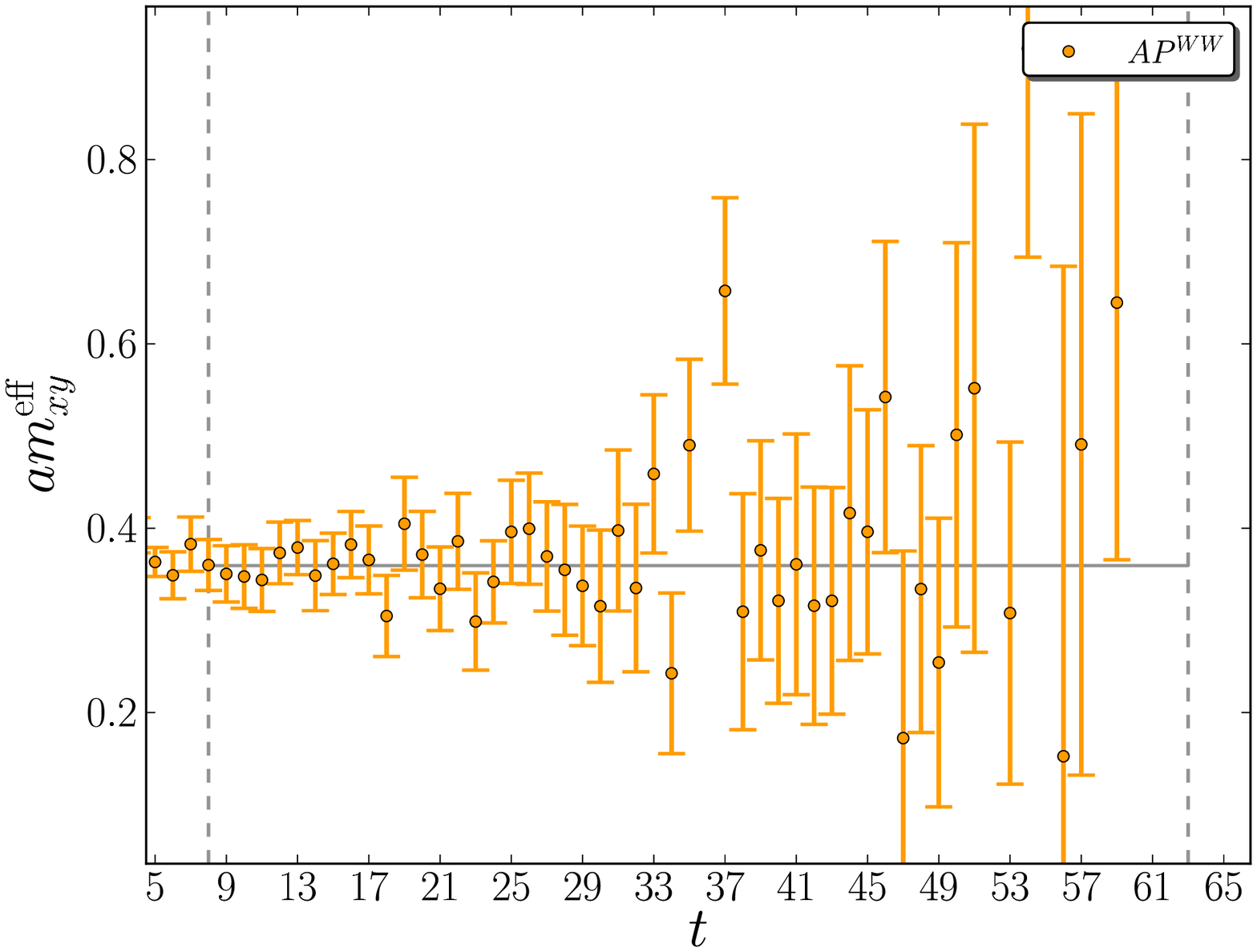}\\[0.1in]
\includegraphics*[width=0.45\textwidth]{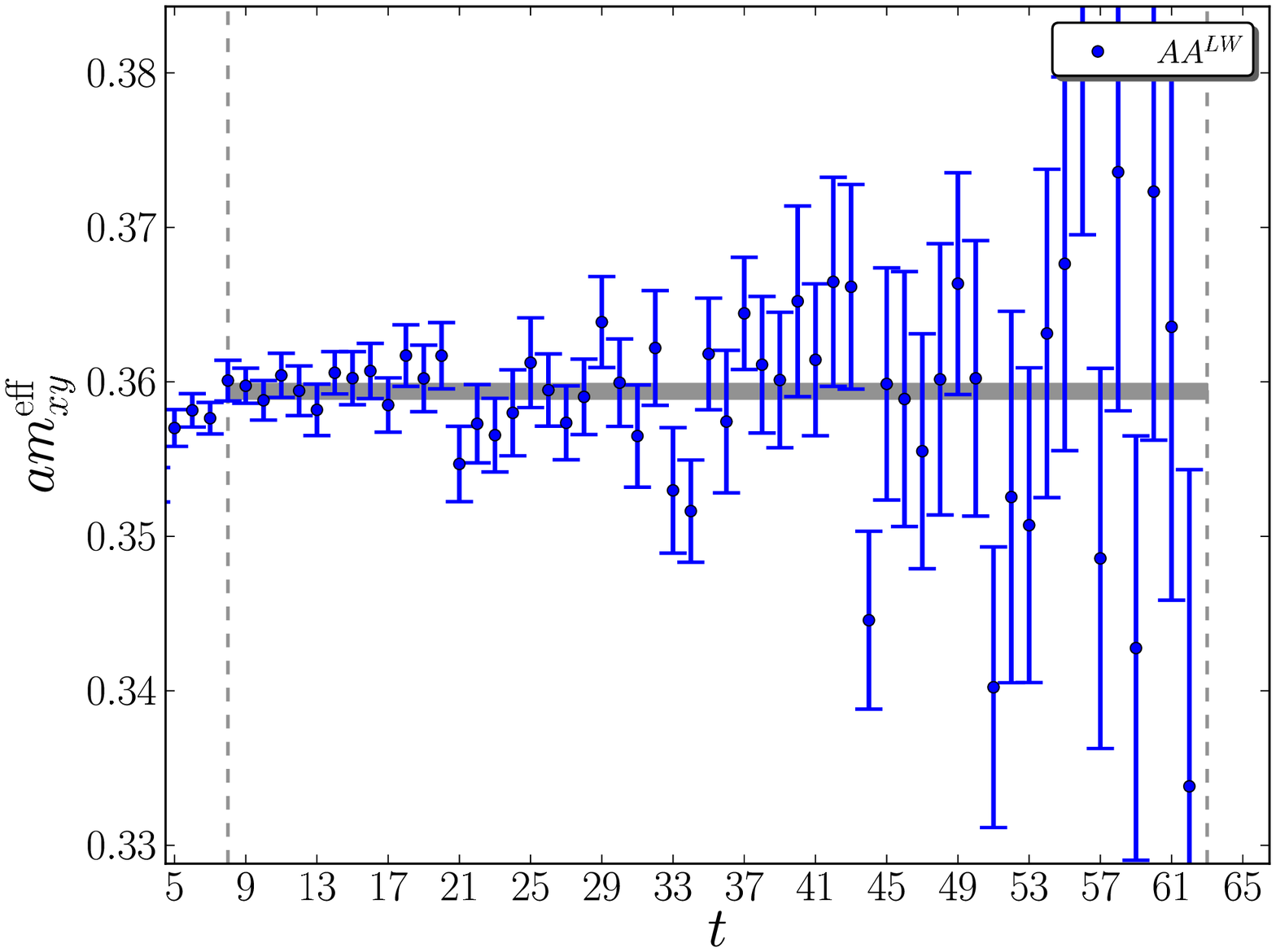}
\caption{
\label{fig:kaonmeff32ID}
Effective unitary kaon masses on the $m_l=0.001$ ensemble from the
PP LW correlator (top left), PP WW correlator (top right),
AP LW correlator (center left), AP WW (center right) and
AA LW correlator (bottom). Note the different vertical scale for the WW correlators. The horizontal bands represent the result for the mass from a simultaneous fit.
}
\end{figure}

We calculated a series of pseudoscalar meson two-point functions of the form:
\begin{equation}
{\cal C}_{O_{1}O_{2}}^{s_{1}s_{2}} (t) = \langle 0| O_1^{s_1}(t)O_2^{s_2}(0) |0\rangle\,.
\end{equation}
Here the subscripts index the interpolating operators and the superscripts denote the operator smearing (wall $W$ or local $L$) at the sink and source respectively. In the following we refer to these by the shorthand $O_1O_2^{s_1s_2}$, for example using $AA^{LW}$ to denote the axial-axial correlator with wall source and point sink. The pseudoscalar masses were determined via a combined fit to the following five correlation functions: $PP^{LW}$, $AP^{LW}$ and $AA^{LW}$, $PP^{WW}$ and $AP^{WL}$. The correlation functions exhibit the following time dependence:
\begin{equation}
        {\cal C}_{O_{1}O_{2}}^{s_{1}s_{2}} (t)
        = \frac{\langle 0 | O_{1}^{s_1} | \pi \rangle \langle \pi | O_{2}^{s_2}
| 0 \rangle }{2 m_{xy} V} \left [ e^{-m_{xy}t} \pm e^{-m_{xy}(2N_t-t)} \right ]\,,
\label{eq:corr-t}
\end{equation}
where the sign in the square brackets is $+$ for the $PP$ and $AA$ correlators and $-$ for the $AP$ correlators. We denote the amplitudes as
\begin{equation}
{\cal N}_{O_1 O_2}^{s_{1}s_{2}} \equiv \frac{\langle 0 |
O_1^{s_1} | \pi \rangle \langle \pi | O_2^{s_2} | 0 \rangle }{2 m_{xy} V}\,.
\end{equation}
Taking full advantage of the doubled time-extent of the lattice, we performed our fits over the time range 8--63 on both ensembles, obtaining the masses listed in table~\ref{tab:mxy_fxy_32ID_mhsim}. The values at the (unitary) physical strange-quark mass are given in tables~\ref{tab:mll_fll_32ID_mhphys} and~\ref{tab:mhl_fhl_32ID_mhphys} for the light-light (pion-like) and strange-light (kaon-like) quark mass combinations respectively. In figures~\ref{fig:pionmeff32ID} and~\ref{fig:kaonmeff32ID} we show example effective mass plots for the data at the simulated strange-quark mass on the $m_l=0.001$ ensemble.

\subsubsection{Pseudoscalar decay constants}

The pseudoscalar decay constants $f_{xy}$ were calculated from the two-point function amplitudes via the following equation:
\begin{equation}
        f_{xy} = Z_A \sqrt{  \frac{2}{m_{xy}V} \frac{{{\cal
        N}_{AP}^{LW}}^2}{{\cal N}_{PP}^{WW} } }\,. \label{eq:fpi-calc}
\end{equation}
Here $Z_A$ relates the local four-dimensional axial current $A_\mu^a$ -- formed with the domain wall surface fields -- to the Symanzik-improved axial current $A_\mu^{Sa}$, and thus renormalizes the local current into the continuum normalization. The indices $a$ and $\mu$ correspond to the flavor and Euclidean direction respectively. 

For domain wall fermions, a partially-conserved five-dimensional axial current $\mathcal{A}_\mu^a$ can also be defined, which is related to the Symanzik improved current by a different renormalization coefficient $Z_\mathcal{A}$. Prior to the 2010 analysis, it was typically assumed that the difference between $Z_\mathcal{A}$ and unity was negligible, hence $Z_A$ was assumed equal to $Z_A/Z_\mathcal{A}$. This can be obtained using the improved ratio~\cite{Blum:2001xb} of the partially-conserved five-dimensional (5D) axial current matrix element $\langle \mathcal{A}_4(t)P(0)\rangle$ to the local axial current matrix element $\langle A_4(t)P(0)\rangle$. As discussed in ref.~\cite{32cubedpaper}, the assumption that $Z_\mathcal{A}=1$ is only true up to terms $\mathcal{O}(a\mres)$, leading to an additional $\mathcal{O}(1\%)$ systematic error in our earlier results.

\begin{figure}[t]
 \centering
\includegraphics*[width=0.45\textwidth]{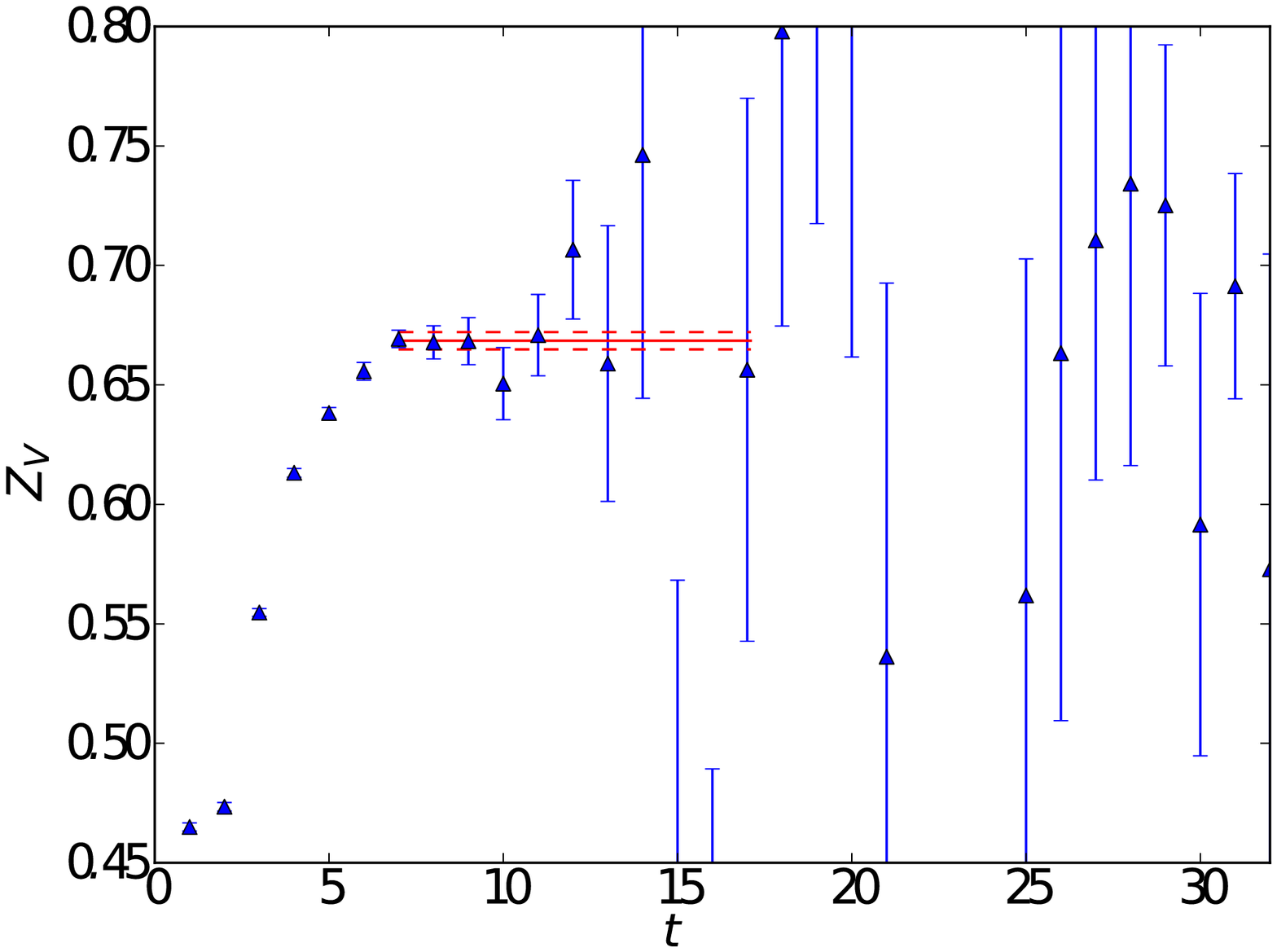}
\includegraphics*[width=0.45\textwidth]{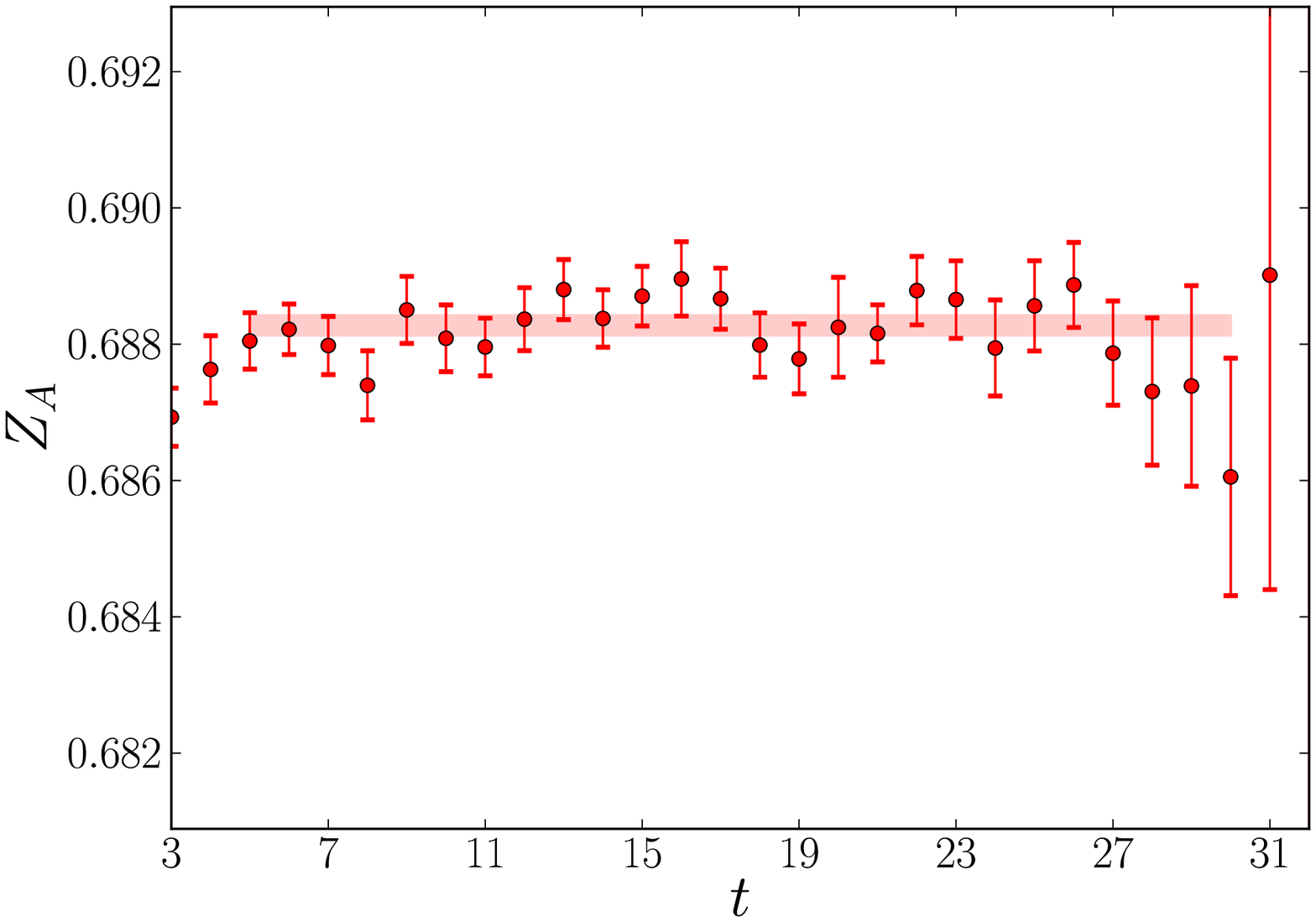}\\
\includegraphics*[width=0.40\textwidth]{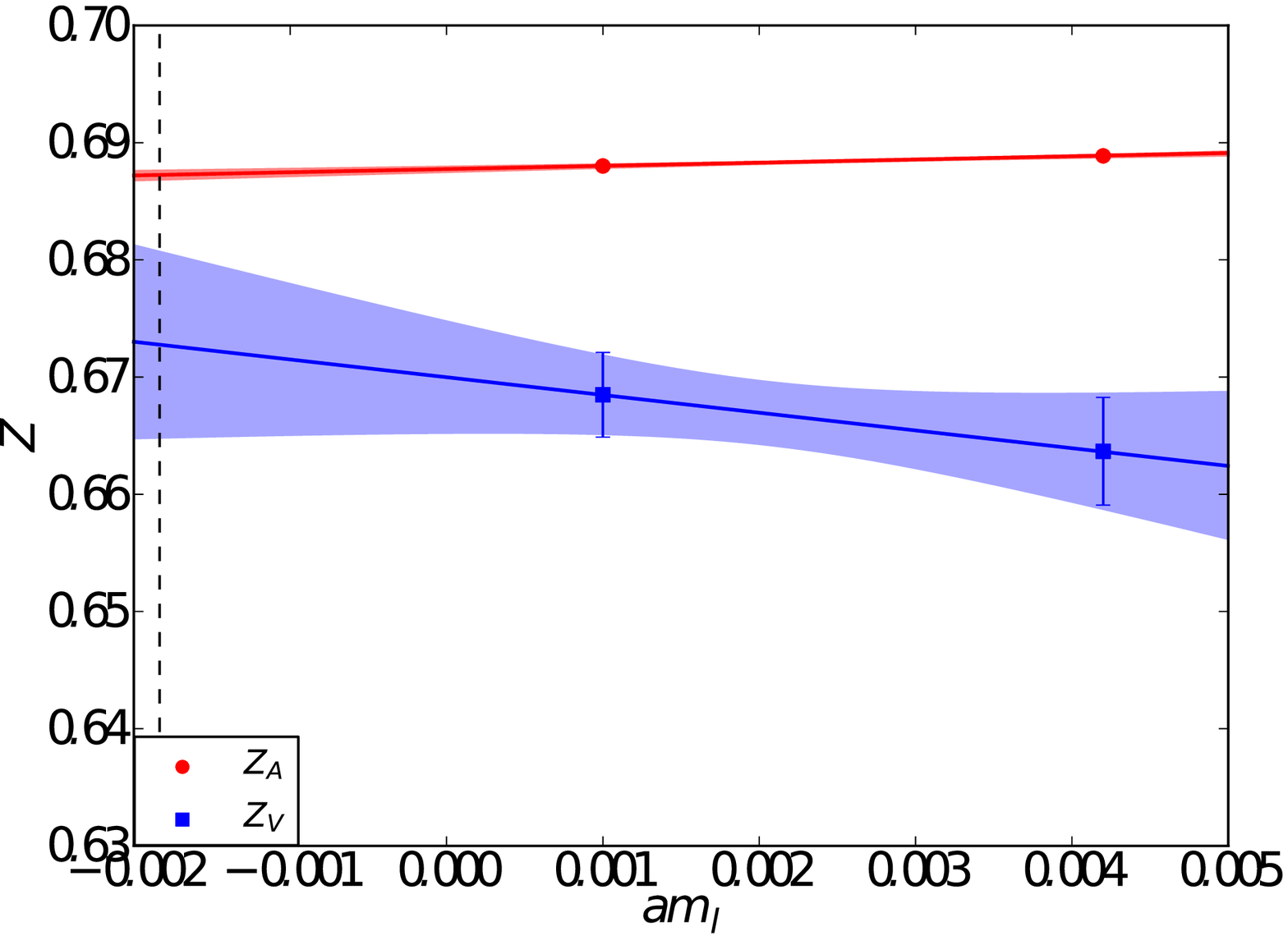}
\caption{\label{fig:ZVZA32ID} $Z_V/Z_\mathcal{V}$ (top left) and $Z_A/Z_\mathcal{A}$ (top right) as a function of time, calculated with unitary quarks on the $m_l=0.001$ ensemble. The bottom figure shows the chiral extrapolation of $Z_V/Z_\mathcal{V}$ and $Z_A/Z_\mathcal{A}$. In these plots the ratios have been abbreviated to $Z_V$ and $Z_A$.}
\end{figure}

However, in refs.~\cite{Sharpe:2007yd} and~\cite{32cubedpaper} it was shown that $Z_A$ is approximately equal to $Z_V$ - the ratio of the Symanzik-improved \textit{vector} current $V_\mu^{Sa}$ to the local vector current $V_\mu^a$ - with their difference of order 
$m_\mathrm{res}^2$. Since the ratio $Z_\mathcal{V}$ of the conserved 5D domain wall vector current $\mathcal{V}_\mu^a$ to its Symanzik-improved current $\mathcal{V}_\mu^{Sa}$ is unity up to terms $\mathcal{O}(a^2)$, this led to the observation that $Z_A$ can be determined much more accurately via the ratio of the local and 5D vector currents, $Z_V/Z_\mathcal{V}$, calculated using the following expression:
\begin{equation}
\frac{Z_V}{Z_\mathcal{V}} = \frac{\sum_{i=1}^3 \sum_{\vec x}\mathcal{V}_i^a(\vec x,t)V_i^a(\vec 0,0)}{\sum_{i=1}^3 \sum_{\vec x}V_i^a(\vec x,t)V_i^a(\vec 0,0)}
\end{equation}
in the limit $t\gg a$. We calculated $Z_V/Z_\mathcal{V}$ on 192 and 93 configurations of the $m_l=0.001$ and $0.0042$ ensembles respectively, and fit to folded data over the time intervals 8--12 and 7--17. Figure~\ref{fig:ZVZA32ID} shows $Z_V/Z_\mathcal{V}(m_x=m_l=0.001)$ as a function of time, illustrating the quality of our data. In the same figure we also show the chiral extrapolation of the results to $m_l=-m_\mathrm{res}$. In table~\ref{tab-zazv} we give the fit results on both ensembles and the chirally extrapolated values. For completeness we also calculate the ratio $Z_A/Z_\mathcal{A}$ using the aforementioned ratio~\cite{Blum:2001xb}, fitting over the time interval 5--30 to folded data. The values of this quantity on each ensemble and in the chiral limit are also given in table~\ref{tab-zazv}, and we show an example correlation function in figure~\ref{fig:ZVZA32ID} alongside a plot of the chiral extrapolation to $m_l=-\mres$. The value of $Z_A/Z_\mathcal{A}$ at the physical strange-quark mass is 
indistinguishable from the value at the simulated mass. Currently we have not measured $Z_V/Z_\mathcal{V}$ on reweighted configurations, however the lack of measurable strange-quark mass dependence of $Z_A/Z_\mathcal{A}$ suggests this will not have any effect on our conclusions.

We calculated the normalized decay constants using the above ratios. The values at the simulated strange-quark mass are listed in the second column of table~\ref{tab:mxy_fxy_32ID_mhsim}, and the pion-like and kaon-like decay constants at the physical strange-quark mass are given in the second columns of tables~\ref{tab:mll_fll_32ID_mhphys} and~\ref{tab:mhl_fhl_32ID_mhphys} respectively. For these quantities, the statistical uncertainty on $Z_V/Z_\mathcal{V}$ is considerably larger than that of the bare decay constant. For example, the bare unitary value on the $m_l=0.001$ ensemble has a 0.3\% error compared to 1.2\% on the normalized quantity. The error on $Z_A/Z_\mathcal{A}$ is much smaller, and if used to normalise the bare decay constants has virtually no effect on the relative error. However we chose to continue using $Z_V/Z_\mathcal{V}$ to normalize the decay constants in order to eliminate the systematic error associated with using the axial currents.

\begin{table}
\begin{center}
\begin{tabular}{lllll}
\hline
$m_q$ & $Z_A/Z_{\mathcal{A}}$ & $Z_V/Z_{\mathcal{V}}$\\
\hline\hline
\hline
0.0042 & 0.68901(9) & 0.6637(46)\\
0.001 & 0.68828(15) & 0.6685(36)\\
\hline
-$m_{\mathrm{res}}$ & 0.68778(34) & 0.6728(80)
\end{tabular}
\caption{Results for $Z_A/Z_{\mathcal{A}}$ and $Z_V/Z_{\mathcal{V}}$ at the simulated strange-quark mass.}
\label{tab-zazv}
\end{center}
\end{table}
\subsubsection{Omega baryon mass}

We determined the Omega baryon masses using box-source propagators with antiperiodic boundary conditions. In order to improve our statistics we averaged the degenerate upper and lower spin-components of the correlation functions prior to fitting. Our fits were performed over the interval $t=3$--10 on both ensembles, giving the values listed in table~\ref{tab:omega32ID}. In figure~\ref{fig:omegabxyeff32ID} we show the effective mass of the Omega baryon on the $m_l=0.001$ ensemble with $m_h=m_x=0.045$, demonstrating the quality of our data.

\subsubsection{Neutral kaon mixing parameter}

\begin{figure}[tp]
\centering
\includegraphics*[width=0.45\textwidth]{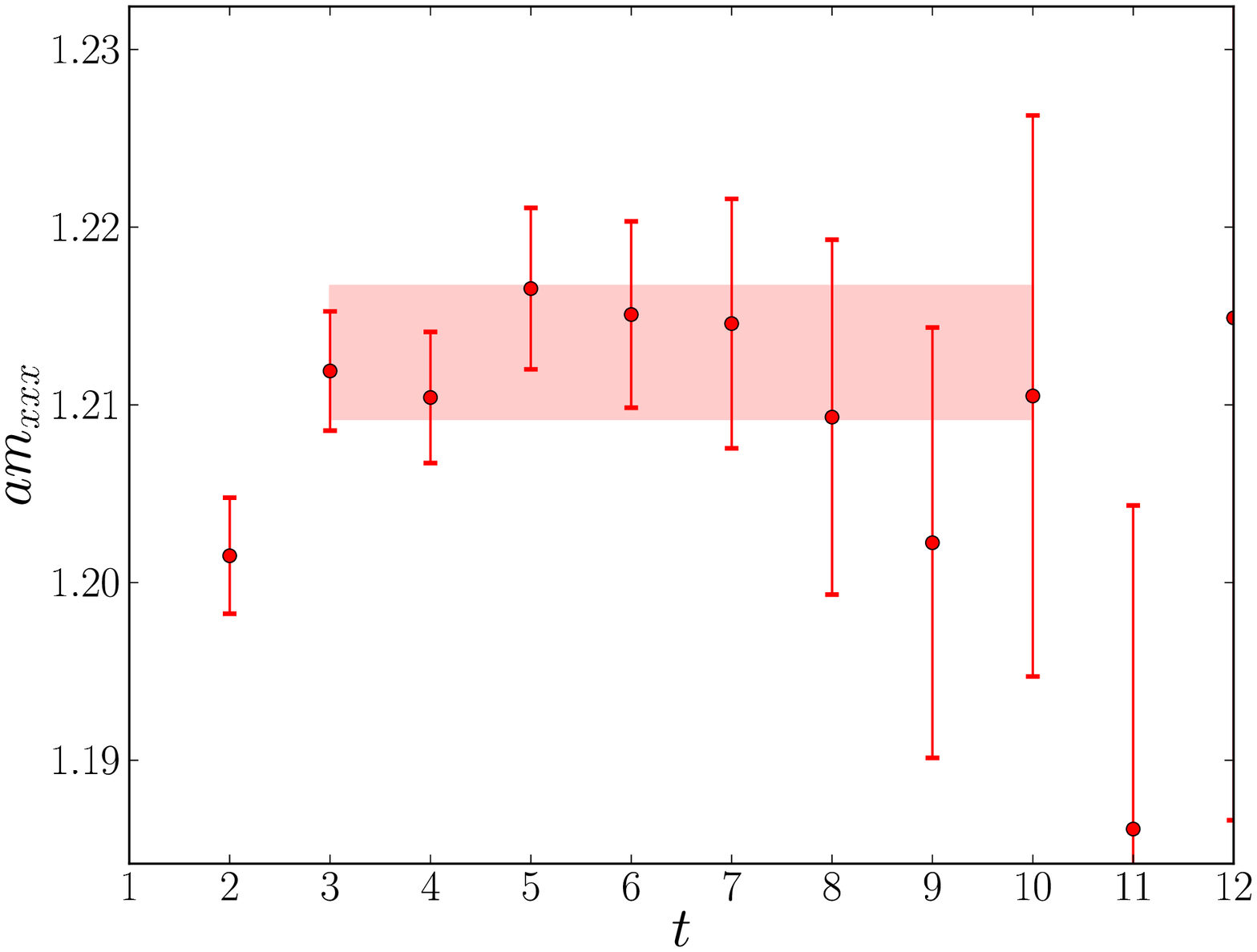}
\includegraphics*[width=0.45\textwidth]{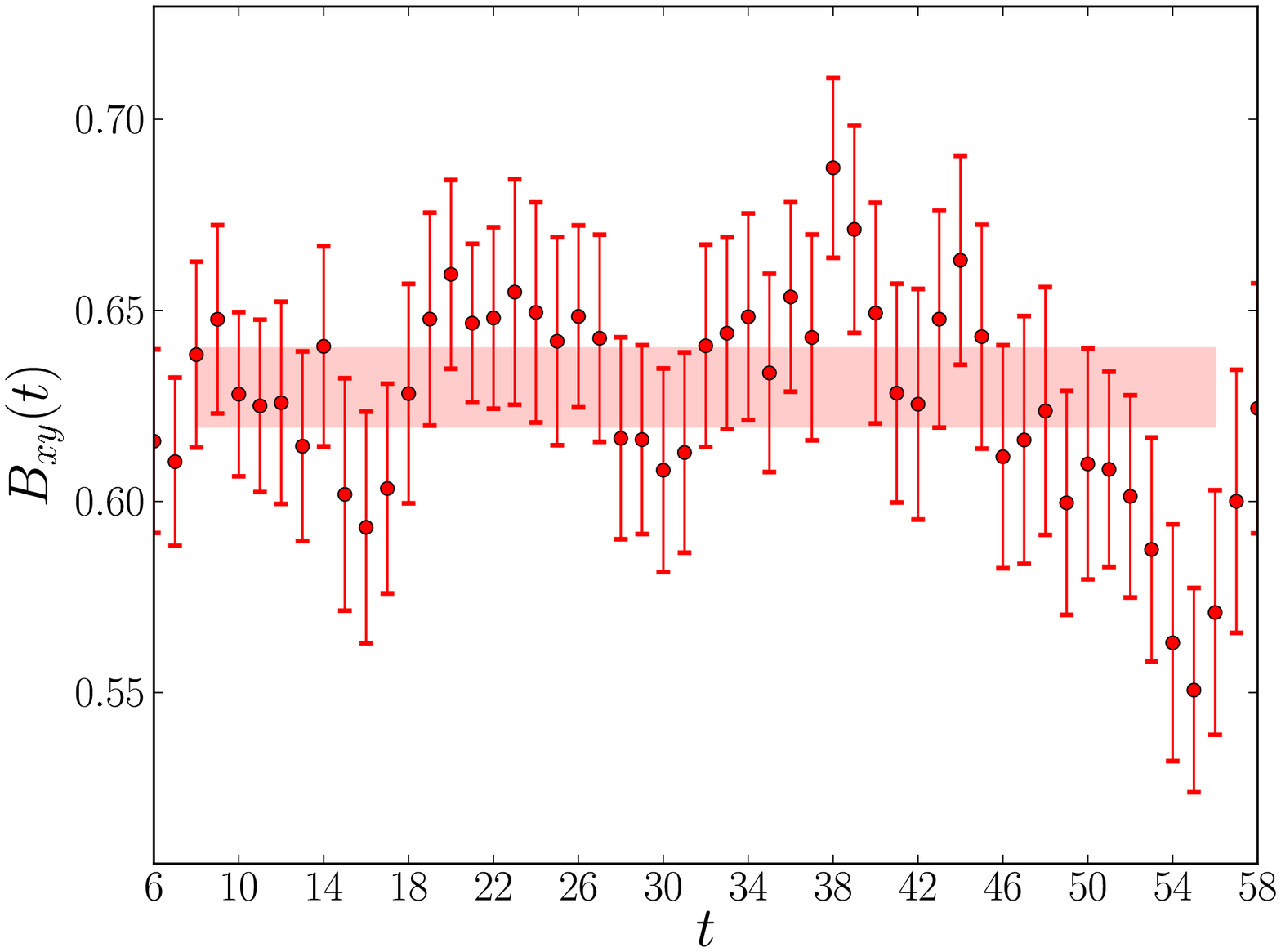}
\caption{
\label{fig:omegabxyeff32ID}
The left panel displays the fit to the $\Omega$ baryon mass with valence strange mass $m_x=0.045$ on the $m_l=0.001$, $m_h=0.045$ ensemble on the 32ID lattice, showing the quality of the fit with our box source. The right panel shows the $B_{xy}$ matrix element with $m_x=m_y=0.001$ as a function of time on the same ensemble.}
\end{figure}

The neutral-kaon mixing parameter $B_{xy}$ was obtained by fitting the time dependence of the following correlation function to a constant:
\begin{equation}
B_K^{\rm lat}(t) = \frac{\langle K^0(t_1)|\mathcal{O}_{\rm VV+AA}(t)|\bar K^0(t_2)\rangle}{\frac{8}{3}\langle K^0(t_1)|A_0(t)\rangle\langle A_0(t)|\bar K^0(t_2)\rangle}\,,\label{eqn-bkoperator}  
\end{equation}
where $\mathcal{O}_{\rm VV+AA}$ is the $\Delta S=2$ four-quark operator responsible for the mixing. This operator is inserted at all times $t$ between $t_1$ and $t_2$. We form the forwards-propagating $K^0$ state using the $p+a$ combination of propagators, and the backwards-propagating $\bar K^0$ state using the $p-a$ combination; in effect this sets $t_1=0$ and $t_2=64$ and reduces the round-the-world effects associated with the kaons propagating through the temporal boundaries. We performed our fits over the time interval 8--56, giving the values listed in table~\ref{tab:bk_32ID_mhsim}. We show an example matrix element in figure~\ref{fig:omegabxyeff32ID} and list the values of $B_{xy}$ at the physical strange-quark mass in table~\ref{tab:bk_32ID_mhphys}. Note that $B_K$ is a renormalization-scheme dependent quantity and must therefore be renormalized into a common scheme prior to being included in our simultaneous fits; this is discussed in more detail in section~\ref{sec:BK}. 

\subsubsection{The Sommer scales}

\begin{figure}[tp]
\centering
\includegraphics*[width=0.45\textwidth]{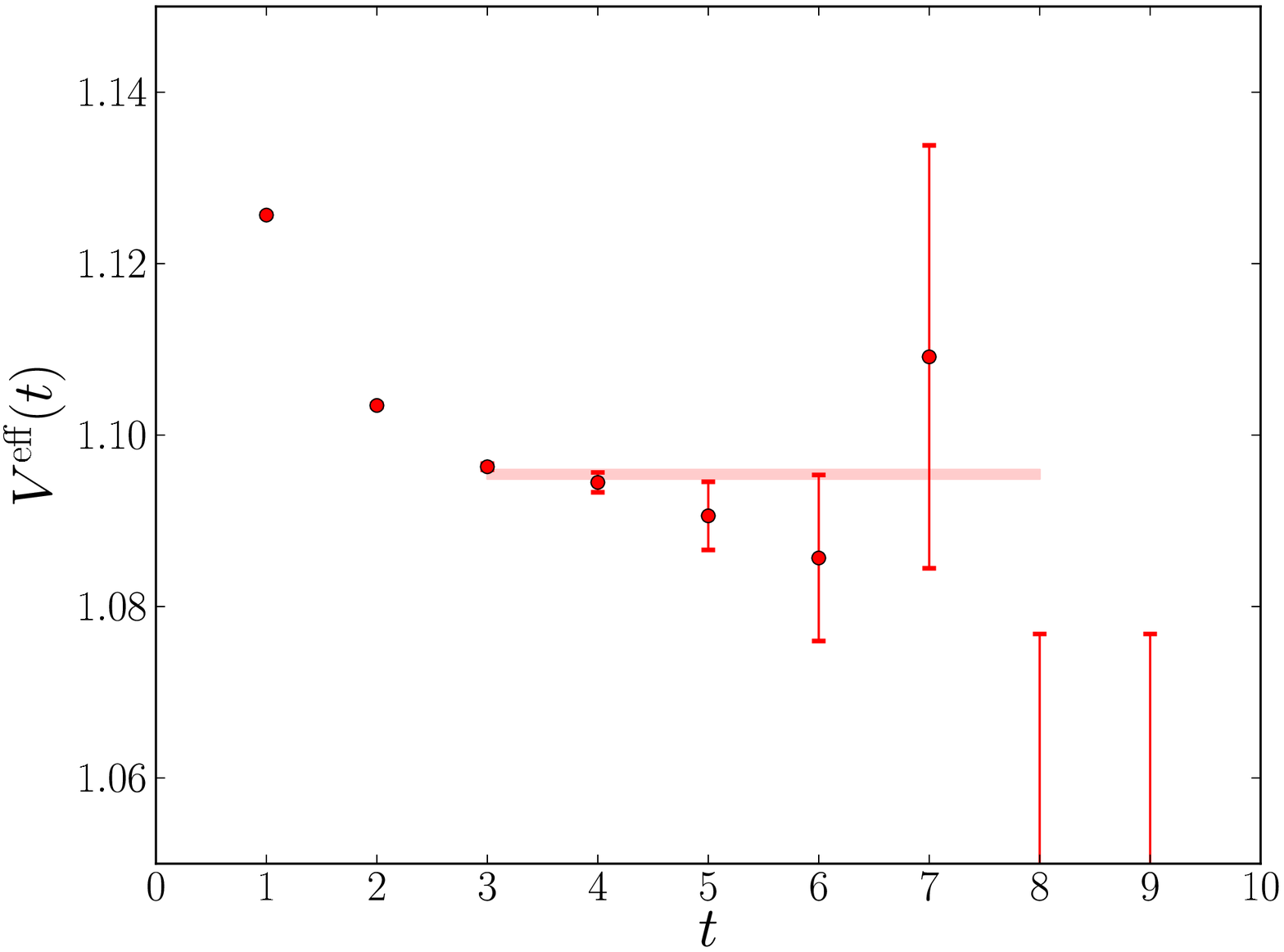}
\includegraphics*[width=0.45\textwidth]{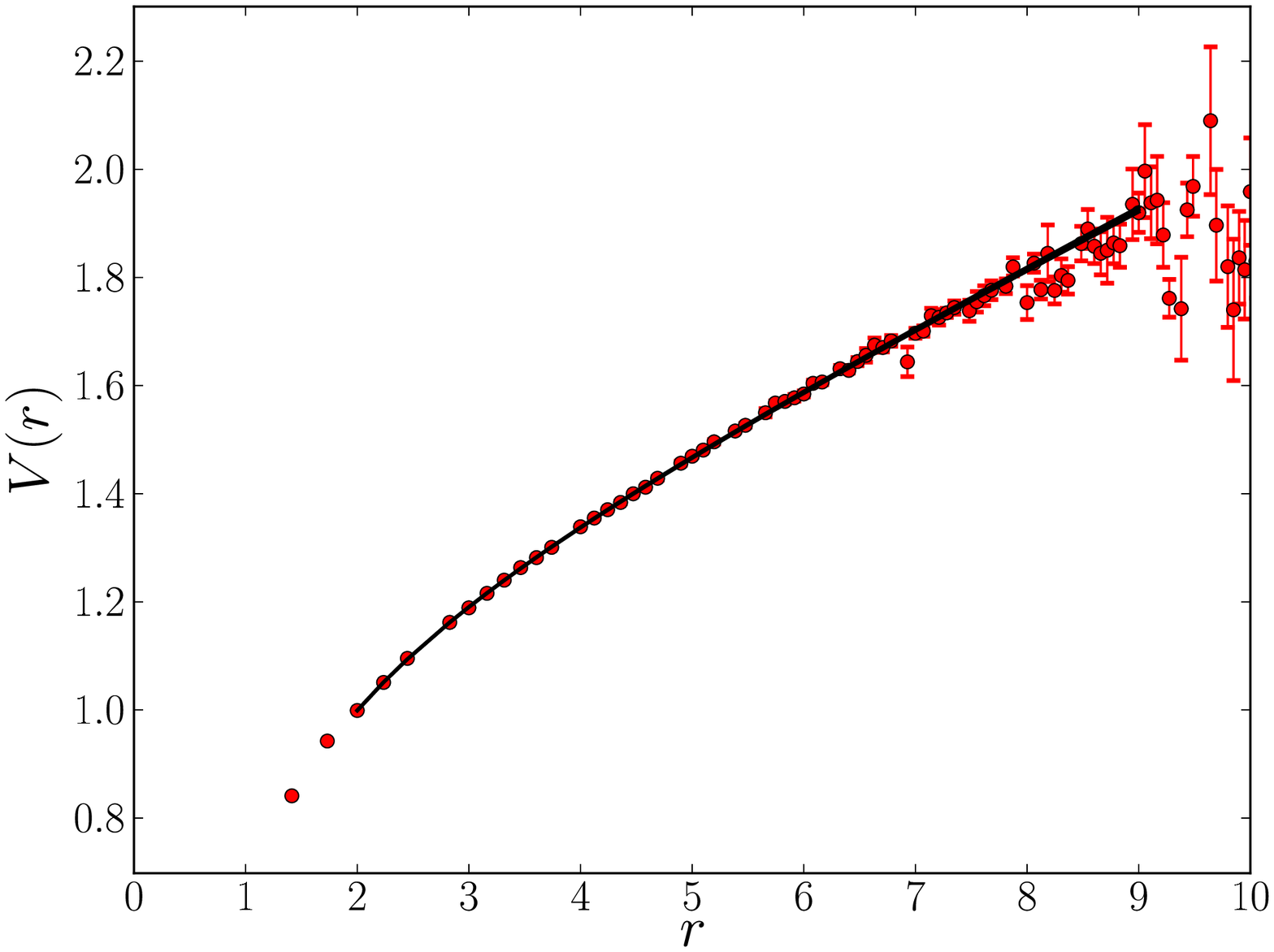}

\caption{
\label{fig-staticpot32ID}
The left panel shows the effective potential of the Wilson loops with a spatial extent of $r=2.45$ on the $m_l=0.001$ ensemble at the simulated strange-quark mass, overlaid by the fit to the range $t=3$--8. The right panel shows the static inter-quark potential $V(r)$ on this ensemble, again at the simulated strange-quark mass, as a function of the spatial extent of the Wilson loops, overlaid by the fit to the Cornell form over the range $r=2.00$--9.00.}
\end{figure}

Finally, we obtain the Sommer scales $r_0$ and $r_1$ using Wilson loops formed from products of time-directed gauge links, for which closure is not required due to Coloumb gauge-fixing. The time dependence of the Wilson loop $W(r,t)$ was fit from $t=3$ to $8$ for each value of the spatial separation $r$, and the resulting potential $V(r)$ then fit over the range $r = 2.00-9$ to the Cornell potential\,\cite{Eichten:1978tg}
\begin{equation}\label{eq:cornell}
V(r)=V_0-\frac{\alpha}{r}+\sigma\,r\,,
\end{equation}
where $V_0$, $\alpha$ and $\sigma$ are constants. The Sommer scales are determined directly from the potential:
\begin{equation}
r_i = \sqrt{\frac{A_i - \alpha}{\sigma} }\,, 
\end{equation}
where $A_0=1.65$ and $A_1=1.00$ for $r_0$ and $r_1$ respectively. In figure~\ref{fig-staticpot32ID} we show an example of the effective potential $V_\mathrm{eff}(t)$ at $r=2.45$ on the $m_l=0.001$ ensemble and the resulting fit to the potential $V(r)$ using the Cornell form. In table~\ref{tab:sommer32ID} we give the values of $r_1$ and $r_0$, as well as their ratios, at the simulated and physical strange-quark masses.

\begin{table}
\begin{tabular}{cc|cc|cc}
\hline
$m_x$ & $m_y$ &$m_{xy}(0.001)$ &$m_{xy}(0.0042)$ & $f_{xy}(0.001)$ &$f_{xy}(0.0042)$  \\
\hline
0.055 & 0.055 &  0.5463(2) & 0.5476(2) &  0.1354(16) & 0.1363(16)  \\
0.045 & 0.055 &  0.5207(2) & 0.5220(2) &  0.1324(16) & 0.1334(16)  \\
0.035 & 0.055 &  0.4941(2) & 0.4954(2) &  0.1291(16) & 0.1302(16)  \\
0.008 & 0.055 &  0.4159(3) & 0.4174(3) &  0.1183(14) & 0.1200(15)  \\
0.0042 & 0.055 &  0.4041(4) & 0.4055(4) &  0.1164(14) & 0.1184(15)  \\
0.001 & 0.055 &  0.3942(6) & 0.3955(6) &  0.1151(14) & 0.1173(15)  \\
0.0001 & 0.055 &  0.3915(7) & 0.3928(7) &  0.1149(14) & 0.1173(15)  \\
0.045 & 0.045 &  0.4940(2) & 0.4953(2) &  0.1294(16) & 0.1305(16)  \\
0.035 & 0.045 &  0.4662(2) & 0.4675(2) &  0.1262(15) & 0.1274(15)  \\
0.008 & 0.045 &  0.3831(3) & 0.3846(3) &  0.1156(14) & 0.1174(14)  \\
0.0042 & 0.045 &  0.3703(3) & 0.3718(4) &  0.1137(14) & 0.1158(14)  \\
0.001 & 0.045 &  0.3594(5) & 0.3610(5) &  0.1123(14) & 0.1148(14)  \\
0.0001 & 0.045 &  0.3564(6) & 0.3581(6) &  0.1121(14) & 0.1147(15)  \\
0.035 & 0.035 &  0.4368(2) & 0.4381(2) &  0.1231(15) & 0.1243(15)  \\
0.008 & 0.035 &  0.3476(3) & 0.3491(3) &  0.1126(14) & 0.1144(14)  \\
0.0042 & 0.035 &  0.3334(3) & 0.3350(3) &  0.1107(13) & 0.1129(14)  \\
0.001 & 0.035 &  0.3212(4) & 0.3230(4) &  0.1092(13) & 0.1118(14)  \\
0.0001 & 0.035 &  0.3178(5) & 0.3197(5) &  0.1090(13) & 0.1118(14)  \\
0.008 & 0.008 &  0.2273(2) & 0.2287(3) &  0.1024(12) & 0.1044(13)  \\
0.0042 & 0.008 &  0.2048(2) & 0.2063(3) &  0.1005(12) & 0.1028(13)  \\
0.001 & 0.008 &  0.1839(2) & 0.1854(3) &  0.0988(12) & 0.1015(12)  \\
0.0001 & 0.008 &  0.1775(2) & 0.1791(3) &  0.0984(12) & 0.1013(13)  \\
0.0042 & 0.0042 &  0.1795(2) & 0.1810(2) &  0.0986(12) & 0.1011(12)  \\
0.001 & 0.0042 &  0.1549(2) & 0.1564(2) &  0.0969(12) & 0.0997(12)  \\
0.0001 & 0.0042 &  0.1472(2) & 0.1487(3) &  0.0964(12) & 0.0994(12)  \\
0.001 & 0.001 &  0.1250(2) & 0.1265(2) &  0.0950(12) & 0.0981(12)  \\
0.0001 & 0.001 &  0.1151(2) & 0.1167(3) &  0.0944(12) & 0.0977(12)  \\
0.0001 & 0.0001 &  0.1042(2) & 0.1058(3) &  0.0938(12) & 0.0973(12)  \\
\end{tabular}
\caption{Pseudoscalar masses $m_{xy}$ ($m_l$) and decay constants $f_{xy}$ ($m_l$) on the 32ID ensembles at the simulated strange-quark mass ($m_h$ = 0.045).}
\label{tab:mxy_fxy_32ID_mhsim}
\end{table}
\begin{table}
\begin{tabular}{cc|cc|cc}
\hline
$m_x$ & $m_y$ &$m_{xy}(0.001)$ &$m_{xy}(0.0042)$ & $f_{xy}(0.001)$ &$f_{xy}(0.0042)$  \\
\hline
0.008 & 0.008 &  0.2272(2) & 0.2286(3) &  0.1027(12) & 0.1048(13)  \\
0.0042 & 0.008 &  0.2048(2) & 0.2063(3) &  0.1008(12) & 0.1031(13)  \\
0.001 & 0.008 &  0.1838(2) & 0.1854(3) &  0.0991(12) & 0.1018(13)  \\
0.0001 & 0.008 &  0.1775(2) & 0.1791(3) &  0.0987(12) & 0.1016(13)  \\
0.0042 & 0.0042 &  0.1794(2) & 0.1809(3) &  0.0989(12) & 0.1014(12)  \\
0.001 & 0.0042 &  0.1548(2) & 0.1563(3) &  0.0972(12) & 0.1000(12)  \\
0.0001 & 0.0042 &  0.1471(2) & 0.1486(3) &  0.0967(12) & 0.0997(12)  \\
0.001 & 0.001 &  0.1249(2) & 0.1265(3) &  0.0953(12) & 0.0984(12)  \\
0.0001 & 0.001 &  0.1151(2) & 0.1166(3) &  0.0947(12) & 0.0981(12)  \\
0.0001 & 0.0001 &  0.1042(2) & 0.1058(3) &  0.0941(12) & 0.0976(12)  \\
\end{tabular}
\caption{Pion masses $m_{xy}$ ($m_l$) and decay constants $f_{xy}$ ($m_l$) on the 32ID ensembles at the physical strange-quark mass ($m_h$ = \msIDbare).}
\label{tab:mll_fll_32ID_mhphys}
\end{table}
\begin{table}
\begin{tabular}{c|cc|cc}
\hline
$m_x$ &$m_{xh}(0.001)$ &$m_{xh}(0.0042)$ & $f_{xh}(0.001)$ &$f_{xh}(0.0042)$  \\
\hline
0.008 &  0.3890(20) & 0.3903(21) &  0.1161(14) & 0.1178(15)  \\
0.0042 &  0.3762(21) & 0.3777(22) &  0.1143(14) & 0.1163(15)  \\
0.001 &  0.3653(21) & 0.3669(23) &  0.1128(14) & 0.1153(15)  \\
0.0001 &  0.3624(21) & 0.3640(23) &  0.1126(14) & 0.1153(15)  \\
\end{tabular}
\caption{Kaon masses $m_{xh}$ ($m_l$) and decay constants $f_{xh}$ ($m_l$) on the 32ID ensembles at the physical strange-quark mass ($m_h$ = \msIDbare).}
\label{tab:mhl_fhl_32ID_mhphys}
\end{table}
\begin{table}
\begin{tabular}{cc|cc}
\hline
$m_y$ & $m_h$ &$m_\Omega(0.001)$ &$m_\Omega(0.0042)$  \\
\hline
0.055 & 0.045 &  1.2641(34) & 1.2735(36) \\
0.045 & 0.045 &  1.2130(37) & 1.2220(41) \\
0.035 & 0.045 &  1.1608(42) & 1.1695(48) \\
\msIDbare & \msIDbare &  1.2248(77) & 1.2326(55) \\
\end{tabular}
\caption{Omega baryon masses on the 32ID ensembles at the simulated strange quark mass $m_h = 0.045$ (first three rows) and at the physical strange quark mass (fourth row).}
\label{tab:omega32ID}
\end{table}

\begin{table}
\begin{tabular}{cc|cc}
\hline
$m_x$ & $m_y$ &$B_{xy}(0.001)$ &$B_{xy}(0.0042)$ \\
\hline
0.008 & 0.055 &  0.645(2) & 0.645(2) \\
0.0042 & 0.055 &  0.643(4) & 0.645(4) \\
0.001 & 0.055 &  0.650(16) & 0.665(16) \\
0.0001 & 0.055 &  0.665(28) & 0.689(28) \\
0.008 & 0.045 &  0.629(2) & 0.628(1) \\
0.0042 & 0.045 &  0.626(3) & 0.625(3) \\
0.001 & 0.045 &  0.630(10) & 0.632(10) \\
0.0001 & 0.045 &  0.639(17) & 0.644(18) \\
0.008 & 0.035 &  0.610(1) & 0.609(1) \\
0.0042 & 0.035 &  0.605(2) & 0.604(2) \\
0.001 & 0.035 &  0.606(6) & 0.602(6) \\
0.0001 & 0.035 &  0.610(10) & 0.605(10) \\
\end{tabular}
\caption{The partially-quenched neutral kaon mixing parameter $B_{xy}$ ($m_l$) on the 32ID ensembles at the simulated strange-quark mass ($m_h$ = 0.045).}
\label{tab:bk_32ID_mhsim}
\end{table}
\begin{table}
\begin{tabular}{c|cc}
\hline
$m_x$ &$B_{xh}(0.001)$ &$B_{xh}(0.0042)$ \\
\hline
0.008 &  0.632(2) & 0.631(2)  \\
0.0042 &  0.630(4) & 0.628(3)  \\
0.001 &  0.638(11) & 0.635(11)  \\
0.0001 &  0.651(17) & 0.649(19)  \\
\end{tabular}
\caption{The partially-quenched neutral kaon mixing parameter $m_{xh}$ ($m_l$) on the 32ID ensembles at the physical strange-quark mass ($m_h$ = \msIDbare).}
\label{tab:bk_32ID_mhphys}
\end{table}
\begin{table}
\begin{tabular}{cc|ccc}
\hline
$m_l$ & $m_h$ &$r_0$ &$r_1$ & $r_1/r_0$ \\
\hline
\multirow{2}{*}{0.0042} & 0.045 & 3.2732(63) & 2.1208(97) & 0.6479(36)\\
                        & \msIDbare  &  3.2616(75) & 2.1270(105) & 0.6521(37)\\
\hline
\multirow{2}{*}{0.001} & 0.045 & 3.2977(62) & 2.1346(98) & 0.6473(34)\\
                        & \msIDbare  &  3.2959(73) & 2.1401(100) & 0.6493(37)\\
\end{tabular}
\caption{The Sommer scales $r_0$ and $r_1$ and their ratio on the 32ID ensembles at the simulated strange quark mass $m_h = 0.045$ (first and third rows) and at the physical strange quark mass (second and fourth row).}
\label{tab:sommer32ID}
\end{table}

\FloatBarrier

%%%%%%%%%%%%%%%%%%%%%%%% Combined Chiral Fits %%%%%%%%%%%%%%%%%%%%%%%%%%%%%%
\section{Simultaneous Chiral/Continuum Fitting Procedure}
\label{sec:CombinedChiralFits}
In order to extrapolate to the continuum limit and physical quark masses we perform a simultaneous \textit{global fit} over our three ensemble sets. In this section we detail the fitting procedure and the subsequent chiral/continuum extrapolation. In addition we discuss the differences between this analysis and the 2010 analysis~\cite{Aoki:2010dy,Aoki:2010pe} of the $24^3$ and $32^3$ DWF+I ensembles.

\subsection{Global fits and scaling}

For a given choice of lattice action and a given bare coupling $\beta$, $2+1$ flavor lattice QCD has two free parameters: the
relevant couplings representing the quark masses. For 2+1 flavor QCD these are  
the average up/down quark mass $am_{u/d}$ and the strange quark mass $m_s$, expressed in lattice units. We can picture taking the continuum limit of the discretized theory as gradually taking $\beta\rightarrow \infty$ while following a curve of $am_{u/d}(\beta)$ and $am_s(\beta)$ that fixes the continuum physics to that of the real world; this curve is known as a \textit{scaling trajectory}. 
Experimental inputs are used to determine the lattice spacing and physical quark masses for
each bare coupling, and this imposes a constraint on each point on this scaling trajectory. (Our standard choice is to require that $m_\Omega$, $m_\pi/m_\Omega$ and 
$m_K/m_\Omega$ take their physical values.) This in turn
allows us to constrain the continuum limit we determine to be the physical point.

We can relate two points $(am_l,am_h,\beta)$ and $(a^\prime m_l^\prime,a^\prime m_h^\prime,\beta^\prime)$ that 
lie on a particular scaling trajectory via two \textit{scaling parameters} $Z_l$ and $Z_h$, defined as~\cite{Aoki:2010dy}--
\begin{equation}
Z_f(\beta,\beta^\prime)= \frac{1}{R_a(\beta,\beta^\prime)} \frac{ a\tilde m_f }{ a^\prime\tilde m_f^\prime}\,,\label{eqn-zfdef}
\end{equation}
where $f\in \{l,h\}$. Here
\begin{equation}
R_a(\beta,\beta^\prime) = \frac{a(\beta)}{a^\prime(\beta^\prime)}
\end{equation}
is the ratio of lattice spacings and $\tilde m_f = m_f + \mres$, where $\mres$ is the residual mass of domain wall QCD. In practice, we define our scaling parameters using the $\beta=2.25$ (32I) ensemble as a reference; we refer to this as the \textit{primary ensemble set}, on which $Z_l$, $Z_h$ and $R_a$ are unity by definition.
We may interpret our matching of quark masses to the bare masses on our primary ensemble set as a convenient, if inelegant,
intermediate renormalization scheme, for which the regularization involves an explicit choice of lattice action and bare coupling,
and whose values are determined by the hadronic inputs. The renormalization scale in this scheme is the scale at which the bare mass is defined: the inverse lattice spacing of the primary ensemble. The renormalized masses are then
\begin{eqnarray}
\tilde{m}_l^{r\prime} &= Z_l R_a[a^\prime \tilde{m}_l^\prime]/a\,,\\
\tilde{m}_h^{r\prime} &= Z_h R_a[a^\prime \tilde{m}_h^\prime]/a\,,
\end{eqnarray}
where unprimed quantities are defined on the primary ensemble set and primed quantities on some other ensemble set, and $a^\prime$ is related to $a$ via $a^\prime = R_a^{-1} a$.

Considering only the unitary observables for simplicity, any observable $Q$ is a function of the bare quark masses and
the bare coupling. We take this as $Q(a^\prime\tilde m_l^\prime,\,a^\prime\tilde m_h^\prime,\,\beta^\prime)$, 
at coupling and quark masses differing from the primary ensemble set.
This can equally be expressed as a function of the renormalized quark masses and the lattice spacing as
\begin{equation}
Q(a^\prime\tilde m_l^\prime,a^\prime\tilde m_h^\prime,\beta^\prime) = f(\tilde{m}_l^{r\prime},\,\tilde{m}_h^{r\prime},\,a^{\prime\,2} )\,.\label{eqn-qconvert}
\end{equation}
The function $f$ depends on the lattice action and on the choice of physical quantities used to determine the scaling trajectory. Since among these input parameters is a quantity with a physical scale (in our case the $\Omega^-$ mass), we choose to view the
function as depending on this scale so its arguments can be expressed in physical units.  The function itself will
have a continuum limit as $\beta$ and $\beta'$ become large. 

Consider a double expansion in quark masses and in lattice spacing around our primary ensemble 
\begin{equation}
\begin{array}{rcl}
f(\tilde{m}_l^{r\prime},\,\tilde{m}_h^{r\prime},\,a^{\prime\,2} )
&=& f(\tilde m_l,\, \tilde m_h,\, a^2) \\
&+&
\frac{\partial f}{\partial\tilde m_l^{r\prime} }(\tilde m_l^{r\prime}-\tilde m_l)\\
&+&
\frac{\partial f}{\partial\tilde m_h^{r\prime} }(\tilde m_h^{r\prime}-\tilde m_h)\\
&+&
\frac{\partial f}{\partial a^{\prime\,2} } (a^{\prime 2}-a^2).\\
&+& O( \tilde m_f^2, a^2 \tilde m_f , a^4)
\end{array}
\end{equation}
where the partial derivatives are evaluated at $R_a = Z_l = Z_h = 1$.

If $f$ is a quantity used to determine the scaling trajectory then we necessarily constrain that
$\frac{\partial f}{\partial a^{\prime\,2} } = 0$ at the match point.
In this paper we introduce a new DSDR term to the effective gauge action. To this order only the 
term $\frac{\partial f}{\partial a^{\prime\,2} }$ depends on the 
the lattice action. We can therefore determine the parameters of $f$ for a given parametrization, accurate to this order,
via a fit to a set of points over multiple ensembles, and including the two different gauge actions.
Even though there is only a single lattice spacing with the DSDR gauge action,
it will usefully contribute to a universality constrained global fit by significantly constraining the mass dependent
terms in a global parametrization of $f(\tilde m_l,\,\tilde m_h,\, a^2)$.

For the purposes of matching ensemble sets with different lattice 
spacing we ignore terms of higher order in $\delta m_l$, $\delta m_h$ and in $a^2$.
Since we allow $Z_h$ for masses in the region of the strange quark
to differ from $Z_l$ for masses in the region of the up/down quarks, in this matching 
context we may consider small variations in the quark masses only.

We can also see immediately that if we instead use a nearby reference point  $(am_l,am_h,\beta) \to (a[m_l+\Delta_l],a[m_h+\Delta_h],\beta)$, 
the ratios $Z_f$ and $R_a$ change only by terms $\mathcal{O}\left(a(\beta)^2 - a^\prime(\beta^\prime)^2\right)$ with 
coefficients that are functions of the mass differences $\Delta_f$ that vanish as $\Delta_f\rightarrow 0$.
This means that there is an \emph{allowed range} overwhich $Z_l$ and $Z_h$ may be simply taken as a constant.
Higher order terms in quark masses are, of course, subsequently introduced in our global chiral-continuum fits,
and we introduce $Z_h$ and $Z_l$ as free fit parameters multiplying quark masses in the allowed range.
In practice we even find $Z_l \sim Z_h$; were the matching and primary ensemble sets taken sufficiently close to the continuum limit, such that lattice artifacts were small and $m_h \ll 1/a$, then we would necessarily find $Z_l = Z_h$. The matching scheme can therefore be considered mass-independent as the mass dependence of the renormalization factors drops out when the renormalization scale becomes large.

In the following subsection we discuss our strategy for determining the scaling parameters $Z_l$, $Z_h$ and $R_a$.

\subsection{Determination of the scaling parameters}
\label{sec:zfactorcalc}
In our analysis~\cite{Aoki:2010dy} of the $24^3$ and $32^3$ DWF+I ensembles, we determined $R_a$, $Z_l$ and $Z_h$ by matching our lattice data at an unphysical light and heavy quark mass within the range of available data on the two simulations. The matching was performed by first choosing a suitable \textit{match point} on one of the ensemble sets (labelled $\bf M$) which can, but does not necessarily have to be, the primary ensemble. On every other ensemble set $\bf e$ (in the 2010 analysis only the 24I ensemble set remained), two dimensionless ratios, $R_l = m_{ll}/m_{hhh}$ and $R_h = m_{lh}/m_{hhh}$, were linearly interpolated in the unitary light and heavy quark masses until their values matched those measured at the match point on ensemble $\bf M$. Here $m_{ll}$, $m_{lh}$ and $m_{hhh}$ are respectively the pion, kaon and Omega baryon masses measured at unphysical light ($l$) and strange ($h$) quark masses. The match point was chosen to minimize the distance of interpolation required on the ensemble 
sets $\bf e$. This procedure provides a pair of equivalent masses (in lattice units), $(a\tilde m_l)^{\bf e}$ and $(a\tilde m_h)^{\bf e}$, for each ensemble set. Using these masses we determined $Z_l$ and $Z_h$ using eqn.~\ref{eqn-zfdef}, calculating $R_a$ from the ratio of the Omega baryon masses at the match point:
%\begin{equation}
\begin{align}
Z_f^{\bf e}= \frac{1}{R_a^{\bf e}}\frac{ (a\tilde m_f)^{\bf 1} }{ (a\tilde m_f)^{\bf e} }\,, &\hspace{1cm} R_a^{\bf e} = \frac{a^{\bf 1}}{a^{\bf e}} = \frac{m_{hhh}^{\bf 1}(\tilde m_l^{\bf 1}, \tilde m_h^{\bf 1})}{m_{hhh}^{\bf e}(\tilde m_l^{\bf e}, \tilde m_h^{\bf e})}\,,\label{eqn-zlralattdef}
\end{align}%\end{equation}
where $f\in \{l,h\}$.

The above procedure defines the scaling parameters such that $m_{ll}$, $m_{lh}$ and $m_{hhh}$ scale perfectly up to terms $\mathcal O(m a^2)$ within the allowed region around the match point. Note that this choice is not unique; we could for instance use the pion and kaon decay constants, $f_{ll}$ and $f_{lh}$, and the Sommer scale $r_0$, and match $r_0 f_{ll}$ and $r_0 f_{lh}$. $R_a$ can then be determined from the ratio of $r_0$ measured at the match point on each ensemble set. In this case $r_0$, $f_{ll}$ and $f_{lh}$ would have no $\mathcal O(a^2)$ dependence instead. In ref.~\cite{Aoki:2010dy} we demonstrated that this produces results that are completely consistent.

\begin{figure}[tp]
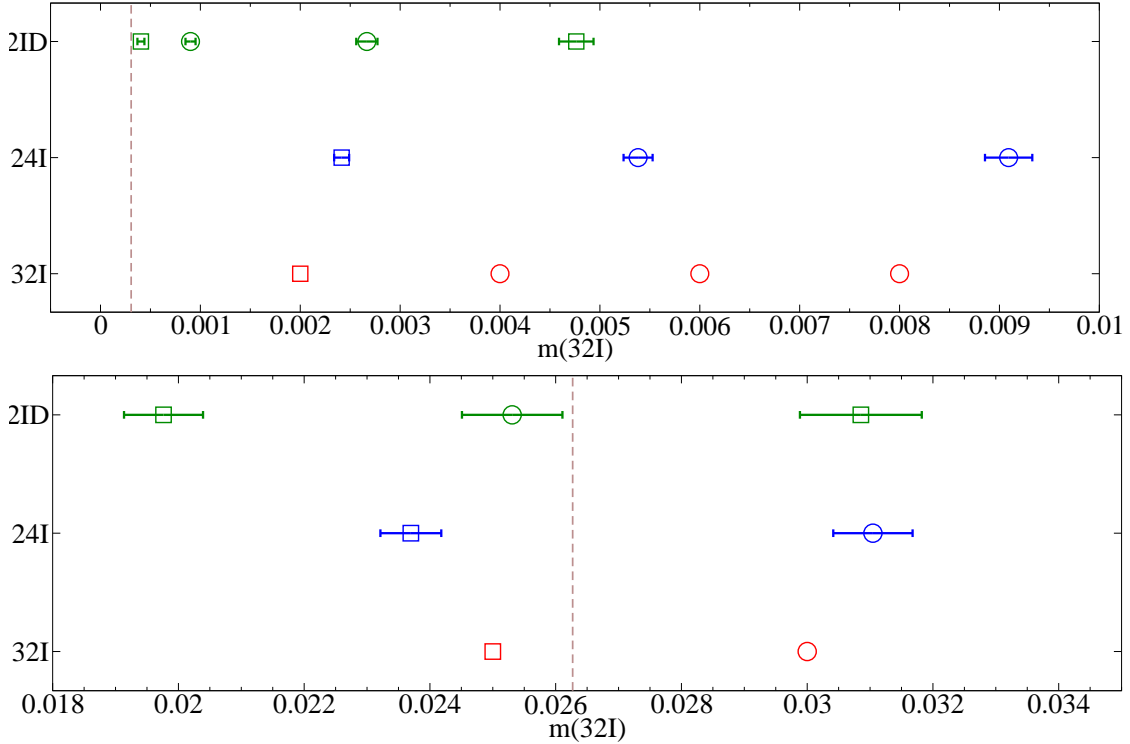

\centering
 \includegraphics[width=0.9\textwidth,clip=true,trim=15 0 0 0]{fig/massrangelight.eps}\\
\vspace{0.1cm}
\includegraphics[width=0.9\textwidth,clip=true,trim=15 0 0 0]{fig/massrangeheavy.eps}
\caption{Simulated quark masses on each of our three ensemble sets brought into a common normalization with the bare quark masses on our 32I ensemble set using the scaling factors determined in section~\ref{sec:FitResults}. The top panel shows the light quark mass regime and the bottom panel the heavy quark mass regime. Circular points are used to mark the unitary masses and square points the partially-quenched masses. The physical up/down and strange quark masses are marked with dashed lines.}
\label{fig-massrangeplot}
\end{figure}

The benefit of this \textit{fixed trajectory} method is that it enables the separation of the matching from the complexities of the subsequent global fits. However, in our combined analysis of the DWF+I and DWF+ID ensemble sets, we find that, apart from the lightest partially-quenched point, the range of light quark masses on the 24I ensemble set does not overlap with that on the 32ID ensemble set (cf. figure~\ref{fig-massrangeplot}). As a result, matching the 24I and 32ID ensemble sets to the 32I primary ensemble set at a single point would require a long extrapolation beyond the unitary mass range. In addition, the use of independent linear interpolations on each ensemble set is more vulnerable to statistical fluctuations than if we were to fit over all data simultaneously. As a result we choose the alternate \textit{generic scaling} method~\cite{Aoki:2010dy}, in which $R_a$, $Z_l$ and $Z_h$ are left as free parameters which are determined, alongside the low-energy constants, in a global fit of $m_\pi$, $m_
K$ and $m_\Omega$ over all ensemble sets. Here the three conditions that define the scaling trajectory are imposed by omitting scaling terms up to $\mathcal O(m a^2)$ from the fit forms describing these quantities, and the values of the ratios are selected as those that minimize the global $\chi^2$. In ref.~\cite{Aoki:2010dy} we demonstrated that this approach gives consistent results with the fixed trajectory approach.

\begin{figure}[t]
\centering
 \includegraphics[width=1.0\textwidth]{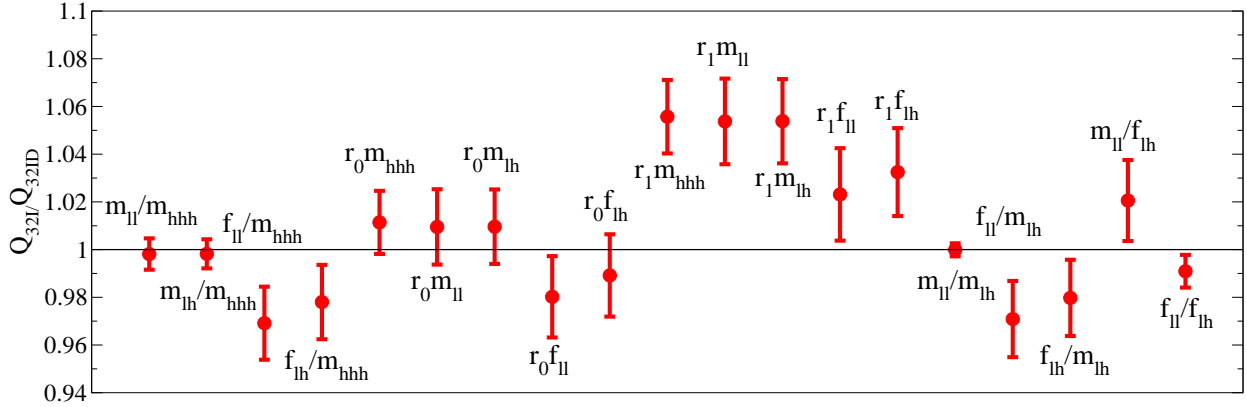}
\caption{Ratios of various dimensionless combinations of observables between the 32I and 32ID ensemble sets. The combination of physical quantities is given above or below the corresponding point. A ratio of unity indicates perfect scaling between the two ensemble sets.}
\label{fig-beijingplot}
\end{figure}

Prior to discussing our fit ans\"{a}tze, it is illustrative to compare the ratios of various dimensionless quantities between the 32I and 32ID ensemble sets at a particular match point, using the scaling parameters determined later in section~\ref{sec:FitResults}. This allows us to visualize the magnitude of the scaling corrections for each quantity. Choosing $[am_l]^{32I} = 0.004$ and the physical strange-quark mass $[am_h]^{32I} = 0.0263(10)$ as a match point, we used the scaling parameters listed in table~\ref{tab-cutlecs} (combining the statistical error with the systematic errors determined using the procedure given in section~\ref{subsec:syserrs}) to determine the corresponding point on the 32ID ensemble as $[am_l]^{32ID} = 0.0066(3)$ and $[am_h]^{32I} = 0.0467(6)$. We then performed linear fits to a range of quantities over each ensemble set independently and interpolated each to the corresponding match point quark masses. In figure~\ref{fig-beijingplot} we plot the ratio of a number of dimensionless 
combinations of these quantities between the two ensemble sets. It is immediately apparent that the scaling parameters do indeed fix $m_\pi$, $m_K$ and $m_\Omega$ to scale between the two ensemble sets, and the errors on the ratios of these quantities are indicative of the size of higher order corrections -- in a fixed trajectory matching at this point those errors would be zero by definition. 

Considering combinations of $m_\pi$, $m_K$ and $m_\Omega$ with other quantities that retain a scale dependence, and for the purpose of making a crude estimate ignoring the small discretization error on the 32I measurements, we can use this plot to read off the rough size of the discretization error for the measurements on the 32ID ensemble set: we estimate $\mathcal{O}(3\%$--$5\%)$ discretization terms for $f_{ll}$ and $f_{lh}$, $\mathcal{O}(1\%$--$2\%)$ for $r_0$, and then a slightly larger $\mathcal{O}(5\%$--$7\%)$ contribution for $r_1$.

As an aid to the reader, we also use the aforementioned scaling parameters to place all of the simulated quark masses on a common scale, and draw a line to indicate the physical point as determined in section~\ref{sec:FitResults}. These plots are shown in figure~\ref{fig-massrangeplot}.

\subsection{Chiral/continuum fitting strategy}

The chiral/continuum fit forms are obtained via a joint expansion in $a^2$ and $\tilde m_f$. As in ref.~\cite{Aoki:2010dy} we consider both an NLO expansion around the SU(2) chiral limit using partially-quenched chiral perturbation theory (PQChPT) and also a leading-order analytic expansion about an unphysical light-quark mass. Including finite-volume effects in the ChPT, this provides three fit ans\"{a}tze, which we label ``analytic'', ``ChPT'' and ``ChPTFV'', where the latter two refer to the chiral perturbation theory forms without and with finite-volume corrections respectively. For each ansatz we expand the heavy-quark mass dependence to linear order in the vicinity of the physical strange-quark mass. We use a power-counting scheme whereby terms of order $\tilde m_f a^2$ and higher are neglected. This truncation leaves only a single $a^2$ term arising from the expansion of the leading order parameter. For example, the analytic form for the pion decay constant $f_{ll}$ in physical units is as follows:
\begin{equation}
f_{ll} = C_0^{f_\pi}\left(1 + C_a^{f_\pi} a^2\right) + C_1^{f_\pi}(m_v^R-m_{l0}^R) + C_2^{f_\pi}(m_l^R-m_{l0}^R) + C_3^{f_\pi}\left(m_h^R - m_{h0}^R\right)\,,\label{eqn-fllana-1}
\end{equation}
where the superscript $R$ indicates a renormalized physical quark mass (in a general scheme), and $m_{l0}^R$ and $m_{h0}^R$ are the expansion points for the light and heavy quark masses respectively. In our power counting scheme, a term in the lattice spacing arises only in the expansion of the leading term $C_0^{f_\pi}$. It is important to note that the $a^2$ coefficients parametrizing the lattice artifacts will differ between the Iwasaki and Iwasaki+DSDR gauge actions, therefore for the remainder of this work we label these coefficients with a superscript denoting the lattice action.

As discussed in ref.~\cite{Aoki:2010dy}, the scaling parameters $Z_l^{\bf e}$ and $Z_h^{\bf e}$ that relate the quark masses between the ensemble ${\bf e}$ and the primary ensemble set can be thought of as ``renormalization coefficients'', removing the ultraviolet divergence and converting the masses into a mass-independent ``matching scheme'' defined with lattice regularization at $\beta=2.25$. It is therefore unnecessary to renormalize the input quark masses into a continuum renormalization scheme such as $\msbar$ prior to performing the fits; we need only convert the input masses into the matching scheme. The predictions for the physical up/down and strange quark masses can be converted into a more conventional scheme \textit{a posteriori}; this is performed in section~\ref{sec:QuarkMasses}.

In the 2010 analysis, we performed our fits to quantities in physical units. However this required us to continually update the lattice spacings and physical quark masses based on the results of the fit, iterating until convergence. For this analysis we instead fit to quantities in lattice units, which removes the need to repeat the global fit multiple times. However, for clarity, we continue to quote our fit forms in dimensionful units; the correctly normalized versions in lattice units for an ensemble $\bf e$ can easily be obtained by inserting powers of $a^{\bf e}$ where appropriate to make the measurement and input quark masses dimensionless; applying factors of $Z_l^{\bf e}$ and $Z_h^{\bf e}$ as before to bring the quark masses into the normalization of the primary ensemble; substituting $a^{\bf e}$ with $a^{\bf 1}/R_a^{\bf e}$; and finally setting $a^{\bf 1}$ to unity.

In the matching scheme the analytic fit form for $f_{ll}$ on the primary ensemble $\bf 1$ becomes:
\begin{equation}
f_{ll}^{\bf 1} = C_0^{f_\pi}\left(1 + C_a^{f_\pi,\,A(\bf 1)}[a^{\bf 1}]^2\right) + C_1^{f_\pi}\tilde m_v^{\bf 1} + C_2^{f_\pi}\tilde m_l^{\bf 1} + C_3^{f_\pi}\left(\tilde m_h^{\bf 1} - m_{h0}\right)\,,
\end{equation}
where $\tilde m_v = \frac{1}{2}(\tilde m_x + \tilde m_y)$ and we have taken advantage of the linearity of the expression to absorb any terms in $\tilde m_l^0$ into the leading coefficient. Here the superscript $A(1)$ denotes the gauge action of the primary ensemble: for our choice of primary ensemble this is the Iwasaki action, labelled $I$. The fit form describing $f_{ll}$ on any other ensemble can then be obtained by applying equation~\ref{eqn-qconvert} to the above and replacing the $a^2$ coefficient with that appropriate to the particular action.

Before continuing, it is illustrative to discuss how the $a^2$ coefficients for the Iwasaki+DSDR gauge action can be determined without having multiple lattice spacings with this action. Let us imagine that we have performed a global fit over the 24I and 32I ensembles as in ref.~\cite{Aoki:2010dy}, and have thus determined the coefficients $C_0^{f_\pi}$ through $C_3^{f_\pi}$ and the Iwasaki scaling coefficient $C_a^{f_\pi,\,I}$. We then perform a fixed trajectory matching between the 32I and 32ID ensemble sets, providing us with $Z_l^{32ID}$, $Z_h^{32ID}$ and $R_a^{32ID}$. The fit form describing $f_{ll}$ on the 32ID ensemble now has only one unknown coefficient, namely $C_a^{f_\pi,\,ID}$, which can be obtained by comparing any single simulated data point with the predicted value or by fitting over several points. In practice we would like the 32ID data to contribute to the determination of the coefficients, thus we perform a combined fit to all three ensemble sets and allow $C_a^{f_\pi,\,ID}$ to be 
determined by minimizing the global $\chi^2$.

Recall that our choice of scaling trajectory defines the pion, kaon and Omega baryon masses to have no lattice spacing dependence up to terms $\mathcal{O}(ma^2)$ arising from the match-point dependence of $Z_l$, $Z_h$ and $R_a$. These terms are neglected by our power counting, hence the fit forms for these quantities contain no discretization terms. For example, the form for the Omega baryon with the analytic ansatz is:
\begin{equation}
m_{hhh} = C_0^{m_\Omega} + C_1^{m_\Omega}\tilde m_l + C_2^{m_\Omega}\left(\tilde m_y-m_{h0}\right) + C_3^{m_\Omega}\left(\tilde m_h -  m_{h0}\right)\,.
\end{equation}

The remaining analytic and ChPT fit forms can be found in section V-B of ref.~\cite{Aoki:2010dy}. Note that as we now measure the strange-quark dependence in the global fit rather than linearly interpolating to the physical strange mass prior to fitting, we include additional parameters for the heavy valence-quark dependence (where appropriate) and the heavy sea-quark dependence, in this order. For the analytic fit forms these coefficients are labelled following the existing sequence, for example the heavy valence and sea quark dependences of $m_{hhh}$ are $C_2^{m_\Omega}$ and $C_3^{m_\Omega}$ respectively. For the ChPT fit forms we label the parameters $c_{Q,m_y}$ and $c_{Q,m_h}$ for the valence and sea dependence of the quantity $Q$ respectively.

We perform our fits with the strange-quark mass expansion point $m_{h0}$ set initially to the un-reweighted strange sea-quark mass on the $32^3$ DWF+I ensemble set. This is then corrected to the physical strange quark mass \textit{a posteriori}; with our power counting this requires only a redefinition of the leading order coefficient (e.g. $C_0^{m_\Omega}$). For the ChPT forms we must also adjust the LECs in order to absorb the effect of adjusting the chiral scale $\Lambda_\chi$ to the conventional 1 GeV once the lattice scale has been determined. 

Once the fits have been performed, we determine the physical up/down and strange quark masses (normalized to the units of the 32I primary ensemble) by numerically adjusting the quark masses in our fit functions such that $m_\pi/m_\Omega$ and $m_K/m_\Omega$ match their physical values in the continuum limit. Here, as in ref.~\cite{Aoki:2010dy}, we use $m_\pi = 135$ MeV, $m_K= 495.7$ MeV and $m_\Omega = 1672.45$ MeV. The primary lattice spacing can then be extracted by dividing the predicted continuum value for $m_\Omega$ in lattice units by its physical value. Using these results and the values of $R_a$, $Z_l$ and $Z_h$ found by fitting the data, the lattice spacings and physical quark masses for the other ensemble sets can be determined; we discuss this in more detail in section~\ref{sec:QuarkMasses}.

%%%%%%%%%%%%%%%%%%%%%%%% Combined Chiral Fit Results %%%%%%%%%%%%%%%%%%%%%%%%%%%%%%
\section{Fit Results and Systematic Error Determination}
\label{sec:FitResults}
\FloatBarrier

Following the 2010 analysis strategy, we split the chiral/continuum fits into three parts. In the first part, to which this section is dedicated, we performed simultaneous fits to $m_\pi$, $m_K$, $m_\Omega$, $f_\pi$ and $f_K$ over the three ensemble sets, from which we determined the physical quark masses (in matching scheme normalization), the lattice spacings and the scaling parameters, along with predictions for the physical pseudoscalar decay constants. The second set of fits were performed to $B_K$ and the third to the Sommer scales $r_0$ and $r_1$; these are documented in sections~\ref{sec:BK} and~\ref{sec:r0r1} respectively. We also separate out the discussion of the determination of the physical quark masses in the $\msbar$-scheme into section~\ref{sec:QuarkMasses}.

\subsection{Fit results}

In the 2010 analysis we did not attempt to correct for finite-volume effects in our analytic fits, as the magnitude of the change was small with respect to the systematic error arising from the chiral extrapolation. However on the 32ID ensemble set we have data reaching down almost to the physical point, hence we might expect that the chiral systematic error will be reduced and that the finite-volume error may begin to dominate (as we discuss below, this does indeed seem to be the case). As a result, in anticipation of our later discussion, we perform our analytic fits to data corrected using ChPT to the infinite-volume limit. Although we do not have multiple volumes from which to measure the size of the correction directly, we expect that the finite-volume terms in NLO chiral perturbation theory will provide a somewhat reliable estimate now that we are so deep in the chiral regime; we therefore estimate the finite-volume correction for each simulated data point as the fractional difference between the 
ChPTFV fit value for that point with and without the finite-volume terms applied. The analytic fits presented in this section are all performed to the finite-volume corrected data. Note that despite the near-physical pion masses on our 32ID ensembles, the smallest $m_\pi L$ is roughly 3.3, which is in fact larger than the value of 3.1 obtained for the lightest pion on the 32I ensembles, due to the greater physical volume of the 32ID lattice. As a result we do not expect our new data involving lighter quark masses to further enhance the finite-volume errors.

In order to prevent accidental correlations between independent data from influencing the fit, while retaining the correlations between data measured on the same ensemble, we make use of the superjackknife technique to propagate the errors through our fits. A superjackknife distribution for a measurement is essentially a collection of independent jackknife distributions, each containing the fluctuations from a particular ensemble. As for the standard jackknife, any procedure, such as a fit or binary operation, is performed sequentially to each jackknife sample in all distributions. The total error on the superjackknife is obtained by evaluating the errors on each of its component jackknife distributions and adding these in quadrature. This technique was also used for our 2010 analysis.

As discussed in the previous section, our use of the strange-quark mass reweighting in the chiral/continuum fits differs from the 2010 strategy. Previously, each quantity was independently interpolated to the physical strange-quark mass prior to fitting; after the fit the values were updated to the new mass and the fit repeated, with this process iterated until convergence. We now constrain the heavy sea-quark dependence of each quantity to be the same on all ensemble sets and include multiple reweighted data points in the fit. As the number of reweighted masses differs between the ensemble sets, and considering that there are likely to be strong correlations between the reweighted data points, we might worry that the $\chi^2$ contributions of the data on the ensemble sets with more reweighted masses will be incorrectly enhanced in our uncorrelated fits. In order to avoid this we used only four reweighted strange-quark masses on each ensemble set, spread uniformly across the range.

Upon performing the fits, we discovered significant (up to $4\sigma$) tensions between the fits and the pion and kaon data on the 32ID ensembles at the upper end of the reweighted mass range. However, the upper limit of this mass range ($m_h=0.052$) is considerably larger than the physical strange quark mass of $\sim 0.047$, which is actually very close to the directly simulated mass of $m_h=0.045$. From the effective number of configurations calculated in section~\ref{sec:SimulationDetails}, we estimated that reweighting to the physical strange quark mass introduces a 10\%--15\% increase in our statistical errors. As we go further from the simulated point we expect the accuracy of the reweighting procedure to further decrease due to the reduced overlap of the reweighted path integral and the original. At $m_h=0.052$ we found that the effective number of configurations was reduced to only 15 on the lighter ensemble (down from 180) and 24 on the heavier (down from 148). This suggests that the measurements at 
the far end of the reweighting range are dominated by only a very small number of configurations and are therefore unreliable. As a result, the tension we observed between the fits and the data at the upper end of the reweighting range is likely to be an artifact of the reweighting procedure. With this in mind, we repeated the fits again using four reweighted masses this time spread only over the range beginning at the simulated strange quark mass and ending at the estimated physical strange quark mass. In doing so we found that the tension disappeared.

The inclusion of the 32ID ensembles greatly enhances the mass range over which our fits are performed. This should reduce the systematic error on the extrapolation to the physical light quark mass, and also allows us to consider removing some of the heavier ensembles from the Iwasaki data sets which may lie near the limits of convergence of NLO chiral perturbation theory. We removed the 24I $m_l=0.01$ ensemble and the 32I $m_l=0.008$ ensemble, as well as the partially-quenched data points on the lighter ensembles containing quarks with these masses. In performing this cut, we restrict our fits to pion masses smaller than \mpicut MeV, where previously the upper bound was $420$ MeV. This amounts to a $\sim 30\%$ reduction in the largest unitary light-quark mass. Note that this is not a straight cut on the partially-quenched pion mass as the elimination of these heavy ensembles also removes a number of partially-quenched pions containing these now ``heavy'' quarks ranging down to $\sim 230$ MeV. 

With the cut data set we were able to obtain excellent fits using the ChPT and ChPTFV ans\"{a}tze. For the analytic ansatz we again found excellent fits to the decay constants as well as $m_K$ and $m_\Omega$, but for the pion mass we found a number of outlying data points on the 32ID ensembles that deviated from the fit by up to $4\sigma$, with the typical size of the deviation being $\mathcal{O}(2\%)$. These deviations appear to occur due to nonlinearities in the light data. The fact that no corresponding deviations appear for the ChPT fits suggests that these nonlinearities are consistent with the NLO chiral logarithms. However, the discrepancies are also of the size expected for NLO terms in the Taylor expansion that are beyond the range of our power counting, hence we cannot draw any strong conclusions about their nature within our modest range of masses. As the linear ansatz must be locally correct around the physical point, we sought to reduce these discrepancies by further lowering the cut for these 
fits, first by eliminating the 32I $m_l=0.006$ ensemble, then by systematically removing the data corresponding to the heaviest partially-quenched pions. The limit to which this bound can be pushed in our analysis is dictated by the stability of the fits and the necessity to retain some data on the remaining 24I ensemble such that the $a^2$ coefficients of the decay constants can be determined; the latter implies that a 240 MeV bound is the lowest that we can currently reach. In practice we reached a similar level of agreement between the analytic fits and the data as found in the ChPT fits by lowering the bound to 260 MeV. Although this removes a large amount of data, we found that the fit remained very stable and that the effect of the cut on the values and precision of the fit parameters and predictions was surprisingly small; the typical change was of the order of a few percent, with the only large, statistically significant change being a 15\% increase in the valence light-quark dependence of $f_\pi$. 
With this in mind, we chose the 260 MeV cut for our analytic fits. The mass combinations of the data points remaining after performing this cut are listed in table~\ref{tab:260MeVcutmasscombs}.

As a side note, we also repeated the ChPT and ChPTFV fits to the full data set, for which the upper bound on the pion mass is 420 MeV. We found that, even over this large range, the NLO SU(2) ChPT fits were able to describe all of our data with only a few points on the 32ID ensembles deviating by between 2 and 3$\sigma$.

\begin{table}
\footnotesize{
\begin{tabular}{c|c|c|cc}
\hline\hline
Ensemble set & $m_l$ & $\{m_x\}$ & $\{m_y\}$\\
\hline
\multirow{3}{*}{32I} & 0.008    & - & -\\
                     & 0.006    & - & -\\
                     & 0.004    & 0.002 & 0.002, 0.004\\
\hline
\multirow{2}{*}{24I} & 0.01    & - & -\\
                     & 0.005   & 0.001 & 0.001\\
\hline
\multirow{2}{*}{32ID} & 0.0042    & 0.0001, 0.001, 0.0042 & 0.0001, 0.001, 0.0042, 0.008 & \multirow{2}{*}{$\Bigg{\}}$(excl. [0.0042,0.008])}\\
                      & 0.001     & 0.0001, 0.001, 0.0042 & 0.0001, 0.001, 0.0042, 0.008

\end{tabular}
}
\caption{Sea and valence quark masses of the data included in the analytic fit with a \mpicutanalytic MeV cut on the pion mass. The third and fourth columns give the set of partially-quenched valence quark masses; the mass combinations of light-light quantities ($m_\pi$ and $f_\pi$) are found by combining each choice of $m_x$ with each choice of $m_y$ from the appropriate columns, with the exception of the $[m_x,m_y]=[0.0042,0.008]$ points on the 32I ensemble set, for which the partially-quenched pion masses are above the cut. For heavy-light data ($m_K$, $f_K$) the light valence-quarks are chosen from the $\{m_x\}$ column, and the heavy valence-quarks from the full set of simulated heavy-quark values. For $m_\Omega$ and the Sommer scales, all data are included on those ensembles not marked with a dash (-).}
\label{tab:260MeVcutmasscombs}
\end{table}

\begin{figure}[t]
\centering
\includegraphics*[width=0.49\textwidth,clip=true,trim=5 0 5 5]{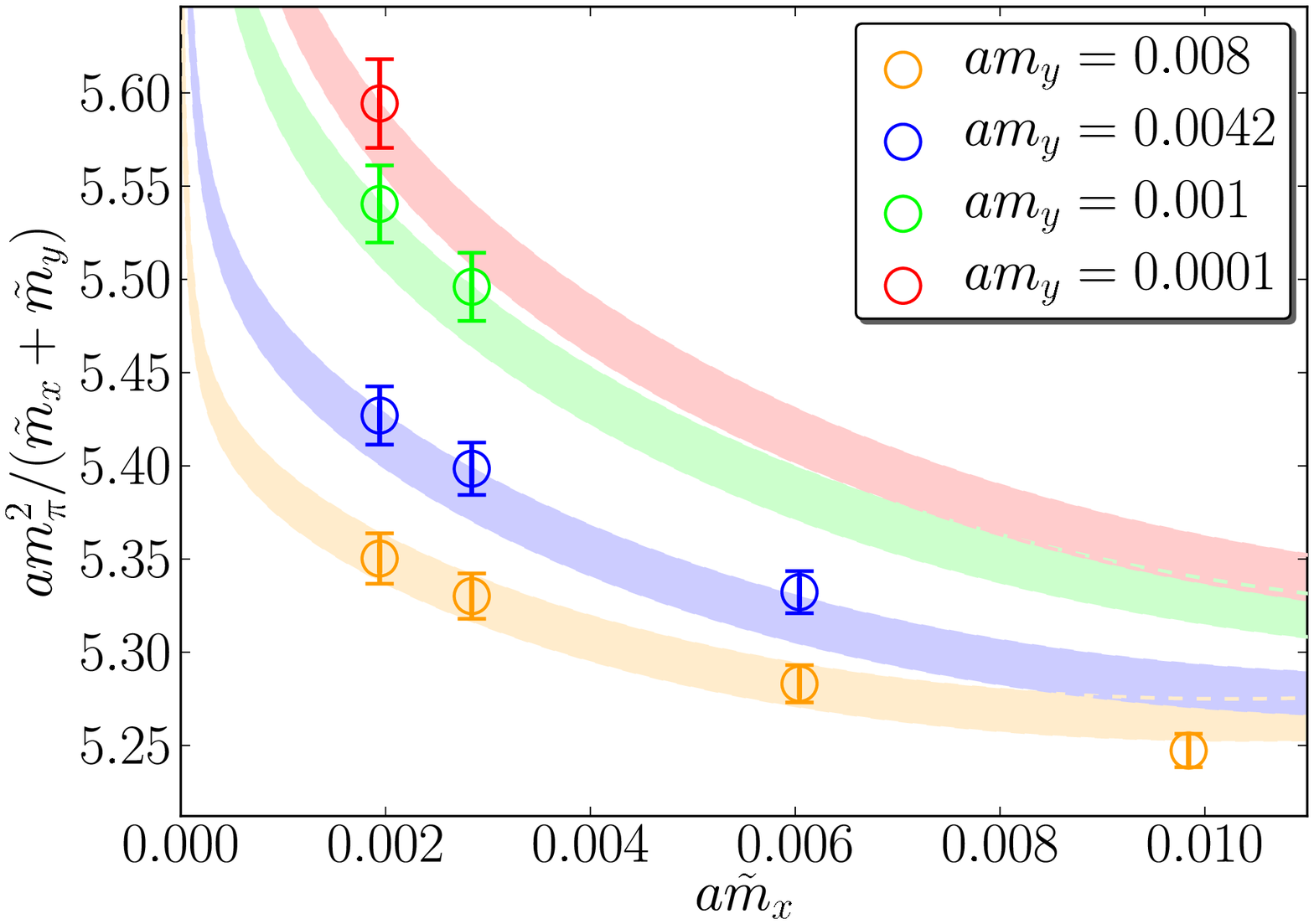}
\includegraphics*[width=0.49\textwidth,clip=true,trim=5 0 5 5]{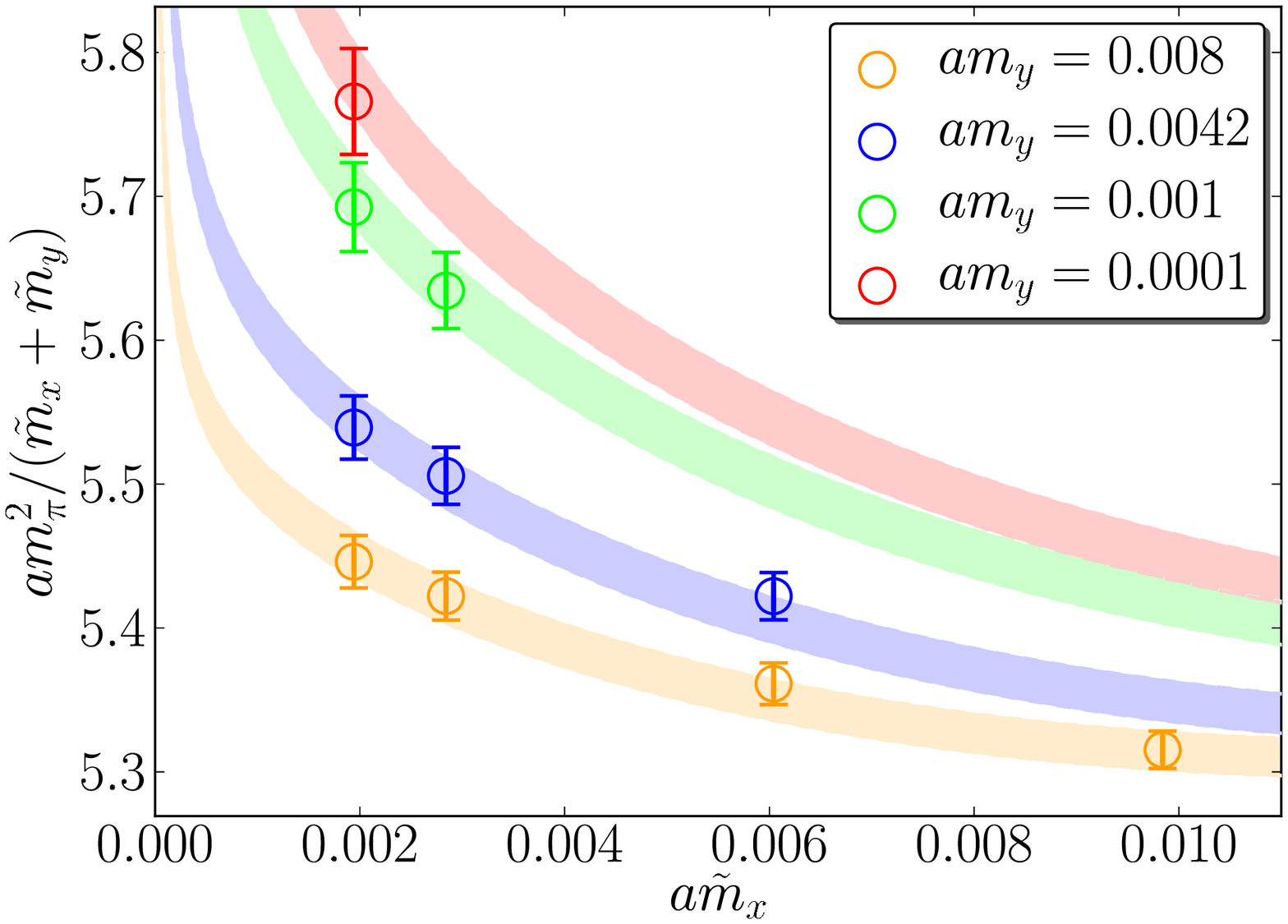}\\[0.1in]
\includegraphics*[width=0.49\textwidth,clip=true,trim=5 0 5 5]{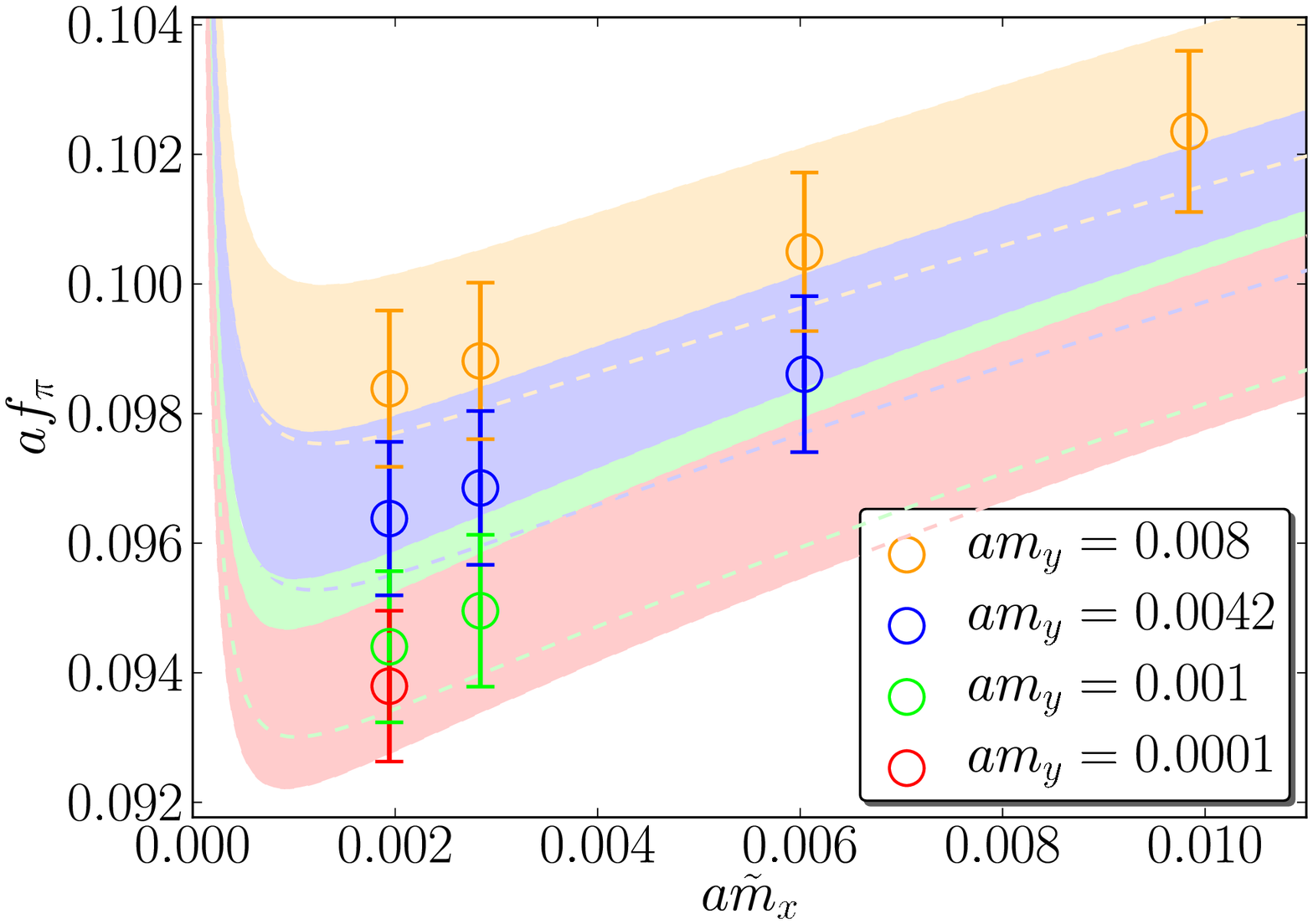}
\includegraphics*[width=0.49\textwidth,clip=true,trim=5 0 5 5]{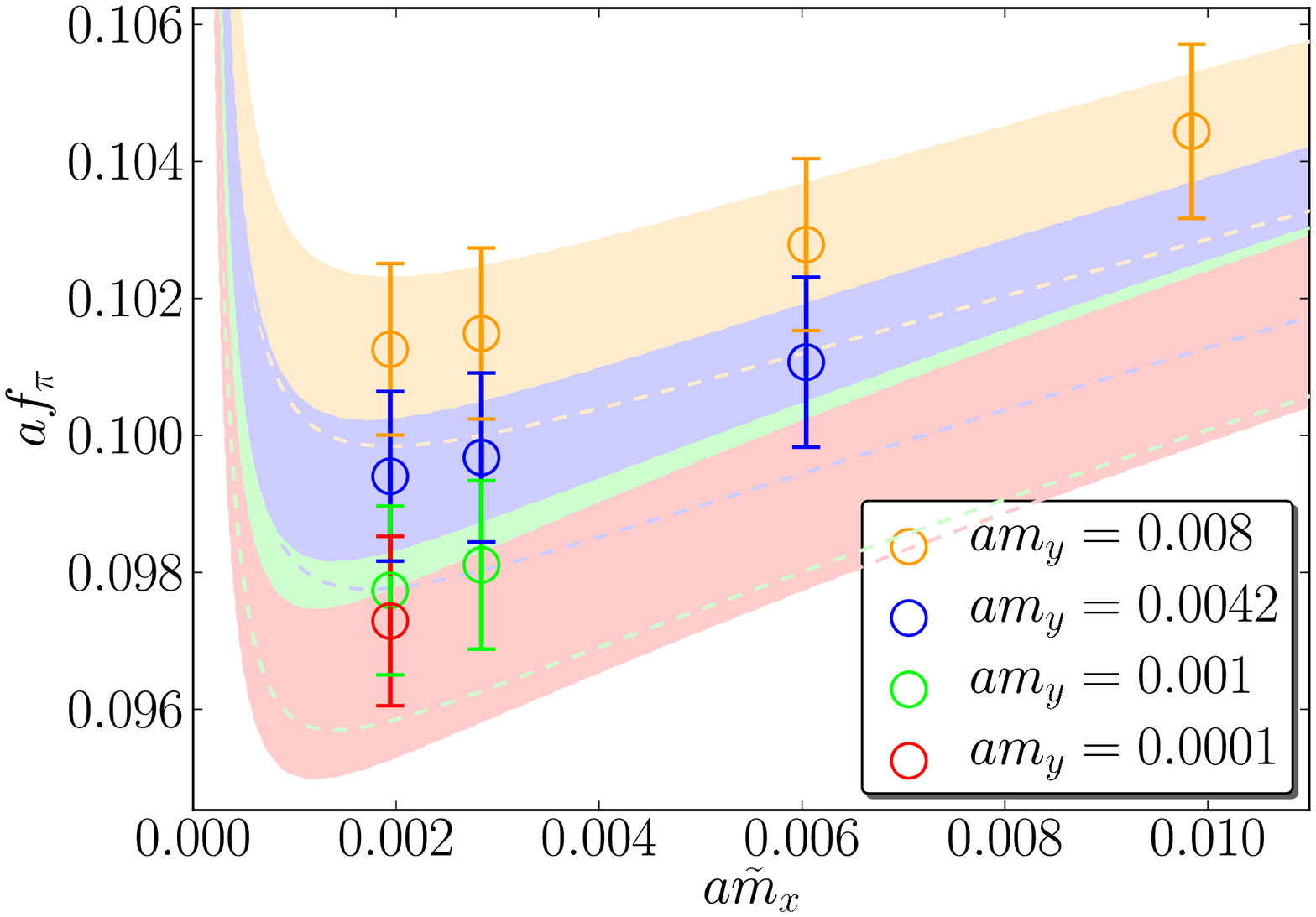}
\caption{
\label{fig:mpifpi:ChPTFV:cut}
Global fits obtained using NLO SU(2) chiral perturbation theory with finite-volume corrections for the pion mass (top) and $f_\pi$ (bottom) on the 32ID ensembles. Here the left-hand plot of each pair show the data at the simulated strange-quark mass and the corresponding fit curves on the $m_l=0.001$ ensemble, and the right-hand plots those on the $m_l=0.0042$ ensemble. The plots of the pion mass have $m_\pi^2/(\tilde m_x+\tilde m_y)$ on the ordinate axis, a quantity used traditionally to emphasize the chiral curvature of the data.}
\end{figure}

\begin{figure}[t]
\centering
\includegraphics*[width=0.49\textwidth,clip=true,trim=5 0 5 5]{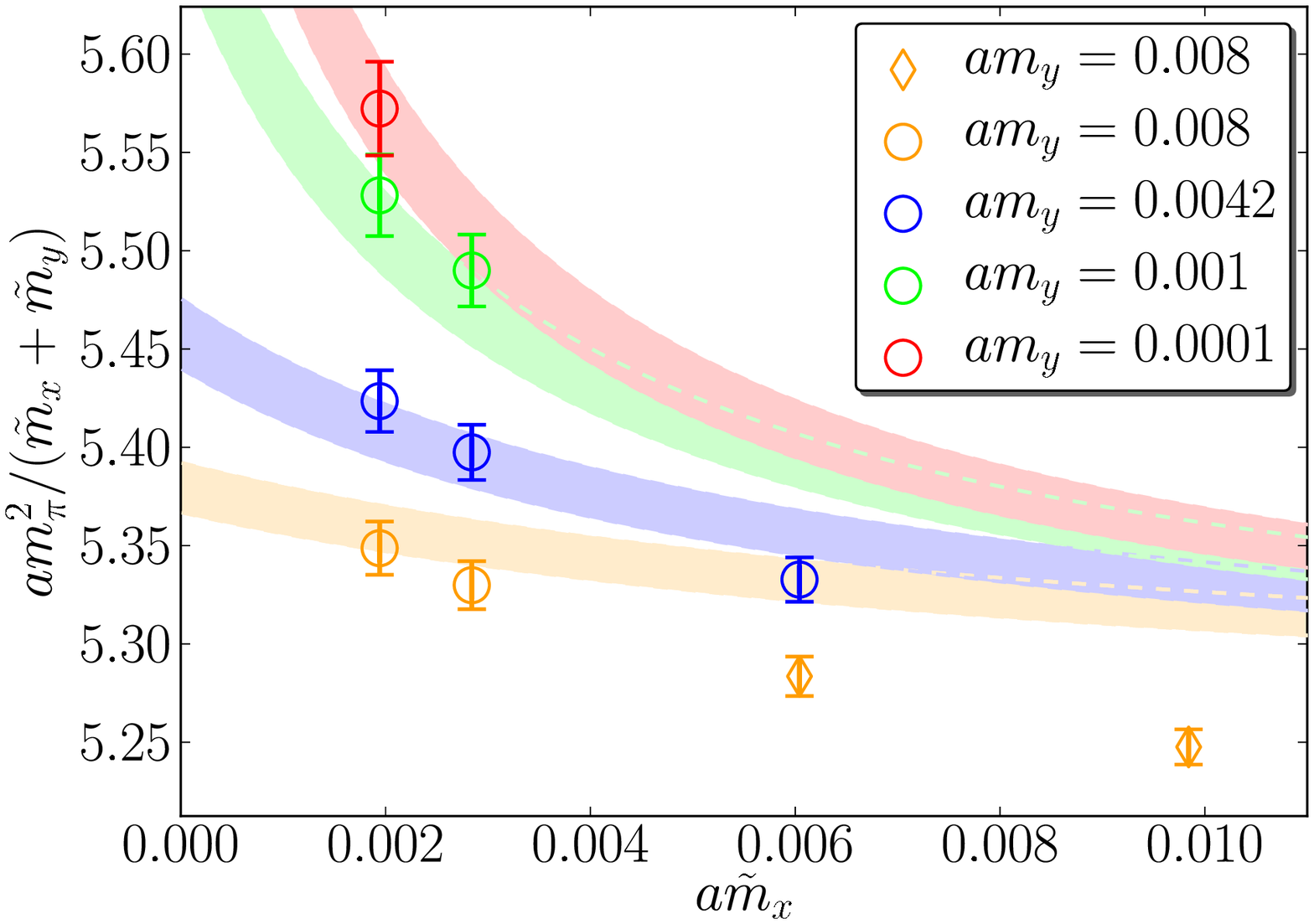}
\includegraphics*[width=0.49\textwidth,clip=true,trim=5 0 5 5]{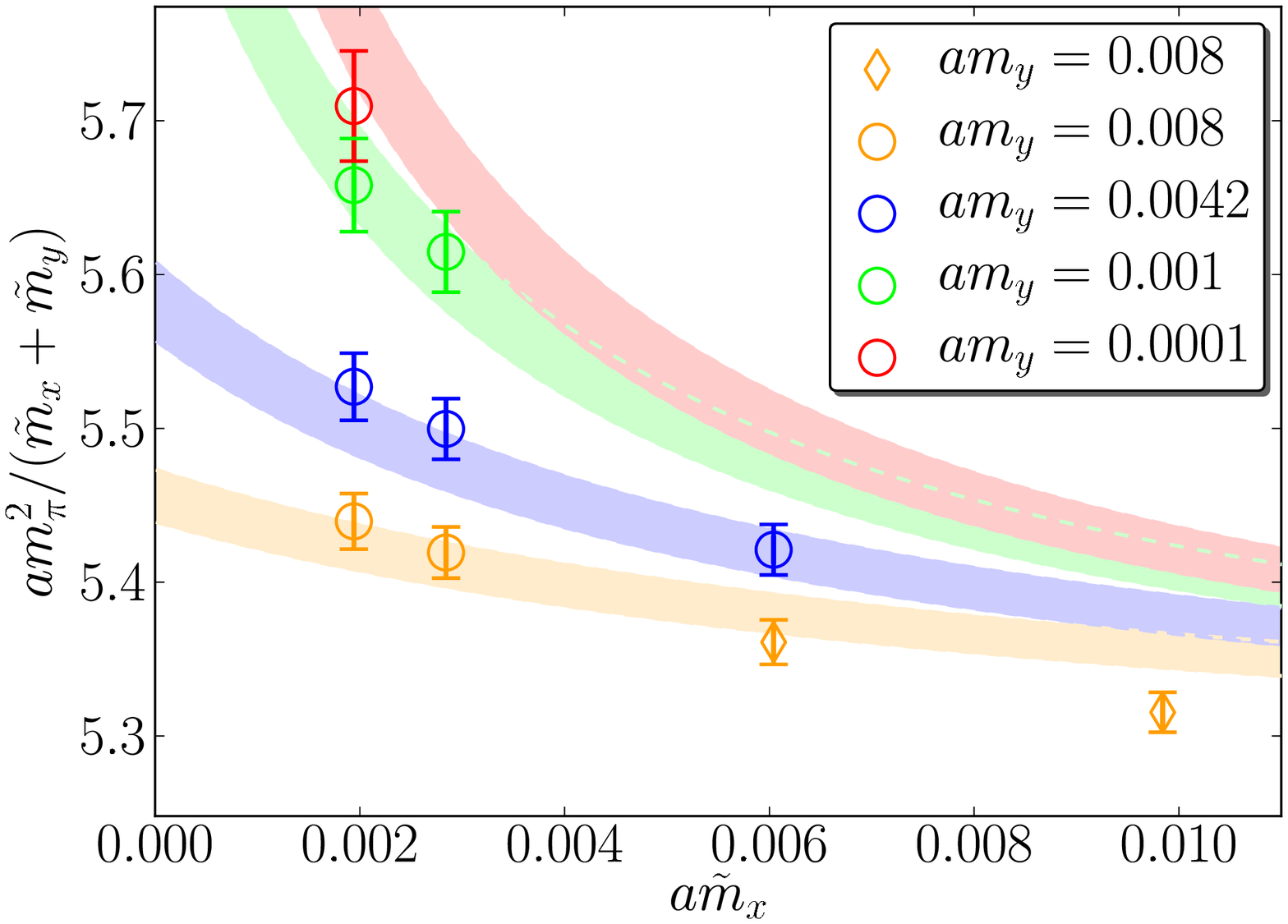}\\[0.1in]
\includegraphics*[width=0.49\textwidth,clip=true,trim=5 0 5 5]{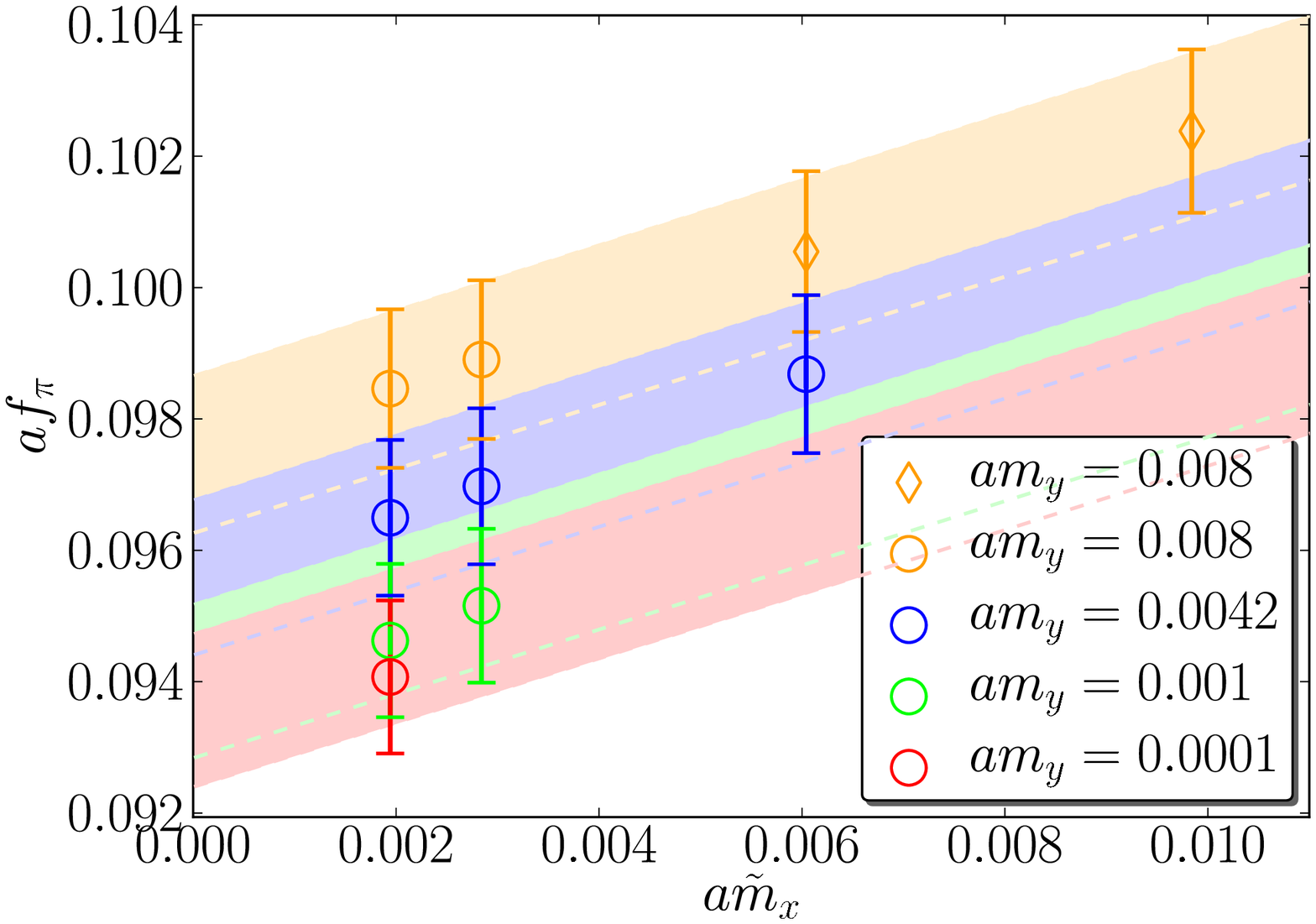}
\includegraphics*[width=0.49\textwidth,clip=true,trim=5 0 5 5]{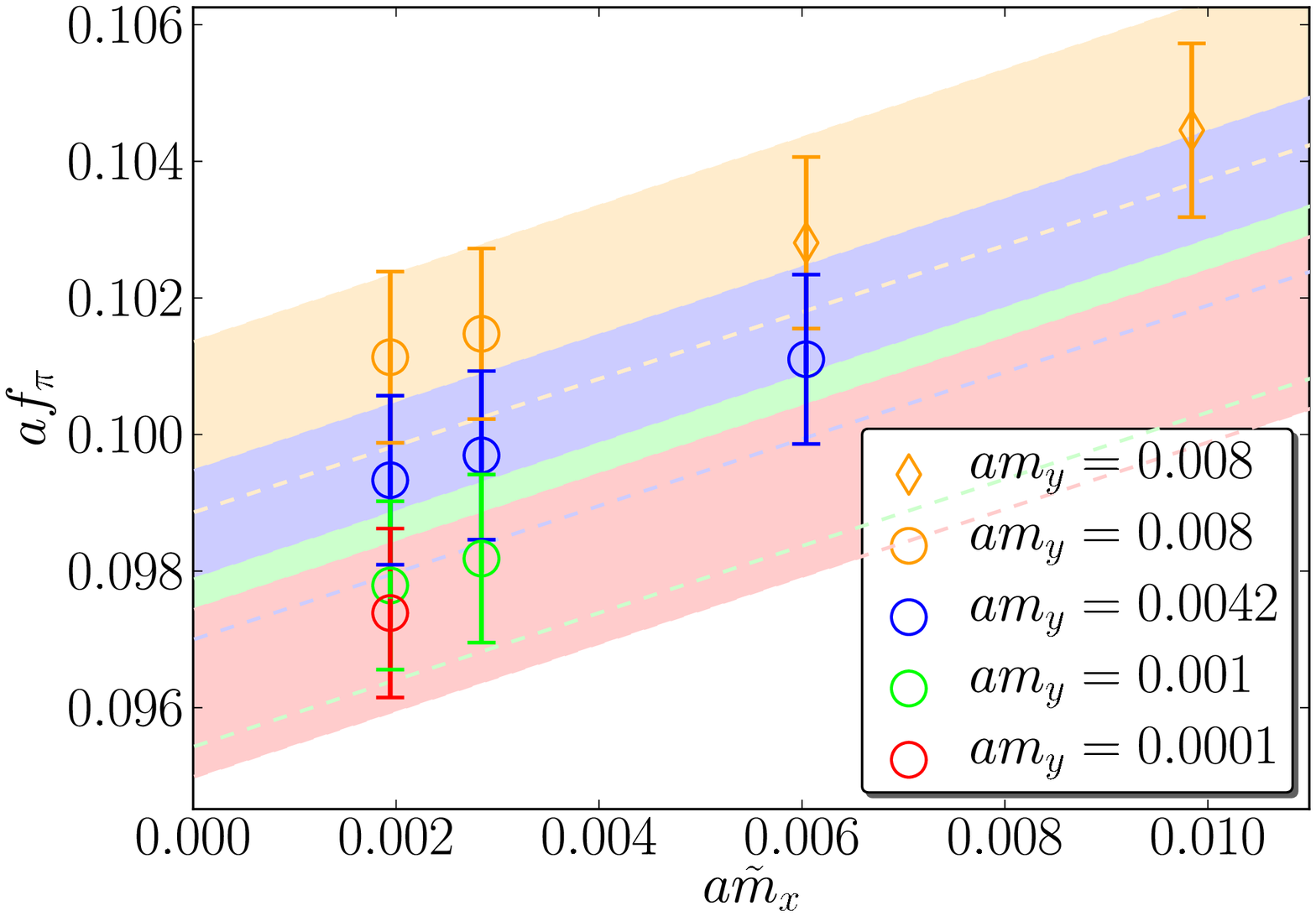}
\caption{
\label{fig:mpifpi:analytic:cut}
Global fits using the analytic ansatz with finite-volume corrected data for the pion mass (top) and $f_\pi$ (bottom) on the 32ID ensembles. Here the left-hand plot of each pair show the data at the simulated strange-quark mass and the corresponding fit curves on the $m_l=0.001$ ensemble, and the right-hand plots those on the $m_l=0.0042$ ensemble. The plots of the pion mass have $m_\pi^2/(\tilde m_x+\tilde m_y)$ on the ordinate axis, a quantity used traditionally to emphasize the chiral curvature of the data. The circular points are those included in the fit, and the diamond points those excluded by the cut on data with $m_\pi \geq \mpicutanalytic$ MeV.}
\end{figure}

\begin{figure}[t]
\centering
\includegraphics*[width=0.49\textwidth,width=0.49\textwidth,clip=true,trim=5 0 5 5]{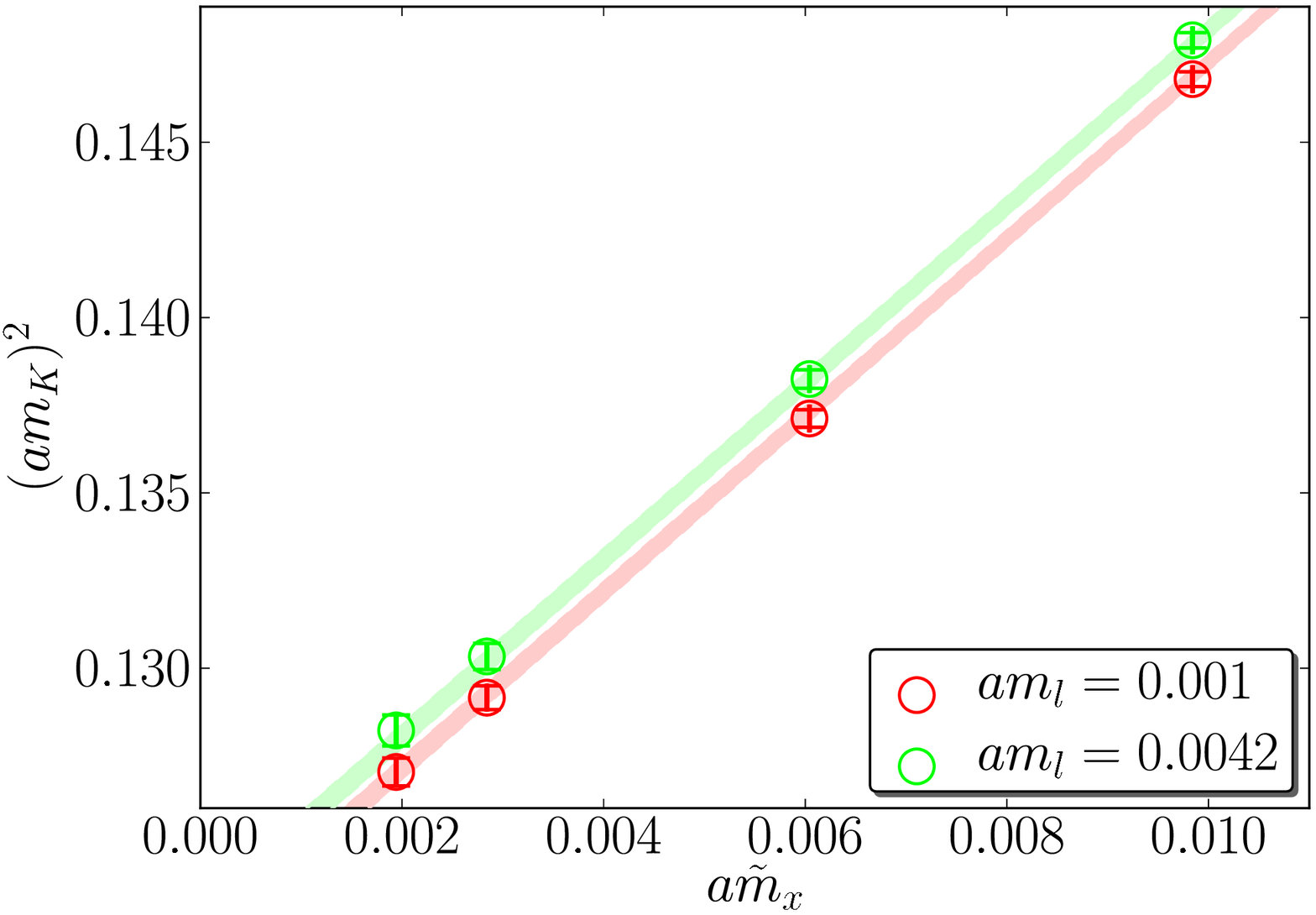}
\includegraphics*[width=0.49\textwidth,width=0.49\textwidth,clip=true,trim=5 0 5 5]{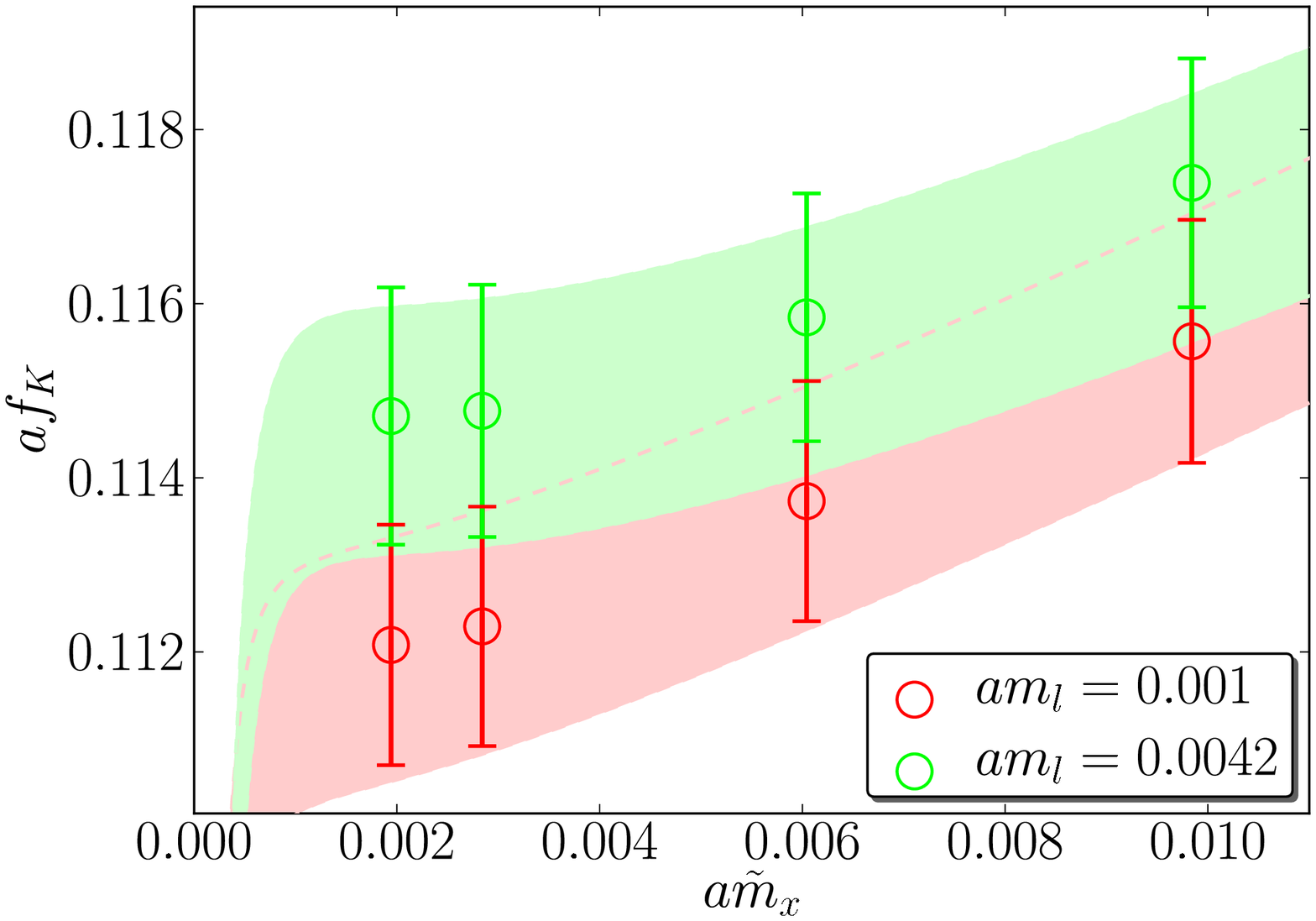}
\caption{
\label{fig:mkfk:ChPTFV:cut}
Global fits obtained using NLO SU(2) chiral perturbation theory with finite-volume corrections for the square of the kaon mass (left) and $f_K$ (right) on the 32ID ensembles.}
\end{figure}

\begin{figure}[t]
\centering
\includegraphics*[width=0.49\textwidth]{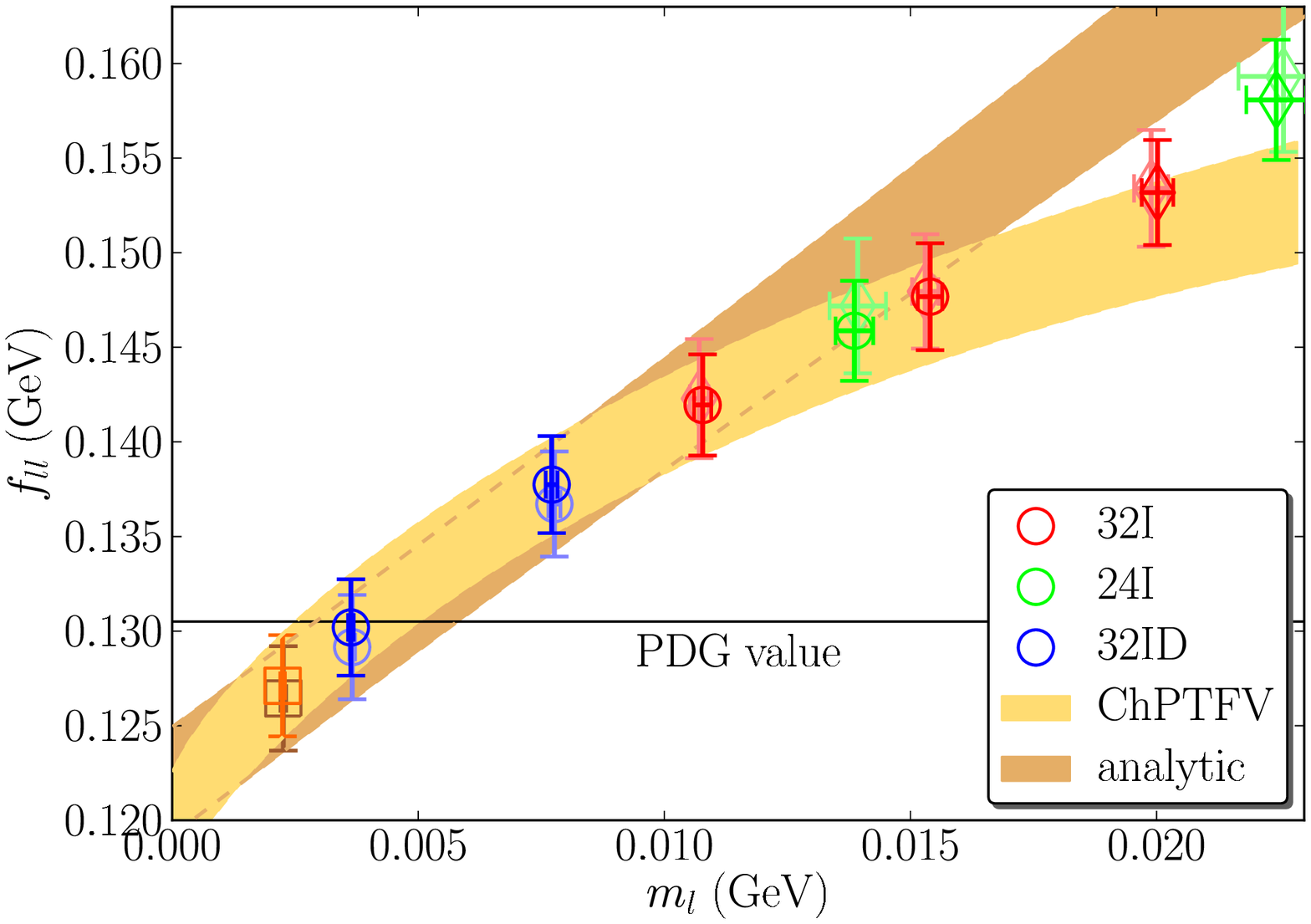}
\includegraphics*[width=0.49\textwidth]{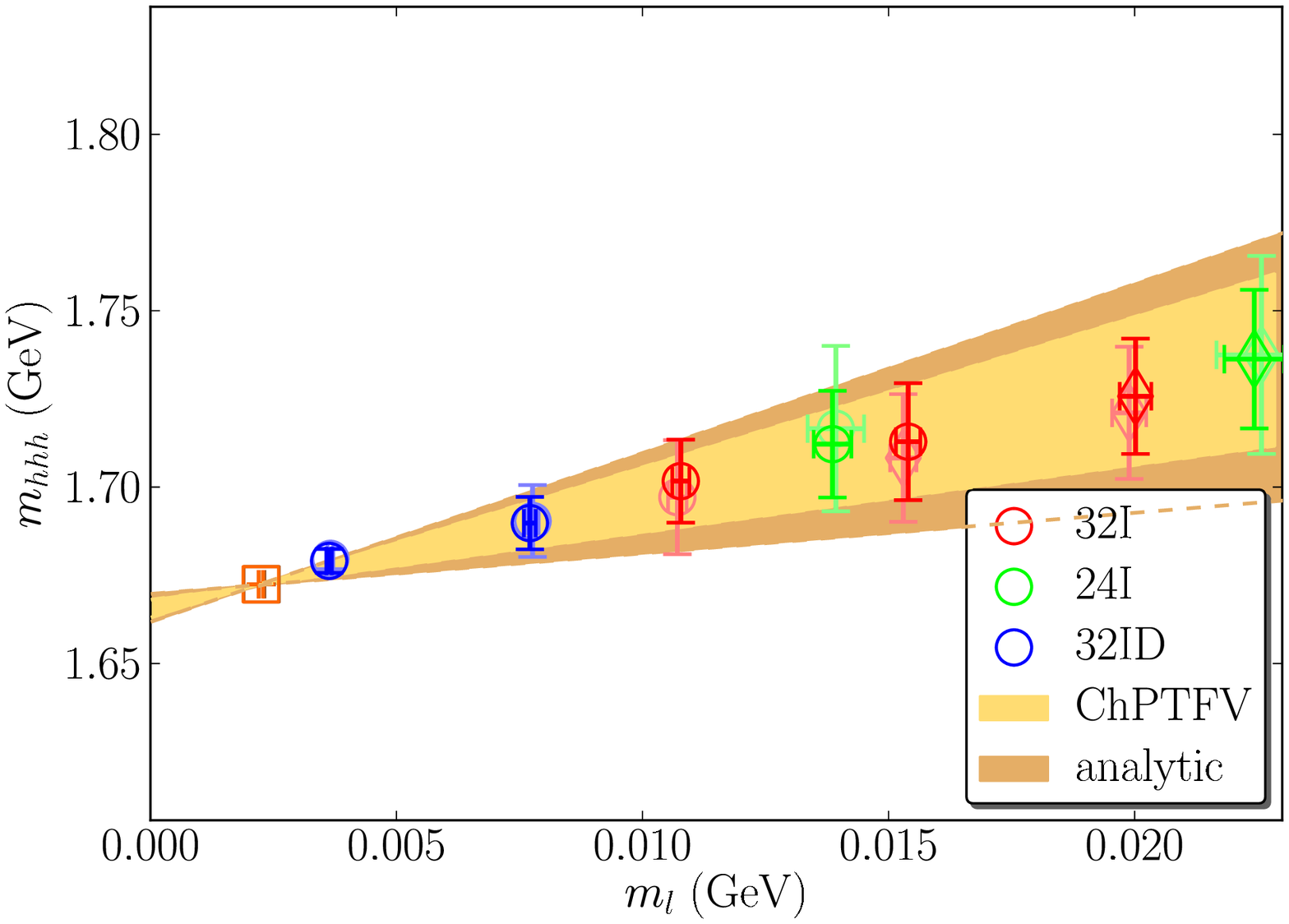}
\caption{
\label{fig:fpi:momega:contanalyticNLOcomparison}
The chiral extrapolation of the pion decay constant (left) and Omega baryon mass (right) using the analytic and ChPTFV ans\"{a}tze. Overlaying these curves we have plotted the unitary data extrapolated to the continuum limit using the $a^2$ dependence of our fit forms. The lighter-shaded points were corrected using the analytic fit form, and the darker points by the ChPTFV form. Here the circular points are those included in the fit, and the diamond points are those excluded by the cuts at \mpicut MeV (ChPTFV) and \mpicutanalytic MeV (analytic). The upper and lower square points show the continuum predictions obtained using the ChPTFV and analytic ans\"{a}tze respectively. Note that for $f_\pi$, the analytic fit does not include any unitary data points on the 32I and 24I ensembles as they lie above the pion mass cut (cf. table~\protect\ref{tab:260MeVcutmasscombs}). Note also that the physical limit of the $\Omega^-$ mass shows no statistical errors and agrees precisely with its physical value because it is 
this quantity that we use to determine the lattice scale.}
\end{figure}

In table~\ref{tab-cutchisqmassesainv} we list the results for the inverse lattice spacings and quark masses obtained using each fit ansatz, alongside the associated uncorrelated $\chi^2$/dof. The results are completely consistent, which suggests that the extrapolation to the physical quark masses is under control. A similar degree of consistency can be seen between the fit parameters (where applicable) given in table~\ref{tab-cutlecs}. Here, as mentioned in the previous section, we have adjusted the chiral scale $\Lambda_\chi$ of the ChPT LECs to the conventional 1 GeV. In figures~\ref{fig:mpifpi:ChPTFV:cut} and ~\ref{fig:mpifpi:analytic:cut} we overlay our simulated data for $m_\pi$ and $f_\pi$ on the 32ID ensembles with the ChPTFV and analytic fit curves respectively, and in figure~\ref{fig:mkfk:ChPTFV:cut} we present similar plots for $m_K$ and $f_K$ overlaid with the ChPTFV fit curves. We list the individual predictions for $f_\pi$, $f_K$ and their ratio at the simulated lattice spacings and the 
continuum limit in table~\ref{tab-fpifkpredindi}. In figure~\ref{fig:fpi:momega:contanalyticNLOcomparison} we plot the chiral extrapolations of $f_\pi$ and $m_\Omega$ overlaying the data corrected to the continuum limit. (Note that the Omega baryon mass data requires no correction due our choice of scaling trajectory.)

\begin{figure}[p]
\includegraphics*[width=0.49\textwidth,clip=true,trim=5 0 5 5]{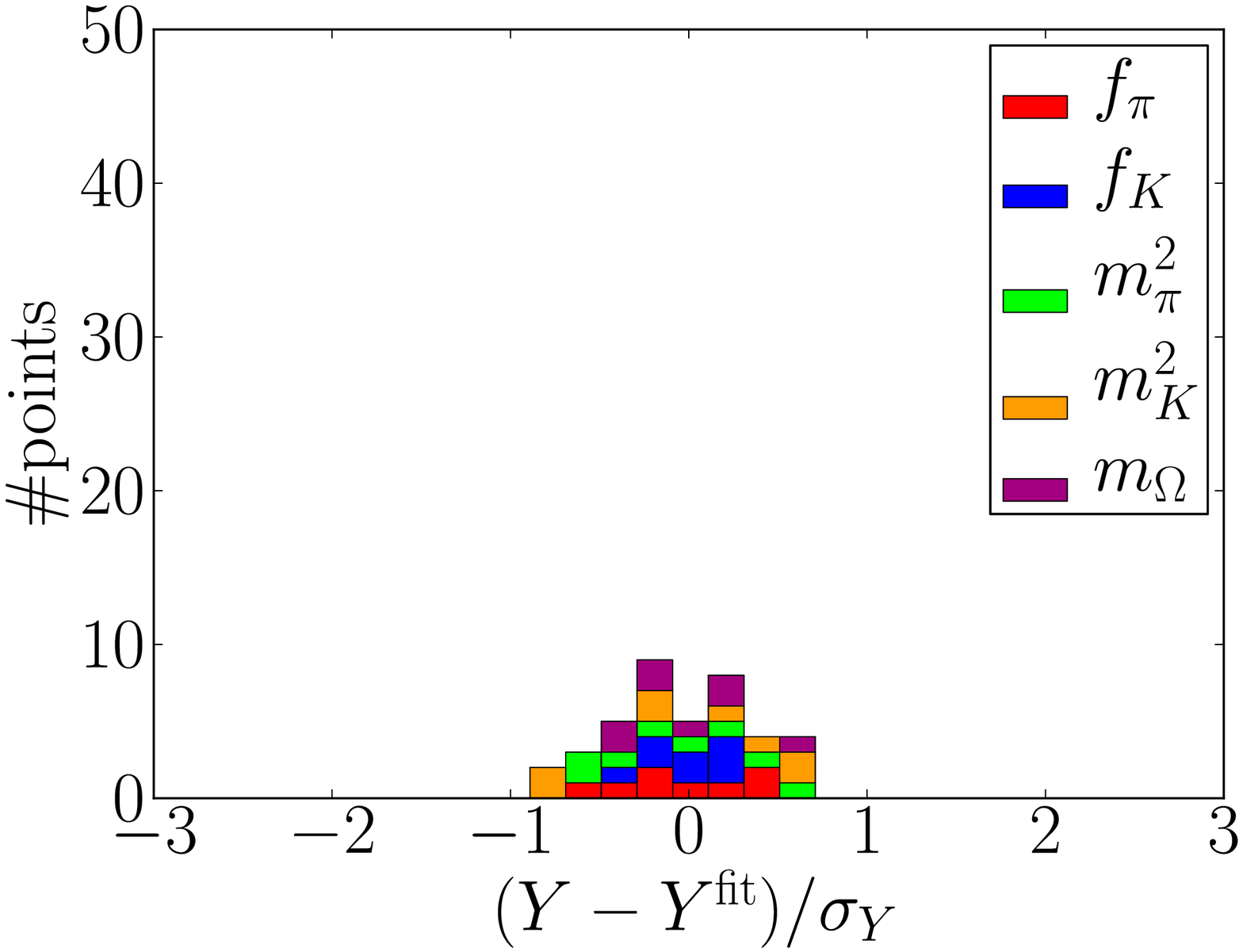}
\includegraphics*[width=0.49\textwidth,clip=true,trim=5 0 5 5]{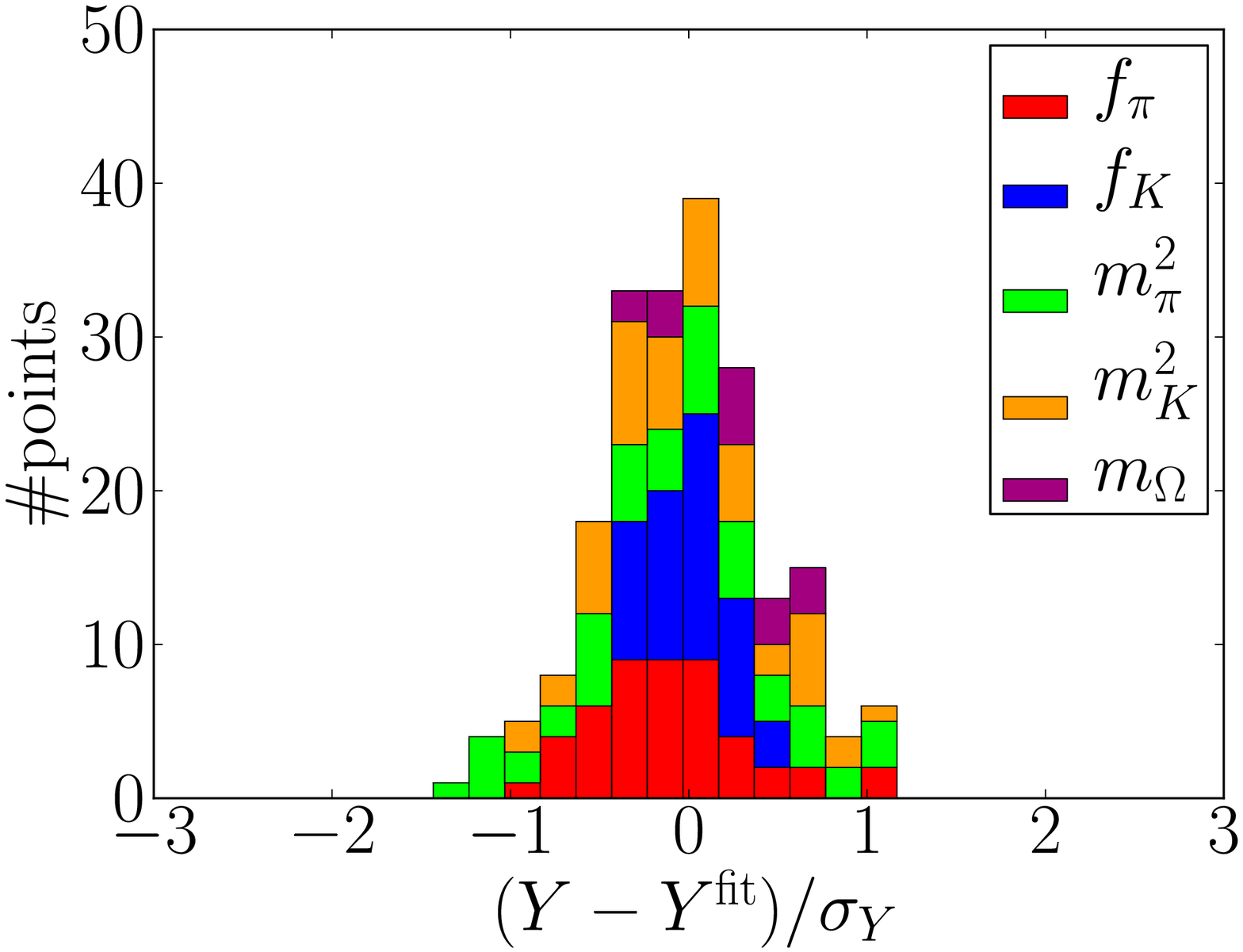}\\
\includegraphics*[width=0.49\textwidth,clip=true,trim=5 0 5 5]{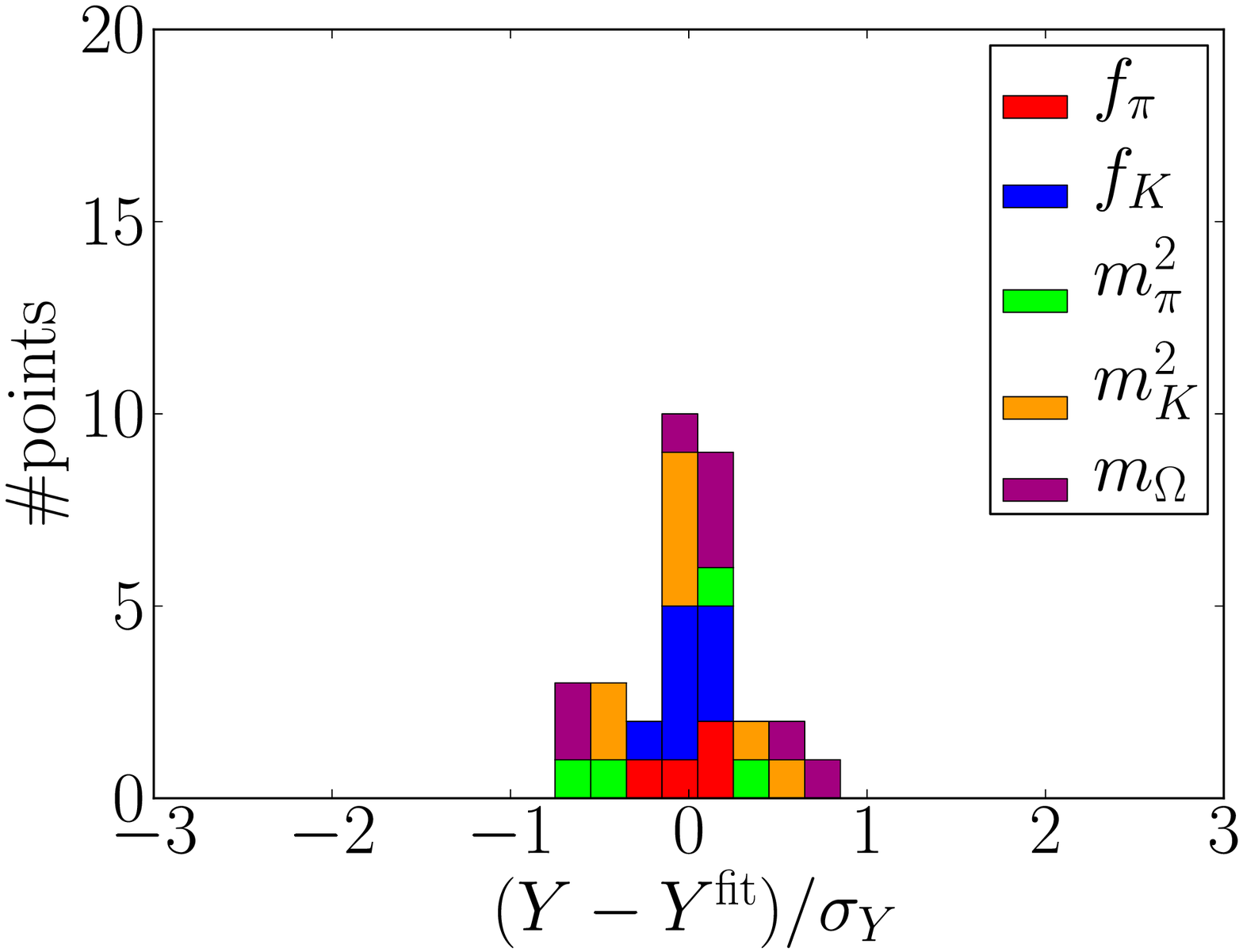}
\includegraphics*[width=0.49\textwidth,clip=true,trim=5 0 5 5]{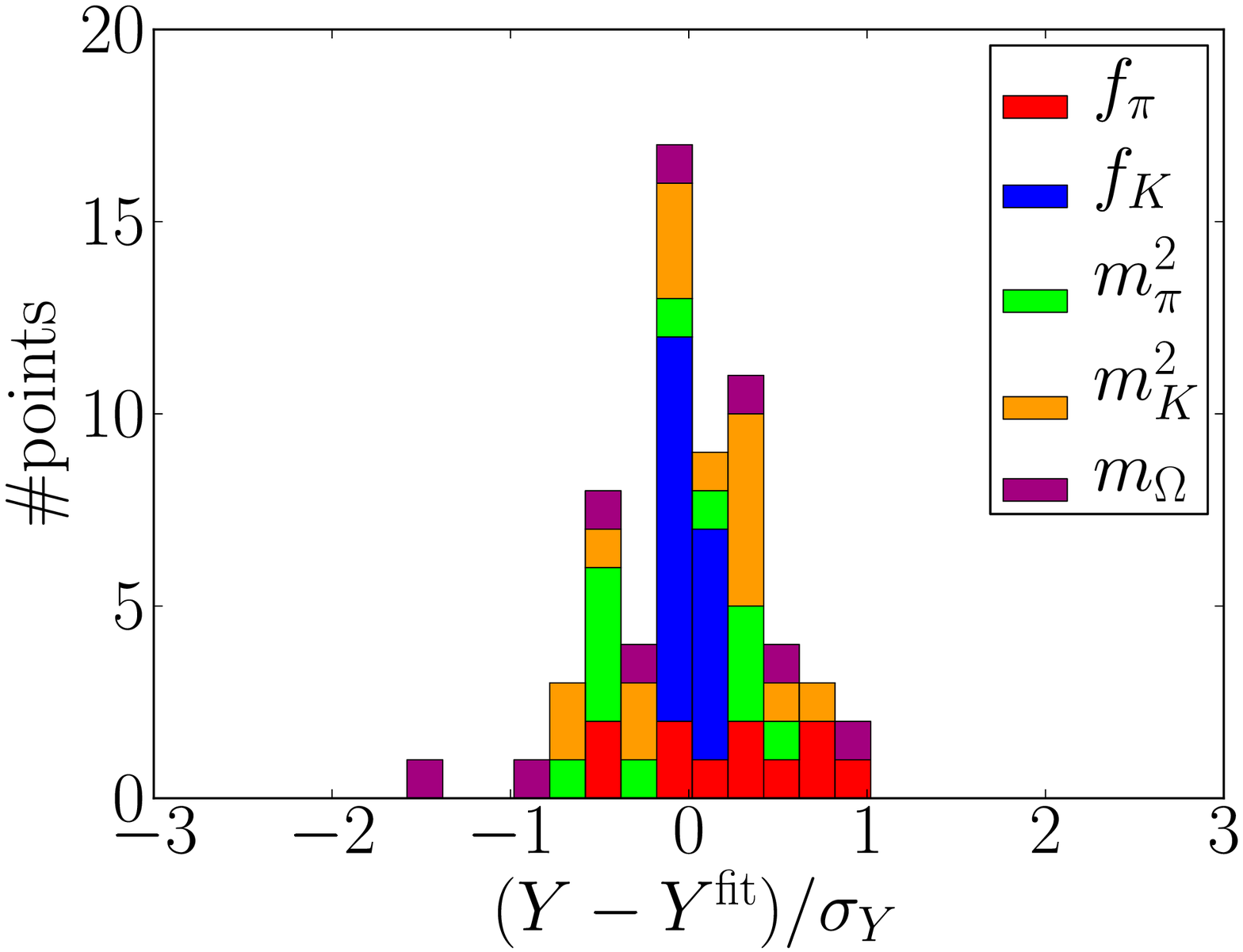}\\
\includegraphics*[width=0.49\textwidth,clip=true,trim=5 0 5 5]{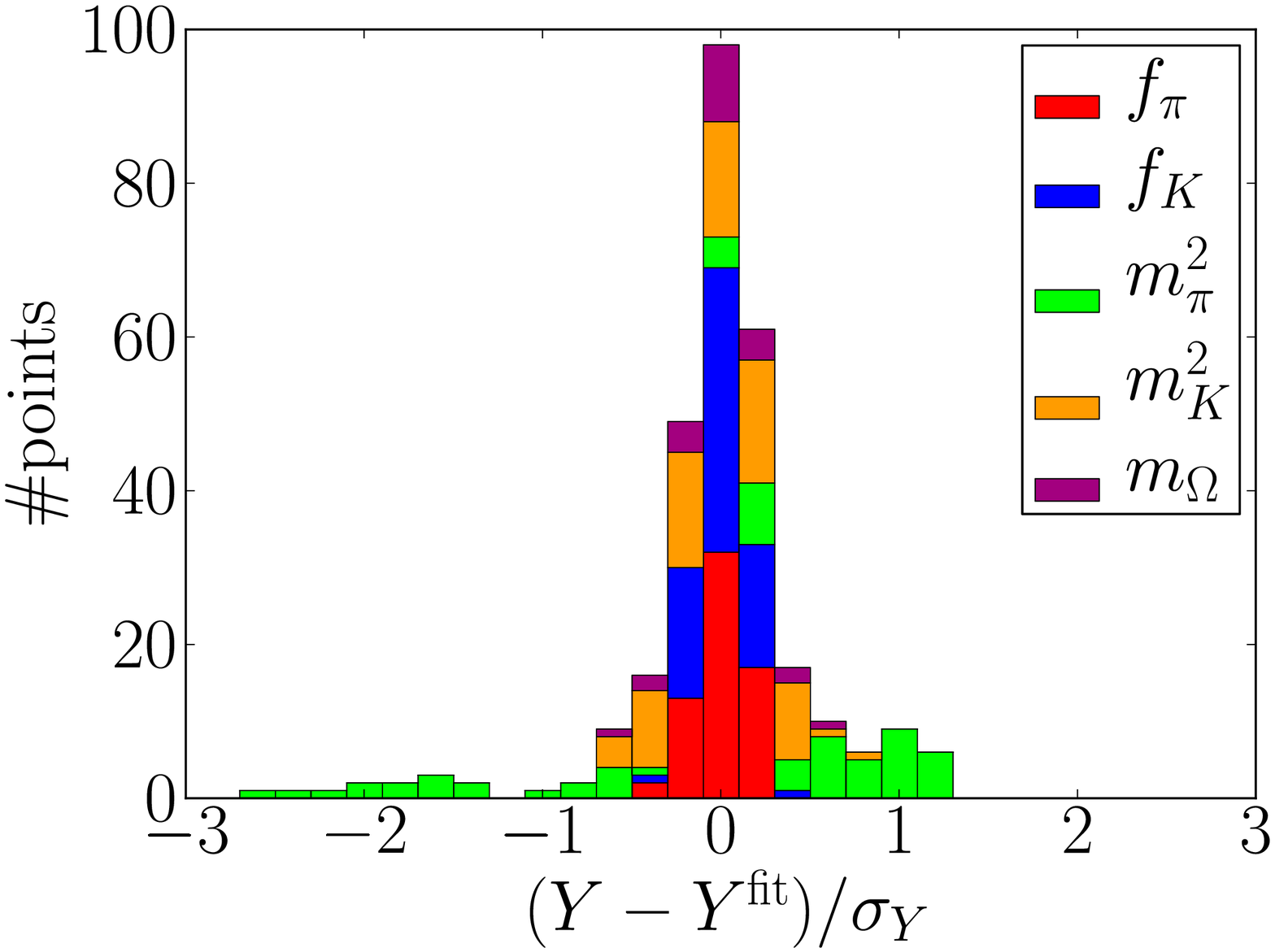}
\includegraphics*[width=0.49\textwidth,clip=true,trim=5 0 5 5]{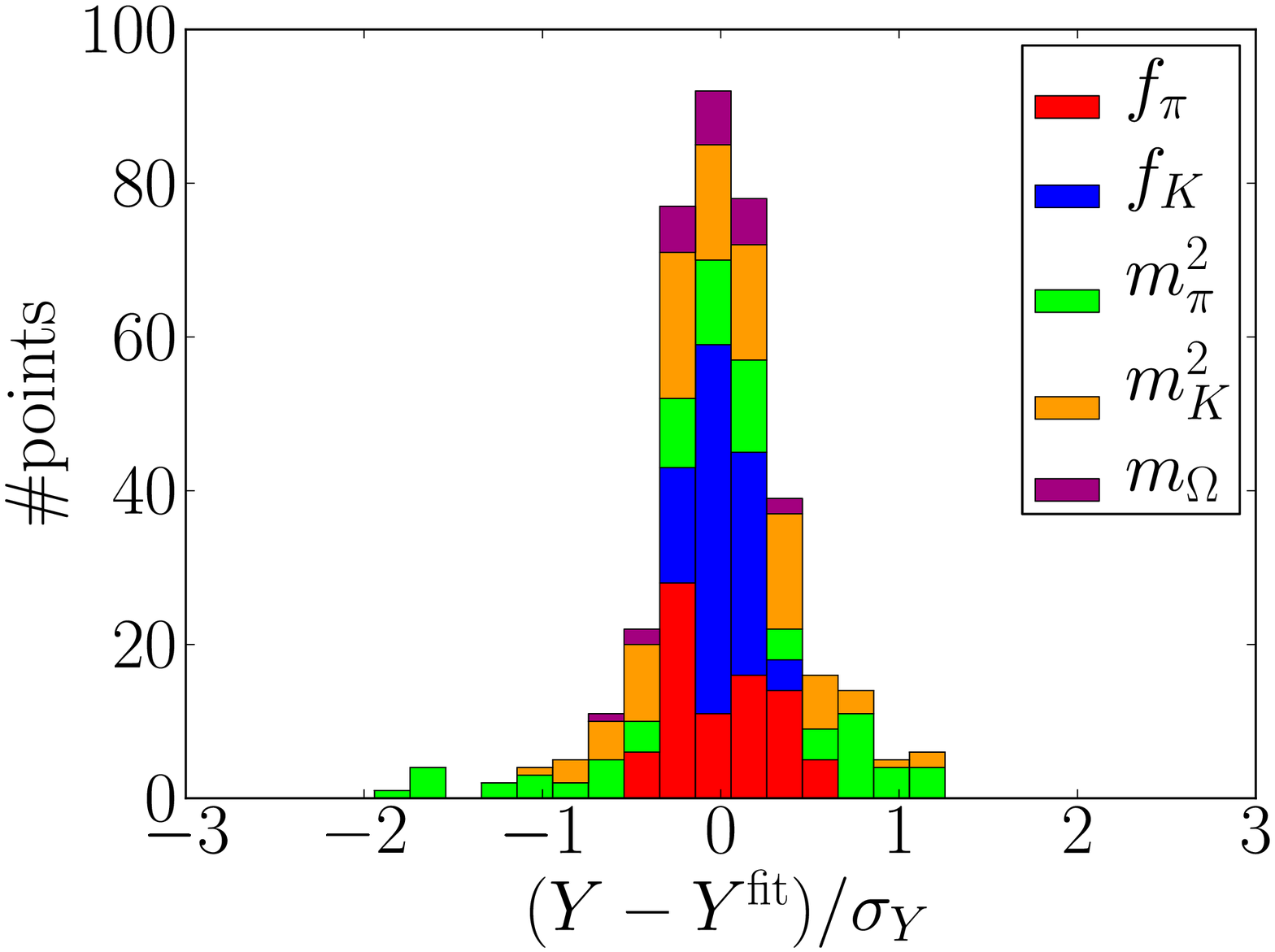}
\vspace{-0.5cm}
\caption{
\label{fig:globalhist_cut}
Histograms of the deviation of the fit from the data for each quantity on each of the three ensemble sets (32I top, 24I middle and 32ID bottom) with the analytic (left) and ChPTFV (right) ans\"{a}tze.}
\end{figure}

The uncorrelated $\chi^2$/dof are all less than unity, suggesting that the fits are behaving. In order to demonstrate the quality of the fits in greater detail, we present histograms of the deviation of the fit from the data in units of the statistical error in figure~\ref{fig:globalhist_cut}.

In the remainder of this section, we discuss how we combine the results of our fits into predictions for $f_\pi$ and $f_K$ and final results for the lattice spacings and physical quark masses (in the matching scheme).

\newpage

\begin{table}[tb]
\centering
\begin{tabular}{c|lll}
\hline
 & Analytic & ChPT & ChPTFV\\
\hline
 \rule{0cm}{0.4cm}$\chi^2/$dof(32IW) & $ 0.279(64) $  & $ 0.191(55) $  & $ 0.221(57) $  \\
\hline
 $am_l$(32I) & $ 0.000320(42) $  & $ 0.000307(34) $  & $ 0.000308(35) $  \\
 $am_s$(32I) & $ 0.02660(98) $  & $ 0.02650(85) $  & $ 0.02627(89) $  \\
 $a^{-1}$(32I) & $ 2.295(40) $ GeV & $ 2.302(35) $ GeV & $ 2.310(37) $ GeV \\
\hline
 $am_l$(24I) & $ -0.001754(83) $  & $ -0.001757(75) $  & $ -0.001749(78) $  \\
 $am_s$(24I) & $ 0.0337(18) $  & $ 0.0338(13) $  & $ 0.0336(13) $  \\
 $a^{-1}$(24I) & $ 1.743(43) $ GeV & $ 1.743(30) $ GeV & $ 1.747(31) $ GeV \\
\hline
 $am_l$(32ID) & $ -0.000090(34) $  & $ -0.000096(21) $  & $ -0.000090(22) $  \\
 $am_s$(32ID) & $ 0.04667(76) $  & $ 0.04674(60) $  & $ 0.04671(61) $  \\
 $a^{-1}$(32ID) & $ 1.372(10) $ GeV & $ 1.371(8) $ GeV & $ 1.371(8) $ GeV \\
\end{tabular}
\caption{The $\chi^2$/dof, unrenormalized physical quark masses in bare lattice units (without $\mres$ included) and the values of the inverse lattice spacing $a^{-1}$ obtained by fitting to data with $m_\pi \leq \mpicut$ for the ChPT and ChPTFV fits, and $m_\pi \leq \mpicutanalytic$ MeV for the analytic fit.}
\label{tab-cutchisqmassesainv}
\end{table}
\FloatBarrier
\setlength{\LTcapwidth}{\textwidth}

\begin{longtable}[p]{l|rr||l|r}
\hline
Parameter & ChPT & ChPTFV & Parameter & Analytic\\
\hline
\endhead
 $Z_l^{\scriptscriptstyle I}$ & $ 0.983(14) $ & $ 0.981(14) $ &  & $ 0.992(21) $\\
 $Z_l^{\scriptscriptstyle ID}$ & $ 0.929(15) $ & $ 0.930(15) $ &  & $ 0.936(16) $\\
 $Z_h^{\scriptscriptstyle I}$ & $ 0.9730(94) $ & $ 0.9719(95) $ &  & $ 0.976(14) $\\
 $Z_h^{\scriptscriptstyle ID}$ & $ 0.939(13) $ & $ 0.935(13) $ &  & $ 0.940(14) $\\
 $R_a^{\scriptscriptstyle I}$ & $ 0.7571(65) $ & $ 0.7562(66) $ &  & $ 0.7595(90) $\\
 $R_a^{\scriptscriptstyle ID}$ & $ 0.5955(72) $ & $ 0.5934(76) $ &  & $ 0.5976(74) $\\
\hline
 $B$ & $ 4.174(83) $ & $ 4.148(86) $ & $C^{m_\pi}_0$ & $ 0.00043(23) $\\
 $L_8^{(2)}$ & $ 0.000616(22) $ & $ 0.000610(23) $ & $C^{m_\pi}_1$ & $ 7.70(16) $\\
 $L_6^{(2)}$ & $ -0.000131(69) $ & $ -0.000159(72) $ & $C^{m_\pi}_2$ & $ 0.173(40) $\\
 $c_{m_\pi,m_h}$ & $ -2.1(2.5) $ & $ -2.5(2.5) $ & $C^{m_\pi}_3$ & $ -0.041(26) $\\
 $f$ & $ 0.1167(31) $ & $ 0.1196(31) $ & $C^{f_\pi}_0$ & $ 0.1221(30) $\\
 $c_f^{\scriptscriptstyle I}$ & $ -0.021(70) $ & $ -0.031(68) $ & $C^{f_\pi,\,\scriptscriptstyle I}_a$ & $ -0.064(88) $\\
 $c_f^{\scriptscriptstyle ID}$ & $ 0.040(45) $ & $ 0.014(43) $ & $C^{f_\pi,\,\scriptscriptstyle ID}_a$ & $ 0.030(47) $\\
 $L_5^{(2)}$ & $ 0.000560(51) $ & $ 0.000524(51) $ & $C^{f_\pi}_1$ & $ 1.054(32) $\\
 $L_4^{(2)}$ & $ -0.00014(13) $ & $ -0.00020(14) $ & $C^{f_\pi}_2$ & $ 0.88(16) $\\
 $c_{f_\pi,m_h}$ & $ 0.422(90) $ & $ 0.484(89) $ & $C^{f_\pi}_3$ & $ 0.120(49) $\\
 $m^{(K)}$ & $ 0.2365(77) $ & $ 0.2364(80) $ & $C^{m_K}_0$ & $ 0.2364(88) $\\
 $\lambda_2$ & $ 0.01907(77) $ & $ 0.02028(76) $ & $C^{m_K}_1$ & $ 3.637(99) $\\
 $\lambda_1$ & $ 0.00220(75) $ & $ 0.00233(80) $ & $C^{m_K}_2$ & $ 0.47(20) $\\
 $c_{m_K,m_y}$ & $ 3.811(61) $ & $ 3.828(64) $ & $C^{m_K}_3$ & $ 3.802(71) $\\
 $c_{m_K,m_h}$ & $ 0.033(43) $ & $ 0.031(43) $ & $C^{m_K}_4$ & $ 0.001(64) $\\
 $f^{(K)}$ & $ 0.1466(36) $ & $ 0.1484(37) $ & $C^{f_K}_0$ & $ 0.1500(35) $\\
 $c_{f^{(K)}}^{\scriptscriptstyle I}$ & $ -0.034(57) $ & $ -0.040(57) $ & $C^{f_K,\,\scriptscriptstyle I}_a$ & $ -0.075(69) $\\
 $c_{f^{(K)}}^{\scriptscriptstyle ID}$ & $ 0.020(38) $ & $ 0.008(38) $ & $C^{f_K,\,\scriptscriptstyle ID}_a$ & $ 0.013(38) $\\
 $\lambda_4$ & $ 0.00622(22) $ & $ 0.00601(22) $ & $C^{f_K}_1$ & $ 0.349(44) $\\
 $\lambda_3$ & $ -0.0034(19) $ & $ -0.0032(20) $ & $C^{f_K}_2$ & $ 0.76(19) $\\
 $c_{f_K,m_y}$ & $ 0.2917(42) $ & $ 0.2923(42) $ & $C^{f_K}_3$ & $ 0.2967(64) $\\
 $c_{f_K,m_h}$ & $ 0.118(40) $ & $ 0.118(40) $ & $C^{f_K}_4$ & $ 0.144(57) $\\
 $m^{(\Omega)}$ & $ 1.6659(100) $ & $ 1.666(11) $ & $C^{m_\Omega}_0$ & $ 1.6657(99) $\\
 $c_{m_\Omega,m_l}$ & $ 2.9(1.2) $ & $ 3.1(1.2) $ & $C^{m_\Omega}_1$ & $ 3.0(1.8) $\\
 $c_{m_\Omega,m_v}$ & $ 5.439(58) $ & $ 5.462(63) $ & $C^{m_\Omega}_2$ & $ 5.441(65) $\\
 $c_{m_\Omega,m_h}$ & $ 0.74(29) $ & $ 0.87(31) $ & $C^{m_\Omega}_3$ & $ 0.35(39) $\\

\caption{The fit parameters of each of our chiral ansatz\"{e} obtained by fitting to data with $m_\pi < \mpicut$ MeV for the ChPT and ChPTFV fits, and $m_\pi \leq \mpicutanalytic$ MeV for the analytic fit. The parameters are given in GeV$^n$ for the appropriate power of $n$, and with the heavy quark mass expansion point adjusted to the physical strange quark mass. We have ordered the table such that the equivalent parameters of the ChPT and analytic fits lie on the same line. The coefficients of the chiral logarithms have also been adjusted so that they are defined at the conventional chiral scale $\Lambda_\chi=1$ GeV.}

\label{tab-cutlecs}
\end{longtable}
\FloatBarrier

\begin{table}[tp]
\centering
\begin{tabular}{l|lll|}
\hline
 & Analytic & ChPT & ChPTFV\\
\hline
 \rule{0cm}{0.4cm}$f_\pi^{\scriptscriptstyle 32IW}$ & $ 0.1249(21) $ & $ 0.1242(22) $ & $ 0.1264(22) $ \\
 $f_\pi^{\scriptscriptstyle 24IW}$ & $ 0.1238(28) $ & $ 0.1238(25) $ & $ 0.1258(26) $ \\
 $f_\pi^{\scriptscriptstyle 32ID}$ & $ 0.1284(18) $ & $ 0.1273(18) $ & $ 0.1280(18) $ \\
 $f_\pi^{\mathrm{\tiny continuum}}$ & $ 0.1264(28) $ & $ 0.1247(27) $ & $ 0.1271(27) $ \\
\end{tabular}\quad
\begin{tabular}{l|lll|}
\hline
 & Analytic & ChPT & ChPTFV\\
\hline
 \rule{0cm}{0.4cm}$f_K^{\scriptscriptstyle 32IW}$ & $ 0.1502(23) $ & $ 0.1499(23) $ & $ 0.1512(24) $ \\
 $f_K^{\scriptscriptstyle 24IW}$ & $ 0.1485(30) $ & $ 0.1491(26) $ & $ 0.1503(27) $ \\
 $f_K^{\scriptscriptstyle 32ID}$ & $ 0.1536(21) $ & $ 0.1526(21) $ & $ 0.1531(21) $ \\
 $f_K^{\mathrm{\tiny continuum}}$ & $ 0.1525(28) $ & $ 0.1509(29) $ & $ 0.1524(30) $ \\
\end{tabular}\\
\begin{tabular}{l|lll|}
\hline
 & Analytic & ChPT & ChPTFV\\
\hline
 \rule{0cm}{0.4cm}$(f_K/f_\pi)^{\scriptscriptstyle 32IW}$ & $ 1.202(12) $ & $ 1.207(9) $ & $ 1.197(9) $ \\
 $(f_K/f_\pi)^{\scriptscriptstyle 24IW}$ & $ 1.199(18) $ & $ 1.205(11) $ & $ 1.195(11) $ \\
 $(f_K/f_\pi)^{\scriptscriptstyle 32ID}$ & $ 1.196(4) $ & $ 1.199(4) $ & $ 1.196(4) $ \\
 $(f_K/f_\pi)^{\mathrm{\tiny continuum}}$ & $ 1.206(14) $ & $ 1.211(12) $ & $ 1.199(12) $ \\
\end{tabular}

\caption{Predictions for $f_\pi$ (top left) and $f_K$ (top right) in GeV as well as their ratio (bottom) for each global fit ansatz at each simulated lattice spacing and in the continuum limit obtained by fitting to data with $m_\pi \leq \mpicut$ for the ChPT and ChPTFV fits, and $m_\pi \leq \mpicutanalytic$ MeV for the analytic fit.}\label{tab-fpifkpredindi}
\end{table}

\subsection{Combining results and estimating systematic errors}
\label{subsec:syserrs}

Prior to discussing our method of estimating the systematic error contributions arising from the chiral extrapolation and finite-volume effects, it may be appropriate to detail two of the contributions that we neglect in our final predictions: those arising from the explicit chiral symmetry breaking due to simulating with finite $L_s$, and those from the truncation of the combined Symanzik-chiral expansion that we discussed in the previous section. We have addressed the explicit chiral symmetry breaking at leading order by additively renormalizing the quark masses in our fit forms with $m_{\rm res}$. However, up to operators of dimension-6, the chiral symmetry breaking also introduces a dimension-5 clover term that potentially introduces ${\cal O}(a\Lambda_\mathrm{QCD})$ discretization errors; we discuss this issue in Appendix~\ref{appendix-cloverterm} and conclude that this can be neglected in our calculations. We may therefore treat our domain wall simulations as ${\cal O}(a)$-improved, which allows us to 
also neglect higher-order terms involving the lattice spacing raised to an odd power, e.g. ${\cal O}(a^3\Lambda_\mathrm{QCD}^3 )$ terms. Regarding the truncation of the Symanzik-chiral expansion, we stated in the previous section that we ignore terms ${\cal O}(m_q a^2 \Lambda_\mathrm{QCD})\sim{\cal O}(m_\pi^2 a^2)$ and higher. These include terms of magnitude ${\cal O}(a^4\Lambda_\mathrm{QCD}^4 )$, ${\cal O}(m_\pi^4\Lambda_\mathrm{QCD}^{-4})$, ${\cal O}(m_{\rm res}m_\pi^2\Lambda_\mathrm{QCD}^{-3})$, ${\cal O}(m_{\rm res}a^2\Lambda_\mathrm{QCD})$, etc. These are expected to be on the scale of a fraction of a percent or less, considerably smaller than the percent-scale chiral and finite-volume errors in our calculation. For example we find that our ${\cal O}(a^2\Lambda_\mathrm{QCD}^2)$ terms are typically $\leq 3\%$, from which we can estimate the ${\cal O}(a^4\Lambda_\mathrm{QCD}^4)$ error as $(0.03)^2\sim 0.1\%$. There are also effects arising from higher-order terms in the Symanzik expansion that are 
typically ignored in lattice calculations: The coefficients of $a^2$ in the expansion are themselves dependent on the lattice spacing through loop corrections, giving rise to terms like $\alpha_s(a)\ln(a\Lambda_{\rm QCD})a^2$. We discuss these in detail in Appendix~\ref{appendix-alphascorrections} and conclude that they can be expected to be of a similar magnitude to the ${\cal O}(a^4\Lambda_\mathrm{QCD}^4)$ errors for our range of lattice spacings. We now proceed to the discussion of the finite-volume and chiral extrapolation errors.

The method of combining the results obtained using our three chiral ans\"{a}tze into a final prediction was discussed at length in ref.~\cite{Aoki:2010dy}. The main issues were first deciding which result or combination of results to use for the central value and second deciding how to estimate the systematic errors arising from finite-volume corrections and the extrapolation to the physical quark masses. The discussion was focussed on the predicted decay constants as they are known to high precision. We observed that our predicted value for $f_\pi$ from the ChPTFV fit was $7(2)\%$ too low, and $4(2)\%$ in the analytic case, where the quoted errors are obtained from the statistical error on the result. Smaller discrepancies were also found in the kaon decay constants. We concluded that these are of the size expected for NNLO terms in the chiral expansion, as obtained by squaring the difference between our data and $f$ -- the leading order term in the ChPT chiral expansion. Noting that both the analytic 
fits and ChPTFV fits appeared to describe our data equally well, we decided to average the two results and take their full difference as our estimate for the chiral extrapolation systematic. We estimated the size of the finite-volume systematic error from the full difference of the ChPTFV and ChPT results.

Now that we have data ranging down almost to the physical point, we are able to revisit the issue of estimating the systematic errors. We first note that the differences between the ChPTFV and analytic results for $f_\pi$ and $f_K$ are now very small, smaller in fact than the formerly sub-dominant finite-volume contributions estimated from the difference between the ChPT and ChPTFV results. By comparing the above results with those obtained by fitting to all available data we observed that this reduction is mainly due to our removal of the data corresponding to heavier pion masses from the fits. 

As discussed in the previous section, we performed our analytic fits to finite-volume corrected data in anticipation of the increased importance of these effects on our results. Here we investigate how large an effect the finite-volume corrections have on the analytic fits by repeating the latter with uncorrected data. The resulting fit parameters and predictions are compared to the original fits in table~\ref{tab-anafvcorrcomparison}. In the table we also provide the superjackknife ratios of the fit results with and without finite-volume corrections. We notice that in taking the ratio, many of the correlated fluctuations cancel, exposing underlying changes that were formerly masked by the statistical error. We observe that in many cases the deviation of the ratio from unity is statistically significant but is only $\mathcal{O}(2\%)$ or less; these changes are of the order expected for higher-order (mass-squared or $a^2 m$) effects that are beyond the range of our power counting, hence we cannot draw any 
conclusions from these results. The only quantities that change significantly are the slopes of $f_\pi$, $f_K$ and $m_\pi$ with respect to the light-quark masses; this behavior is expected as the finite-volume corrections will be larger in the light quark-mass regime, in which the physical length scales are greater. We observe a $1.7$ MeV upwards shift in the continuum prediction for $f_\pi$, which is consistent with the $2.4$ MeV difference between the ChPT and ChPTFV results.

Although we now correct for the finite-volume using NLO chiral perturbation theory, we note that resummation techniques~\cite{Colangelo:2005gd} may lead to somewhat larger estimates of the finite-volume effects. As we lack the ability to repeat our calculations on a larger volume, we choose to continue to include a conservative finite-volume systematic error in our final results, obtained, as before, from the full difference of the ChPTFV and ChPT results.

In the previous section we demonstrated that the ChPTFV fit forms describe our data reliably over a considerably larger range of pion masses than the linear ansatz. For the final predictions given in the following sections we therefore take the ChPTFV results for our central values and use the analytic ansatz only to estimate the chiral systematic. However, we continue to find it surprising that a linear ansatz appears capable of describing QCD at the 1\% level from the \mpicutanalytic MeV pion-mass regime down to the physical point, and at the 2\% level if that range is extended to 350 MeV.

In some cases we observed that the superjackknife errors on the differences between results obtained using the three parametrizations were larger than the differences between the central values. In these cases we chose to be conservative and took the statistical error on the difference for our estimate of the systematic error.

\clearpage
\begin{longtable}{l|ll|ll}
\hline
Quantity & Original data & FV corrected data & Ratio $R$ & $|R-1|/\sigma$\\
\hline
\endhead
$\chi^2$/dof & 0.219(54) & 0.274(65) & 1.254(33) & 7.672\\
$m_l$ & 0.002230(58) & 0.002259(58) & 1.01293(53) & 24.379\\
$m_h$ & 0.0627(12) & 0.0626(12) & 0.99857(31) & 4.603\\
$Z_l(24I)$ & 0.996(22) & 0.992(21) & 0.99619(41) & 9.377\\
$Z_l(32ID)$ & 0.927(16) & 0.936(16) & 1.00932(42) & 22.131\\
$Z_h(24I)$ & 0.975(14) & 0.976(14) & 1.00073(20) & 3.580\\
$Z_h(32ID)$ & 0.942(14) & 0.940(14) & 0.99876(29) & 4.366\\
$R_a(24I)$ & 0.7595(91) & 0.7595(90) & 1.00004(17) & 0.218\\
$R_a(32ID)$ & 0.5977(75) & 0.5976(74) & 0.99980(22) & 0.911\\
$a^{-1}(32I)$ & 2.295(40) & 2.295(40) & 1.00025(24) & 1.032\\
$a^{-1}(24I)$ & 1.743(43) & 1.743(43) & 1.00029(41) & 0.710\\
$a^{-1}(32ID)$ & 1.372(10) & 1.372(10) & 1.000048(21) & 2.282\\
$f_\pi$ & 0.1247(27) & 0.1264(28) & 1.01359(72) & 18.879\\
$f_K$ & 0.1515(28) & 0.1525(28) & 1.00627(31) & 19.916\\
$f_K/f_\pi$ & 1.213(12) & 1.202(12) & 0.99143(36) & 24.076\\
$C^{m_\pi}_0$ & -0.00011(16) & -0.00014(16) & 1.28(51) & 0.563\\
$C^{m_\pi}_1$ & 3.378(30) & 3.355(30) & 0.99334(28) & 24.172\\
$C^{m_\pi}_2$ & 0.084(18) & 0.075(18) & 0.892(20) & 5.490\\
$C^{m_\pi}_3$ & -0.016(11) & -0.018(11) & 1.084(74) & 1.146\\
$C^{f_\pi}_0$ & 0.0539(13) & 0.0547(13) & 1.01534(81) & 18.970\\
$C^{f_\pi,\,\scriptscriptstyle I}_a$ & -0.013(17) & -0.012(17) & 0.91(13) & 0.702\\
$C^{f_\pi,\,\scriptscriptstyle ID}_a$ & 0.0093(91) & 0.0057(89) & 0.61(38) & 1.032\\
$C^{f_\pi}_1$ & 1.121(32) & 1.054(32) & 0.9404(21) & 28.102\\
$C^{f_\pi}_2$ & 0.94(16) & 0.88(16) & 0.9414(88) & 6.662\\
$C^{f_\pi}_3$ & 0.120(48) & 0.120(49) & 1.001(10) & 0.110\\
$C^{m_K}_0$ & 0.06589(63) & 0.06597(62) & 1.00113(11) & 10.556\\
$C^{m_K}_1$ & 1.600(30) & 1.585(30) & 0.99058(37) & 25.545\\
$C^{m_K}_2$ & 0.208(86) & 0.206(85) & 0.99130(46) & 18.716\\
$C^{m_K}_3$ & 1.6544(97) & 1.6561(97) & 1.00101(11) & 9.208\\
$C^{m_K}_4$ & -0.000(28) & 0.000(28) & 0(1500) & 0.009\\
$C^{f_K}_0$ & 0.0705(14) & 0.0710(14) & 1.00633(32) & 19.911\\
$C^{f_K,\,\scriptscriptstyle I}_a$ & -0.014(13) & -0.014(13) & 1.011(22) & 0.510\\
$C^{f_K,\,\scriptscriptstyle ID}_a$ & 0.0041(73) & 0.0024(72) & 0.60(76) & 0.529\\
$C^{f_K}_1$ & 0.378(44) & 0.349(44) & 0.9246(84) & 9.007\\
$C^{f_K}_2$ & 0.77(20) & 0.76(19) & 0.9793(31) & 6.752\\
$C^{f_K}_3$ & 0.2965(65) & 0.2967(64) & 1.00060(22) & 2.749\\
$C^{f_K}_4$ & 0.144(57) & 0.144(57) & 0.9982(45) & 0.395\\
$C^{m_\Omega}_0$ & 0.7992(100) & 0.7994(99) & 1.00023(11) & 2.056\\
$C^{m_\Omega}_1$ & 3.0(1.8) & 3.0(1.8) & 0.9937(29) & 2.158\\
$C^{m_\Omega}_2$ & 5.436(65) & 5.441(65) & 1.00093(22) & 4.281\\
$C^{m_\Omega}_3$ & 0.35(39) & 0.35(39) & 1.013(29) & 0.454\\

\caption{A comparison of the results of analytic fits to the simulated data and the data corrected to the infinite volume using the ChPTFV fit forms. The quantity in the fourth column is the jackknife ratio of the results, $R$, and the quantity in the fifth column is the statistical significance of the deviation of this ratio from unity.}
\label{tab-anafvcorrcomparison}
\end{longtable}

\subsection{Global fit predictions}
\label{sec:globalfitpredictions}
Applying the procedure detailed above, we present our predictions for the pion and kaon decay constants:
\begin{align}
f_\pi & = 127.1(2.7)(0.9)(2.5)\ {\rm MeV},\\
f_K & = 152.4(3.0)(0.7)(1.5)\ {\rm MeV},\\
f_K/f_\pi & = 1.1991(116)(69)(116)\,.
\end{align}
Here the errors are statistical, chiral and finite-volume respectively. Note that by restricting the ChPTFV fit to $m_\pi<\mpicut$ MeV rather than $m_\pi<420$ MeV used in the 2010 analysis (a 30\% cut in the light quark mass), we obtain a value for $f_\pi$ that is now consistent with the known physical value, justifying our assertion that the previously observed deviation was mainly due to the influence of higher order terms in the chiral expansion.

For the inverse lattice spacings we obtain--
\begin{align}
a^{-1}({\rm 32I}) &= 2.310(37)(17)(9)\ {\rm GeV},\\
a^{-1}({\rm 24I}) &= 1.747(31)(24)(4)\ {\rm GeV},\\
a^{-1}({\rm 32ID}) &= 1.3709(84)(56)(3)\ {\rm GeV}. 
\end{align}
For comparison, in the 2010 analysis we obtained $a^{-1}({\rm 32I}) = 2.282(28)(1)(1)\ {\rm GeV}$ and $a^{-1}({\rm 24I}) = 1.730(25)(1)(0)\ {\rm GeV}$ by fitting only to the Iwasaki data. These results are statistically consistent, although we find a considerable enhancement in the systematic errors. Upon detailed investigation we determined that these differences arise almost entirely because the scaling factors $Z_l$, $Z_h$ and $R_a$ are now allowed to vary between the fits (generic scaling), as opposed being fixed to the values obtained at some unphysical mass point (fixed trajectory) as in the 2010 analysis: In the fixed trajectory case the prediction for the physical Omega baryon mass, which we use to set the overall scale, can vary only through the values of the physical light and strange quark masses, whereas in the generic scaling case the scaling parameters are those that contribute to the minimization of the global $\chi^2$, and can thus introduce larger variations in the predicted Omega mass. This 
does not, however, suggest that generic scaling is worse than the fixed trajectory approach, as the shifts in the scaling parameters between the three ans\"{a}tze in the former approach would simply be absorbed elsewhere in the latter, increasing the systematic error on some other quantities.

Using the NLO SU(2) ChPT fits we can obtain values for the effective couplings $\bar l_3$ and $\bar l_4$. For the ChPTFV and ChPT fits on their own, we find--
\begin{equation}\begin{array}{ll}
\bar{l_3} = 2.91(23), & \bar{l_4} = 3.99(16)\;({\rm ChPTFV})\\
\bar{l_3} = 2.98(22), & \bar{l_4} = 3.90(16)\;({\rm ChPT})\,.
\end{array}\end{equation}
As before we take the ChPTFV result for our central value. Although we cannot obtain a chiral extrapolation error without a corresponding analytic fit result, we can continue to estimate a finite-volume error from the difference between the two ChPT results. Therefore, our final values for the effective couplings are as follows:
\begin{equation}\begin{array}{lr}
\bar{l_3} = 2.91(23)(7), & \bar{l_4} = 3.99(16)(9)\,,
\end{array}\end{equation}
where the errors are statistical and finite-volume respectively. In the 2010 analysis (applying the same procedure to obtain the finite-volume error), we found $\bar l_3 = 2.57(18)(25)$ and $\bar l_4 = 3.83(9)(7)$. Comparing to our fits without the reduced pion mass cuts, we determined that the inflation of the statistical error and the rises in the central values over the 2010 analysis results derive mostly from the lowering of the cut from 420 to 350 MeV. However the values for $\bar l_3$ and $\bar l_4$ agree more closely in our current analysis even without the reduced cut, suggesting that the inclusion of the 32ID ensembles has some stabilizing influence upon the fit. For comparison, the FLAG working group obtained~\cite{Colangelo:2010et} an estimate of $\bar l_3 = 3.2(8)$, which was chosen to cover a large number of independent lattice results for this quantity, among which there are some discrepancies between the values. Our result is entirely consistent with this estimate. For $\bar l_4$, the 
inconsistencies between the results were considered too large to make a meaningful estimate. For both of these quantities, recent results include 2+1f determinations by the MILC collaboration~\cite{Bazavov:2010hj,Bazavov:2010yq} and our 2010 analysis paper~\cite{Aoki:2010dy}, and a 2+1+1f determination by the ETM collaboration~\cite{Baron:2010bv}.

Finally we give our predictions for the physical quark masses on the primary ensemble set:
\begin{equation}
\begin{array}{lr}
\tilde m_{ud}({\rm 32I}) = 2.243(46)(24)(10)\ {\rm MeV}, & \tilde m_s({\rm 32I}) = 62.2(1.1)(0.5)(0.3)\ {\rm MeV}
\end{array}
\end{equation}
In the 2010 analysis we obtained $\tilde m_{ud}({\rm 32I}) = 2.355(81)(79)(42)$ and $\tilde m_s({\rm 32I}) = 63.7(9)(1)(1)$. These numbers are again consistent, although here it appears that the enhanced control over the chiral extrapolation afforded by the 32ID ensembles has decreased the statistical error on the average up/down quark mass in spite of our exclusion of a large number of data points. We also observe a vastly improved chiral extrapolation systematic and a substantially reduced finite-volume error on this quantity. In the next section we discuss how these masses are renormalized into the $\msbar$ scheme.

%%%%%%%%%%%%%%%%%%%%%%%%% Quark Masses %%%%%%%%%%%%%%%%%%%%%%%%%%%%%%%%
\section{Physical Results For The Light- And Heavy-Quark Masses}
\label{sec:QuarkMasses}
\FloatBarrier

In the previous section we determined the physical quark masses in lattice units in the matching scheme defined in section~\ref{sec:FitResults}. In this section we discuss how we convert these into the conventional $\msbar$-scheme.

\subsection{Nonperturbative renormalization for the quark masses}

We cannot simulate with a noninteger number of dimensions, hence we must match our lattice results to perturbation theory in order to quote a result in the $\msbar$-scheme. Rather than matching using lattice perturbation theory, which is often poorly convergent, we obtain the renormalization coefficients $Z_m^{\msbar}$ nonperturbatively at each lattice spacing via several intermediate renormalization schemes -- the so-called RI/SMOM schemes -- that are variants of the Rome-Southampton RI/MOM scheme. In these schemes the renormalization coefficients are calculated by fixing the values of appropriate amputated vertex functions, constructed using quark propagators on Landau-gauge fixed configurations, at a renormalization scale defined by the quark momenta. These schemes are defined without reference to a particular regularization, hence they can easily be formulated in continuum perturbation theory with dimensional regularization, and the matching coefficients between them and the $\msbar$ scheme can be 
determined without reference to the lattice regularization. The matching is performed at a sufficiently high energy scale to be within the perturbative regime.

We have shown~\cite{Aoki:2010dy} that renormalizing at 3 GeV rather than the conventional 2 GeV results in a significant improvement in the contribution to the systematic error from the truncation of the perturbative series. The quark masses in ref.~\cite{Aoki:2010dy} were calculated at the 2 GeV scale; in this analysis we update the procedure to use the higher scale, and use twisted boundary conditions to gain better control of the discretization effects on the off-shell amplitudes entering the renormalization~\cite{Arthur:2010ht}. The RI/SMOM$\rightarrow \msbar$ matching coefficients at one-loop~\cite{Sturm:2009kb} and two-loops are known~\cite{Gorbahn:2010bf,Almeida:2010ns}.

For the lattice calculation of the RI/SMOM renormalization coefficients, we are constrained in our choice of renormalization scale only by the desire to avoid large discretization and finite-volume effects. Therefore for a lattice of spatial extent $L$ and lattice spacing $a$, we must choose a scale $\mu$ in the window: 
\begin{equation}
L^{-2} \ll \mu^2 \ll (\pi /a)^2\,.\label{eq-nprwindow}
\end{equation}
However, if we wish to match to the $\msbar$ scheme, this window is further constrained to the typically much smaller regime in which both the discretization and nonperturbative effects are small:
\begin{equation}
\Lambda_{\rm QCD}^2 \ll \mu^2 \ll (\pi /a)^2\,.\label{eq-rome-southampton}
\end{equation}
This is known as the Rome-Southampton window~\cite{Martinelli:1994ty}. For the 32I and 24I lattices, with $a^{-1}\sim 2.3$ GeV and $1.75$ GeV respectively, our target of 3 GeV is accessible directly within this window. However, for the 32ID lattice, with $a^{-1}\sim 1.37$ GeV, we cannot calculate the lattice renormalization conditions in the perturbative regime without incurring large discretization errors. The 32ID renormalization factors are not needed for the analysis of the quark masses in this section (see below), but this is an issue for $B_K$; we discuss this further in section~\ref{sec:BK}.

The need to calculate the RI/SMOM coefficients within the perturbative regime can be circumvented via the use of off-shell step-scaling functions~\cite{Arthur:2010ht,Arthur:2011cn} determined through a continuum extrapolation of the scale dependence (with a fixed lattice action) -- in this limit the dependence on the action disappears and the scale dependence becomes universal. Similar step-scaling functions were used in our recent analysis of the $K\rightarrow \pi \pi$ $\Delta I=3/2$ amplitudes~\cite{KtopipiPRD}. In that analysis, performed only on the 32ID ensemble set, we used the following strategy:
\begin{itemize}
\item[1.]
We evaluated the Z-factors (or the matrix of Z-factors in the case of operator-mixing) at a low energy scale $\mu_0$ on the 32ID lattice and computed the relevant renormalized matrix elements. The scale $\mu_{\rm 0}$ was chosen within the region given in equation~\ref{eq-nprwindow}, in which the finite-volume and discretization effects are small. In practice we chose $\mu_0\sim 1.1$ GeV.
\item[2.]
We computed the scale evolution between $\mu_0 \sim 1.1$ GeV and $\mu=3 \GeV$ of these operators on the finer Iwasaki (IW) lattices, upon which the high scale lies within the usual Rome-Southampton window. At finite lattice spacing $a_{\rm IW}$, if $Z^{\schemeS}(\mu,a_{\rm IW})$ is the renormalization factor of the operator under consideration in a (lattice) scheme $\schemeS$, the corresponding scale evolution is given by
\begin{equation}
\Sigma^\schemeS (\mu,\mu_0,a_{\rm IW}) = %\lim_{a_{\rm IW}\to 0} 
Z^\schemeS(\mu,a_{\rm IW}) \,{(Z^\schemeS(\mu_0,a_{\rm IW}))}^{-1}\,.
\end{equation}
The result was extrapolated to the continuum limit, giving the universal running in this energy range for this given scheme ${\schemeS}$:
\begin{equation}
\sigma^{\schemeS} (\mu,\mu_0) = \lim_{a_{\rm IW}\to 0} \Sigma^{\schemeS} (\mu,\mu_0,a_{\rm IW})\,.\label{eqn-stepscalingfactor}
\end{equation}
\item[3.]
We multiplied the Z-factors obtained in step 1 at the scale $\mu_0$ by the continuum nonperturbative running obtained in step 2 to obtain the desired Z-factors at 3 GeV. We then converted these to the $\msbar$ scheme using one-loop perturbation theory~\cite{Lehner:2011fz}.
\end{itemize}
Further details of the renormalization strategy used in the aforementioned analysis can be found in ref.~\cite{Boyle:2011cc}.

It is conceptually cleaner to divide our determination of the $\msbar$-scheme quark masses in a similar way to the above, separating the calculation of the non-perturbative renormalization coefficients and their subsequent continuum extrapolation from the perturbative matching stage. We therefore first calculate the RI/SMOM coefficients at a low energy scale $\mu_0$ and then calculate the step-scaling functions from this scale to 3 GeV. As discussed in section~\ref{sec:BK}, the choice $\mu_0 = 1.4$ GeV is optimal for the $B_K$ analysis -- we use this scale in the quark-mass analysis for consistency. Providing the jackknife/bootstrap errors are propagated correctly, the value of $Z_m^{\schemeS}(3\ {\rm GeV})$ obtained after applying the step-scaling function to the 1.4 GeV result will be exactly the same as if we had performed the continuum extrapolation directly at 3 GeV, due to the fact that the step-scaling functions are calculated using the same data.

\subsubsection{Determination of the lattice renormalization coefficients}
\label{sec-zmmethod}
Before presenting the results of our analysis, we summarize our measurement strategy, highlighting several important improvements over the original RI/MOM methods. 

The original RI/MOM scheme, defined in ref.~\cite{Martinelli:1994ty}, was shown~\cite{Aoki:1997xg} to suffer from greatly enhanced chiral symmetry breaking errors. These were found to occur due to the use of so-called exceptional kinematics, for which the vertex has channels along which the momentum transfer is zero; these allow quark and gluon loops with momenta below the spontaneous chiral symmetry breaking scale to exist even when the external momenta are moderately hard. The persistence of nonperturbative effects at high energy gives rise to large uncertainties in the perturbative matching. In order to avoid this problem we follow the 2010 analysis procedure in using non-exceptional ``symmetric'' kinematics~\cite{Aoki:1997xg} for which no exceptional channels exist. With these kinematics the nonperturbative effects fall off much faster as the virtuality is increased.

The quark mass renormalization coefficient $Z_m$, which is taken in product with the bare quark mass to obtain the renormalized quantity, is determined from the flavor nonsinglet scalar and pseudoscalar vertex renormalization coefficients, $Z_S$ and $Z_P$ respectively, via the relation $Z_m = 1/Z_S = 1/Z_P$. The equivalence of $Z_S$ and $Z_P$ is not exact if the chiral symmetry is broken; this occurs due to the low-energy spontaneous chiral symmetry breaking of QCD and to a much lesser degree from finite-$L_s$ effects. With nonexceptional kinematics, the former vanishes as $1/p^6$~\cite{Aoki:2007xm}, and is therefore small at the 3 GeV scale at which we perturbatively convert to the $\msbar$-scheme. In ref.~\cite{Aoki:2007xm} and during the present analysis we found that the effect of the difference between $Z_S$ and $Z_P$ on our final $\msbar$ scheme quark masses was considerably smaller than the error associated with the truncation of the perturbative series; as a result we do not need to include a 
systematic error for this effect. For the central values we arbitrarily chose to take the average of the scalar and pseudoscalar renormalization factors to determine $Z_m$, as was performed in ref.~\cite{Aoki:2007xm}.

The scalar and pseudoscalar vertex functions $\Pi_S$ and $\Pi_P$ were constructed at all sink locations of two quark propagators with momenta $p_1$ and $p_2$. The symmetric kinematics require that the momenta are chosen such that $p_1^2 = p_2^2 = (p_1-p_2)^2 = q^2 = -\mu^2$ for a renormalization scale $\mu$. As before we used volume momentum source propagators as these have been shown~\cite{Aoki:2010pe} to significantly reduce the statistical error on the NPR coefficients.

The renormalization conditions for the scalar and pseudoscalar vertex functions, applied at the scale $\mu$ in the three-flavor chiral limit, are: $\displaystyle \frac{Z_S}{Z_q}\Lambda_S = 1$ and $\displaystyle\frac{Z_P}{Z_q}\Lambda_P = 1$, where
\begin{align}
\Lambda_S=\frac{Z_S}{Z_q}\frac{1}{12}\mathrm{tr}\left[\Pi_S\cdot I\right]\,, & \hspace{1cm}\Lambda_P=\frac{1}{12}\mathrm{tr}\left[\Pi_P\cdot \gamma^5\right]\,.
\end{align}
Here $I$ is the identity matrix and $Z_q$ is the wave-function renormalization factor. $Z_m$ in the nonexceptional schemes is thus calculated as
\begin{equation}
Z_m Z_q = \frac{1}{2}(\Lambda_S + \Lambda_P)\,.
\end{equation}

The wave-function renormalization factor is determined from the renormalization condition on the vector current: $\displaystyle\frac{Z_V}{Z_q} \Lambda_V =1$, where
\begin{equation}
\Lambda_V = \frac{1}{12}\mathrm{tr}\left[\Pi_{V_\mu}\cdot \Gamma_\mu\right]
\end{equation}
for the vector bilinear vertex $\Pi_{V_\mu}$. With symmetric kinematics, the momentum transfer $q^2$ is nonzero, hence we have two choices for the projection matrix $\Gamma_\mu$, namely $\gamma_\mu/4$ and $\slashed {\hat q} \hat q_\mu/\hat q^2$; these define two different renormalization schemes which we label RI/SMOM$_{\gamma_\mu}$ and RI/SMOM respectively. Here we have used $\hat q_\mu = \sin(q_\mu)$ following ref.~\cite{Aoki:2010dy}. In the remainder of this work we refer to the two schemes collectively as the ``SMOM schemes''.

In the above, the vector renormalization coefficient $Z_V$ is identical to the factor relating the four-dimensional vector current to the corresponding Symanzik current. In section~\ref{sec:DSDRresults} we discussed how this quantity can be calculated independently using the ratio of the local four-dimensional and conserved five-dimensional vector currents. (The values for this quantity on the Iwasaki ensemble sets were determined in ref.~\cite{Aoki:2010dy}.) As $Z_V$ is known, we can combine its measurement with the ratio $\frac{Z_V}{Z_q}$, obtained from the vector-vertex renormalization condition, in order to determine $Z_q$.

In principle a separate measurement of $Z_q$ could be obtained using the axial-vector vertex. As was the case for the scalar and pseudoscalar vertex functions, this measurement can differ from that calculated via the vector vertex due to the residual effects of the low-energy spontaneous chiral symmetry breaking and small finite-$L_s$ effects. However, in refs.~\cite{Aoki:2010dy} and~\cite{Aoki:2010pe} we found that the effect of the difference between the vector and axial-vector vertex functions on our final result is again negligable compared to the perturbative truncation error.

As mentioned above, the renormalization conditions are applied in the three-flavor chiral limit. In practice we generate data on each ensemble with quark masses set equal to the dynamical light-quark mass; the chiral extrapolation is then performed using a linear fit over the unitary light-quark mass-dependence. The vertex functions are flavor-independent, hence this extrapolation also takes the valence strange-quark, but not the sea strange-quark, to the chiral limit. As we have only a single simulated dynamical strange-quark mass and reweight over only a short range, we cannot reliably take this final mass to the chiral limit. In refs.~\cite{Aoki:2010dy} and~\cite{Aoki:2010pe} we estimated the effect of not taking the strange sea-quark to zero using the slope of the unitary light-quark extrapolation, reduced by a factor of two to obtain the contribution of a single flavor. For the RI/MOM scheme, the two-flavor mass-dependence was found to be significant, resulting in a large systematic effect comparable in 
size to the truncation systematic. However, for the RI/SMOM schemes we found a very benign mass-dependence that was statistically indistinguishable from zero. Note that this estimate is highly conservative as the slope is likely dominated by the valence mass dependence; this suggests that we can ignore this systematic effect in our present analysis, for which we use only the nonexceptional schemes.

In ref.~\cite{Aoki:2010dy} we calculated the renormalization factors over a range of momentum scales. The scales at which we could perform our lattice measurements were limited by the need to form a symmetric momentum configuration with spatial momentum components that are discretized in units of $2\pi/L$ by the periodic boundary conditions. The resulting momentum configurations were typically distinct under the hypercubic group, hence the measurements were susceptible to lattice artifacts that vary under $O(4)$ rotations. These induced a scatter in the data, breaking the expected smooth scale dependence; as a result we were forced to artificially inflate our errors by a factor of $\sqrt{\chi^2/{\rm dof}}$, taken from a straight-line fit to the data. In ref.~\cite{Arthur:2010ht} we showed that the scatter can be eliminated entirely using twisted boundary conditions to induce quark momenta with a fixed direction; the remaining lattice artifacts can be removed by a continuum extrapolation. This approach was 
used for the renormalization of $B_K$ in the second of the 2010 analysis papers~\cite{Aoki:2010pe}. In the current analysis we also adopt this technique for the quark mass renormalization.

\subsubsection{Perturbative matching to the $\msbar$ scheme}

The conversion factors $C_m^{{\rm RI/SMOM}\rightarrow \msbar}$ between the RI/SMOM and $\msbar$ schemes were first computed at one loop in ref.~\cite{Sturm:2009kb} and the two-loop corrections for both RI/SMOM and RI/SMOM$_{\gamma_\mu}$ are known from refs.~\cite{Gorbahn:2010bf,Almeida:2010ns}. 
Regarding our notation, we write the running of the renormalized quark mass (in a given scheme $\schemeS$) between the scale $\mu_0$ and $\mu$ in the form 
\begin{equation}
\label{mscheme}
m^\schemeS(\mu) = m^\schemeS(\mu_0) \,  \exp\left( \int_{a_s(\mu_0)}^{a_s(\mu)} {\rm d}x \frac{\gamma_m^S(x)}{\beta(x)} \right) \,,
\end{equation}
where, following~\cite{Chetyrkin:1999pq}, we use $a_s=(\alpha_s/\pi)$. We expand the anomalous dimension $\gamma_m^{\schemeS}$ and the $\beta$-function (dropping the superscript $S$ for clarity)
\begin{eqnarray}
\gamma_m^{\schemeS} (a_s)&=& - \gamma^{(0)} \, a_s - \sum_{i\ge1} \gamma_{\schemeS}^{(i)} \, a_s^{i+1} \,,\\
\beta(a_s)    &=& - \sum_{i\ge0} \beta_i \, a_s^{i+2} \,,
\end{eqnarray}
where we have made explicit the fact that $\gamma^{(0)}$ is scheme-independent (we do not discuss here the scheme dependence of $\alpha_s$, which cancels in equation~\ref{mscheme}).
We can then express the result of eqn.~\ref{mscheme} with the help of 
\begin{equation}
\exp\left( \int_{a_s(\mu_0)}^{a_s(\mu)} {\rm d}x \frac{\gamma_m^S(x)}{\beta(x)} \right) 
= \frac{c^{\schemeS}(\mu)}{c^{\schemeS}(\mu_0)} \,,
\end{equation}
where
\begin{equation}
\label{cscheme}
c^{\schemeS}(\mu) = {a_s(\mu)}^{\frac{\gamma^{(0)}}{\beta_0}} \, 
\left( 1 + 
\left( \frac{\gamma_\schemeS^{(1)}}{\beta_0}  - \frac{\beta_1\gamma^{(0)}}{{\beta_0}^2}\right) a_s(\mu) 
+\mathcal{O}(a_s^2) \right) \,.
\end{equation}
Still following~\cite{Chetyrkin:1999pq}, we then define the renormalization-group-invariant (RGI) mass $\hat m$ by
\begin{equation}
\hat m =\lim_{\mu\to \infty} m^{\schemeS}(\mu) \,  {a_s(\mu)}^{-\frac{\gamma^{(0)}}{\beta_0}} \,.
\end{equation}
Using equation~\ref{mscheme} and~\ref{cscheme}, this gives
\begin{equation}
\label{toRGI}
\hat m  = \frac{m^\schemeS(\mu_0)}{c^\schemeS(\mu_0)}\, \quad \forall \mu_0\,.
\end{equation}
In particular, since $\hat m$ is renormalization group invariant, we can use the ratio of $c$'s to 
change scheme: for example, the conversion factor between a scheme $\schemeS_1$ and a scheme $\schemeS_2$ at the scale $\mu$ 
is given by
\begin{equation}
C_m^{{\schemeS_1} \to {\schemeS_2}}(\mu) = \frac{c^{{\schemeS_2}}(\mu)}{c^{{\schemeS_1}}(\mu)} \,.
\end{equation}
The RGI mass is obtained from $m^S(\mu)$ using equation~\ref{toRGI}, which implies that 
\begin{equation}
C_m^{\schemeS \to \rm RGI}(\mu) = \frac{1}{c^{\schemeS}(\mu)} \,.
\end{equation}
With some simple linear algebra we are able to convert the numerical results
of ref.~\cite{Sturm:2009kb} to our conventions and evaluate eqn.~\ref{cscheme}.

Finally, to obtain $\alpha_s$ at 3 GeV in the three-flavor theory, we used the four-loop running of refs.~\cite{vanRitbergen:1997va} and~\cite{Chetyrkin:1997sg} and took $\alpha_s(m_Z)=0.1184$~\cite{Beringer:1900zz} as an initial condition. We ran this quantity down to the charm mass, changing the number of flavors when crossing each threshold, obtaining $\alpha_s(3\ {\rm GeV})= 0.2454$.

Putting everything together, we found
\begin{eqnarray}
\frac{1}{c^{\rm RI/SMOM} (3\ {\rm GeV})} &=& 3.1052 \,( 1 - 0.0825 - 0.0066  + \mathcal{O}(a_s^3) ) = 2.8283\,,\\
\frac{1}{c^{{\rm RI/SMOM}_{\gamma_\mu}} (3\ {\rm GeV})} &=& 3.1052 \,(1 -0.1086 - 0.0147  + \mathcal{O}(a_s^3) ) = 2.7223\,,\\
\frac{1}{c^{\msbar} (3\ {\rm GeV})} &=& 3.1052 \,( 1 - 0.0699- 0.0035 + \mathcal{O}(a_s^3) ) = 2.8773\,.
\end{eqnarray}
Combining these we obtained for the conversion factors:
\begin{eqnarray}
C_m^{{\rm RI/SMOM}\rightarrow \msbar} (3\ {\rm GeV}) = \frac{ c^{\overline{MS} }(a_s(3\ {\rm GeV})) }{ c^{{\rm RI/SMOM} }(3\, {\rm GeV}) } &=& 0.9830\,,\\
C_m^{{\rm RI/SMOM}_{\gamma_\mu}\rightarrow \msbar} (3\ {\rm GeV}) = \frac{ c^{\overline{MS} }(a_s(3\ {\rm GeV})) }{ c^{{\rm RI/SMOM}_{\gamma_\mu}}(3\ {\rm GeV}) } &=& 0.9462\,,
\label{eqn-msbarconvfact}
\end{eqnarray}
which are correct to order $a_s^3$.

With the four-loop anomalous dimension of ref.~\cite{Chetyrkin:1999pq}, we obtain
\begin{equation}
C_m^{\msbar\rightarrow {\rm RGI}}(3\ {\rm GeV}) =
3.1052 \,( 1 -0.0699 -0.0035 -0.0001 + \mathcal{O}(a_s^4)) = 2.8769\,.\label{eqn-msbar3gev-rgi}
\end{equation}

For completeness we apply the same procedure at a renormalization scale of $\mu=2$ GeV (using $\alpha_s (2 \ {\rm GeV}) = 0.2960$):
\begin{eqnarray}
\frac{1}{c^{{\rm RI/SMOM}} (2\ {\rm GeV})} = 2.5452+ \mathcal{O}(a_s^3)\,,\\
\frac{1}{c^{{\rm RI/SMOM}_{\gamma_\mu}} (2\ {\rm GeV})} = 2.4218+ \mathcal{O}(a_s^3)\,,\\
\frac{1}{c^{\msbar} (2\ {\rm GeV})} = 2.6017+ \mathcal{O}(a_s^3)\,,
\end{eqnarray}
which are again quoted to $\mathcal{O}(a_s^3)$. Thus we find
\begin{eqnarray}
C_m^{{\rm RI/SMOM}\rightarrow \msbar} (2\ {\rm GeV}) = 0.9783\,,\\
C_m^{{\rm RI/SMOM}_{\gamma_\mu}\rightarrow\msbar} (2\ {\rm GeV}) = 0.9309\,,\\
C_m^{\msbar \rightarrow {\rm RGI}} (2\ {\rm GeV}) = 2.6012\,,\label{eqn-msbar2gev-rgi}
\end{eqnarray}
where, as before, the SMOM to $\msbar$ conversion factors are correct up to terms $\mathcal{O}(a_s^3)$, whereas the RGI conversion factor is true up to terms $\mathcal{O}(a_s^4)$ by virtue of using the four-loop anomalous dimension. As expected, these numbers are in very nice agreement with the ones given in ref.~\cite{Gorbahn:2010bf}.

We close this paragraph with a remark about the definition of the RGI quantities:
Our convention is such that, at the first order of perturbation theory, the 
conversion to the RGI quark mass is given by
\begin{equation}
%\frac{1}{c^{S}(\mu)} 
C_m^{\rm A \to RGI} (\mu)= (\alpha_s(\mu)/\pi)^{-4/(11-2 n_f/3)} \;.
\end{equation}
This convention differs from the one used for $B_K$, where $\alpha_s$ is not divided by $\pi$:
\begin{equation}
%\frac{1}{c^{S}(\mu)} 
C_{B_K}^{\rm A \to RGI} (\mu) = (\alpha_s(\mu))^{-2/(11-2 n_f/3)}\,.
\end{equation}
(Here the difference between the anomalous dimensions of the quark mass and $B_K$ accounts for the factor of two in the exponent between the two expressions.) Although the difference in conventions is rather unfortunate, we adopt them in order to match those commonly used in the literature.

\subsubsection{Calculation of $Z_m$ in the RI/SMOM schemes}

\begin{table}[tp]
\centering
\begin{tabular}{cccc}
\hline\hline
24I & $p_1 $ & $p_2$ & $\theta$  \\
\hline
& (-2,0,2,0) & (0,2,2,0)  &  $\{-0.45136, 0.732\}$ \\
& (-3,0,3,0) & (0,3,3,0)  &  $\frac{3}{16} n$ : $n = \{-2,1...,12\}$\\
& (-4,0,4,0) & (0,4,4,0)  &  $\frac{3}{2}$ \\
\hline
32I & $p_1 $ & $p_2$ & $\theta$  \\
\hline
& (-2,0,2,0) & (0,2,2,0)  &  $\{-0.413, 0.783\}$ \\
& (-3,0,3,0) & (0,3,3,0)  &  $\frac{1}{4}$ \\ %\{\frac{1}{4},0.135\}
& (-4,0,4,0) & (0,4,4,0)  &  $\{-\frac{3}{4}$ , $\frac{3}{8}$\}\\
& (-5,0,5,0) & (0,5,5,0)  &  $\{-\frac{5}{8} , \frac{3}{8}\}$ \\ %-0.531292
\hline
32ID & $p_1 $ & $p_2$ & $\theta$  \\
\hline
& (-3,0,3,0) & (0,3,3,0)  &  $\{ 0.0 \}$ \\
& (-4,0,4,0) & (0,4,4,0)  &  $\frac{1}{2} n$ : $n= \{-1,0,1,2\}$ \\
& (-5,0,5,0) & (0,5,5,0)  &  $\frac{1}{2} n$ : $n= \{-1,0\}$ \\
\end{tabular}
\caption{
Nonexceptional momenta and twist angles used for the evaluation of amputated twisted Green's functions in our NPR analyses. The momenta here
are listed in $(x,y,z,t)$ order. The integer Fourier mode numbers $\{n_i\}$ are related to the lattice momenta via $a p_i = \frac{n_i 2\pi}{L_i}$. The momentum added by the twist is determined by the twist angle $\theta$ giving $a p_i = \frac{ (2 n_i  +  \theta) \pi}{L_i}$. The twists that are not multiples of $\frac{1}{8}$ are chosen to match specific momenta on a larger volume lattice that will be described in a forthcoming publication.
\label{tab:tw_momenta}
}
\end{table} 

\begin{figure}[tp]
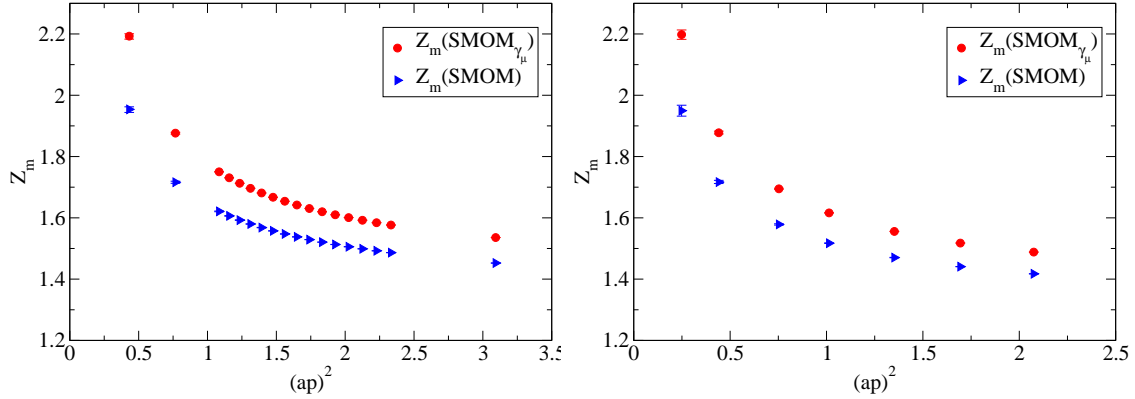

\centering
\includegraphics[width=0.45\textwidth]{fig/Zm/24I/Zm_SMOMq_SMOMg_24cubed.eps}
\includegraphics[width=0.45\textwidth]{fig/Zm/32I/Zm_SMOMq_SMOMg_32cubed.eps}
\caption{
\label{fig-zmap2dep}
$Z_m$ in the two SMOM schemes at a range of scales on the 24I (left) and 32I (right) ensemble sets.}
\end{figure}

We calculated the RI/SMOM and RI/SMOM$_{\gamma_\mu}$ bilinear vertex functions on each ensemble of the Iwasaki lattices, using quark propagators with the (twisted) momenta given in the first two blocks of table~\ref{tab:tw_momenta} (as explained below, the 32ID renormalization factors are not needed). These were then linearly extrapolated to the two-flavor chiral limit. We plot the scale dependence of the resulting chiral-limit renormalization factors in figure~\ref{fig-zmap2dep}; here we clearly see the smooth scale dependence arising from the use of twisted boundary conditions. 

In order to obtain the values at 1.4 GeV we performed an interpolation over several data points in the region surrounding the 1.4 GeV renormalization scale. Using the lattice spacings from section~\ref{sec:FitResults} we find the corresponding values of $(ap)^2$ to be 0.367 and 0.642 on the 32I and 24I lattices respectively. In this region of figure~\ref{fig-zmap2dep} we see a nonlinear scale dependence arising from the (supressed) poles and the renormalization group running, hence we cannot perform our interpolation using a simple linear function. Upon experimenting with several different nonlinear forms, we found that the following parametrization:
\begin{equation}
Z_m[(ap)^2] = C_0 + C_1/(ap)^2 + C_2 (ap)^2\,,
\end{equation}
fit the data well and was stable when the number of points was increased. We present the results of interpolating to $\mu=1.4$ GeV in table~\ref{tab-zmvals}. In order to later obtain the step-scaling factors, we repeated the above with a 3.0 GeV renormalization scale; these results are also given in the table. Note that the error quoted for the results in this table contains only the statistical contributions from the amputated vertex functions; the fluctuations arising from the statistical and systematic uncertainties on the lattice spacings and $Z_V$ are discussed below.

\begin{table}[tp]
\begin{tabular}{c|cc|cc}
\hline\hline
                  & \multicolumn{2}{c|}{32I}                    &  \multicolumn{2}{c}{24I}\\
 & RI/SMOM & $\text{RI/SMOM}_{\gamma_\mu}$     & RI/SMOM & $\text{RI/SMOM}_{\gamma_\mu}$ \\
\hline
$Z_m^{\schemeS}(\mu=1.4\ {\rm GeV})$ & 1.7782(62) & 1.9612(52)      & 1.7763(43) & 1.9558(36)\\
$Z_m^{\schemeS}(\mu=3.0\ {\rm GeV})$ & 1.4414(5)  & 1.5183(2)       & 1.4579(2) & 1.5419(2)

\end{tabular}
\caption{Renormalization factors in the intermediate RI/SMOM scheme $\schemeS$ at the scale $\mu$. Here the quoted error contains only the statistical contributions from the amputated vertex functions, not the fluctuations from the uncertainties on the lattice spacings and $Z_V$.}
\label{tab-zmvals}
\end{table}

\subsubsection{Renormalization of the continuum quark masses}
\label{zm-recont-subsec}
The physical quark masses determined in section~\ref{sec:FitResults} are quoted in the ``matching scheme'', whereas the renormalization factors above act upon the \textit{bare} physical quark masses. Therefore in order to obtain the quark masses in either the $\msbar$ scheme or one of the Rome-Southampton schemes, we must first convert the matching scheme masses into bare masses using equation~\ref{eqn-zlralattdef}.

The matching scheme is a noncontinuum (due to its explicit cutoff dependence), mass-independent scheme in which a bare quark mass in physical units that is determined at a coupling $\beta$ is renormalized by fixing its value to that obtained on a $32^3\times 64\times 16$ domain wall lattice with the Iwasaki gauge action at $\beta = 2.25$:
\begin{equation}
m_f^{\rm match} = Z^{\rm match}(m_f,\beta) m_f(\beta)\,.
\end{equation}
As discussed in section~\ref{sec:CombinedChiralFits}, the renormalization factor $Z^{\rm match}(m_f,\beta)$ at finite lattice-spacing is only weakly dependent upon the mass, hence we require just two factors: one to renormalize heavy quarks near the physical strange quark mass, and one to renormalize the light quarks. We labelled these $Z_h(\beta)$ and $Z_l(\beta)$ respectively, and calculated their values on the 24I and 32ID lattices as part of our global fits in section~\ref{sec:FitResults} (the values on the 32I ensemble are unity by definition).

Given the values of $m_{u/d}^{\rm match}$ and $m_s^{\rm match}$, we can obtain quark masses renormalized in one of our intermediate RI/SMOM schemes $\rm S$ at a given $\beta$ using the nonperturbative renormalization factors calculated above via the following ratios:
\begin{equation}
\begin{array}{ccc}
m_{u/d}^{\rm S}(\beta) = m_{u/d}^{\rm match} \times Z_m^{S}(\beta)/Z_l(\beta)\ \   \rm{and} & m_s^{\rm S}(\beta) = m_s^{\rm match} \times Z_m^{\rm S}(\beta)/Z_h(\beta)\,.
\end{array}
\end{equation}
These quantities still retain lattice artifacts which must be removed via a continuum extrapolation. Since, by definition, $Z_h$ and $Z_l$ absorb the coupling dependence of the quark masses, we need only extrapolate the ratios
\begin{equation}
\begin{array}{ccc}
 Z_{ml}^{\rm S}(\beta) = Z_m^{S}(\beta)/Z_l(\beta) & \rm{and} & Z_{mh}^{\rm S}(\beta) = Z_m^{\rm S}(\beta)/Z_h(\beta)\,.
\end{array}
\label{eqn-zml-zmh-def}
\end{equation}
For this we assume a linear dependence on $a^2$, neglecting the higher order effects. Note that we cannot include the values of $Z_m$ calculated on the 32ID lattice in this extrapolation due to this lattice having a different gauge action, and hence a different scale dependence, than the 24I and 32I lattices. As a result we have not analyzed this quantity in the present analysis.

In order to correctly propagate the statistical errors and the chiral/finite-volume errors on the various quantities we use the superjackknife procedure as before and repeated the analysis using $Z_l$, $Z_h$, the lattice spacings and the quark masses calculated using each of the three chiral ans\"{a}tze separately, taking the differences between these results at the final stage to determine the systematic errors in the usual way. In practice, the determination of the renormalization coefficients was performed using bootstrap resampling and used only the final results for the lattice spacings in determining the renormalization scale. In order to ensure that the systematic and statistical errors were correctly propagated we devised a procedure for generating suitable 'super-jackknife' distributions from these; this procedure is given in Appendix~\ref{appendix-zerrorprop}.

\begin{figure}[tp]
\centering
\includegraphics*[width=0.46\textwidth]{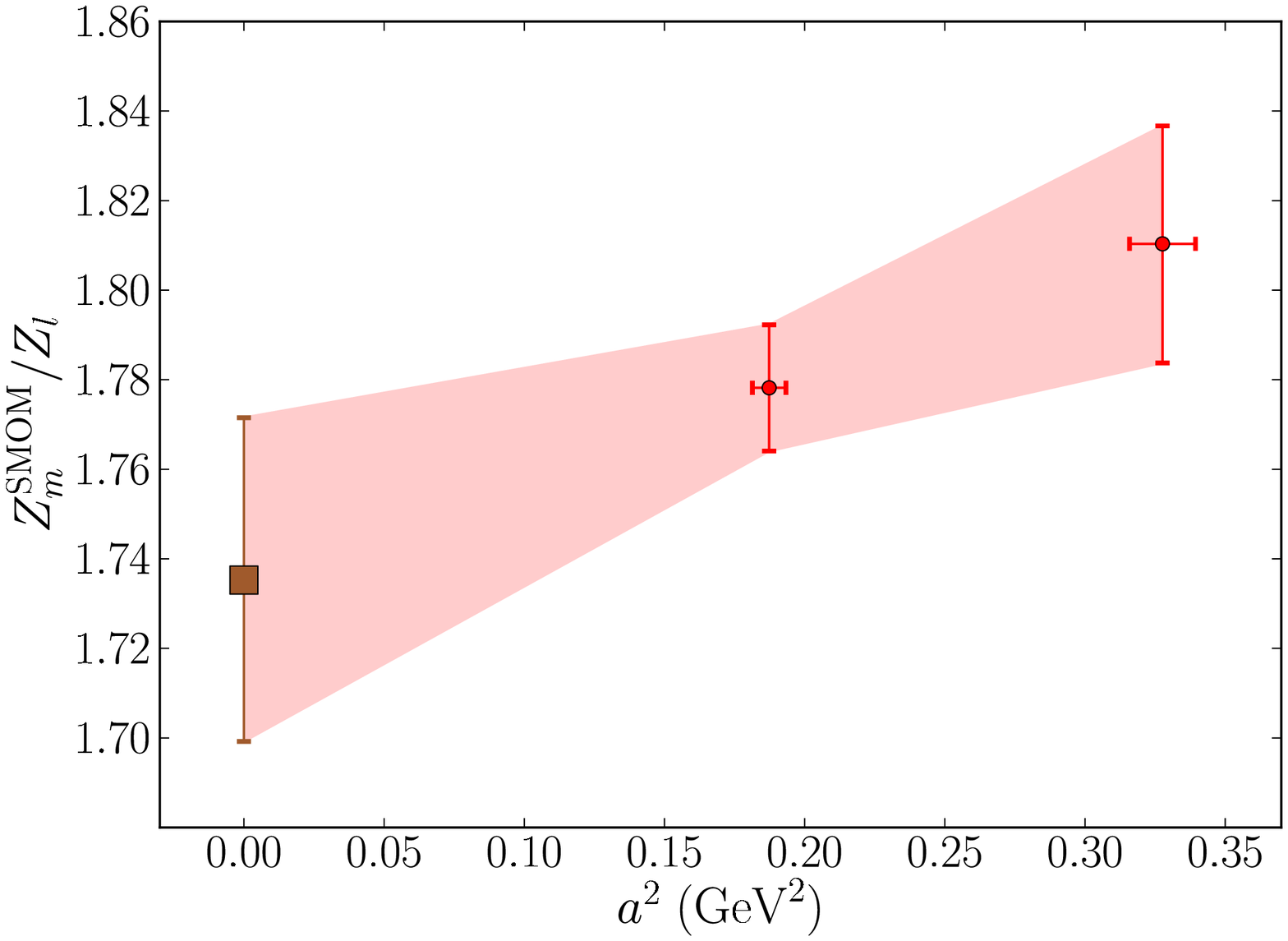}
\includegraphics*[width=0.46\textwidth]{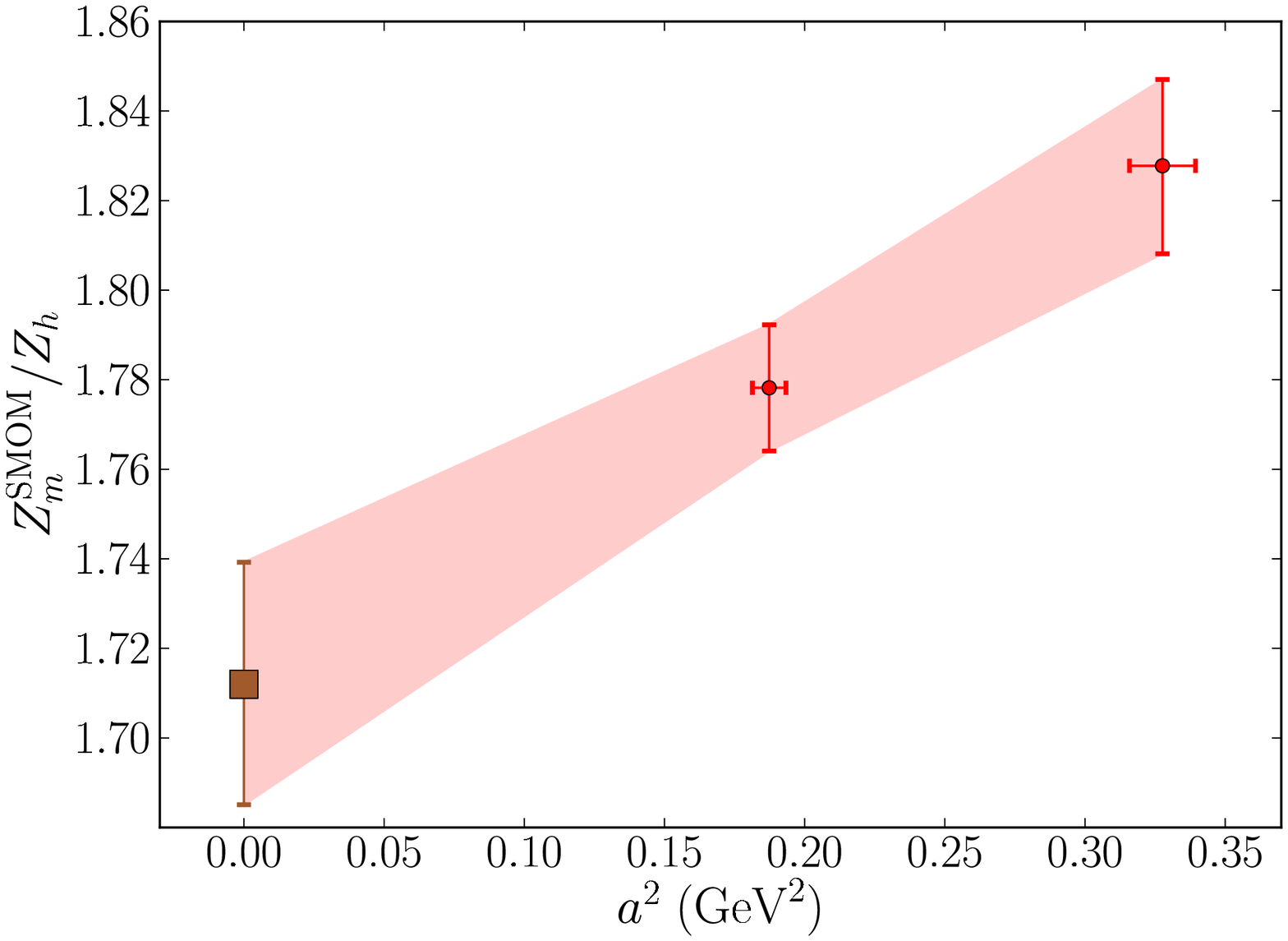}
\vspace{-0.5cm}
\caption{
\label{fig:contextrapZmZlhrat}
The continuum extrapolations of $Z_{ml}$ (left) and $Z_{mh}$ (right) in the RI/SMOM at 1.4 GeV.}
\end{figure}

We performed the continuum extrapolation of $Z_{ml}^{\schemeS,{\rm A}}(\mu=1.4\ {\rm GeV})$ and $Z_{mh}^{\schemeS,{\rm A}}(\mu=1.4\ {\rm GeV})$ for each choice of scheme $\schemeS$ and chiral ansatz $\rm A$, obtaining the values listed in table~\ref{zm-allansatz}. In the table and below we add a superscript ``c'' to denote continuum quantities. An example of the continuum extrapolation is shown in figure~\ref{fig:contextrapZmZlhrat}. 

The step-scaling factors $\sigma^{\schemeS,{\rm A}}(3\ {\rm GeV},1.4\ {\rm GeV})$ were then determined via a continuum extrapolation over the Iwasaki lattices of the ratio $\Sigma$ of renormalization coefficients at 3 GeV and 1.4 GeV (cf. equation~\ref{eqn-stepscalingfactor}). This was repeated for each scheme and chiral ansatz, giving the values also listed in table~\ref{zm-allansatz}. We then applied the step-scaling factors to $Z_{ml}^c$ and $Z_{mh}^c$ at the 1.4 GeV scale to obtain the corresponding values at 3 GeV; these are again listed in table~\ref{zm-allansatz}. Note that there is a quite considerable cancellation between the statistical fluctuations on the step-scaling factors and the 1.4 GeV renormalization coefficients; this cancellation is necessary to reproduce the smaller statistical errors on the 3 GeV factors and justifies the use of superjackknife error propagation. (Similar results might be obtained using bootstrap resampling for all quantities, with a consistent number of bootstrap 
samples, although this risks accidental cancellation between ostensibly uncorrelated fluctuations.)

\begin{table}[p]
\begin{tabular}{c|lc|ccc}
\hline\hline
                             &                                   &                                           & \multicolumn{3}{c}{Ansatz $\rm A$}\\
Quantity                     & Scheme $\schemeS$                 & Scale(s) $\mu$                            &  ChPTFV & ChPT & Analytic\\
\hline
\multirow{2}{*}{$Z_{ml}^c$}  & SMOM                              & \multirow{4}{*}{$1.4\ {\rm GeV}$}         & 1.735(36) &1.735(36) &1.752(51)\\
                             & SMOM$_{\gamma^\mu}$               &                                           & 1.918(39) &1.917(39) &1.935(55)\\
\multirow{2}{*}{$Z_{mh}^c$}  & SMOM                              &                                           & 1.712(27) &1.711(27) &1.712(34)\\
                             & SMOM$_{\gamma^\mu}$               &                                           & 1.893(29) &1.890(29) &1.890(37)\\ 
\hline
\multirow{2}{*}{$\sigma$}    & SMOM                              & \multirow{2}{*}{$1.4\rightarrow 3.0$ GeV} & 0.797(8) &0.798(8) &0.799(8)\\
                             & SMOM$_{\gamma^\mu}$               &                                           & 0.755(7) &0.756(7) &0.758(7)\\
\hline
\multirow{2}{*}{$Z_{ml}^c$}  & SMOM                              & \multirow{8}{*}{$3.0\ {\rm GeV}$}         & 1.383(27) &1.385(27) &1.401(40)\\
                             & SMOM$_{\gamma^\mu}$               &                                           & 1.449(28) &1.450(28) &1.466(42)\\
                             & $\msbar$ (via SMOM)               &                                           & 1.360(26) &1.361(26) &1.377(40)\\
                             & $\msbar$ (via SMOM$_{\gamma^\mu}$)&                                           & 1.371(26) &1.372(26) &1.387(40)\\
\multirow{2}{*}{$Z_{mh}^c$}  & SMOM                              &                                           & 1.365(18) &1.365(18) &1.368(25) \\
                             & SMOM$_{\gamma^\mu}$               &                                           & 1.429(18) &1.429(18) &1.432(26)\\ 
                             & $\msbar$ (via SMOM)               &                                           & 1.341(17) &1.342(17) &1.345(25)\\
                             & $\msbar$ (via SMOM$_{\gamma^\mu}$)&                                           & 1.352(17) &1.353(17) &1.355(25)\\
\end{tabular}
\caption{
\label{zm-allansatz}
The factors $Z_{ml}^c$ and $Z_{mh}^c$ used to convert our matching-scheme physical quark masses into each intermediate NPR scheme at 1.4 and 3.0 GeV, and the step-scaling factors used to run between those scales. We also list the $\msbar$ renormalization factors with $\mu=3.0$ GeV, obtained by applying the perturbative conversion from each of the intermediate RI/SMOM schemes. The superscript `c' on the renormalization factors is used to indicate that these are continuum quantities. The right-most columns correspond to the three choices of chiral ansatz used to obtain the lattice spacings used for the scale-setting and continuum extrapolations.}
\end{table}

\subsubsection{$\msbar$-scheme renormalization factors and systematic errors}
\label{sec-zmmsbarsys}
Applying the perturbative conversion factors to $Z_{ml}^c$ and $Z_{mh}^c$ at 3 GeV, we finally obtain the $\msbar$ renormalization coefficients for the quark masses determined in section~\ref{sec:FitResults}. We list the values in table~\ref{zm-allansatz}. All that remains prior to obtaining the $\msbar$ quark masses is to decide which intermediate scheme to use for the renormalization and to analyze the systematic errors.

In the 2010 analysis we decided that the most reliable $\msbar$ renormalization coefficients were obtained using the SMOM${}_{\gamma^\mu}$ intermediate scheme. This was based on the fact that this scheme showed a considerably smaller scatter from $O(4)$-symmetry breaking lattice artifacts than the SMOM scheme. However, now that the scatter has been eliminated through the use of twisted boundary conditions, we base our choice of ``best'' scheme on the size of the error in the matching of the intermediate scheme to $\msbar$, which we estimate from the size of the two-loop terms in equation~\ref{eqn-msbarconvfact}. We see that the SMOM-scheme conversion factors appear to converge faster than those in the SMOM${}_{\gamma^\mu}$-scheme, with a two-loop term roughly 75\% smaller. As a result we adopt the SMOM scheme for our final numbers.

We expect the main contribution to the systematic error to be associated with the truncation of the perturbative expansion of the $\msbar$ scheme-change factors. In ref.~\cite{Aoki:2010dy} we discussed two suitable methods for estimating this error: The first is to use the size of the two-loop term in the perturbative conversion and the second to take the full difference between the $\msbar$ coefficients calculated at 3 GeV using our two intermediate SMOM schemes. For the 2010 analysis, the most conservative estimate was obtained from the size of the two-loop term, however, now that we have adopted the RI/SMOM scheme for our final result we find that the $0.4\%$ two-loop contribution is smaller than the $0.8\%$ difference between the results obtained via the SMOM and SMOM${}_{\gamma^\mu}$ intermediate schemes. We therefore use the latter as our estimate of the truncation error.

In section~\ref{sec-zmmethod} we detailed several additional sources of error in our renormalization procedure that arise from nonperturbative effects; specifically, we highlighted the effects of the low-energy spontaneous chiral symmetry breaking and those associated with the dynamical strange sea-quark mass-scale. There are also likely to be additional effects at the $\Lambda_{\rm QCD}$ scale that were not considered. Although we concluded that the nonperturbative effects at the 3 GeV matching scale are negligible compared to the truncation error on our final results, it is illustrative to consider at what point they enter into our calculations. The RI/(S)MOM schemes are actually defined in the limit $\mu^2 \gg \Lambda_{QCD}^2$, at which the behavior is purely perturbative. The momentum schemes that we actually implement on our lattice can be therefore be regarded as different schemes that take into account the nonperturbative behavior. We therefore consider the aforementioned errors not as properties 
of the numerical renormalization factors, but rather as additional errors on the perturbative conversion to the $\msbar$-scheme, arising from the fact that the scheme-change factors are calculated using a slightly different scheme than the numerical results.

There are two final sources of systematic error on the renormalization conditions -- those arising from the chiral extrapolation and finite-volume errors on the lattice spacings used in the scale-setting and the continuum extrapolation. In the previous section, we repeated the analysis using the lattice spacings obtained from our global fits with the three different chiral ans\"{a}tze. We can therefore estimate these errors using the procedure discussed in section~\ref{subsec:syserrs}, namely estimating the chiral systematic error as the larger of two values, the first being the difference in central values between the results obtained using the ChPTFV and analytic parametrizations, and the second the superjackknife error on this difference. The same procedure is applied to the ChPTFV and ChPT results to estimate the finite-volume error. We take the central value and statistical error from the ChPTFV ansatz.

The final values for the quark mass renormalization factors are:
\begin{equation}\begin{array}{rl}
Z_{ml}^c(\msbar, 3\ {\rm GeV}) &= 1.360(26)(22)(2)(11)\,,\\
Z_{mh}^c(\msbar, 3\ {\rm GeV}) &= 1.341(17)(15)(1)(11)\,.\\
\end{array}\end{equation}
Here the errors are due to statistical, chiral, finite-volume and truncation effects.

\subsection{Results for the physical quark masses}

Multiplying $Z_{ml}$ and $Z_{mh}$ by the physical quark masses in the matching scheme, we obtain
\begin{equation}\begin{array}{cc}
m_{ud}(\msbar, 3\ {\rm GeV}) = 3.05(8)(6)(1)(2)\, {\rm MeV}, & m_s(\msbar, 3\ {\rm GeV}) = 83.5(1.7)(0.8)(0.4)(0.7)\, {\rm MeV},
\end{array}\end{equation}
where the errors are statistical, chiral, finite-volume and from the perturbative matching. The quark masses obtained in our 2010 analysis were quoted in the $\msbar$ scheme at 2 GeV. In order to facilitate a comparison between these and our new results we must therefore convert to a common scheme; for this we use the Renormalisation-Group invariant (RGI) scheme, for which the conversion factors from $\msbar$ are given in eqns.~\ref{eqn-msbar2gev-rgi} and~\ref{eqn-msbar3gev-rgi} for 2 and 3 GeV respectively. Applying the latter to the results above we find:
\begin{equation}\begin{array}{cc}
\hat m_{ud} = 8.78(24)(17)(3)(7)\, {\rm MeV}, & \hat m_s = 240.1(4.8)(2.4)(1.2)(2.0)\, {\rm MeV},
\end{array}\end{equation}
where the hat is used to label the RGI values. In the 2010 analysis we obtained
\begin{equation}\begin{array}{cc}
\hat m_{ud} = 9.34(34)(31)(16)(21)\, {\rm MeV}, & \hat m_s = 250.2(3.9)(0.5)(0.3)(5.5)\, {\rm MeV}\,.
\end{array}\end{equation}
Our new result appears to be consistent with that of the 2010 analysis, but has a renormalization systematic error that is over a factor of two smaller by virtue of performing the matching to the $\msbar$ scheme at 3 GeV, rather than 2 GeV, at which the perturbation theory is more reliable. For the up/down quark mass we also see a substantial improvement in the chiral and finite-volume systematics, resulting from the lowering of the pion mass cut in the fit and the inclusion of the 32ID data. For the strange quark mass, the 32ID data does not have the same effect because the Iwasaki data were already (after reweighting) at the physical mass, and the light-quark mass dependence of the kaon is small. The larger chiral and finite-volume systematics on this quantity likely arise from allowing the scaling parameter $Z_h$, and also to a lesser extent $Z_l$, to differ between the fit ans\"{a}tze rather than remaining fixed; this allows the larger changes in the quality of the fit for the other fitted quantities to 
influence the kaon fit. A similar effect was observed for the lattice spacings and was discussed in section~\ref{sec:globalfitpredictions}.

For comparison with the above, the FLAG working group give $m_{ud}(\msbar, 2\ {\rm GeV})=3.43(11)$ MeV and $m_{s}(\msbar, 2\ {\rm GeV})=94(3)$ MeV~\cite{Colangelo:2010et}. These values were obtained by combining results from the MILC~\cite{Bazavov:2010yq,Bazavov:2009fk} and HPQCD~\cite{McNeile:2010ji} collaborations, as well as our 2010 analysis results. Converting to the RGI scheme using the conversion factor given above, these become $\hat m_{ud} = 8.92(29)$ MeV and $\hat m_{ud} = 245(8)$ MeV, which both agree very well with our results.

Finally, for completeness we also calculate the ratios of the strange and up/down quark masses:
\begin{equation}
\frac{m_s}{m_{ud}} = 27.36(39)(31)(22)(0)\,,
\end{equation}
where the errors are again as above.

%%%%%%%%%%%%%%%%%%%%%%%%% B_K %%%%%%%%%%%%%%%%%%%%%%%%%%%%%%%%
\section{Chiral/Continuum Fits And Physical Results For $B_K$}
\label{sec:BK}
In this section we present our results for the neutral kaon mixing parameter $B_K$. Continuum results are obtained by performing chiral/continuum fits over our three ensemble sets following the strategy outlined in section~\ref{sec:CombinedChiralFits}. This analysis extends that in ref~\cite{Aoki:2010pe} through the inclusion of the 32ID ensemble set.

As $B_K$ is a scheme-dependent quantity we must perform our fits to renormalized data. We determine the renormalization factors again using variants of the RI/MOM scheme with symmetric kinematics. We first outline this calculation, then discuss the application of our chiral fitting techniques to this quantity. Finally we present the continuum results in the $\msbar$ scheme at 3 GeV.

\subsection{Nonperturbative renormalization factors}

Unlike in the case of the quark mass renormalization, we require renormalization factors for $B_K$ on both the Iwasaki and Iwasaki+DSDR ensemble sets. In this case, the option of calculating our lattice renormalization factors directly at 3 GeV is not an option since we cannot simulate within the perturbative regime without incurring large lattice artifacts. (We remind the reader that perturbation theory is required to match the renormalization factors computed on the lattice to a continuum scheme, typically $\overline{\rm MS}$ in which the Wilson coefficients are computed.) As discussed in section~\ref{sec:QuarkMasses}, our analysis~\cite{KtopipiPRD} of the $\Delta I=3/2$ $K\rightarrow \pi\pi$ amplitudes had a similar issue, which was solved by computing the renormalization factors at a low energy scale, $\mu_0=1.1$ GeV, at which finite-volume effects and lattice artifacts are small (i.e. satisfying eqn.~\ref{eq-nprwindow}), and using the continuum step-scaling factors to evolve this to the perturbative 
matching scale. For this analysis we adopt a similar procedure.

\subsubsection{Determining the NPR factors}
\label{sec-bknprdetr}
We follow ref.~\cite{Aoki:2010pe} in calculating the renormalization factors in four different lattice schemes. First we consider the process
\be
d(p_1)\bar{s}(-p_2)\to\bar{d}(-p_1)u(p_2)
\ee
with a variety of momenta satisfying the symmetric momentum configuration $p_1^2=p_2^2=(p_1-p_2)^2=\mu^2$. We write the corresponding amputated Green's function evaluated on Landau gauge-fixed configurations as $\Lambda_{\alpha \beta, \gamma \delta}^{ij, kl}$ (the color indices $i,j,\ldots$
and Dirac indices $\alpha,\beta,\ldots$ correspond to the external states). We have to project these Green's functions onto their Dirac-color structure, where, as before we, define two projectors using both the $\gamma$-matrices and $\slashed{\hat q}$ (where $N_C$ is the number of colors and $\hat q_\mu = \sin(q_\mu)$)
:
\bea
P^{(\gamma^{\mu})\,ij, kl}_{\alpha \beta, \gamma \delta} 
&=&\frac{1}{128N_c(N_c+1)}\,
\left[ (\gmuL)_{\beta\alpha} (\gmuL)_{\delta\gamma} \right]
 \delta^{ij} \delta^{kl}
\\
P^{(\qslashs)\,ij, kl}_{\alpha \beta, \gamma \delta} 
&=& \frac{1}{32\hat q^{2}N_c(N_c+1)}
\left[ (\qslashL)_{\beta\alpha} (\qslashL)_{\delta\gamma} \right]
\delta^{ij} \delta^{kl} \,.
\eea
These act on $\Lambda$ in the following way:
\be
\label{eq:proj}
M\equiv P\{\Lambda\} \equiv P^{ij, kl}_{\alpha \beta, \gamma \delta} \Lambda_{\alpha \beta, \gamma \delta}^{ij, kl}
\ee

As before we can renormalize the quark field, and hence obtain $Z_q$, in both the RI/SMOM and RI/SMOM$_{\gamma^\mu}$ schemes; we therefore have four independent renormalization schemes for $Z_{B_K}$:
\be
\label{Zbk}
Z_{(27,1)}^{(A,B)} = 
(Z^{(B)}_q)^2 \big[ P^{(A)}\{\Lambda\} \big]^{-1}
\ee
where $A$ and $B$ can be either $\gamma^{\mu}$ or $\qslash$. Here the label $(27,1)$ refers to the $SU(3)_L\times SU(3)_R$ transformation properties of the $VV+AA$ four-quark operator that forms the numerator of equation~\ref{eqn-bkoperator}. Motivated by~\cite{Aoki:2010pe}, we focus only on two schemes :
the $(A,B)={(\gamma^{\mu}, \gamma^{\mu})}$ and $(\qslash, \qslash)$ combinations.

The renormalization factor for $B_K$ is then
\begin{equation}
Z_{B_K}^{(A,B)} = {Z_{(27,1)}^{(A,B)} \over Z_A^2} \,. 
\end{equation}
We obtain $Z_q^2/Z_A^2$ from the renormalization conditions on the vector and axial-vector vertices:
\begin{equation}
\frac{Z_q}{Z_A} = \half(\Lambda_A+\Lambda_V)\,.
\end{equation}
As discussed in section~\ref{sec:QuarkMasses}, the difference between these vertices in the SMOM schemes is tiny and can be ignored; we used their average only such that the same procedure can be applied for the exceptional schemes.

\subsubsection{Perturbative conversion factors}
\label{sec:bk-pertconv}

The one-loop perturbative conversion factors for converting to the $\msbar$-scheme from the SMOM schemes are obtained using the expressions in ref.~\cite{Aoki:2010pe}, resulting in the following:
\begin{equation}\begin{array}{rccl}
C_{B_K}^{(\slashed q,\slashed q)} &= 1 &-&  0.45465\left(\frac{\alpha_s}{4\pi}\right) = 0.99112\ \ {\rm and}\\
C_{B_K}^{(\gamma^\mu,\gamma^\mu)} &= 1 &+&  0.21197\left(\frac{\alpha_s}{4\pi}\right) = 1.00414\,,
\end{array}\end{equation}
where
\begin{equation}
\alpha_s(3\;\GeV) = 0.24544\,.
\end{equation}

As discussed in the following section, we do not use the SMOM$(\slashed q,\gamma^\mu)$ or SMOM$(\gamma^\mu,\slashed q)$ schemes for our final predictions, hence we have not listed the corresponding conversion factors above.

\subsubsection{Renormalization scales}

As the 3 GeV matching scale lies within the Rome-Southampton windows for the two Iwasaki lattices, we need only compute the 32ID renormalization factors at the low energy scale and subsequently use the continuum step-scaling factors to run these up to the same scale as the Iwasaki coefficients. However in practice we found that the statistical errors on the step-scaling factors were quite large, which resulted in considerably larger errors on the 3 GeV renormalization factors than their Iwasaki counterparts. Note that contrary to the case of the mass renormalization, no cancellation occurs between the statistical fluctuations on $Z_{B_K}(\mu_0)$ and $\sigma(3\ {\rm GeV},\mu_0)$ as the data sets from which they were determined are entirely independent. 

The disparity in the statistical errors between the renormalization factors has the effect of weakening the constraints that the 32ID data imposes on the simultaneous chiral/continuum fit under the global $\chi^2$ minimization. As a na\"{i}ve test of the impact of this disparity, we repeated our fits with the errors on the 32ID renormalization factors artificially reduced to match those on the Iwasaki lattices. We found that the central value of the continuum prediction for $B_K$ shifted by an amount comparable to the chiral and finite-volume systematics; an effect too large to be ignored. As we pointed out in section~\ref{sec:FitResults} when discussing the number of reweighting samples to use on each lattice, it is important to treat each ensemble set uniformly such that the weight of each of the ensemble sets in the fit depends only on the statistics of the data. We therefore calculate the renormalization factors for all three lattices at the same scale, chosen within the regime in which the 
discretization effects are under control. The 1.1 GeV scale used in ref.~\cite{KtopipiPRD} meets this criteria, although we found a noticable reduction in the statistical errors by raising this to 1.4 GeV (actually 1.426 GeV, the nearest scale at which we had a simulated point). Of course, using a larger scale increases the size of the discretization effects on the 32ID lattice, however, as we ultimately perform a universality-constrained continuum extrapolation, only the $\mathcal{O}[(ap)^4]$ terms and higher remain in the final result for $B_K$. Only after performing the continuum limit do we apply the step-scaling factor to evolve the continuum prediction to 3 GeV, at which the matching to $\msbar$ is performed.

\subsubsection{Results}
\label{sec:zbk-results}
Following the above strategy we calculated $Z_{B_K}$ at $\mu_0=1.426$ GeV on each of the three ensemble sets. In addition, we recalculated the Iwasaki renormalization factors at 3 GeV such that we could obtain the continuum step-scaling functions. The quark momenta used in these measurements are listed in table~\ref{tab:tw_momenta}, and we present the values at both renormalization scales in table~\ref{tab-zbkrawvals}. We used the central values of the lattice spacings given in section~\ref{sec:globalfitpredictions} to set the physical scales in these determinations. 

In order to correctly propagate errors on the lattice spacings, we formed superjackknife distributions for the renormalization factors that include the fluctuations on the lattice spacings, following the procedure in section~\ref{zm-recont-subsec}. As before, separate distributions were obtained for each of the three chiral ans\"{a}tze, with the central values shifted appropriately, allowing us to later separate the chiral and finite-volume systematic errors. The formation of the superjackknife distributions requires the derivatives of $Z_{B_K}$ with respect to the lattice spacings, which we again determined by measuring the differences in the central values as the lattice spacings are varied by their total error. We use the full superjackknife distributions to renormalize $B_K$ in the following sections.

We determined the step-scaling factors by taking the continuum limit of the ratio of $Z_{B_K}$ at 3 GeV and 1.4 GeV in each of the four schemes. The results are given in table~\ref{tab-bkstepscaling}.

\begin{table}[tp]
\begin{tabular}{clc|ccc}
\hline\hline
Quantity                   & Projector P                 & Scale $\mu$ & \multicolumn{3}{c}{Value}\\
                           &                             &             & 32I       &   24I      & 32ID\\
\hline
\multirow{4}{*}{$Z_{B_K}$} & $(\slashed q,\slashed q)$  &\multirow{4}{*}{1.426 GeV} & 1.0608(12) &1.0320(11) & 0.9992(11)\\
                           & $(\gamma^\mu,\gamma^\mu)$  &                           & 0.9788(9)  &0.9527(3)  & 0.9210(8)\\
                           & $(\slashed q,\gamma^\mu)$  &                           & 0.8758(25) &0.8554(17) & 0.8187(13) \\
                           & $(\gamma^\mu,\slashed q)$  &                           & 1.1865(38) &1.1496(32) & 1.1241(24)\\

\hline
\multirow{4}{*}{$Z_{B_K}$} & $(\slashed q,\slashed q)$ &\multirow{4}{*}{3 GeV}      & 0.9765(1) &0.9549(1) & --\\
                           & $(\gamma^\mu,\gamma^\mu)$ &                            & 0.9396(2) &0.9153(1) & --\\
                           & $(\slashed q,\gamma^\mu)$ &                            & 0.8795(4) &0.8537(2) & --\\
                           & $(\gamma^\mu,\slashed q)$ &                            & 1.0432(4) &1.0238(2) & --\\
\end{tabular}
\caption{$B_K$ renormalization factors in the four intermediate RI/SMOM schemes at the scales $\mu$. Here the quoted error contains only the statistical contributions from the amputated vertices, not the fluctuations from the uncertainties on the lattice spacings. Note that we did not calculate the 32ID renormalization factors at 3 GeV as this point lies beyond the Rome-Southampton window on this lattice.}
\label{tab-zbkrawvals}
\end{table}

\begin{table}[tp]
\begin{tabular}{l|ccc}
\hline\hline
Projector P & \multicolumn{3}{c}{Chiral Ansatz} \\
            & ChPTFV & ChPT & Analytic \\
\hline
$(\slashed q,\slashed q)$ & 0.9140(34) &0.9145(33) &0.9150(34)\\
$(\gamma^\mu,\gamma^\mu)$ & 0.9589(21) &0.9591(21) &0.9593(21)\\
$(\slashed q,\gamma^\mu)$ & 1.0127(74) &1.0127(75) &1.0128(75)\\
$(\gamma^\mu,\slashed q)$ & 0.8641(80) &0.8647(80) &0.8654(81) \\
\end{tabular}
\caption{Nonperturbative step-scaling factors for each intermediate scheme SMOM(P), used {\it a posteriori} to run $Z_{B_K}$ from 1.426 to 3 GeV. A different value is obtained for each determination of the lattice spacings.}
\label{tab-bkstepscaling}
\end{table}

\FloatBarrier
\subsection{Chiral/continuum fits}

\begin{table}
\centering
\begin{tabular}{l|rr||l|r}
\hline
Parameter & ChPT & ChPTFV & Parameter & Analytic\\
\hline
 \rule{0cm}{0.4cm}$\chi^2$/dof &  0.71(45)  &  0.56(40)   &    & 0.49(33)\\
\hline
 \rule{0cm}{0.4cm}$B$ & $ 4.144(89) $ & $ 4.110(93) $ &  & \\
 $f$ & $ 0.1221(29) $ & $ 0.1259(28) $ &  & \\
\hline
 \rule{0cm}{0.4cm}$B_K^0$ & $ 0.580(10) $ & $ 0.584(10) $ & \rule{0cm}{0.4cm}$C_0^{B_K}$ & $ 0.597(11) $\\
 $c_{B_K,a}^{\scriptscriptstyle I}$ & $ 0.073(44) $ & $ 0.072(44) $ & $C_a^{B_K,\,\scriptscriptstyle I}$ & $ 0.059(46) $\\
 $c_{B_K,a}^{\scriptscriptstyle ID}$ & $ 0.099(23) $ & $ 0.095(23) $ & $C_a^{B_K,\,\scriptscriptstyle ID}$ & $ 0.086(23) $\\
 $c_{B_K,m_x}$ & $ 0.00458(72) $ & $ 0.00398(76) $ & $C_1^{B_K}$ & $ 0.33(24) $\\
 $c_{B_K,m_l}$ & $ -0.0079(16) $ & $ -0.0079(17) $ & $C_2^{B_K}$ & $ -0.07(54) $\\
 $c_{B_K,m_y}$ & $ 1.440(39) $ & $ 1.450(40) $ & $C_3^{B_K}$ & $ 1.450(40) $\\
 $c_{B_K,m_h}$ & $ -0.08(13) $ & $ -0.06(13) $ & $C_4^{B_K}$ & $ -0.04(13) $\\
\end{tabular}
\caption{The $\chi^2$/dof and parameters for each of our chiral fit ansatz\"{e} for $B_K$, with the fits performed to data renormalized in the SMOM$(\slashed q,\slashed q)$ scheme with a cut on data with corresponding pion masses $m_\pi > \mpicut$ MeV. The parameters are given in physical units and with the heavy quark mass expansion point adjusted to the physical strange quark mass. For the ChPT and ChPTFV ansatz\"{e} the chiral scale $\Lambda_\chi$ has been adjusted to 1 GeV.}
\label{tab-bklecs}
\end{table}

\begin{table}
\centering
\begin{tabular}{l|cc}
\hline
        & \multicolumn{2}{c}{Scheme} \\
 Ansatz & SMOM$(\slashed q,\slashed q)$ & SMOM$(\gamma^\mu, \gamma^\mu)$\\
\hline
\rule{0cm}{0.4cm}Analytic  & $ 0.5978(87) $ & $ 0.5506(77) $ \\
ChPT  & $ 0.5871(84) $ & $ 0.5410(75) $ \\
ChPTFV  & $ 0.5904(85) $ & $ 0.5436(75) $ \\
\end{tabular}
\caption{Predictions for $B_K$ in the continuum limit in the SMOM$(\slashed q,\slashed q)$ and SMOM$(\gamma^\mu, \gamma^\mu)$ schemes at $\mu=1.426$ GeV for each global fit ansatz. These results were obtained using simultaneous/chiral continuum fits to renormalized data with a pion mass cut of \mpicut MeV.}
\label{tab:bkcontinuum}
\end{table}

The determination of $B_K$ on the 32ID ensemble set was discussed in section~\ref{sec:DSDRresults} and the values listed in tables~\ref{tab:bk_32ID_mhsim} and~\ref{tab:bk_32ID_mhphys}. These data and those on the Iwasaki ensemble sets were renormalized into the RI/SMOM intermediate schemes at $\mu = 1.426\GeV$ using the results of the previous section. Anticipating the discussion in the following section, we present only the results of fitting to data renormalized in the SMOM$(\gamma^\mu,\gamma^\mu)$ and SMOM$(\slashed q,\slashed q)$ intermediate schemes.

As before, we obtain our chiral/continuum fit forms by performing an expansion in the quark masses and $a^2$ to NLO, with the light-quark mass expanded about both the chiral limit -- using chiral perturbation theory -- and about a fixed mass via a Taylor expansion. For example, for the analytic ansatz we obtain the following:
\begin{equation}
B_{xy}^{\bf 1} = C_0^{B_K}\left(1 + C_a^{B_K,\,A(\bf 1)}[a^{\bf 1}]^2\right) + C_1^{B_K}\tilde m_x^{\bf 1} + C_2^{B_K}\tilde m_l^{\bf 1} + C_3^{B_K}\left(\tilde m_y^{\bf 1} - m_{h0}\right) + C_4^{B_K}\left(\tilde m_h^{\bf 1} - m_{h0}\right)\,,
\end{equation}
and for the ChPT ansatz:
\begin{equation}\begin{array}{rl}
\displaystyle B_{xy}^{\bf 1} &= \displaystyle B_K^0\left\{1 + c_{B_K,a}^{A(\bf 1)}[a^{\bf 1}]^2 + \frac{c_{B_K,m_l}\chi_l^{\bf 1}}{f^2} + \frac{c_{B_K,m_x}\chi_x^{\bf 1}}{f^2} - \frac{\chi_l^{\bf 1}}{32\pi^2f^2}\log\left(\frac{\chi_x^{\bf 1}}{\Lambda_\chi^2}\right)\right\}\\
&\displaystyle\hspace{1cm}+ c_{B_K,m_y}\left(\tilde m_y^{\bf 1} - m_{h0}\right) + c_{B_K,m_h}\left(\tilde m_h^{\bf 1} - m_{h0}\right)\,,
\end{array}\end{equation}
where $\chi_q = 2B\tilde m_q$ and the chiral scale $\Lambda_\chi$ is set to 1 GeV. These fit forms apply specifically to the primary lattice $\bf 1$; the forms for any other ensemble set $\bf e$ can be obtained by inserting factors of $Z_l^{\bf e}$ and $Z_h^{\bf e}$ and selecting the $a^2$ coefficient appropriate to the lattice action. The finite-volume correction terms for the ChPT fit form can be found by applying the rules given in Appendix C of ref.~\cite{Allton:2008pn}. 

Following the 2010 analysis strategy, we fixed the leading order LECs $B$ and $f$ in the ChPT fits to those obtained in section~\ref{sec:FitResults}, reducing the number of free parameters. We also fix the the scaling factors $Z_l$, $Z_h$ and $R_a$, as well as the physical quark masses and the overall scale to those obtained using the corresponding ansatz in section~\ref{sec:FitResults}.

\begin{figure}[t]
\centering
\includegraphics*[width=0.48\textwidth]{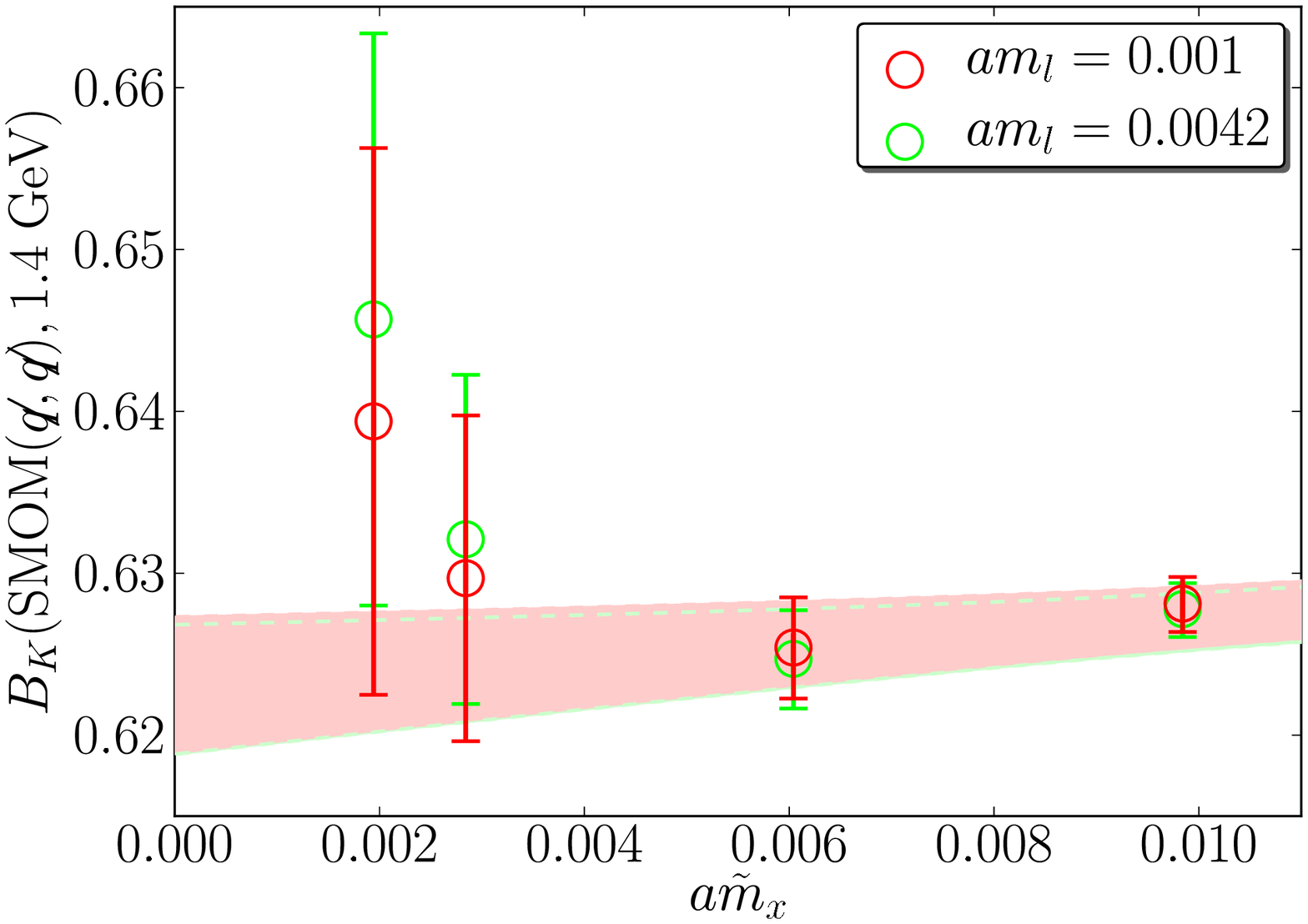}\quad
\includegraphics*[width=0.48\textwidth]{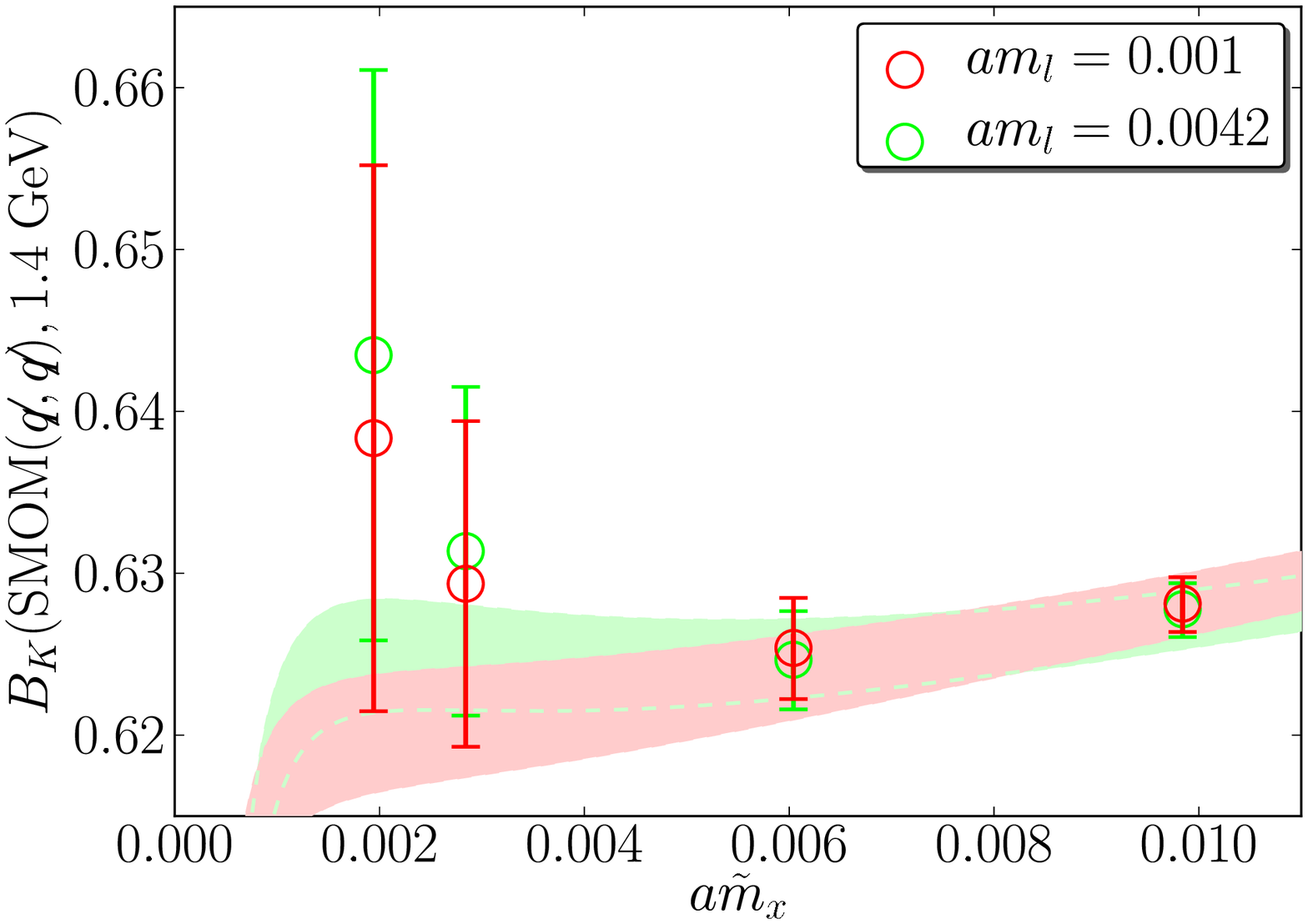}
\caption{
\label{fig:bk_analytic_ChPTFV_cut}
The analytic (left) and ChPTFV (right) fit curves overlaying the partially-quenched data on the 32ID ensembles at the simulated strange quark mass. The fits were performed to the data set with corresponding pion masses $m_\pi < \mpicut$ MeV, with the data renormalized in the SMOM$(\slashed{q},\slashed{q})$ intermediate scheme.}
\end{figure}

We once again performed cuts to the data set used in the ChPT and ChPTFV fits, reducing the largest pion mass to $\mpicut$MeV. In the main analysis we performed our analytic fits with a lower pion mass cut of $\mpicutanalytic$MeV in order to obtain a better fit to the data. When using this cut for the analyic fits to $B_K$, we found that we lost almost all statistical precision on our continuum prediction because the statistical errors on the 32ID ensembles become very large in the light-mass regime (cf. figure~\ref{fig:bk_analytic_ChPTFV_cut}), hence the effective number of points contributing to the fit after the cut is smaller than in the case of $m_K$ or $f_K$. Raising the cut to \mpicut MeV produced much more reliable results, hence we adopt this higher cut for the analytic fits in this section. This is justified by the fact that we observed no statistically significant deviations of the fit functions from the data over this expanded range, hence we have no reason to believe that this will lead to an 
incorrect estimate for the chiral systematic error. This was not the case for the fits to $m_\pi$, where we observed significant deviations. 

The analytic fits were again performed to data corrected to the infinite-volume using the ChPTFV fit form.

\begin{figure}[p]
\includegraphics*[width=0.49\textwidth,clip=true,trim=5 0 0 5]{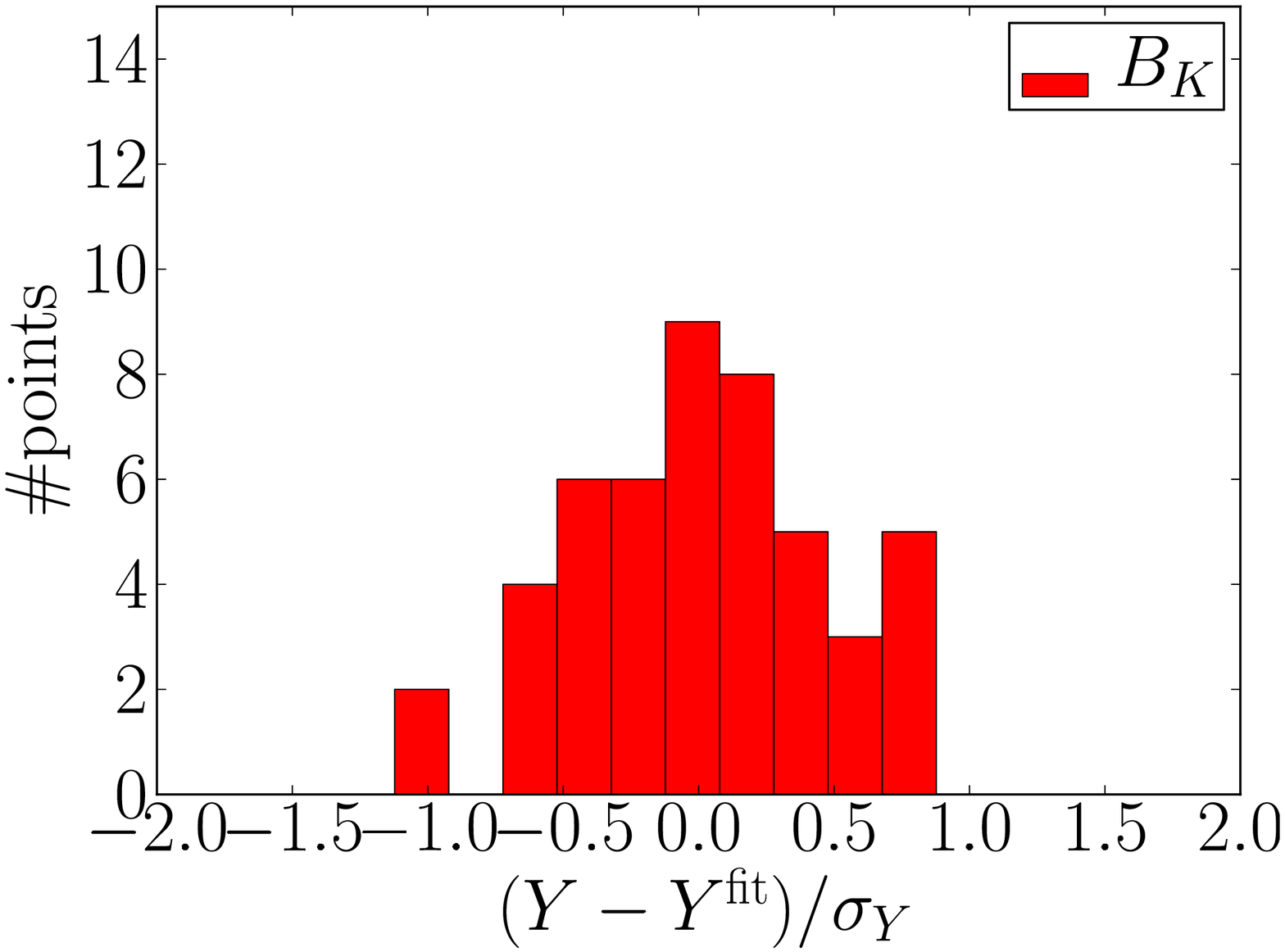}
\includegraphics*[width=0.49\textwidth,clip=true,trim=5 0 0 5]{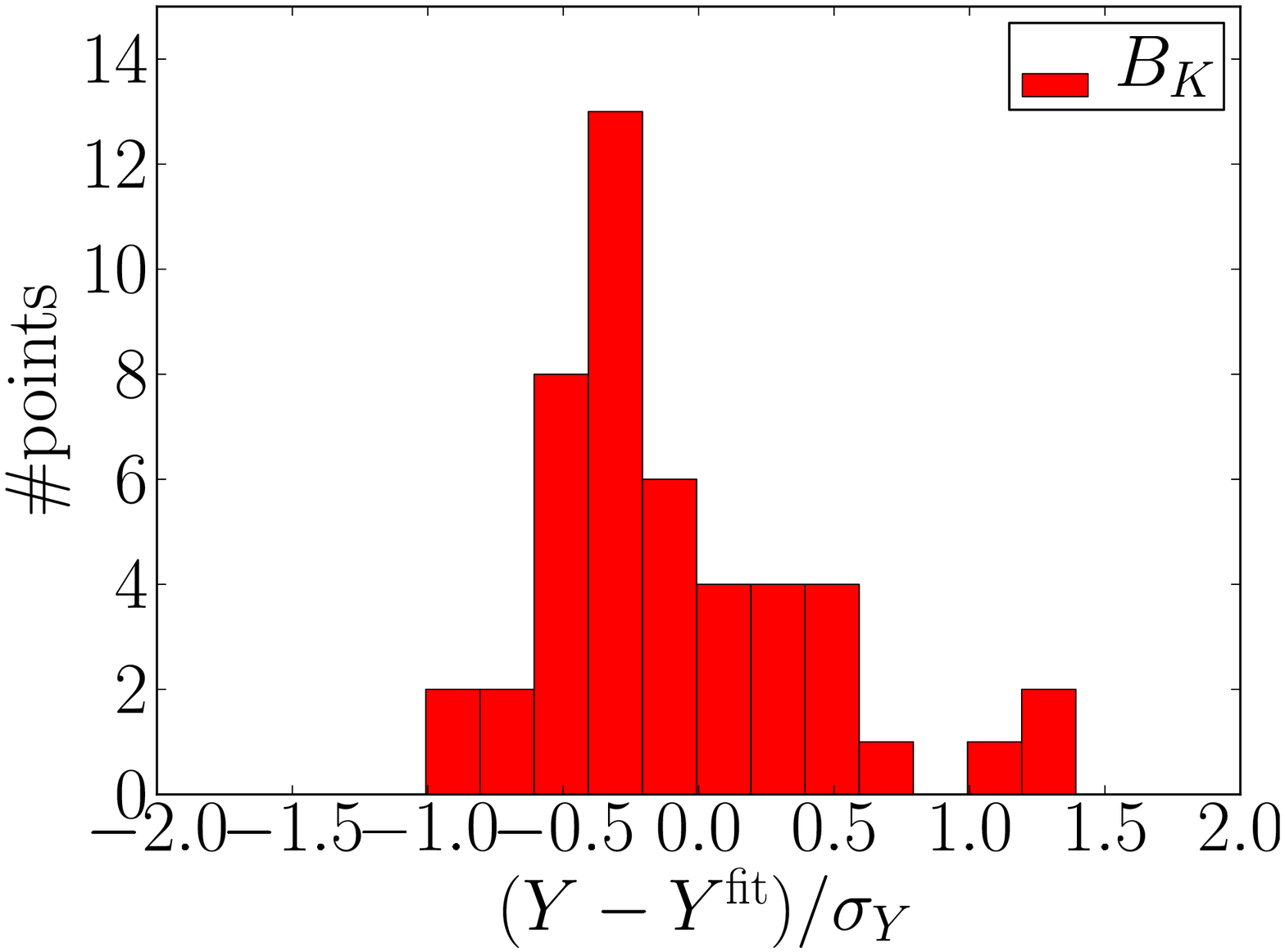}\\
\includegraphics*[width=0.49\textwidth,clip=true,trim=5 0 0 5]{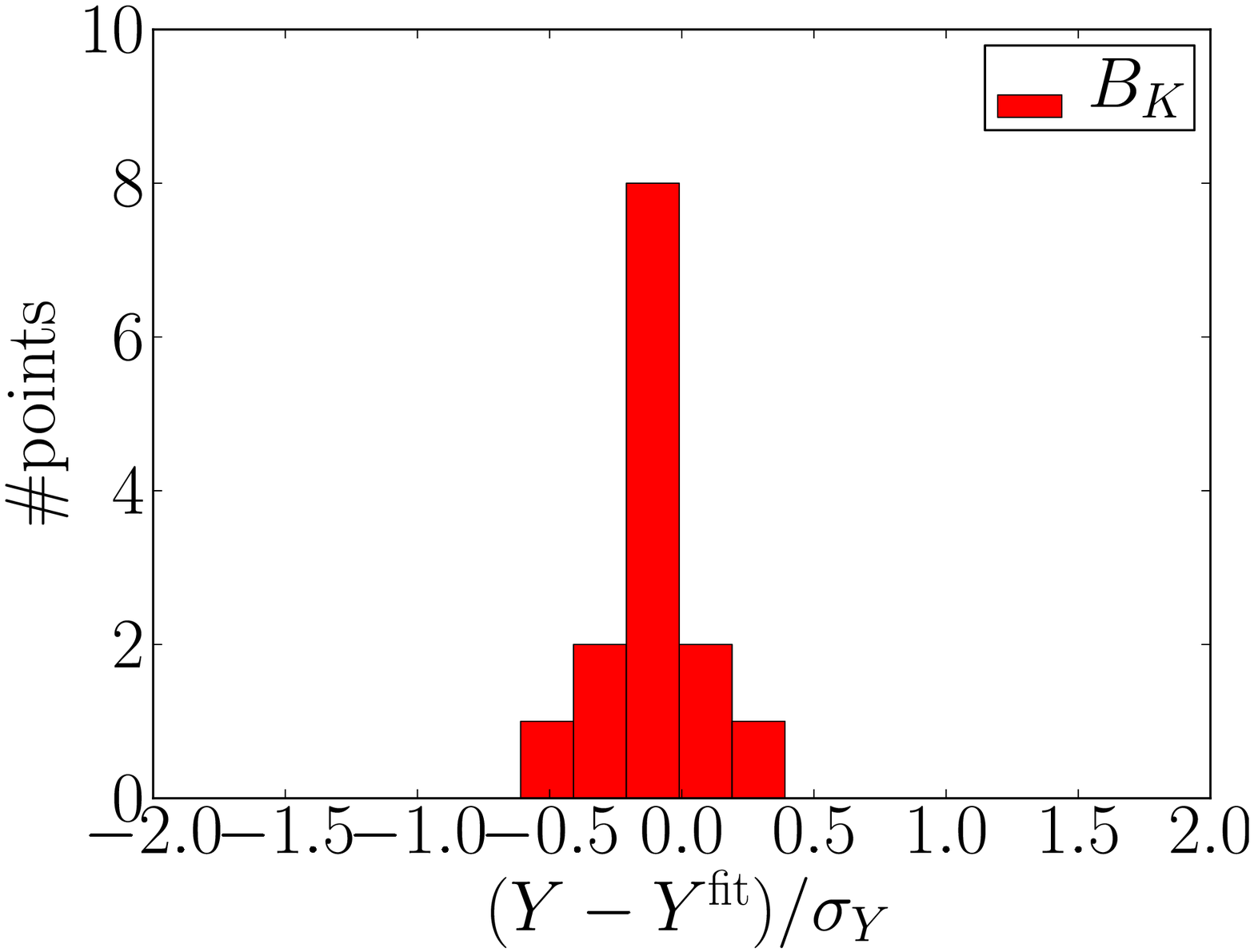}
\includegraphics*[width=0.49\textwidth,clip=true,trim=5 0 0 5]{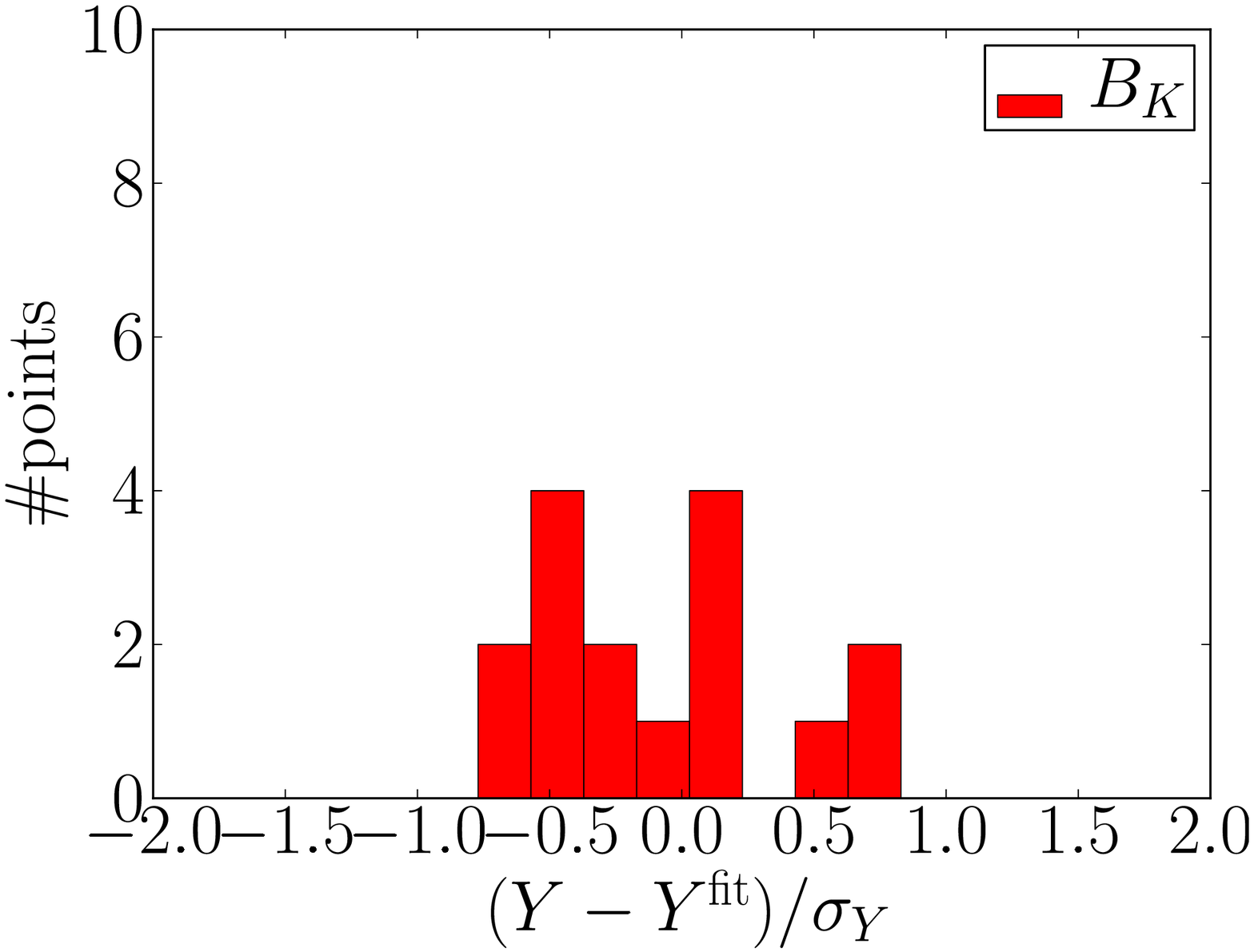}\\
\includegraphics*[width=0.49\textwidth,clip=true,trim=5 0 0 5]{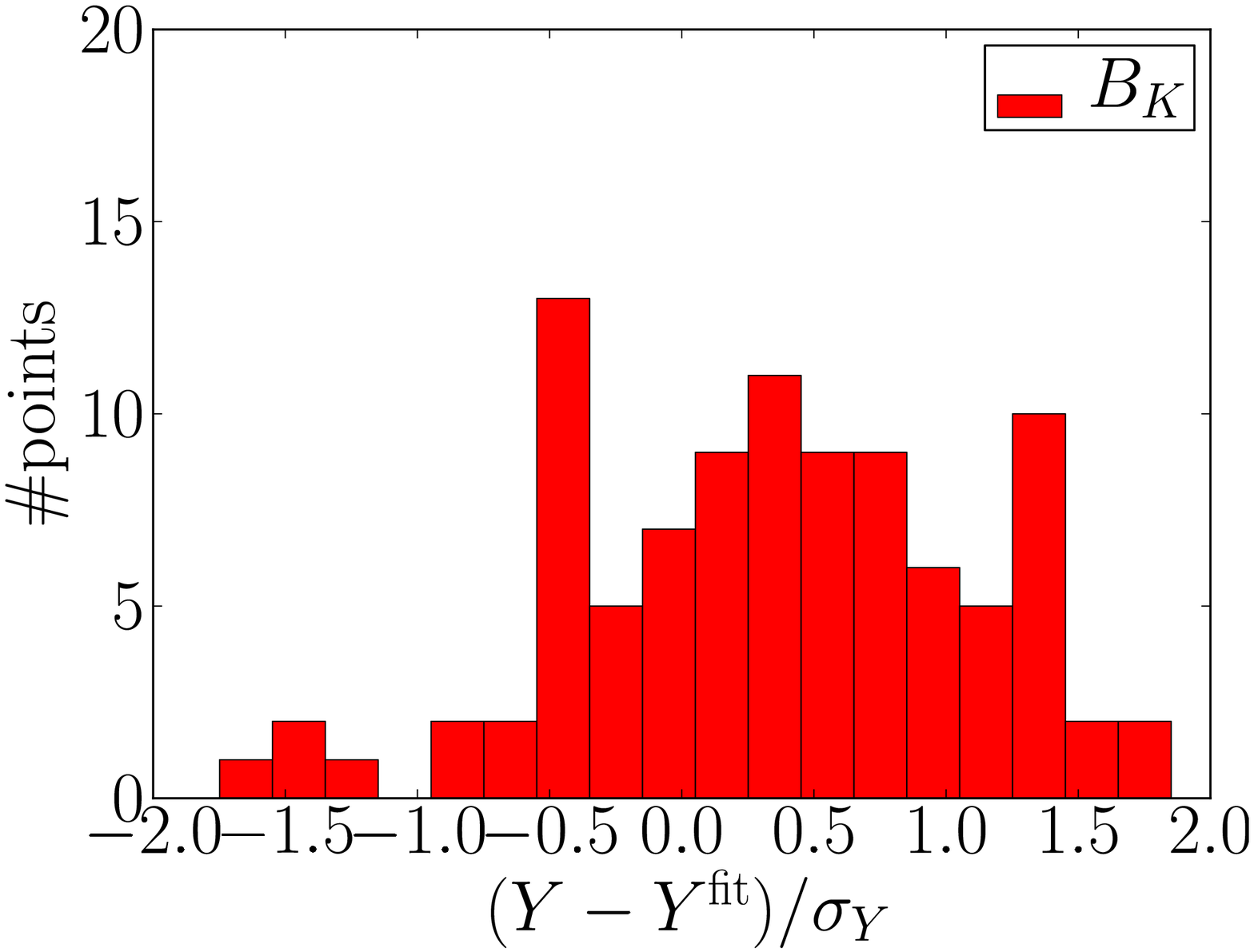}
\includegraphics*[width=0.49\textwidth,clip=true,trim=5 0 0 5]{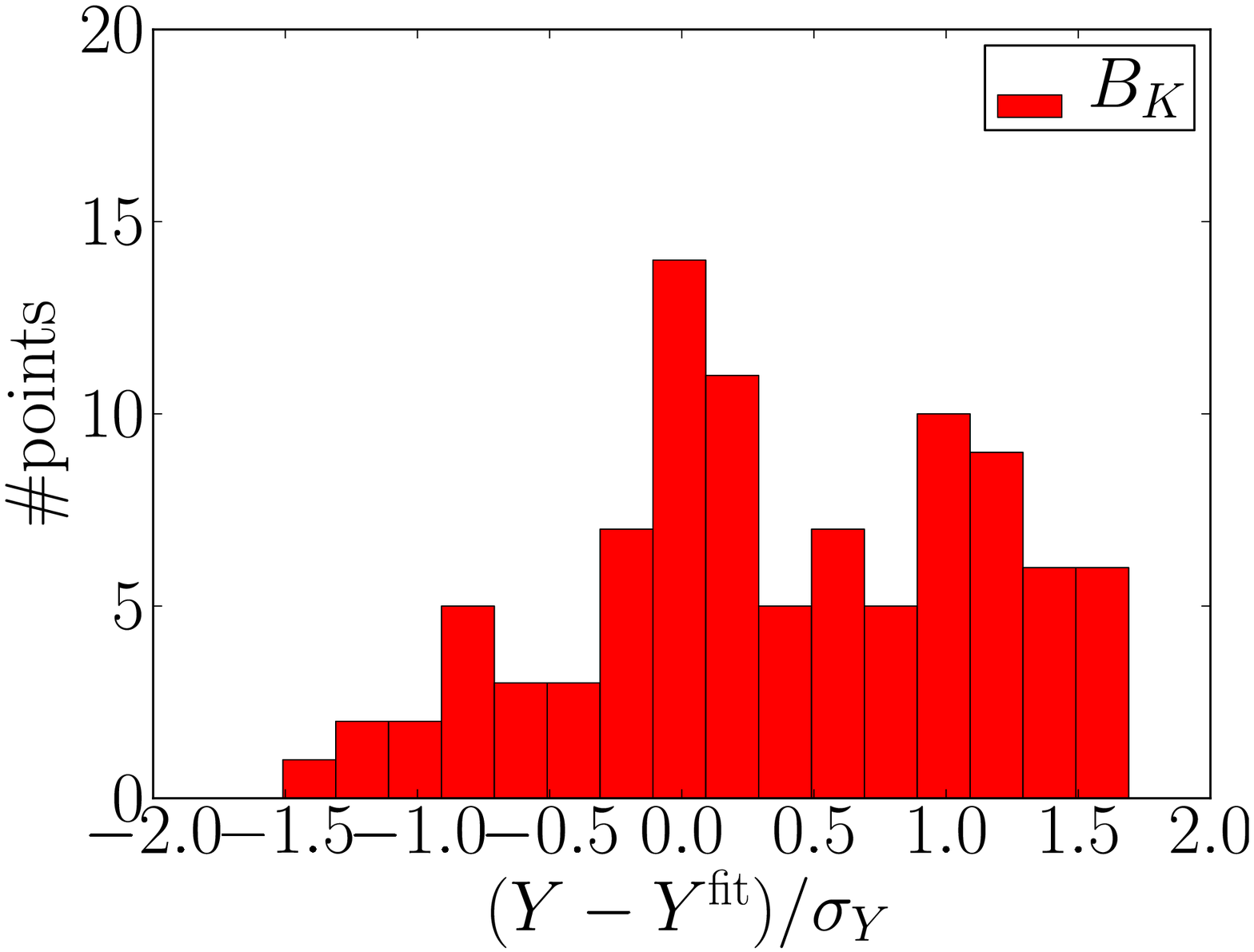}
\vspace{-0.5cm}
\caption{
\label{fig:hist_bk_cut}
Histograms of the deviation of the fit from the data for $B_K$ on each of the three ensemble sets (32I top, 24I middle and 32ID bottom) with the analytic (left) and ChPTFV (right) ans\"{a}tze.}
\end{figure}

The parameters and uncorrelated $\chi^2$/dof obtained by fitting to data renormalized in the SMOM$(\slashed q, \slashed q)$ are listed in table~\ref{tab-bklecs} and we give histograms showing the deviation of the data from the fits in figure~\ref{fig:hist_bk_cut}. We list the continuum predictions in both the SMOM$(\slashed q, \slashed q)$ and SMOM$(\gamma^\mu, \gamma^\mu)$ schemes in table~\ref{tab:bkcontinuum}.

\begin{figure}[t]
\centering
\includegraphics*[width=0.49\textwidth]{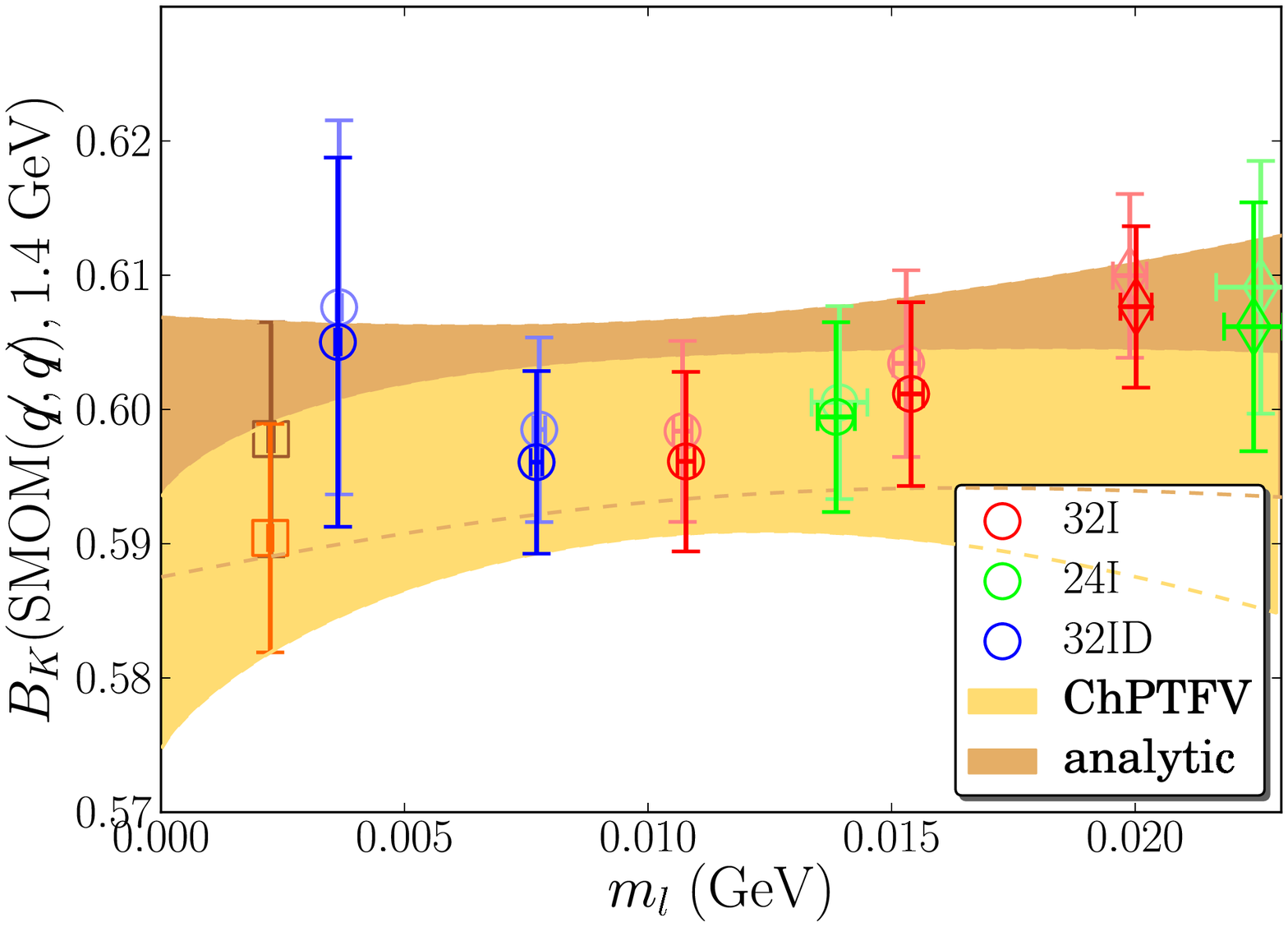}
\includegraphics*[width=0.49\textwidth]{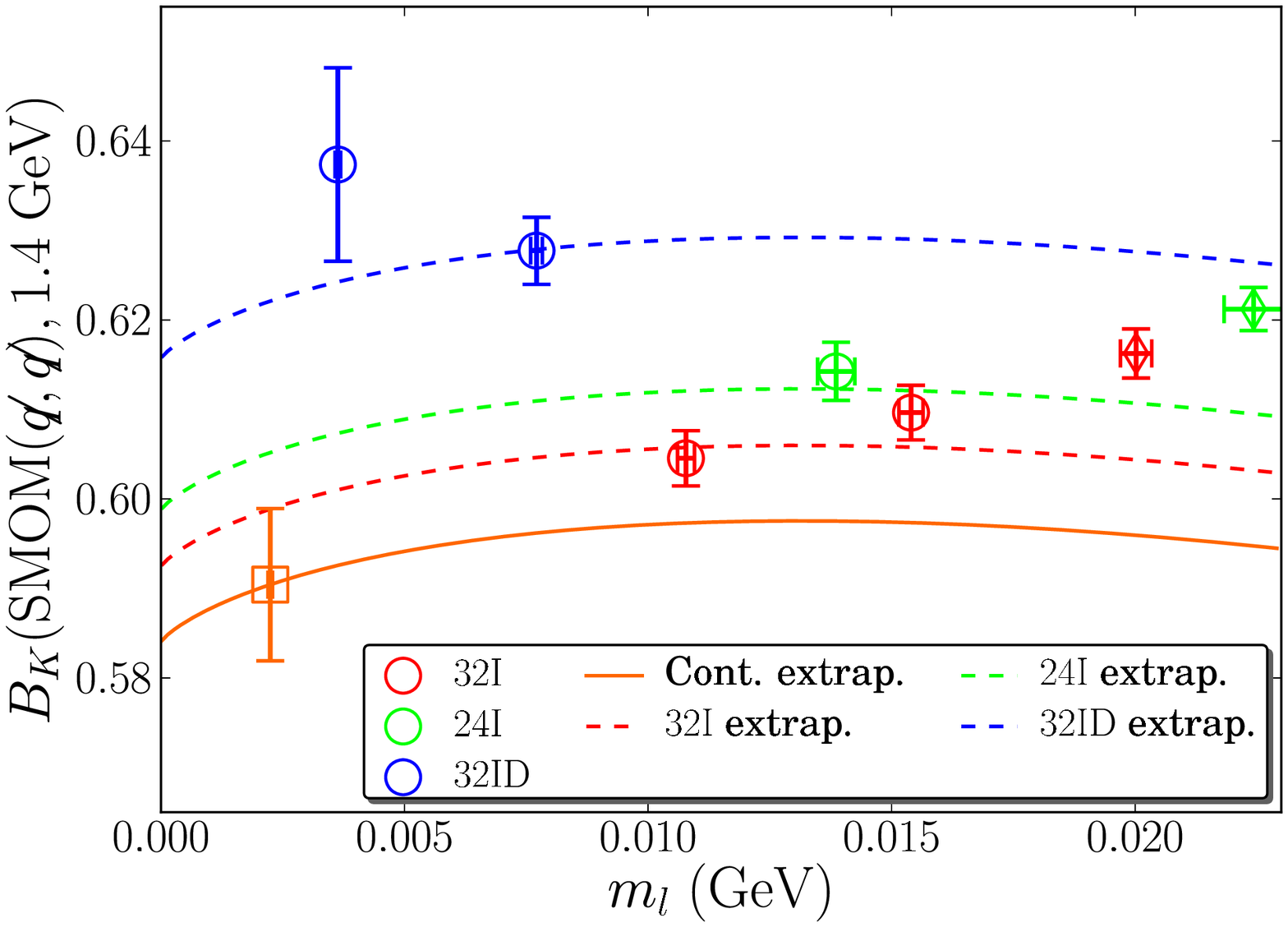}
\caption{
\label{fig:bk:contanalyticNLOcomparison}
The left figure shows the chiral extrapolation of $B_K$ in the continuum limit, renormalized in the SMOM$(\slashed{q},\slashed{q})$ scheme at a scale of $1.4$ GeV. The circular and diamond-shaped data points in darker shades show the data corrected to the continuum limit using the ChPTFV fit form, and those in lighter shades via the analytic form. The circular points indicate those data included in the fits, and the diamond points those that were not. The upper and lower curves show the analytic and ChPTFV chiral fit forms and the corresponding square data points the extrapolated values at the physical up/down quark mass. All data and curves are shown at the physical strange quark mass. The right figure shows the data at finite-$a$, adjusted to the infinite volume limit and the physical strange quark mass, overlaid by the ChPTFV fit curves at finite-$a$ and the continuum curve shown in the previous plot (shown without error bands for clarity).}
\end{figure}

In figure~\ref{fig:bk_analytic_ChPTFV_cut} we overlay the data with the fit curves on the 32ID ensembles, and in figure~\ref{fig:bk:contanalyticNLOcomparison} we show the chiral extrapolation overlaying data corrected to the continuum and infinite-volume limits as well as the physical strange quark mass via the ChPTFV and analytic parameterzations. In the latter we also plot the data at finite lattice spacing (adjusted to the infinite-volume limit and physical strange quark mass as before) and the corresponding finite-$a$ fit curves. The separation of the points at the physical up/down quark mass in the former is used as a measure of the error on the chiral extrapolation. In these figures we see that the statistical errors increase substantially as we approach the chiral limit. The central values also appear to trend upwards, although this apparent curvature is in the opposite direction to that suggested by chiral perturbation theory and is therefore likely to be simply due to the low resolution on these 
data points.

\begin{table}[t]
\centering
\begin{tabular}{l|cc}
\hline
        & \multicolumn{2}{c}{Scheme} \\
 Ansatz & SMOM$(\slashed q,\slashed q)$ & SMOM$(\gamma^\mu, \gamma^\mu)$\\
\hline
\rule{0cm}{0.4cm}Analytic  &  0.5213(72) &0.5397(76)\\
ChPT  & 0.5188(72) &0.5369(76) \\
ChPTFV  & 0.5282(73) &0.5470(78) \\
\end{tabular}
\caption{Predictions for $B_K$ in the continuum limit in the SMOM$(\slashed q,\slashed q)$ and SMOM$(\gamma^\mu, \gamma^\mu)$ schemes at $\mu=3$ GeV for each global fit ansatz. These results were obtained by applying the continuum step-scaling factors to the values in table~\protect\ref{tab:bkcontinuum}.}
\label{tab:bkcontinuum3gev}
\end{table}

\subsection{Final results for $B_K$}

Applying the step-scaling factors given in table~\ref{tab-bkstepscaling} to the continuum predictions in table~\ref{tab:bkcontinuum}, we obtained $B_K$ in the SMOM$(\slashed q, \slashed q)$ and SMOM$(\gamma^\mu, \gamma^\mu)$ schemes at a 3 GeV renormalization scale. These results are listed in table~\ref{tab:bkcontinuum3gev}. Once again we see some cancellation between the statistical fluctuations on the step-scaling factor and the 1.4 GeV quantity.

Finally, we apply the $\msbar$ conversion factors given in section~\ref{sec:bk-pertconv} to convert our results into the $\msbar$ scheme for the convenience of the reader. Before quoting our final results, we first discuss the various contributions to the systematic error. 

\subsubsection{Systematic errors}

For our central values and statistical errors of our final $\msbar$ prediction, we follow the 2010 analysis in taking the results obtained using the SMOM$(\slashed q, \slashed q)$ intermediate scheme, which is best described by one-loop perturbation theory. Following section~\ref{sec:FitResults} we estimate the finite-volume and chiral extrapolation systematics on this quantity from the differences between the ChPTFV result (which we take as our central value) and the ChPT and analytic results respectively, taking for our estimate the larger of the superjackknife error on the difference or the difference in central values. As we propagated the differences between the lattice spacings through our analysis in section~\ref{sec:zbk-results}, the aforementioned systematics on the renormalization factors are automatically included in the differences above. 

The remaining systematic errors are associated with the perturbative conversion into the $\msbar$ scheme. The largest of these is the perturbative truncation error. To determine this we again follow the 2010 analysis strategy of taking the difference between the values of $B_K$ in the $\msbar$-scheme at 3 GeV obtained using the SMOM$(\slashed q, \slashed q)$ and SMOM$(\gamma^\mu, \gamma^\mu)$ intermediate schemes, the latter of which is also well-described by perturbation theory. As discussed in section~\ref{sec-zmmsbarsys} and above, there are nonperturbative effects associated with the spontaneous chiral symmetry breaking and the presence of additional energy-scales ($\Lambda_{\rm QCD}$, $m_s$, etc.), that contribute to the perturbative systematic. In ref.~\cite{Aoki:2010pe} we found that in the nonexceptional schemes these effects are tiny compared to the truncation systematic, therefore we do not include these effects in our systematic error budget.

\subsubsection{Final results}

Using the ChPTFV result in the SMOM$(\slashed q, \slashed q)$ for the central value and statistical error, and obtaining the chiral and finite-volume systematic errors as above, we find:
\begin{equation}
B_K({\rm SMOM}(\slashed q, \slashed q),3\;\GeV) = 0.540(8)(7)(3)\,.
\end{equation}
where the errors are associated with the statistical, chiral, and finite-volume respectively. Converting this to the $\msbar$-scheme at 3 GeV using one-loop perturbation theory we obtain
\begin{equation}
B_K(\msbar,3\;\GeV) = 0.535(8)(7)(3)(11)\,,
\end{equation}
where the first three errors are as before, and the final error is that associated with the truncation of the perturbative series. Converting to the Renormalisation-Group invariant (RGI) scheme, we find
\begin{equation}
\hat B_K = 0.758(11)(10)(4)(16)\,.
\end{equation}

In the 2010 analysis we obtained:
\begin{equation}
B_K(\msbar,3\;\GeV) = 0.529(5)(15)(2)(11)\,.
\end{equation}
This is highly consistent with the result of the present analysis. In our new result we see a large improvement in the chiral extrapolation systematic, which results from lowering the pion mass cut to 350 MeV from the 420 MeV used in the previous analysis. 

For comparison, the FLAG working group give $\hat B_K = 0.738(20)$~\cite{Colangelo:2010et} for $B_K$ in the RGI scheme with 2+1 quark flavors, which was determined by combining our 2010 analysis result~\cite{Aoki:2010pe} with the value calculated by Aubin {\it et al}~\cite{Aubin:2009jh}, which used domain wall valence quarks on the 2+1 flavor staggered fermion lattices produced by the MILC collaboration. The result of $\hat B_K = 0.758(22)$ obtained in the current analysis is consistent with this value. Other calculations performed since the publication of the FLAG 2010 paper include refs.~\cite{Durr:2011ap},~\cite{Bae:2011ff} and~\cite{Laiho:2011dy}.

%%%%%%%%%%%%%%%%%%%%%%%%% B_K %%%%%%%%%%%%%%%%%%%%%%%%%%%%%%%%
\section{Chiral/Continuum Fits And Physical Results For The Sommer Scales}
\label{sec:r0r1}
\FloatBarrier
In this section we present the results of applying our global fit technique to the Sommer scales, $r_0$ and $r_1$. In ref.~\cite{Aoki:2010dy} we determined continuum values for these parameters using global fits to our Iwasaki ensemble sets. In this paper we extend these fits to include the 32ID ensemble set and observe the effect of lowering the pion mass cut. The values of $r_0$ and $r_1$ measured on the 32ID ensemble sets can be found in section~\ref{sec:DSDRresults}.

Assuming a linear dependence on the quark masses and on $a^2$, we performed our chiral/continuum fits using the following form:
\begin{equation}
r_i^{\bf 1} = c_{r_i,0}(1 + c_{r_i,a}^{A(\bf 1)}[a^{\bf 1}]^2) + c_{r_i,m_l}\tilde m_l^{\bf 1} + c_{r_i,m_h}(\tilde m_h^{\bf 1} - m_{h0})
\end{equation}
on the primary lattice $\bf 1$. As always the fit form describing another ensemble set, $\bf e$, is obtained by inserting factors of $Z_l^{\bf e}$ and $Z_h^{\bf e}$ to convert the simulated quark masses on ensemble $\bf e$ into the matching scheme, and selecting the $a^2$ coefficient for the lattice action of the ensemble set.

\begin{table}
\centering
\begin{tabular}{c|c|c}
\hline
\rule{0cm}{0.4cm}Ansatz & $\chi^2/\mathrm{dof}$ & $\chi^2/\mathrm{dof}$\\
 & Uncut & Cut \\ 
\hline
 \rule{0cm}{0.4cm}Analytic & $ 1.45(66) $ & $ 0.141(71) $ \\
 ChPT & $ 1.47(67) $ & $ 0.41(40) $ \\
 ChPTFV & $ 1.47(67) $ & $ 0.42(40) $ \\
\end{tabular}
\caption{Fit ansatze and the associated uncorrelated $\chi^2/\mathrm{dof}$ obtained by fitting to $r_0$ and $r_1$ over the full data set (second column) and to the cut data set (third column). The upper bounds on the pion mass in the cut data sets are $m_\pi = \mpicut$ MeV for the ChPT and ChPTFV fits and $m_\pi < \mpicutanalytic$ MeV for the analytic fit.}
\label{tab-r0fitchisq}
\end{table}

\begin{table}
\centering
\begin{tabular}{l|rrr|rrr}
\hline
\rule{0cm}{0.4cm} & \multicolumn{3}{c|}{Uncut} & \multicolumn{3}{c}{Cut}\\
Parameter & Analytic & ChPT & ChPTFV & Analytic & ChPT & ChPTFV\\
\hline
 \rule{0cm}{0.4cm}$c_{r_0,0}$ (GeV$^{-1}$) & $ 2.479(34) $ & $ 2.445(36) $ & $ 2.438(38) $ & $ 2.462(49) $ & $ 2.453(51) $ & $ 2.441(52) $\\
 $c_{r_0,a}^{\scriptscriptstyle I}$ (GeV$^2$) & $ -0.065(53) $ & $ -0.013(46) $ & $ -0.008(47) $ & $ -0.008(85) $ & $ -0.018(64) $ & $ -0.010(65) $\\
 $c_{r_0,a}^{\scriptscriptstyle ID}$ (GeV$^2$) & $ -0.055(24) $ & $ -0.028(26) $ & $ -0.023(28) $ & $ -0.032(35) $ & $ -0.030(33) $ & $ -0.021(34) $\\
 $c_{r_0,m_l}$ (GeV$^{-2}$) & $ -1.67(87) $ & $ -1.65(88) $ & $ -1.64(87) $ & $ -5.0(1.7) $ & $ -3.6(1.4) $ & $ -3.6(1.4) $\\
 $c_{r_0,m_h}$ (GeV$^{-2}$) & $ -0.83(42) $ & $ -0.83(42) $ & $ -0.83(42) $ & $ -0.27(64) $ & $ -0.56(52) $ & $ -0.56(51) $\\
\hline
 $c_{r_1,0}$ (GeV$^{-1}$) & $ 1.697(24) $ & $ 1.675(26) $ & $ 1.671(27) $ & $ 1.662(41) $ & $ 1.650(40) $ & $ 1.642(40) $\\
 $c_{r_1,a}^{\scriptscriptstyle I}$ (GeV$^2$) & $ -0.099(64) $ & $ -0.050(58) $ & $ -0.045(58) $ & $ 0.00(11) $ & $ 0.014(91) $ & $ 0.023(92) $\\
 $c_{r_1,a}^{\scriptscriptstyle ID}$ (GeV$^2$) & $ -0.148(25) $ & $ -0.123(26) $ & $ -0.118(28) $ & $ -0.110(38) $ & $ -0.097(38) $ & $ -0.088(39) $\\
 $c_{r_1,m_l}$ (GeV$^{-2}$) & $ -1.84(60) $ & $ -1.82(59) $ & $ -1.81(59) $ & $ -2.6(2.4) $ & $ -2.2(1.1) $ & $ -2.2(1.1) $\\
 $c_{r_1,m_h}$ (GeV$^{-2}$) & $ -1.02(20) $ & $ -1.02(20) $ & $ -1.01(20) $ & $ -0.88(37) $ & $ -0.73(24) $ & $ -0.73(24) $\\
\end{tabular}
\caption{The $a^2$ and mass dependences of $r_0$ and $r_1$ obtained by fitting to the full and cut data sets. We repeat the fits for each choice of chiral ansatz used for the determination of the scaling parameters. The upper bounds on the pion mass in the cut data sets are $m_\pi = \mpicut$ MeV for the ChPT and ChPTFV fits and $m_\pi < \mpicutanalytic$ MeV for the analytic fit. The parameters are given in physical units and with the heavy quark mass expansion point adjusted to the physical strange quark mass}
\label{tab:r0r1fit_lecs}
\end{table}

\begin{table}
\centering
\begin{tabular}{l|lll|lll}
\hline
\rule{0cm}{0.4cm} & \multicolumn{3}{c|}{Uncut} & \multicolumn{3}{c}{Cut}\\
 & Analytic & ChPT & ChPTFV & Analytic & ChPT & ChPTFV\\
\hline
 $r_0^{\mathrm{\tiny continuum}}$ & $ 2.475(33) $ & $ 2.441(35) $ & $ 2.435(37) $ & $ 2.451(48) $ & $ 2.445(49) $ & $ 2.433(50) $ \\
 $r_1^{\mathrm{\tiny continuum}}$ & $ 1.693(23) $ & $ 1.671(24) $ & $ 1.666(25) $ & $ 1.657(38) $ & $ 1.645(38) $ & $ 1.637(39) $ \\
 $(r_1/r_0)^{\mathrm{\tiny continuum}}$ & $ 0.684(8) $ & $ 0.684(8) $ & $ 0.684(8) $ & $ 0.676(11) $ & $ 0.673(11) $ & $ 0.673(11) $ \\
\end{tabular}
\caption{Continuum predictions for $r_0$ and $r_1$ in GeV$^{-1}$ as well as their ratio, using scaling parameters obtained from each of the three global fit ansatz\"{e}. The first set of columns contain the values obtained by fitting to the full data set, and the second set those obtained by fitting to the cut data set. The upper bounds on the pion mass in the cut data sets are $m_\pi = \mpicut$ MeV for the ChPT and ChPTFV fits and $m_\pi < \mpicutanalytic$ MeV for the analytic fit.}
\label{r0r1continuum}
\end{table}

\begin{figure}[tp]
\centering
\includegraphics*[width=0.49\textwidth,clip=true,trim=5 0 5 5]{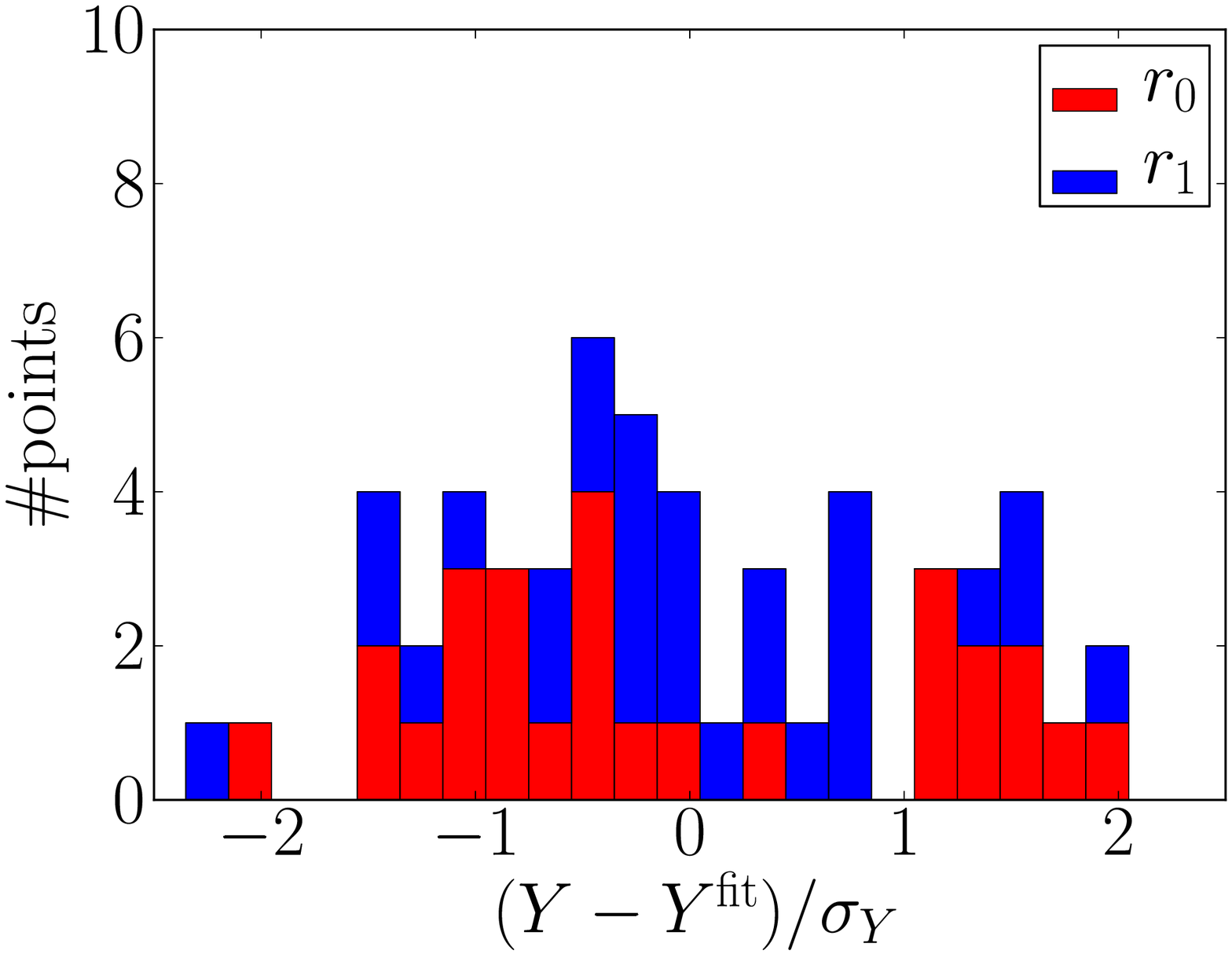}
\includegraphics*[width=0.49\textwidth,clip=true,trim=5 0 5 5]{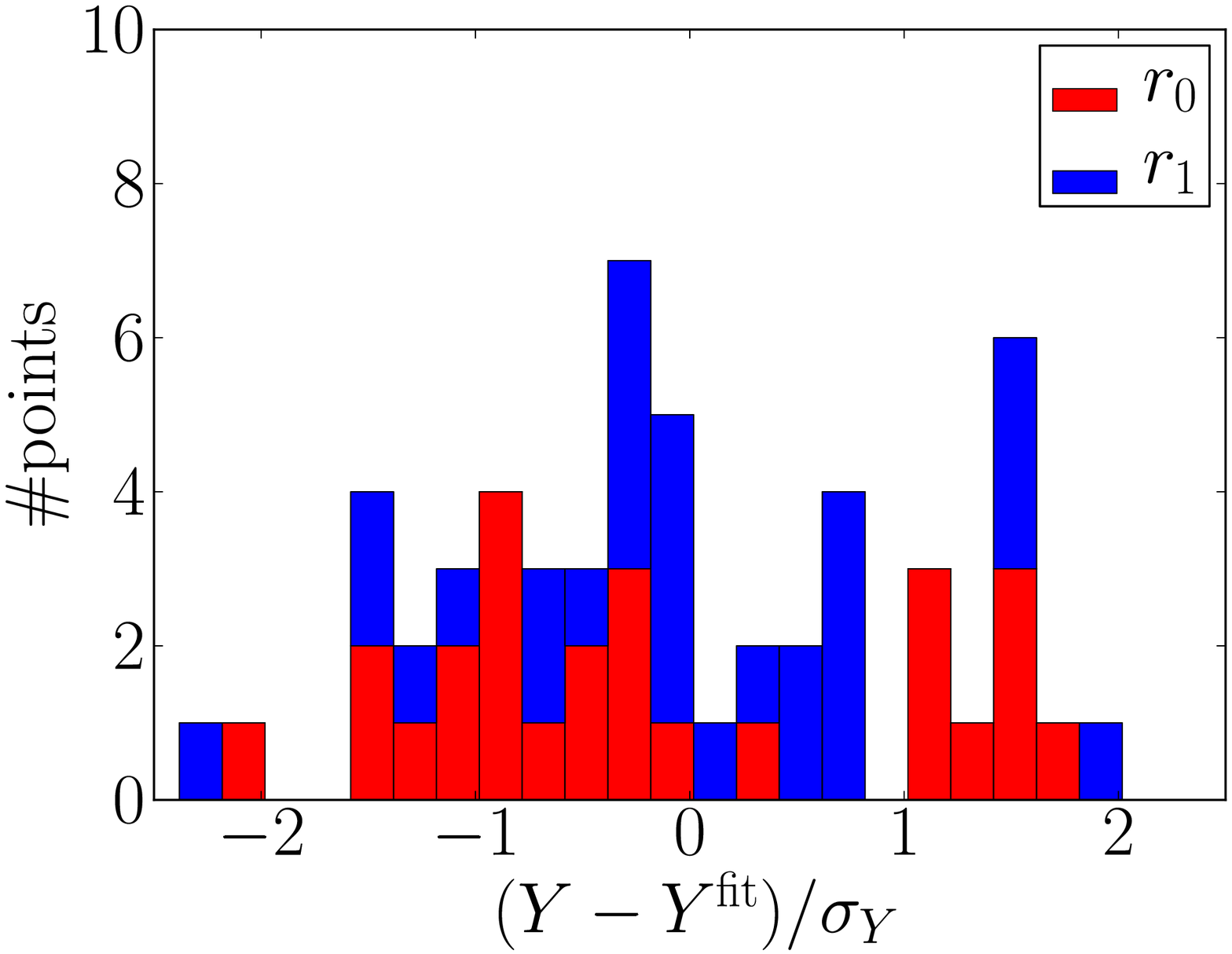}\\
\includegraphics*[width=0.49\textwidth,clip=true,trim=5 0 5 5]{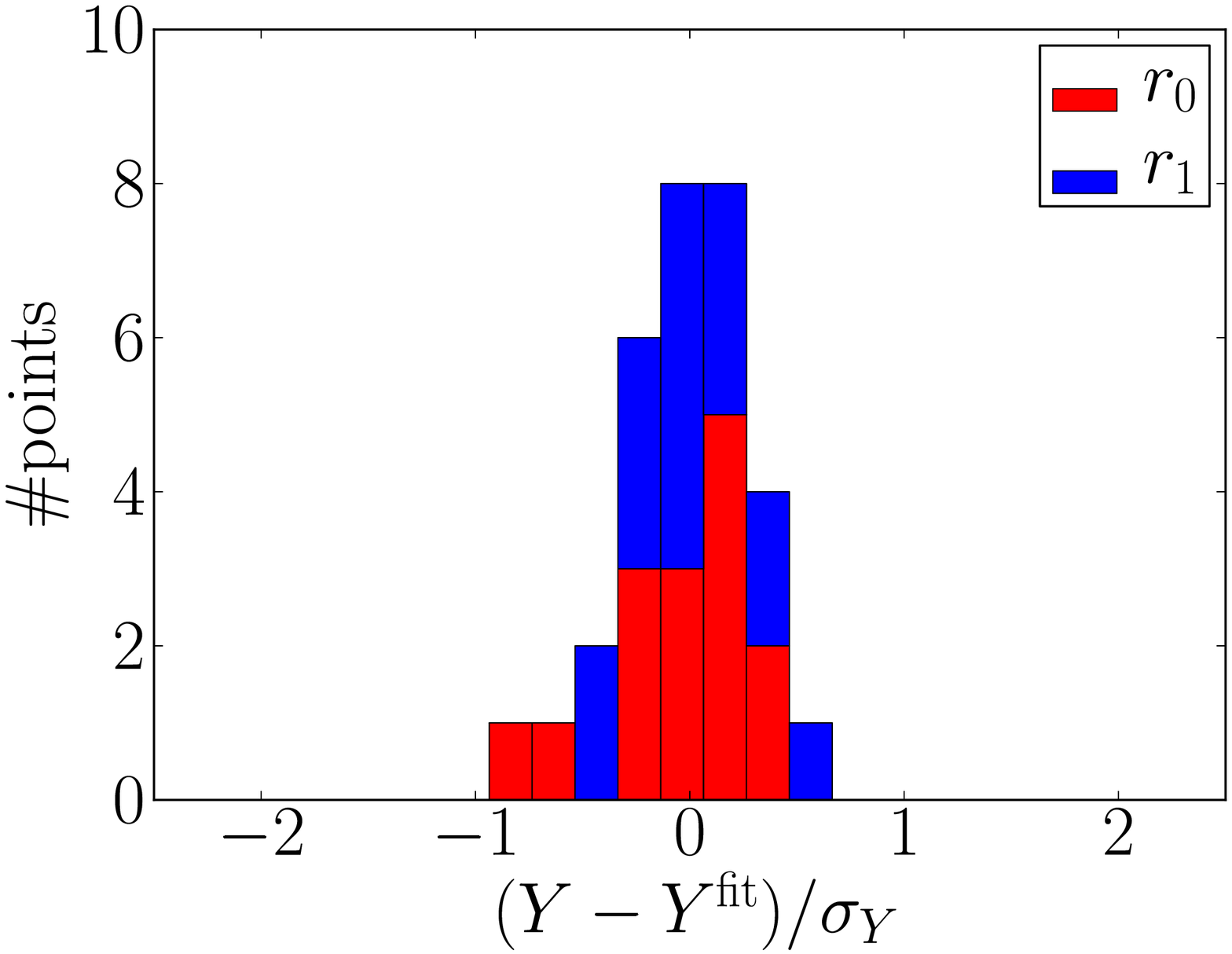}
\includegraphics*[width=0.49\textwidth,clip=true,trim=5 0 5 5]{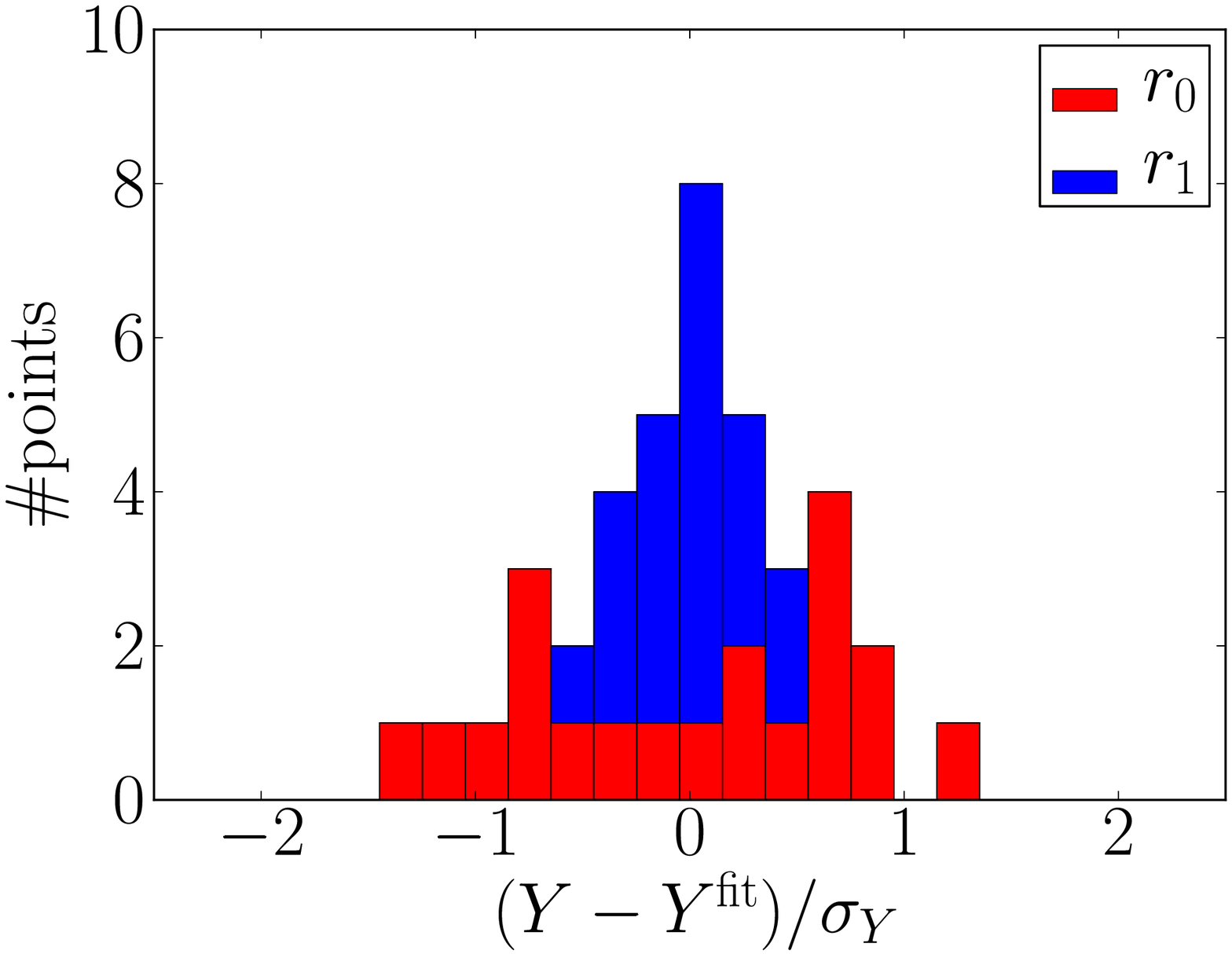}\\
\vspace{-0.5cm}
\caption{
\label{fig:globalhistr0}
Histograms of the deviation of the fit from the data for $r0$ and $r_1$ over all three ensemble sets, fitting with the analytic (left) and ChPTFV (right) ans\"{a}tze to the uncut (top) and cut (bottom) data sets.}
\end{figure}

For convenience, we simultaneously fit both $r_0$ and $r_1$, even though they do not share any common parameters other than the scaling factors, $Z_l$ and $Z_h$. The lattice spacings and scaling factors were fixed to those obtained in the main analysis, with the fits repeated for each of the three chiral ans\"{a}tze. For each fit we applied the same cuts as were performed to the data in section~\ref{sec:FitResults}; this corresponds to removing the data points on the 32I, $m_l=0.008$ and 24I, $m_l=0.01$ ensembles, and also in the analytic fit, the data point on the 32I, $m_l=0.006$ ensemble (cf. table~\ref{tab:260MeVcutmasscombs}). For later comparison we also quote the results of fitting to the full data set in this section, although as previously discussed these results are flawed due to the poor fit to several of the pion mass data points on the 32ID ensembles. In table~\ref{tab-r0fitchisq} we give the uncorrelated $\chi^2$/dof of our fits and in figure~\ref{fig:globalhistr0} we show histograms of the 
deviations of the data from unity for the fits. We list the fit parameters in table~\ref{tab:r0r1fit_lecs} and the continuum predictions for $r_1$, $r_0$ and their ratio in table~\ref{r0r1continuum}.

\begin{figure}[tp]
\centering
\includegraphics*[width=0.49\textwidth,clip=true,trim=5 0 5 0]{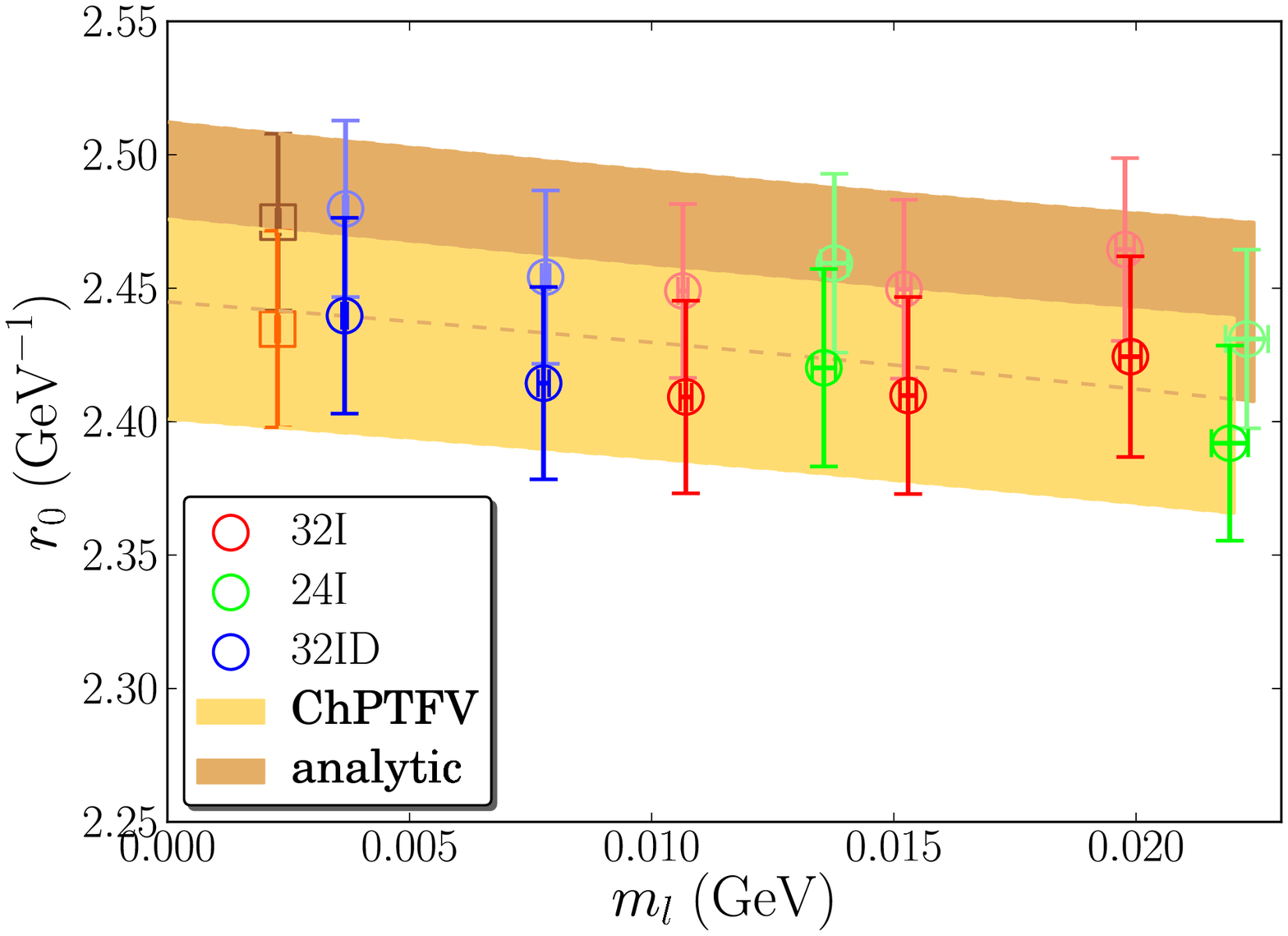}
\includegraphics*[width=0.49\textwidth,clip=true,trim=5 0 5 0]{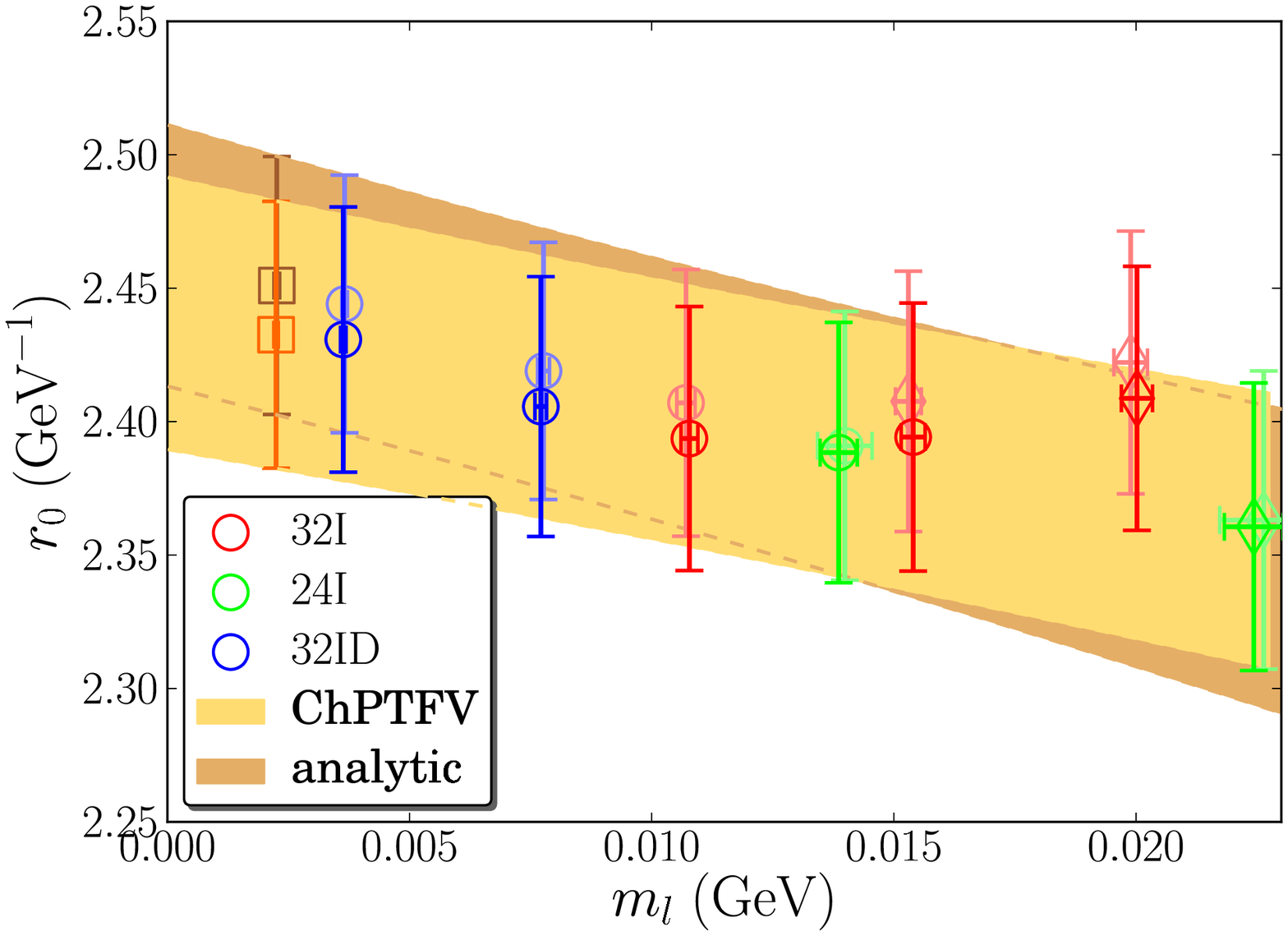}\\
\includegraphics*[width=0.49\textwidth,clip=true,trim=5 0 5 0]{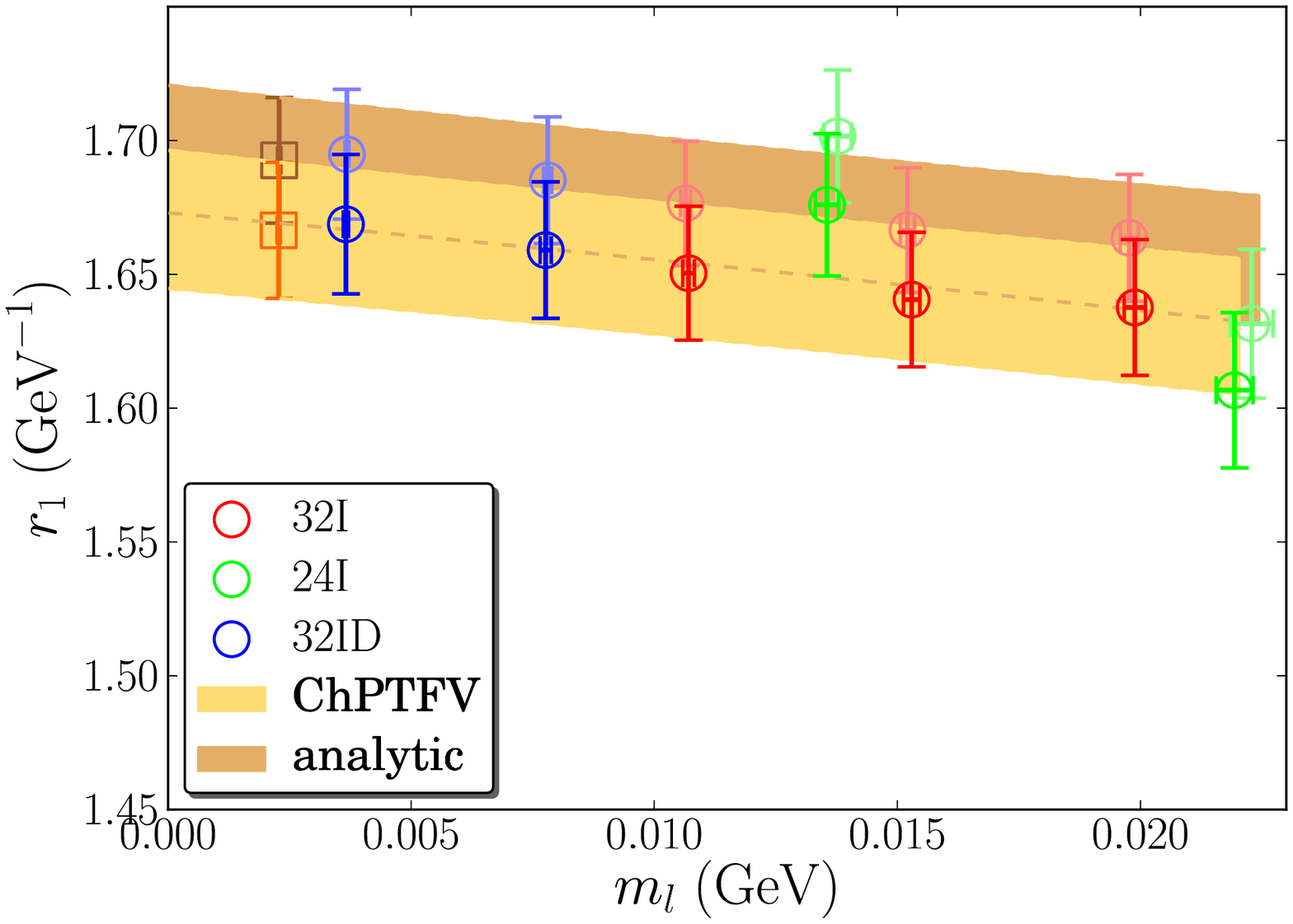}
\includegraphics*[width=0.49\textwidth,clip=true,trim=5 0 5 0]{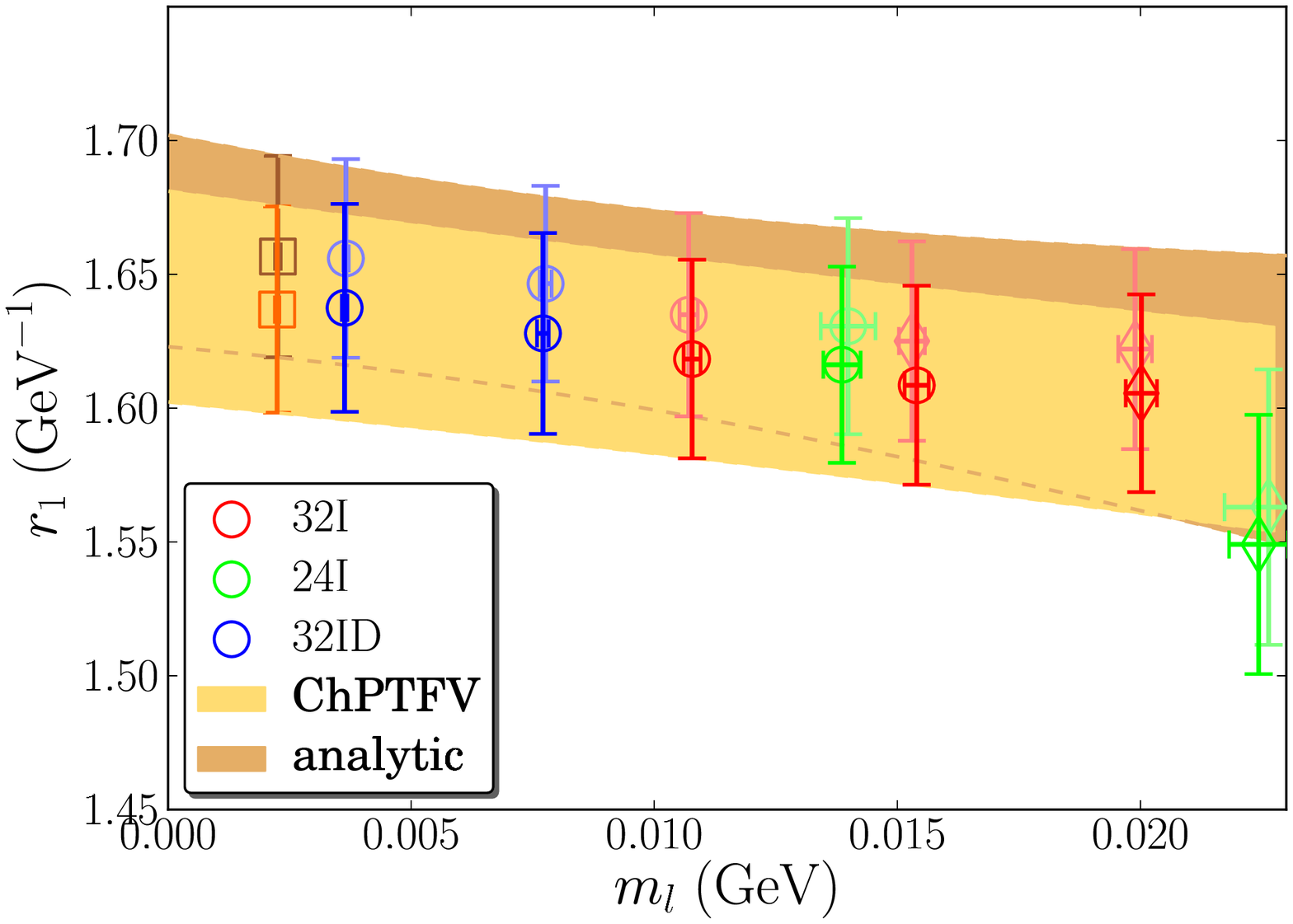}\\
\vspace{-0.5cm}
\caption{
\label{fig:chiralextrapr0r1}
The chiral extrapolation of $r_0$ (top) and $r_1$ (bottom) using the analytic and ChPTFV ans\"{a}tze. The plots on the left show the fits to the full data set and those on the right to the cut data sets. We have overlayed the fit curves with the data points corrected to the continuum limit and physical strange quark mass using each of the aforementioned fit functions; those points shown in bold colors were corrected using the ChPTFV fits and those in pastel colors using the analytic fits. The circular data points are those included in the fits and the diamond points those that were not. The square points show the predicted value at the physical point.}
\end{figure}

\begin{figure}[tp]
\centering
\includegraphics*[width=0.49\textwidth,clip=true,trim=5 0 5 0]{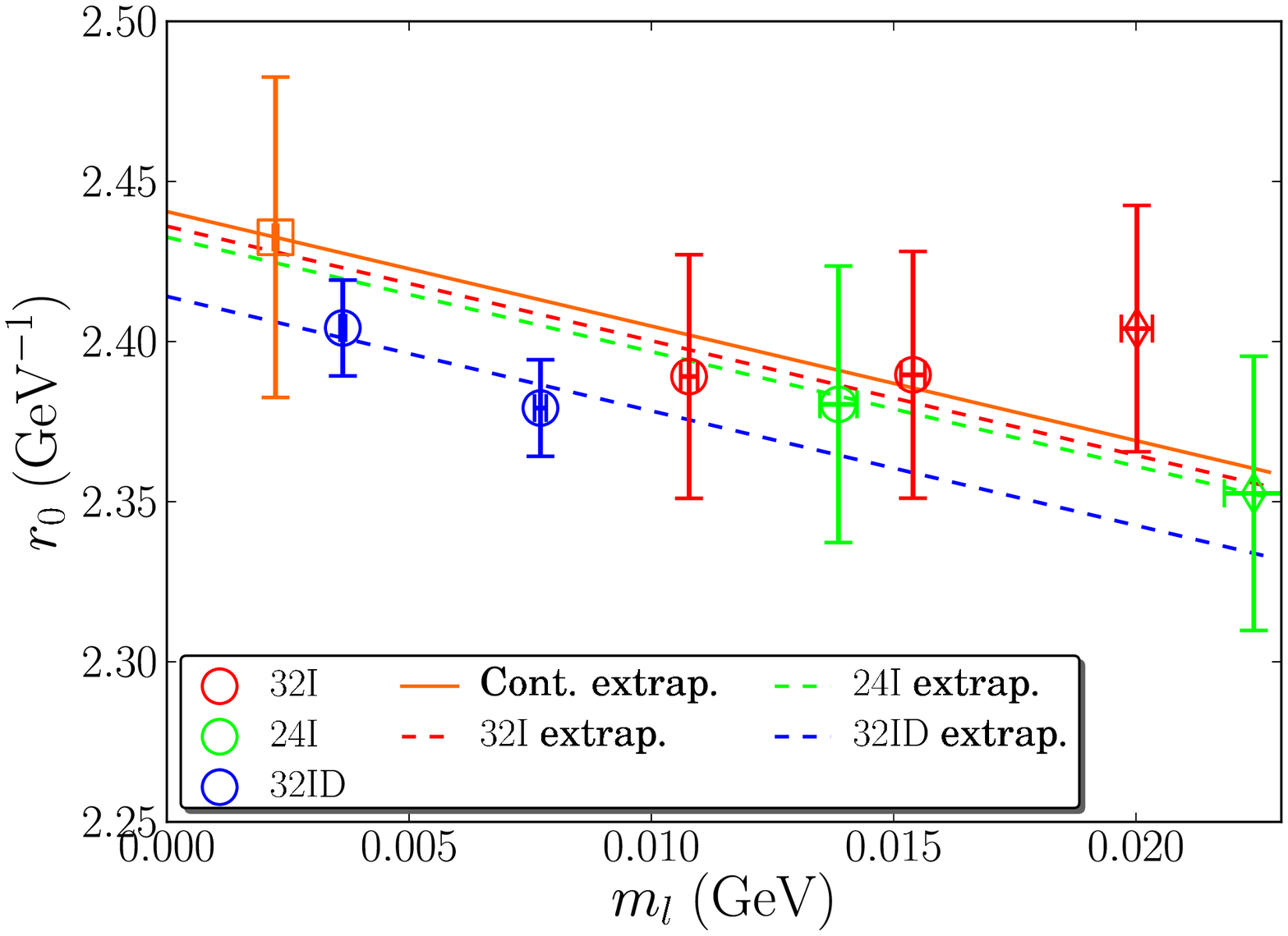}
\includegraphics*[width=0.49\textwidth,clip=true,trim=5 0 5 0]{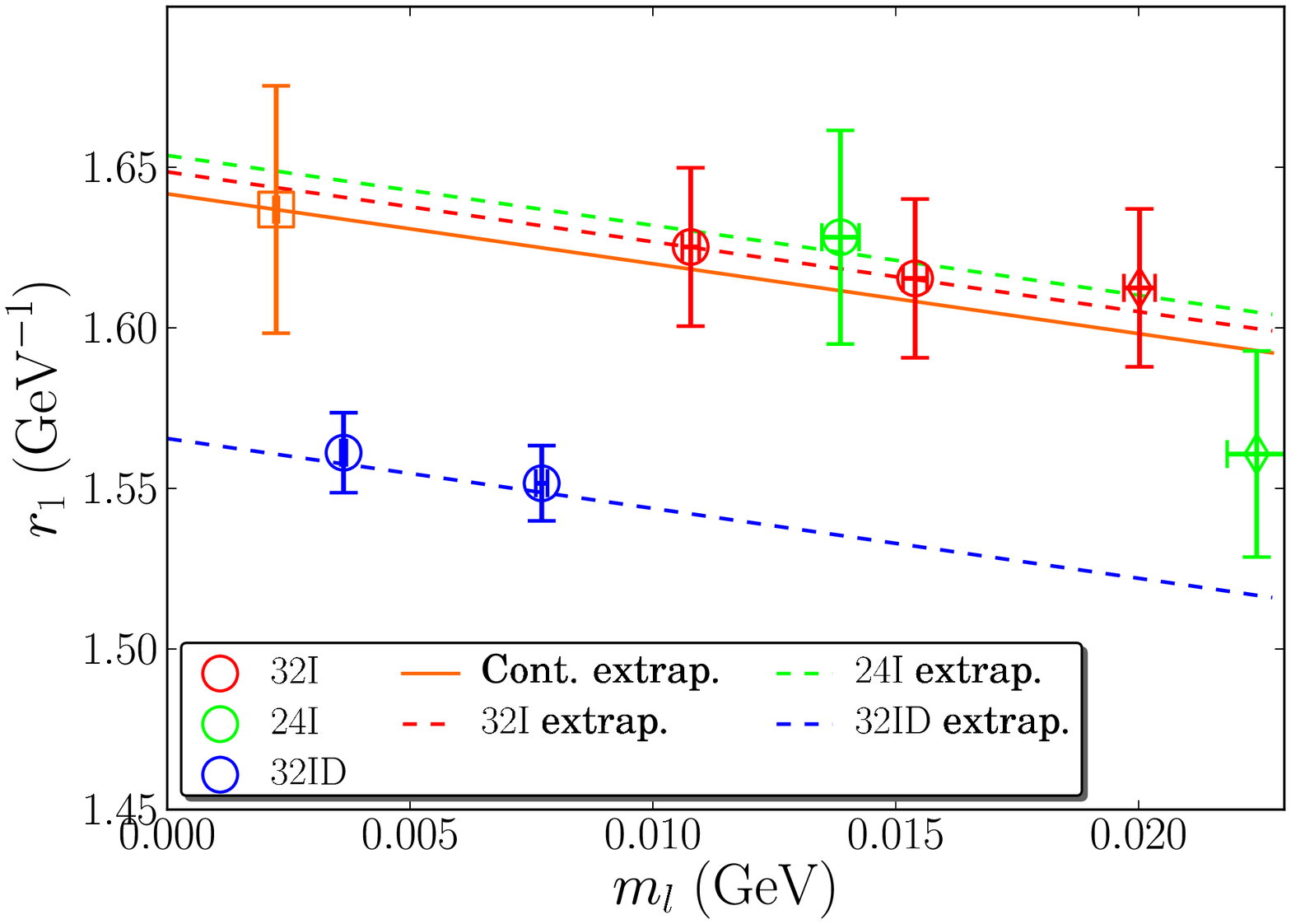}
\vspace{-0.5cm}
\caption{
\label{fig:chiralextrapr0r1finitea}
The chiral extrapolation of $r_0$ (left) and $r_1$ (right) using the ChPTFV ansatz applied to the cut data set. Here we have overlayed the fit curves at finite lattice spacing (dashed lines) with the raw data points corrected to the physical strange quark mass. We also show the continuum fit curve (solid line) and the physical point (square). As before the circular data points are those included in the fits and the diamond points those that were not.}
\end{figure}

In the 2010 analysis we remarked on a tension between the fit and the value of $r_1$ on the heaviest 24I ensemble, which led to us inflating the error on the prediction for this quantity. In figures~\ref{fig:chiralextrapr0r1} and~\ref{fig:chiralextrapr0r1finitea} we plot the chiral extrapolation in the continuum limit and at finite lattice spacing respectively. In these figures we see the large apparent difference in the slopes of $r_1$ with respect to $m_l$ between the two Iwasaki ensemble sets that was responsible for this tension. It appears however that the slopes of $r_1$ agree very well between the 32I and 32ID ensemble sets, which has led to a substantially better fit to $r_1$ upon including the 32ID data. Restricting the fits to lighter data markedly improves our fits, reducing the $\chi^2/$dof by at least a factor of three. The chiral behavior of the data, as illustrated in the right-hand plots in figure~\ref{fig:chiralextrapr0r1}, is now very linear. As a result of these observations, we decided 
that inflating the error on $r_1$ is no longer necessary.

We obtain continuum predictions for $r_1$, $r_0$ and their ratio from the cut fit results using the strategy detailed in section~\ref{subsec:syserrs}. We find
\begin{equation}\begin{array}{lll}
r_1 & = 1.637(39)(20)(8)\ {\rm GeV}^{-1} & = 0.3230(77)(39)(16)\ {\rm fm},\\
r_0 & = 2.433(50)(18)(13)\ {\rm GeV}^{-1} & = 0.4795(99)(35)(26)\ {\rm fm},\\
r_1/r_0 & = 0.6729(109)(30)(2)\,,
\end{array}\end{equation}
where the errors are statistical, chiral and finite-volume respectively. The values determined in ref.~\cite{Aoki:2010dy} were $r_1 = 0.3333(93)(2)(1)$ fm, $r_0 = 0.4870(89)(2)(2)$ fm and $r_1/r_0 = 0.6844(97)(1)(0)$. By comparing the results in table~\ref{r0r1continuum} with those obtained in the 2010 analysis we find that, as with the Omega mass, the use of the generic scaling procedure for determining the scaling factors leads to considerably larger chiral and finite-volume systematic errors than the fixed trajectory approach. In the case of $r_1$, we see a reduced systematic error in the continuum prediction due to the improved control over the chiral extrapolation. However for $r_0$ -- which formerly did not display any tensions with the linear ansatz requiring error inflation -- this is offset by the reduction in the amount of data. For comparison, the MILC collaboration recently obtained $r_1 = 0.3106(17)\ {\rm fm}$~\cite{Bazavov:2010hj} and in an earlier work $r_0 = 0.462(12)\ {\rm fm}$~\cite{Aubin:2004wf}, both of which appear to be consistent with our results.

%%%%%%%%%%%%%%%%%%%%%%%%% Conclusions %%%%%%%%%%%%%%%%%%%%%%%%%%%%%%%%
\FloatBarrier
\section{Conclusions}
\label{sec:Conclusions}

Using the Iwasaki gauge action with the addition of the DSDR term we were able to simulate with domain wall fermions (DWF) at a relatively strong coupling ($\beta=1.75$, $a^{-1}= \ainvtwo$ GeV) while retaining good chiral symmetry and topological tunneling; this enabled us to work with a large enough physical volume ($[4.61\ {\rm fm}]^3$) to accomodate pions as light as $143(1)$ MeV without suffering from large finite-volume effects ($m_\pi L \approx 3.2$ for the lightest partially-quenched point and $m_\pi L \approx 4$ for the lightest unitary point) and without having to simulate with a large number of lattice sites; the dimensionless lattice volume is $32^3\times 64\times 32$, where the final number is the length of the fifth dimension $L_s$ that governs the size of the chiral symmetry breaking in the domain wall formulation. 

The aim of this paper was to combine these data in a simultaneous chiral/continuum fit with our $24^3\times 64\times 16$ and $32^3\times 64\times 16$ DWF ensembles with the Iwasaki gauge action at $\beta = 2.13$ ($a^{-1}= \ainvone$ GeV) and $\beta = 2.25$ ($a^{-1} = \ainvzero$ GeV) respectively, and under the constraint of universality obtain continuum predictions for various quantities. In this we broadly followed the strategy of our 2010 analysis~\cite{Aoki:2010dy,Aoki:2010pe}.

The fits were performed assuming three forms for the mass dependence: the ChPTFV and ChPT forms were obtained from NLO SU(2) chiral perturbation theory with and without finite-volume corrections respectively, and the analytic ansatz from a linear Taylor expansion about an unphysical mass point.

The largest change from our 2010 analysis strategy was the use of the ``generic scaling'' method to obtain the scaling parameters $Z_l$ and $Z_h$ that relate the physical quark masses between our ensemble sets, and $R_a$ that relates the lattice scales. In this approach (which was discussed in ref.~\cite{Aoki:2010dy} but not used in the final analysis) the scaling parameters are left as free parameters in our fits and the results are those that, along with the mass dependences and $a^2$ dependence, minimize the global $\chi^2$. In the 2010 analysis we used the ``fixed trajectory'' approach in which the ensemble sets were matched at an unphysical mass point prior to performing the fits. Changing to the generic scaling approach allows for differences between the scaling parameters $Z_l$, $Z_h$ and $R_a$, which relate the physical quark masses and lattice spacings between the ensemble sets, as we go between the three chiral ans\"{a}tze. We associated these with chiral and finite-volume systematic errors; in the 
fixed trajectory approach these differences would have been absorbed by other parameters in the fits. These differences gave rise to larger systematic errors on the lattice spacing predictions due to their influence on the fit form for the Omega baryon mass, which we used to set the overall scale. 

In these fits we were able to determine the $a^2$ dependence of the 32ID ensembles even without a second lattice spacing using this action. This is because, within our power counting, the choice of action affects only the coefficient of the $a^2$ term. As all other parameters are shared with the Iwasaki ensemble sets, this only introduces one additional parameter per quantity. This parameter could in principle be determined by comparing a single data point to the continuum value predicted using the Iwasaki ensemble sets alone, however we choose to maximize the use of our data by including it in the global fit.

We investigated removing data associated with the heavier pions, constraining our fits to a smaller range. For the ChPT and ChPTFV fits, we lowered the pion mass cut to 350 MeV, down from the 420 MeV used in the 2010 analysis. For the analytic fits, we found large deviations of our fits from the data when fitting to this range, necessitating a further reduction in the largest pion mass to 260 MeV. With this cut the analytic fit produced results with errors only slightly larger than the ChPT determinations. The necessity of lowering the cut for the analytic fits hints at the presence of nonlinearity in our combined data set, which appears to be consistent with NLO SU(2) ChPT, although we cannot rule out other higher-order terms such as $m^2$ with our present statistics.

We presented the results of simultaneously fitting $m_\pi$, $m_K$, $f_\pi$, $f_K$ and $m_\Omega$ in section~\ref{sec:FitResults}. As in the 2010 analysis, the pion, kaon and Omega baryon masses were used to set the up/down quark mass, strange quark mass and lattice scale respectively. We were then able to make predictions for the other physical quantities. For the pseudoscalar decay constants, we obtained $f_\pi = 127.1(3.8)\ {\rm MeV}$ and $f_K  = 152.4(3.4)\ {\rm MeV}$. These agree very well with the known continuum values of~\cite{Nakamura:2010zzi} $f_{\pi^-} = 130.4(2)$ and $f_{K^-} = 156.1(8)$, which is a marked improvement from the 2010 analysis, in which the predictions for these quantities were considerably lower. The improvement stems mainly from our removal of data associated with the heavier pions.

Combining our ChPT and ChPTFV fit results, we obtained values for the effective chiral couplings $\bar{l_3} = 2.91(24)$ and $\bar{l_4} = 3.99(18)$, which we found to be highly consistent with our 2010 analysis results and with other lattice calculations.

In section~\ref{sec:QuarkMasses} we discussed the renormalization of the physical quark masses into the $\msbar$ scheme. We used variants of the Rome-Southampton RI/MOM scheme with symmetric kinematics as intermediate nonperturbative schemes, which were applied at 1.4 GeV and the results run to 3 GeV using continuum step-scaling factors. These were then converted into $\msbar$ using perturbation theory. This analysis improved on the 2010 result in the use of twisted-boundary conditions to remove O(4)-breaking lattice artifacts in our measurements. We also increased the renormalization scale from 2 GeV to 3 GeV, as this considerably reduces the systematic error arising from the truncation of the perturbative series. We obtained $m_{ud}(\msbar, 3\ {\rm GeV}) = 3.05(10)\, {\rm MeV}$ and $m_s(\msbar, 3\ {\rm GeV}) = 83.5(2.0)\, {\rm MeV}$ for the average up/down quark mass and strange-quark mass respectively.

In section~\ref{sec:BK} we applied our chiral/continuum fits to the neutral kaon mixing parameter. This analysis improved on the 2010 result through the inclusion of the Iwasaki+DSDR ensembles. We found a marked improvement in the chiral extrapolation systematic due to the inclusion of these data. For our final result we obtained $B_K(\msbar,3\;\GeV) = 0.535(16)$.

Finally, in section~\ref{sec:r0r1} we performed chiral/continuum fits to the Sommer scales $r_0$ and $r_1$, for which we obtained $r_0 = 0.480(11)\ {\rm fm}$ and
$r_1 = 0.323(9)\ {\rm fm}$. Here the inclusion of the 32ID ensembles provided considerably greater stability to the fits than in the 2010 analysis, resulting in much reduced errors, particularly for $r_1$, for which we were formerly forced to inflate the errors due to the poor $\chi^2$/dof on the fits.

Although the inclusion of the 32ID ensembles resulted in considerable improvements in the chiral extrapolation systematic error in most cases, there is still room for improvement. Our collaboration has recently gained access to IBM Blue Gene/Q computers, which have performances in the region of several hundred Teraflops per rack. Particularly when used with the improved techniques that we and others have developed (some of which are discussed in section~\ref{sec:SimulationDetails} and Appendix~\ref{appendix-integrators}), these computers have the capability of generating domain wall fermion ensembles with physical quark masses and large enough $L_s$ and physical volumes to maintain small chiral symmetry breaking and finite-volume corrections. With such ensembles the necessity of extrapolating to the physical point will be removed and only the continuum extrapolation will remain. However, in the meantime the results of this analysis, particularly the physical quark masses and lattice spacings, will be 
essential for any physics measurements performed on the Iwasaki and Iwasaki+DSDR ensembles.

%%%%%%%%%%%%%%%%%%%%%%%%%%%% Acknowledgments %%%%%%%%%%%%%%%%%%%%%%%%%
\section*{Acknowledgments}
The generation of the $32^3\times 64$ Iwasaki+DSDR ensembles and the measurements on them were performed using the IBM Blue Gene/P machines at the Argonne Leadership Class Facility (ALCF) provided under the Incite Program of the U.S. DOE. Much of the computation for the $32^3\times 64$ Iwasaki ensembles was also performed at this facility, with the remainder, along with the computation for the $24^3\times 64$ Iwasaki ensembles, performed using the QCDOC computers \cite{Boyle:2005qc,Boyle:2003mj,Boyle:2005fb} at Columbia University, Edinburgh University, and at the Brookhaven National Laboratory (BNL). At BNL, the QCDOC computers of the RIKEN-BNL Research Center and the USQCD
Collaboration were used. The software used includes the CPS QCD code (http://qcdoc.phys.columbia.edu/cps.html), supported in part by the USDOE SciDAC program; the BAGEL (http://www2.ph.ed.ac.uk/~paboyle/bagel) assembler kernel generator for many of the high-performance optimized kernels~\cite{Boyle:2009bagel}; and the UKHadron codes. Renormalization was performed using STFC funded DiRAC resources.

C.T.S and A.T.L are funded by STFC Grant No. ST/J000396/1. The University of Southampton's Iridis cluster is funded by STFC Grant No. ST/H008888/1. T.B is funded by the U.S. DOE Grant No. \#DE-FG02-92ER40716. R.A is supported by a SUPA prize studentship. N.G is supported by the STFC Grant No. ST/G000522/1 and acknowledges the EU Grant No. 238353 (STRONGnet). P.A.B, N.G and R.J.H are supported by STFC Grants No. ST/K000411/1, No. ST/J000329/1 and No. ST/H008845/1. C.K, N.H.C and R.D.M are partially supported by U.S. DOE Grant No. \#DE-FG02-92ER40699. A.S, C.J and T.I are partially supported by DOE Contract No. \#AC-02-98CH10886 (BNL). T.I is also partially supported by JSPS Kakenhi Grant No. 22540301 and No. 23105715. J.M.Z is supported by the Australian Research Council through a Future Fellowship (FT100100005).

%%%%%%%%%%%%%%%%%%%%%%%%%%%%%% APPENDICES %%%%%%%%%%%%%%%%%%%%%%%%%%
\appendix

\section{Numerical integration scheme}
\label{appendix-integrators}

Integrators used in lattice simulation must be both reversible and
symplectic. Consider a general Hamiltonian with both a kinetic
($T$) and potential ($S$) term:
\begin{equation}\label{lat_hamiltonian}
  H=T(p)+S(U).
\end{equation}
In general this Hamiltonian cannot be integrated exactly, as the corresponding time
evolution operator,
\begin{equation}
  \exp\left(\tau \widehat{H}\right)
  =\exp\left(\tau\left(\widehat{T}+\widehat{S}\right)\right)\,,
\end{equation}
involves noncommuting operators $\widehat{T}$ and
$\widehat{S}$. However, by making use of the Baker-Campbell-Hausdorff
(BCH) formula one can separate $\widehat{T}$ and $\widehat{S}$ and
integrate them at different steps. 

One of the simplest integrators
that can be constructed in this way is the leapfrog integrator,
\begin{equation}
  U_{QPQ}(\tau)=
  \exp\left(\frac{1}{2}\tau \widehat{T}\right)
  \exp\left(\tau \widehat{S}\right) 
  \exp\left(\frac{1}{2}\tau \widehat{T}\right).
\end{equation}
Using the BCH formula it can be shown that
\begin{equation}
  U_\textrm{QPQ}(\tau)=\exp\left(
  \tau\left(\widehat{T}+\widehat{S}\right)
  +\mathcal{O}(\tau^3)
  \right)\,.
\end{equation}
The $\mathcal{O}(\tau^3)$ error is accumulated over the integration such that the total error is $\mathcal{O}(\tau^2)$, hence the leapfrog integrator is a second-order integrator. Another popular second-order integrator is the Omelyan integrator,
\begin{equation}
  U_\textrm{QPQPQ}(\tau)=
  \exp\left(\alpha\tau \widehat{T}\right)
  \exp\left(\frac{1}{2} \tau \widehat{S}\right) 
  \exp\left(\left(1-2\alpha\right)\tau \widehat{T}\right)
  \exp\left(\frac{1}{2} \tau \widehat{S}\right) 
  \exp\left(\alpha\tau \widehat{T}\right)\,,
\end{equation}
where $\alpha$ is a tunable parameter.

Recent development on integrators has introduced
the force gradient integrator (FGI)~\cite{Kennedy:2009fe} as a fourth order
integrator. The force gradient integrator is constructed by
introducing the ``force gradient term'' into the integration
steps. This extra force evaluation helps to eliminate the second order
errors and makes the force gradient integrator a fourth order
integrator. One choice of the force gradient integrator is
\begin{equation}
  \begin{split}
    U_\textrm{FGI}(\tau)=&
    \exp\left(\frac{3-\sqrt{3}}{6}\tau \widehat{T}\right)
    \exp\left(\frac{1}{2} \tau
    \widehat{S}-\frac{2-\sqrt{3}}{48}\tau^3\widehat{\left\{S,\left\{S,T\right\}\right\}}\right)
    \cdot \\
    &\exp\left(\frac{\sqrt{3}}{3}\tau \widehat{T}\right)
    \exp\left(\frac{1}{2} \tau \widehat{S}-\frac{2-\sqrt{3}}{48}\tau^3\widehat{\left\{S,\left\{S,T\right\}\right\}}\right) 
    \exp\left(\frac{3-\sqrt{3}}{6}\tau \widehat{T}\right)
  \end{split}
\end{equation}

\subsection{Sexton-Weingarten integration}

In practice the action contain contributions from both the
gauge fields and the fermions,
\begin{equation}
  H=T(p)+S_\textrm{G}(U)+S_\textrm{F}(U).
\end{equation}
It is usually the case that the gauge force is larger than the fermion
force by a factor of 10 or more. If both the gauge action and the
fermion action are integrated in the same step then the step size
$\tau$ has to be chosen to accommodate the larger gauge force. This
approach incurs an extra cost on the fermion part, which usually
dominates the computing time.

The Sexton-Weingarten integration scheme can be used to mitigate the
issue. Define
\begin{align}
  H=&T'+S_\textrm{F}(U) \\
  T'=&T(p)+S_\textrm{G}(U),
\end{align}
then $T'$ and $S_\textrm{F}(U)$ can be fit into one integrator. When
integrating $T'$, its 2 parts $T(p)$ and $S_\textrm{G}(U)$ can be fit
into another integrator. For example, when using the leapfrog QPQ
integrator for both levels one has the following
\begin{align}
  \exp\left(\tau \widehat{H}\right)\approx&
  \exp\left(\frac{1}{2}\tau \widehat{T'}\right)
  \exp\left(\tau \widehat{S_\textrm{F}}\right) 
  \exp\left(\frac{1}{2}\tau \widehat{T'}\right) \\
  \exp\left(\frac{1}{2}\tau \widehat{T'}\right)\approx&
  \left(\exp\left(\frac{1}{4n}\tau \widehat{T}\right)
  \exp\left(\frac{1}{2n}\tau \widehat{S_\textrm{G}}\right) 
  \exp\left(\frac{1}{4n}\tau \widehat{T}\right)\right)^n,\label{SW_integration_n}
\end{align}
where $n$ can be chosen as any positive integer. In this way different
time steps are assigned to $S_\textrm{G}(U)$ and $S_\textrm{F}(U)$,
which can be tuned to minimize the cost.

\subsection{Hasenbusch mass splitting}

Hasenbusch mass splitting breaks a single fermion action into a few
parts and offers a fine control on distributing fermion forces among
them.

The fermion action is derived from the following fermion determinant
\begin{equation}
  \det\left(\frac{M^\dag(m)M(m)}{M^\dag(1)M(1)}\right)=
  \int \mathcal{D}\phi^\dag \mathcal{D}\phi \;
  \exp\left(-\phi^\dag M(1)\frac{1}{M^\dag(m)M(m)}M^\dag(1)\phi\right).
\end{equation}
The Hasenbusch factorization~\cite{Hasenbusch:2001ne} rewrites the above
quotient action as a product of quotient actions by introducing
intermediate masses
\begin{align}
  &\det\left(\frac{M^\dag(m)M(m)}{M^\dag(1)M(1)}\right)=
  \prod^{k+1}_{i=1}\det\left(
  \frac{M^\dag(m_{i-1})M(m_{i-1})}{M^\dag(m_i)M(m_i)}
  \right) \\
  =&\prod_{i=1}^{k+1}\int \mathcal{D}\phi_i^\dag \mathcal{D}\phi_i\;
  \exp\left(
  -\phi_i^\dag
  M(m_i)\frac{1}{M^\dag(m_{i-1})M(m_{i-1})}M^\dag(m_i)\phi_i
  \right),
\end{align}
where $m=m_0<m_1<\cdots<m_{k+1}=1$. 

This method offers fine grained control on the sizes of the fermion
forces since all intermediate masses $m_i (i=1,2,\cdots,k)$ can be
tuned continuously. In what follows the symbol $S_\textrm{Q}(m_a,
m_b)$ will be used to represent the quotient fermion action
\begin{equation}
  S_\textrm{Q}(m_a,m_b)
  =\phi^\dag M(m_b)\frac{1}{M^\dag(m_a)M(m_a)}M^\dag(m_b)\phi,
\end{equation}
The Q in $S_\textrm{Q}$ means ``quotient''. Note that each quotient
action has a different pseudofermion field $\phi$. This fact is not
represented in the above symbol.

\subsection{Final scheme}

The quotient action discussed above accounts for 2 types of
fermions. This is used to simulate the 2 light quarks in our
simulation. For simulating strange quark, the rational approximation needs
to be used:
\begin{equation}
  \det\left(\frac{M^\dag(m)M(m)}{M^\dag(1)M(1)}\right)^{1/2}
  =\int \mathcal{D}\phi^\dag \mathcal{D}\phi\;
  \exp\left(
  -\phi^\dag
  \left(M^\dag(1)M(1)\right)^{1/4}
  \frac{1}{\left(M^\dag(m)M(m)\right)^{1/2}}
  \left(M^\dag(1)M(1)\right)^{1/4}\phi
  \right),
\end{equation}
where rational approximations of function $x^{1/4}$ and $x^{-1/2}$ are
used to evaluate the noninteger powers of matrices. In what follows
we will use the symbol $S_\textrm{R}(m_1,m_2)$ to represent this
rational action
\begin{equation}
  S_\textrm{R}(m_1,m_2)=
  \phi^\dag
  \left(M^\dag(m_2)M(m_2)\right)^{1/4}
  \frac{1}{\left(M^\dag(m_1)M(m_1)\right)^{1/2}}
  \left(M^\dag(m_2)M(m_2)\right)^{1/4}\phi,
\end{equation}
where power functions such as $x^{1/4}$ and $x^{-1/2}$ are understood
to be shorthand notations of their corresponding rational
approximations, the ``R'' in $S_\textrm{R}$ means ``rational''.

The final action used in the evolution contains the following components:
\begin{equation}
  H=T(p)+S_\textrm{G}+\sum_i S_\textrm{Q}(m_{i-1},m_i)+S_\textrm{R}(m_s,
  1)+S_\textrm{DSDR},
\end{equation}
where $m_0=m_l$, $m_{k+1}=1$, $m_l$ and $m_s$ represents the light
quark mass and strange quark mass respectively. It is also possible
to replace the quotient action $S_\textrm{Q}(m, 1)$ with two copies of the
same rational action $S_\textrm{R}(m,1)$.

When evolving the above action, we use multiple levels of nested
integrators to separate the different parts of the action. A general
multilevel Sexton-Weingarten Integration scheme can be written as
follows:
\begin{align}
  H=T'_0=&T'_1+S_1 \\
  T'_i=&T'_{i+1}+S_{i+1}\;\;\; i=1,2,\cdots,k-1,
\end{align}
where $T'_k=T(p)$. The above equations separate the entire action into
$k$ levels.

The details of the evolution schemes for the 2 ensembles are listed
in tables~\ref{tab:integrators0.0042} and~\ref{tab:integrators0.001}. The second column specifies which component of the
action is used in $S_i$. The value given in the fourth column, $n_i$, denotes the number of
integration steps for $T'_i$. This quantity is equivalent to $n$ in equation~\ref{SW_integration_n}.

\begin{table}[!htbp]
  \begin{tabular}{c|cccc}
    \hline
    Level($i$) & $S_i$ & Integrator type & $n_i$ & Step size \\
    \hline
    \hline
    1 & $S_\textrm{Q}(0.0042,0.015)+S_\textrm{Q}(0.015, 0.045)$ &
    Omelyan QPQPQ & 1 & $1/8$ \\
    \hline
    2 & $S_\textrm{R}(0.045, 1)+S_\textrm{R}(0.045,
    1)+S_\textrm{R}(0.045, 1)$ & Omelyan QPQPQ & 2 & - \\
    \hline
    3 & $S_\textrm{DSDR}$ & Omelyan QPQPQ & 4 & - \\
    \hline
    4 & $S_\textrm{G}$ & Omelyan QPQPQ & 1 & - \\
    \hline
  \end{tabular}
  \caption{$m_l=0.0042$, $m_s=0.045$ ensemble evolution details, with
    total 4 levels of nested integrators. Also note that 2 copies of
    rational action $S_\textrm{R}(0.045, 1)$ are used to replace a
    single quotient action $S_\textrm{Q}(0.045, 1)$. We use $\alpha=0.22$ for the Omelyan integrators.\label{tab:integrators0.0042} }
\end{table}

\begin{table}[!htbp]
  \begin{tabular}{c|cccc}
    \hline
    level($i$) & $S_i$ & integrator type & $n_i$ & step size \\
    \hline
    \hline
    \raisebox{12pt}{1} & 
    \shortstack{$S_\textrm{Q}(0.001,0.01)+S_\textrm{Q}(0.01, 0.04)$\\
      $+S_\textrm{Q}(0.04, 0.12)+S_\textrm{Q}(0.12, 0.31)$\\
      $+S_\textrm{Q}(0.31, 0.62)+S_\textrm{Q}(0.62, 1)+S_\textrm{R}(0.045, 1)$} &
    \raisebox{12pt}{FGI QPQPQ} & \raisebox{12pt}{3} & \raisebox{12pt}{$1/9$} \\
    \hline
    2 & $S_\textrm{DSDR}$ & FGI QPQPQ & 1 & - \\
    \hline
    3 & $S_\textrm{G}$ & FGI QPQPQ & 1 & - \\
    \hline
  \end{tabular}
  \caption{$m_l=0.001$, $m_s=0.045$ ensemble evolution details, with
    total 3 levels of nested integrators.\label{tab:integrators0.001}}
\end{table}
\section{Error propagation in the quark mass renormalization}
\label{appendix-zerrorprop}

In section~\ref{zm-recont-subsec} we performed the continuum extrapolation of the ratios of quark mass renormalization factors, $Z_{ml}$ and $Z_{mh}$, which we defined in equation~\ref{eqn-zml-zmh-def}. These ratios combine the scaling parameters $Z_l$ and $Z_h$ that represent the renormalization factors in the intermediate mass-independent ``matching scheme'' used during the fits and the nonperturbative renormalization factor $Z_m$ in the SMOM schemes calculated using the Rome-Southampton method. In this calculation, the propagation of statistical and systematic errors through the extrapolation and the subsequent application of the step-scaling factors is non-trivial. First, we note that the 32I and 24I lattice spacings are very strongly correlated through $R_a$ (recall that $a^{24I}$ is obtained as $a^{32I}/R_a^{24I}$). As the errors on these quantities give rise to uncertainties on both the renormalization scale and on the coordinates used in the continuum extrapolation, naively treating them as 
independent between the lattices could potentially give rise to unrealistically large errors on the final renormalized quark masses. In the earlier parts of this analysis, the propagation of finite-volume and chiral extrapolation effects was performed by repeating the global fit with each of the three chiral/continuum fit ans\"{a}tze (analytic, ChPT and ChPTFV) separately, taking the difference between these only at the final stage to estimate the corresponding systematic errors. All correlations were taken into account through the use of the superjackknife method to propagate the statistical errors. However, the determination of the nonperturbative renormalization coefficients was performed using bootstrap resampling for the error propagation. Therefore, in order to propagate the effects of the statistical and systematic errors on the lattice spacings in a fashion consistent with the main analysis, we created `superjackknife' distributions from the bootstrap distributions via the following procedure:
\begin{enumerate}
 \item On each Iwasaki lattice, we calculated $Z_m^{\schemeS}$ in each of the RI/SMOM schemes $\schemeS$ at 1.4 GeV and 3 GeV using bootstrap resampling to propagate the statistical errors. The results of these calculations were given in the previous section. We used only the central values of $Z_V$ and the lattice spacings during this procedure such that the statistical error contains only the fluctuations from the measurements of the amputated vertex functions. For the lattice spacings we used the central values from the ChPTFV determination, which we previously chose as our ``best'' ansatz.\label{enum-zm1}

 \item We repeated the previous step once again, only this time we shifted the lattice spacings by their total error. From the change in $Z_m^{\schemeS}$ we obtained its slope with respect to $a$. (The slopes are negative for all of our schemes, as can be seen in figure~\ref{fig-zmap2dep}).

 \item Using $dZ_m^{\schemeS}/da$ we shifted $Z_m^{\schemeS}$ to the values we would have obtained if we had repeated step~\ref{enum-zm1} using the lattice spacings obtained with the ChPT and analytic ans\"{a}tze. Along with the original measurement we then had values of $Z_m$ with the physical scales set using the results of each of the three global fit ans\"{a}tze. We henceforth refer to these with an additional superscript $\rm A$ denoting the chiral ansatz.

 \item For each fit ansatz we placed the corresponding bootstrap distribution on a fictitious ``superjackknife'' ensemble, ensuring that the statistical fluctuations remain independent from others in the analysis. (Our code is able to include both bootstrap and jackknife distributions within the same framework.) The remaining superjackknife samples were modified to account for the statistical fluctuations in the lattice spacings by setting each sample $i$ to the following:
$$(Z_m^{\schemeS, {\rm A}})_i = \langle Z_m^{\schemeS, {\rm A}} \rangle + \frac{dZ_m^{\schemeS}}{da} (a_i - \langle a \rangle)\,.$$
Here $\langle ...\rangle$ denotes the central value of the distribution.

 \item For the final step we take into account the fluctuations on $Z_V$ by dividing the ``superjackknife'' distributions for $Z_m^{\schemeS, {\rm A}}$ by $Z_V/\langle Z_V \rangle$, where the quantity in the numerator is the superjackknife distribution used to normalise $f_\pi$ in the main analysis. 
\end{enumerate}

These superjackknife distributions were used for the analysis documented in section~\ref{zm-recont-subsec}.
\section{$O(a)$ errors and chiral symmetry breaking}
\label{appendix-cloverterm}

In the Symanzik effective theory, explicit chiral symmetry breaking manifests as a dimension-3 term corresponding to the residual mass as well as a dimension-5 clover term. The clover term introduces ${\rm O}(a)$ discretization errors that make it difficult to perform continuum extrapolations with traditional Wilson fermions. For domain wall fermions however, both terms are suppressed due to the separation of the left- and right-handed chiral modes in the fifth dimension. As we discussed in section~\ref{sec:SimulationDetails}, dislocations in the gauge fields, that manifest more frequently at stronger coupling, can allow fermion modes to tunnel between the walls, breaking the usual exponential suppression; it is these that the DSDR factor was designed to suppress. For the ensembles used in this paper, the DSDR parameters were tuned to minimize the residual mass while retaining sufficient levels of topological tunneling. In this appendix we present evidence that this procedure has also heavily suppressed the 
clover term contributions, and hence that it is not necessary to consider ${\rm O}(a)$ discretization errors in our continuum extrapolations.

\begin{figure}[tp]
\centering
\includegraphics[width=0.4\textwidth]{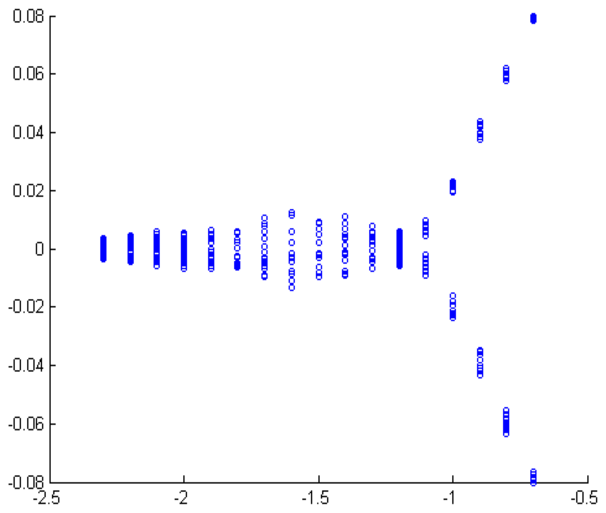}\\
\includegraphics[width=0.4\textwidth]{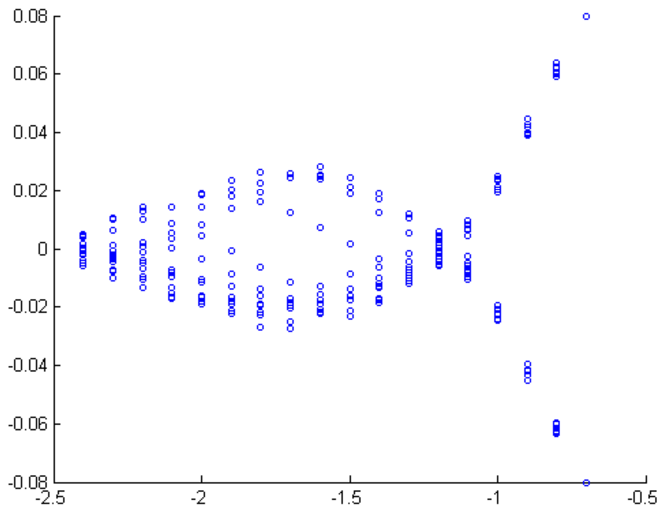}\quad
\includegraphics[width=0.4\textwidth]{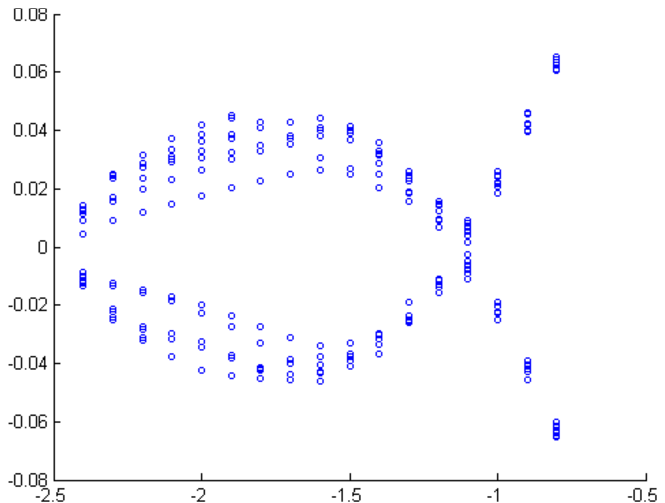}
\caption{The 12 lowest eigenmodes of the 4D Wilson-Dirac operator as a function of the domain wall mass $-M_5$ for positive $M_5$, measured on a single representative configuration of each of three $16^3\times 8\times 32$ ensembles; the upper plot on a $\beta=1.95$ ensemble with the Iwasaki gauge action, the lower-left on a $\beta=1.95$ ensemble with the Iwasaki+DSDR gauge action, and the lower-right on a $\beta=1.75$ ensemble with the Iwasaki+DSDR gauge action.\label{fig-eyediag}}
\end{figure}

Both the residual mass and the clover term are expected to be enhanced by dislocations in the gauge fields corresponding to zero modes of the four-dimensional (4D) Wilson-Dirac operator. In figure~\ref{fig-eyediag} we reproduce plots from ref.~\cite{Renfrew:2009wu} that show the effect of the DSDR factor on the 12 lowest eigenmodes of the 4D Wilson-Dirac matrix as a function of the 4D mass, $-M_5$ (for positive $M_5$), measured on a single representative configuration of each of three $16^3\times 8\times 32$ domain wall ensembles, including one with a different gauge coupling. At the mass $M_5=1.8$ that we used in our simulation we can clearly see that the DSDR factor provides a strong suppression of the lowest modes. Due to the common origin of both the clover term and the residual mass term, we expect both to be similarly suppressed by the reduction in the number of dislocations such that the observed reduction in $m_{\rm res}$ in our simulation is accompanied by a corresponding reduction in size of the clover term.

\begin{figure}[tp]
\centering
\includegraphics[width=0.4\textwidth]{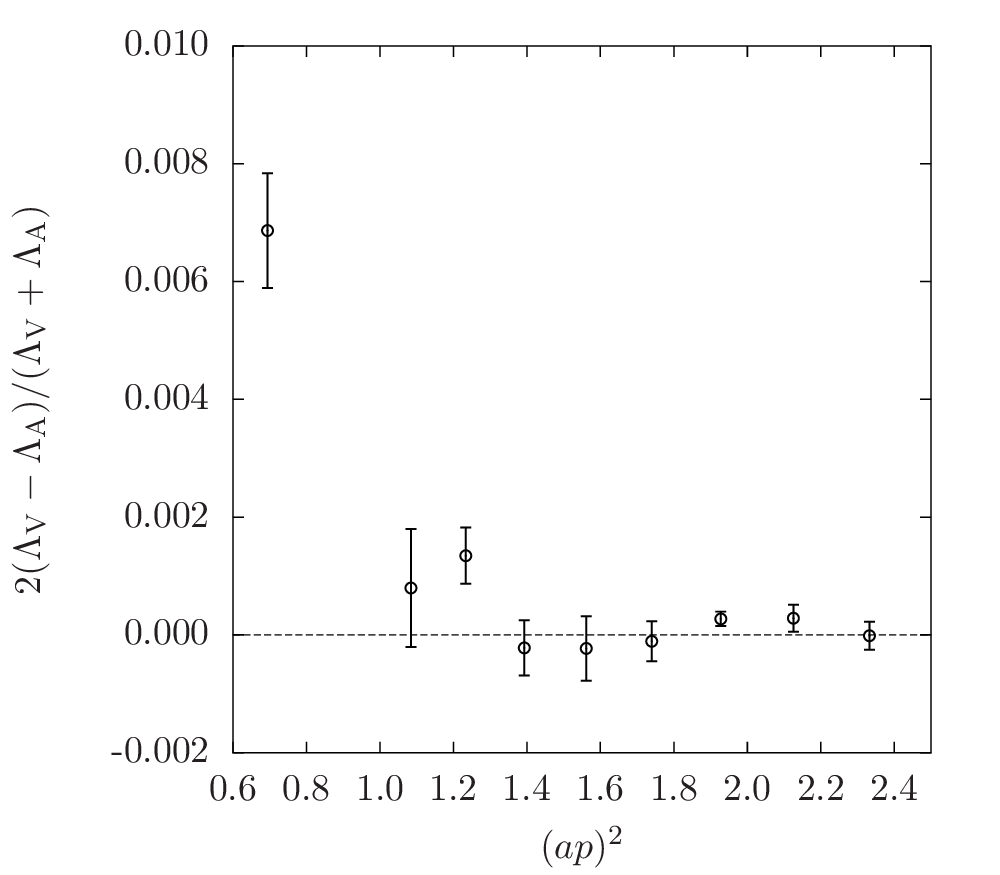}\\
\includegraphics[width=0.4\textwidth]{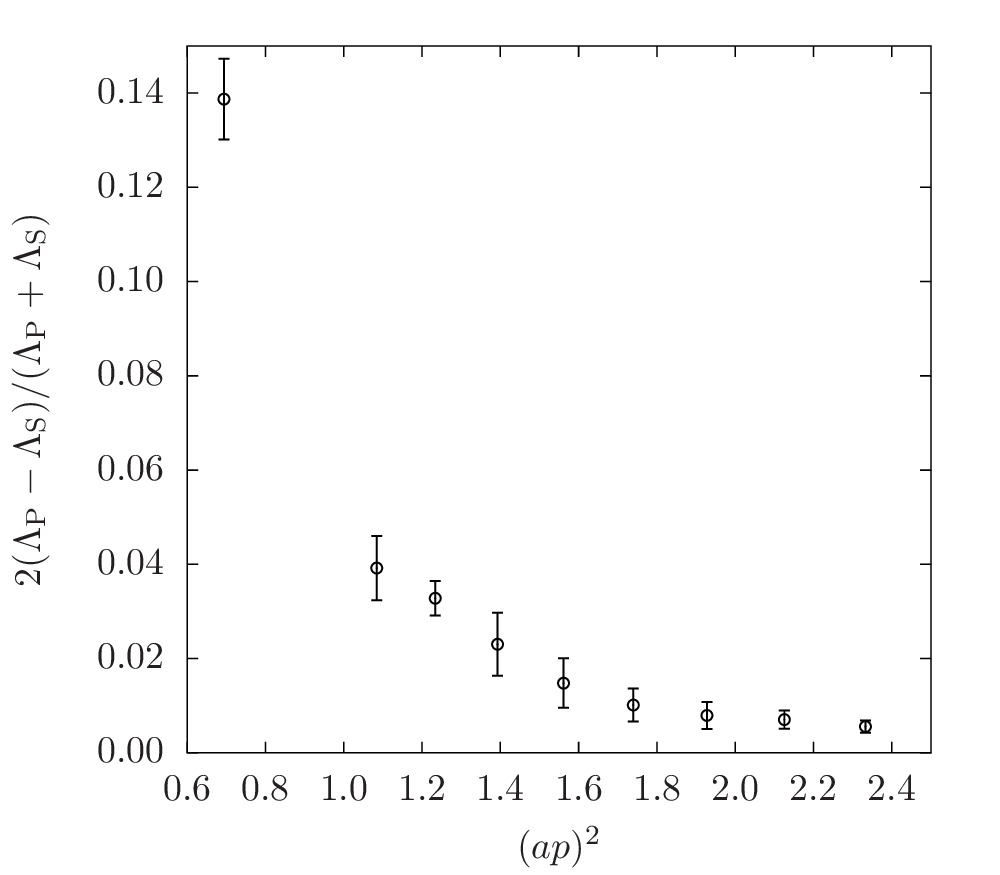}\quad
\includegraphics[width=0.4\textwidth]{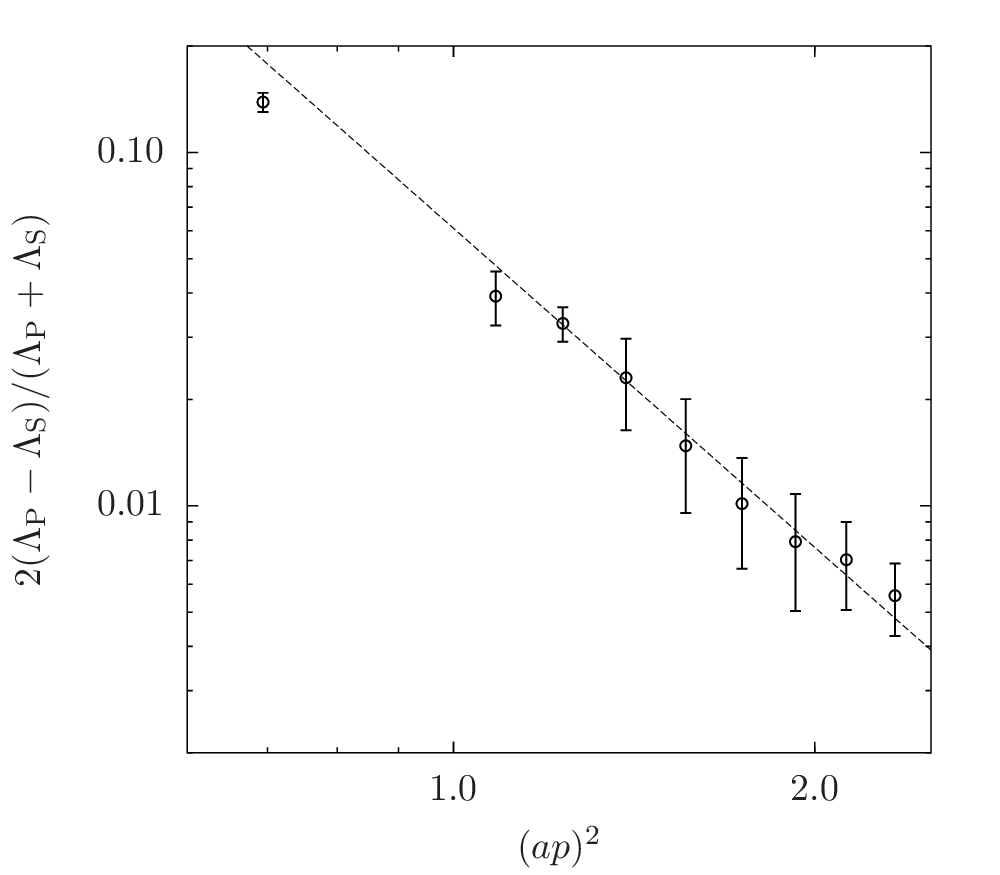}
\caption{The fractional difference between the local vector and axial-vector operators, $\Lambda_V$ and $\Lambda_A$ (top), and also between the scalar and pseudoscalar operators, $\Lambda_S$ and $\Lambda_P$ (bottom), as a function of the square of the momentum in lattice units. These values were obtained by measuring amputated bilinear vertex functions at a scale defined by the momentum of the incoming quark propagators. The lower-right figure is plotting with logarithmic axes and is overlayed by a line with the expected $1/(ap)^6$ dependence.\label{fig-vma-pms}}
\end{figure}

Additional evidence for the absence of large clover term contributions can be obtained by measuring the size of the explicit chiral symmetry breaking in our simulation beyond the effects of $m_{\rm res}$. One place where this should be apparent is in a larger than expected difference between the local vector and axial-vector vertex functions, $\Lambda_V$ and $\Lambda_A$ respectively, and similarly between those of the scalar and pseudoscalar operators, $\Lambda_S$ and $\Lambda_P$, evaluated at the chiral limit (including $m_{\rm res}$) in large-momentum Green's functions which might be expected to have a greater sensitivity to a dimension-5 operator. In the infinite-$L_s$ limit these quantities differ only through the dynamical chiral symmetry breaking at low energies, an effect that diminishes as $1/(ap)^6$ as the momentum-scale $p$ is increased. These are obtained with very high precision using the nonperturbative renormalization techniques discussed in section~\ref{sec:QuarkMasses} of this paper. In 
figure~\ref{fig-vma-pms} we plot the fractional differences as a function of the square of the momentum in lattice units. We see that the difference $\Lambda_V - \Lambda_A$ is consistent with zero at high energies, and $\Lambda_S-\Lambda_P$, while falling more slowly, demonstrates the expected $1/(ap)^6$ dependence and at the largest measured momentum is only a fraction of a percent. Note also that the behavior of the latter is very similar to that observed on our two finer DWF+Iwasaki lattices in ref.~\cite{Aoki:2010dy} (pg. 90) which do not use the DSDR factor. We have also published the results~\cite{KtopipiPRD} (pg. 42) of a similar analysis, performed on our Iwasaki+DSDR ensembles, of the off-diagonal components of the operator mixing matrix between the 4-quark operators used in our $K\rightarrow \pi\pi$ calculation, where we reached the same conclusion regarding the size of the explicit chiral symmetry breaking.

\section{Higher order corrections to symanzik coefficients}
\label{appendix-alphascorrections}

The standard treatment of the continuum limit is based on the usual
Symanzik analysis and assumes that the dominant discretization
errors can be described by an effective theory given by continuum
QCD with extra, dimension-six operators whose coefficients are 
proportional to $a^2$.  Higher order corrections arise from
dimension eight operators with $a^4$ coefficients.  (Here we are 
exploiting the chiral symmetry of the DWF formalism and considering correction
terms with only even dimensions.)  Using our two Iwasaki ensembles with 
$1/a=1.73$ and 2.28 GeV, we extrapolate linearly in $a^2$, assuming the 
$a^2$ term dominates, to obtain continuum limit results.  Since the 
results on the $1/a=1.73$ ensemble differ from the continuum limit 
values by typically $\le 3\%$, we estimate the systematic errors 
resulting from the $a^4$ terms as $(0.03)^2 \sim 0.1\%$, much smaller 
than the systematic errors from other sources.  

While this is presently the standard approach to evaluating the 
continuum limit in a lattice QCD calculation, we should recognize 
that in the Symanzik theory the coefficients of the $O(a^2)$ operators 
are actually not constant but will themselves contain logarithms of 
the lattice spacing, having the form: 
\begin{equation}
c(a) = c_0 + c_1 \alpha_s(a) \ln(a\Lambda_\mathrm{QCD})+\ldots,
\label{eq:a^2log}
\end{equation}
where ``\ldots'' represents terms with higher powers of the QCD
coupling $\alpha_s(a)$ evaluated at the lattice scale and more powers 
of the logarithm $\ln(a\Lambda_\mathrm{QCD})$.   The 
logarithms in $c(a)$ result from loop corrections and appear both 
explicitly and implicitly through the dependence of $\alpha_s$ on $a$.  
Let us examine how such $a$-dependence of $c(a)$ affects the 
determination of the continuum limit.

Consider a physical quantity $A(a)$ whose lattice spacing dependence 
is determined by the Symanzik coefficient $c(a)$:
\begin{equation}
A(a) = A_0 +a^2 c(a) A_1\,.
\label{eq:cont_limit}
\end{equation}
Here $A_0$ is the matrix element of the operator which gives the 
continuum value of $A$ while $A_1$ is the associated matrix element 
of the dimension-6 Symanzik correction operator.  We must determine 
numerically $A(a)$ in a range of accessible lattice spacings and then 
remove the unphysical $a^2c(a)A_1$ term.  To the extent that the 
logarithms appearing in Symanzik coefficient $c(a)$ are constant over 
the range of $a$ explored in the lattice calculation, they have no 
effect and a simple linear extrapolation will remove the entire 
$a^2 c(a) A_1$ term.  Note this procedure will give the correct continuum
limit (assuming that $c(a)$ is constant in the region in which the calculation
is performed) even if $c(a)$ has a strong dependence on $a$ as 
$a\to0$~\cite{Balog:2009yj} provided the Symanzik expansion is valid and 
the product $a^2 c(a)$ vanishes as $a\to0$.

However, if $c(a)$ does depend on $a$ in the region where a continuum
extrapolation is attempted then an error will result which we can
estimate.   For example, since $c(a)$  is likely slowly varying, we might
assume that it can be approximated by a low-order polynomial that could
be obtained by a simple Taylor expansion about an appropriately chosen
point $a_0$ within the range of the calculation.  Since we are expanding 
around a nonzero value of $a$ we can choose to expand in either $a$ or 
$a^2$.  We find the latter more convenient since it permits an easy 
comparison between this and the usual $a^4$ corrections expected 
in the DWF theory.  Approximating the Symanzik coefficient $c(a)$ as
\begin{equation}
c(a) = c_0 + c_2 a^2
\end{equation}
and using values of $A(a)$ obtained at two lattice spacings $a_1$ and $a_2$
to perform the usual subtraction to remove the $O(a^2)$ term gives our
approximation to the continuum limit:
\begin{eqnarray}
A_\mathrm{approx}^\mathrm{lin} &=&  A(a_1) - \frac{A(a_2)-A(a_1)}{a_2^2-a_1^2}a_1^2 \\
        &=& A_0 - c_2 a_2^2a_1^2 A_1\,.
\label{eq:sys_err_lin}
\end{eqnarray}
The second term represents the systematic error in our evaluation of the
continuum limit.   If instead we assume the logarithmic behavior of $c(a)$
present at one loop and given in Eq.~(\ref{eq:a^2log}) and ignore the $a$
dependence of $\alpha_s$, we find a similar systematic error:
\begin{eqnarray}
A_\mathrm{approx}^\mathrm{log} &=&  A_0
                     - c_1 \alpha_s \frac{\ln(a_2/a_1)}{a^2_2-a^2_1} a_2^2a_1^2 A_1\,.
\label{eq:sys_err_log}
\end{eqnarray}
Of course, Eq.~(\ref{eq:sys_err_log}) reduces to Eq.~(\ref{eq:sys_err_lin})
if we assume $\ln(a_2/a_1) \approx  (a^2_2-a^2_1)/(a_1^2+a_2^2)$ 
and use
\begin{equation}
c_2 = c_1\frac{\alpha_s}{a_1^2+a_2^2}\,.
\end{equation}

We can compare the size of this systematic error in the evaluation
of the continuum limit with the error that results from the neglect of the 
conventional $(a\Lambda_\mathrm{QCD})^4$ corrections.  
Equation~(\ref{eq:sys_err_lin}) can also be used to estimate the systematic
error introduced by the omission of $(a\Lambda_\mathrm{QCD})^4$ corrections:
\begin{equation}
\mathrm{Err}^{(a\Lambda_\mathrm{QCD})^4} 
                 = (a_1^2a_2^2\Lambda_\mathrm{QCD})^4 
                 \approx 0.05\%\,.
\end{equation}
where we assume $c_2=\Lambda_\mathrm{QCD}^4$ and  
$\Lambda_\mathrm{QCD}=300$ MeV and use our two Iwasaki lattice spacings 
$1/a=1.73$ and 2.28 GeV.  We can make a similar estimate of the systematic
error which arises from the neglect of a possible logarithm in the Symanzik
coefficient $c(a)$ by using Eq.~(\ref{eq:sys_err_log}) and assuming 
$c_1=\Lambda_\mathrm{QCD}^2/\pi$ and $\alpha_s=0.3$:
\begin{equation}
\mathrm{Err}^{a^2\ln(a)}
        = a_1^2a_2^2 \Lambda_\mathrm{QCD}^2
             \frac{\alpha_s}{\pi}\frac{\ln(a_2/a_1)}{a_2^2-a_1^2}
       \approx 0.1\%\,.
\end{equation}
Both of these estimates are much smaller than the other systematic
errors present in the calculations described in this paper.  However, it
is important to recognize that the error arising from the neglected 
logarithms of $a$ in the Symanzik coefficients may be as large or
larger than the more familiar $a^4$ errors and that these errors will
become increasingly dominant as $a$ is reduced.  This suggests a
future strategy that uses additional lattice spacings to allow a more
accurate polynomial description of $c(a)$ and a more accurate 
subtraction of this $O(a^2)$ Symanzik term.

We emphasize that it is our use of the Symanzik description of lattice
artifacts which permits this approach to determining the continuum limit.  
Instead of attempting to literally evaluate the limit $a^2 \to 0$, we can 
adopt a procedure to identify (through their $a^2$ dependence) and to 
subtract specific terms in the Symanzik expansion.  This approach may 
be viewed as complementary to the alternative effort to reach as small 
a value of $a$ as possible.  (Of course, smaller $a$ will always be 
required if sufficiently massive quarks are present in the calculation that 
the Symanzik expansion cannot be relied upon.)

In this approach we need not be concerned with possible singular behavior as 
$a^2\to 0$ such as found, for example, by  Balog {\it et al.}~\cite{Balog:2009yj}.
They examine $c(a)$ as $a^2$ varies over many orders of magnitude in 
two dimensional field theories.  For such a large range of values of $a^2$ 
a sum of leading logarithms must be performed and a simple linear or 
logarithmic description of $c(a)$ is inadequate.  Of course, for a calculation in 
four dimensions such a large range of lattice spacings is not available 
and the description of the variation of $c(a)$ by a low order polynomial 
should be very accurate.  

It should be emphasized that the effect of such $a^2\ln^n(a^2)$ terms 
on the evaluation of the continuum limit is very different from the effect 
of the $m_\pi^2\ln^n(m_\pi^2)$ terms that appear in chiral perturbation 
theory.  In the case of a chiral extrapolation we are interested in 
extrapolating these logarithmic terms to a nonzero value of $m_\pi$, 
outside the region in which calculations have been performed.  In the 
case of the continuum limit we need only subtract the unphysical 
$a^2c(a)$ term and need not be interested in its behavior outside the 
region in which lattice results have been obtained.

%%%%%%%%%%%%%%%%%%%%%%%%%%%%  Bibliography %%%%%%%%%%%%%%%%%%%%%%%%%%%
\FloatBarrier
\bibliography{paper}

\begin{thebibliography}{10}
\expandafter\ifx\csname bibnamefont\endcsname\relax
  \def\bibnamefont#1{#1}\fi
\expandafter\ifx\csname bibfnamefont\endcsname\relax
  \def\bibfnamefont#1{#1}\fi
\expandafter\ifx\csname url\endcsname\relax
  \def\url#1{\texttt{#1}}\fi
\expandafter\ifx\csname urlprefix\endcsname\relax\def\urlprefix{URL }\fi
\expandafter\ifx\csname bibinfo\endcsname\relax \def\bibinfo#1#2{#2}\fi
\expandafter\ifx\csname eprint\endcsname\relax \def\eprint#1{#1}\fi

\bibitem{Aoki:2010dy}
\bibinfo{author}{\bibfnamefont{Y.}~\bibnamefont{Aoki}} \emph{et~al.}
  (\bibinfo{collaboration}{RBC and UKQCD Collaborations}),
  \bibinfo{journal}{Phys. Rev.} \textbf{\bibinfo{volume}{D83}},
  \bibinfo{pages}{074508} (\bibinfo{year}{2011}), \eprint{1011.0892}.

\bibitem{Aoki:2010pe}
\bibinfo{author}{\bibfnamefont{Y.}~\bibnamefont{Aoki}},
  \bibinfo{author}{\bibfnamefont{R.}~\bibnamefont{Arthur}},
  \bibinfo{author}{\bibfnamefont{T.}~\bibnamefont{Blum}},
  \bibinfo{author}{\bibfnamefont{P.}~\bibnamefont{Boyle}},
  \bibinfo{author}{\bibfnamefont{D.}~\bibnamefont{Brommel}}, \emph{et~al.}
  (\bibinfo{collaboration}{RBC and UKQCD collaborations}),
  \bibinfo{journal}{Phys.Rev.} \textbf{\bibinfo{volume}{D84}},
  \bibinfo{pages}{014503} (\bibinfo{year}{2011}), \eprint{1012.4178}.

\bibitem{Vranas:1999rz}
\bibinfo{author}{\bibfnamefont{P.~M.} \bibnamefont{Vranas}}, in
  \emph{\bibinfo{booktitle}{Lattice Fermions and Structure of the Vacuum}}
  (\bibinfo{publisher}{Springer}, \bibinfo{address}{New York},
  \bibinfo{year}{1999}), chap. \bibinfo{chapter}{Domain wall fermions in vector
  theories}.

\bibitem{Vranas:2006zk}
\bibinfo{author}{\bibfnamefont{P.~M.} \bibnamefont{Vranas}},
  \bibinfo{journal}{Phys. Rev.} \textbf{\bibinfo{volume}{D74}},
  \bibinfo{pages}{034512} (\bibinfo{year}{2006}), \eprint{hep-lat/0606014}.

\bibitem{Fukaya:2006vs}
\bibinfo{author}{\bibfnamefont{H.}~\bibnamefont{Fukaya}} \emph{et~al.}
  (\bibinfo{collaboration}{JLQCD Collaboration}), \bibinfo{journal}{Phys.Rev.}
  \textbf{\bibinfo{volume}{D74}}, \bibinfo{pages}{094505}
  (\bibinfo{year}{2006}), \eprint{hep-lat/0607020}.

\bibitem{Renfrew:2009wu}
\bibinfo{author}{\bibfnamefont{D.}~\bibnamefont{Renfrew}},
  \bibinfo{author}{\bibfnamefont{T.}~\bibnamefont{Blum}},
  \bibinfo{author}{\bibfnamefont{N.}~\bibnamefont{Christ}},
  \bibinfo{author}{\bibfnamefont{R.}~\bibnamefont{Mawhinney}},
  \bibnamefont{and} \bibinfo{author}{\bibfnamefont{P.}~\bibnamefont{Vranas}},
  \bibinfo{journal}{PoS} \textbf{\bibinfo{volume}{LATTICE2008}},
  \bibinfo{pages}{048} (\bibinfo{year}{2008}), \eprint{0902.2587}.

\bibitem{KtopipiPRD}
\bibinfo{author}{\bibfnamefont{T.}~\bibnamefont{Blum}},
  \bibinfo{author}{\bibfnamefont{P.}~\bibnamefont{Boyle}},
  \bibinfo{author}{\bibfnamefont{N.}~\bibnamefont{Christ}},
  \bibinfo{author}{\bibfnamefont{N.}~\bibnamefont{Garron}},
  \bibinfo{author}{\bibfnamefont{E.}~\bibnamefont{Goode}}, \emph{et~al.}
  (\bibinfo{collaboration}{RBC and UKQCD Collaborations}),
  \bibinfo{journal}{Phys.Rev.} \textbf{\bibinfo{volume}{D86}},
  \bibinfo{pages}{074513} (\bibinfo{year}{2012}), \eprint{1206.5142}.

\bibitem{Antonio:2007tr}
\bibinfo{author}{\bibfnamefont{D.~J.} \bibnamefont{Antonio}} \emph{et~al.}
  (\bibinfo{collaboration}{RBC and UKQCD Collaborations}),
  \bibinfo{journal}{Phys.Rev.} \textbf{\bibinfo{volume}{D77}},
  \bibinfo{pages}{014509} (\bibinfo{year}{2008}), \eprint{0705.2340}.

\bibitem{Aoki:2001su}
\bibinfo{author}{\bibfnamefont{S.}~\bibnamefont{Aoki}} \bibnamefont{and}
  \bibinfo{author}{\bibfnamefont{Y.}~\bibnamefont{Taniguchi}},
  \bibinfo{journal}{Phys.Rev.} \textbf{\bibinfo{volume}{D65}},
  \bibinfo{pages}{074502} (\bibinfo{year}{2002}), \eprint{hep-lat/0109022}.

\bibitem{Aoki:2001dea}
\bibinfo{author}{\bibfnamefont{S.}~\bibnamefont{Aoki}} \emph{et~al.}
  (\bibinfo{collaboration}{CP-PACS Collaboration}),
  \bibinfo{journal}{Nucl.Phys.Proc.Suppl.} \textbf{\bibinfo{volume}{106}},
  \bibinfo{pages}{718} (\bibinfo{year}{2002}), \eprint{hep-lat/0110126}.

\bibitem{Golterman:2003qe}
\bibinfo{author}{\bibfnamefont{M.}~\bibnamefont{Golterman}} \bibnamefont{and}
  \bibinfo{author}{\bibfnamefont{Y.}~\bibnamefont{Shamir}},
  \bibinfo{journal}{Phys.Rev.} \textbf{\bibinfo{volume}{D68}},
  \bibinfo{pages}{074501} (\bibinfo{year}{2003}), \eprint{hep-lat/0306002}.

\bibitem{Golterman:2004cy}
\bibinfo{author}{\bibfnamefont{M.}~\bibnamefont{Golterman}},
  \bibinfo{author}{\bibfnamefont{Y.}~\bibnamefont{Shamir}}, \bibnamefont{and}
  \bibinfo{author}{\bibfnamefont{B.}~\bibnamefont{Svetitsky}},
  \bibinfo{journal}{Phys.Rev.} \textbf{\bibinfo{volume}{D71}},
  \bibinfo{pages}{071502} (\bibinfo{year}{2005}), \eprint{hep-lat/0407021}.

\bibitem{Golterman:2005fe}
\bibinfo{author}{\bibfnamefont{M.}~\bibnamefont{Golterman}},
  \bibinfo{author}{\bibfnamefont{Y.}~\bibnamefont{Shamir}}, \bibnamefont{and}
  \bibinfo{author}{\bibfnamefont{B.}~\bibnamefont{Svetitsky}},
  \bibinfo{journal}{Phys.Rev.} \textbf{\bibinfo{volume}{D72}},
  \bibinfo{pages}{034501} (\bibinfo{year}{2005}), \eprint{hep-lat/0503037}.

\bibitem{Svetitsky:2005qa}
\bibinfo{author}{\bibfnamefont{B.}~\bibnamefont{Svetitsky}},
  \bibinfo{author}{\bibfnamefont{Y.}~\bibnamefont{Shamir}}, \bibnamefont{and}
  \bibinfo{author}{\bibfnamefont{M.}~\bibnamefont{Golterman}},
  \bibinfo{journal}{PoS} \textbf{\bibinfo{volume}{LAT2005}},
  \bibinfo{pages}{129} (\bibinfo{year}{2006}), \eprint{hep-lat/0508015}.

\bibitem{Allton:2008pn}
\bibinfo{author}{\bibfnamefont{C.}~\bibnamefont{Allton}} \emph{et~al.}
  (\bibinfo{collaboration}{RBC and UKQCD Collaborations}),
  \bibinfo{journal}{Phys. Rev.} \textbf{\bibinfo{volume}{D78}},
  \bibinfo{pages}{114509} (\bibinfo{year}{2008}), \eprint{0804.0473}.

\bibitem{Yin:2011sz}
\bibinfo{author}{\bibfnamefont{H.}~\bibnamefont{Yin}} \bibnamefont{and}
  \bibinfo{author}{\bibfnamefont{R.~D.} \bibnamefont{Mawhinney}},
  \bibinfo{journal}{PoS} \textbf{\bibinfo{volume}{LATTICE2011}},
  \bibinfo{pages}{051} (\bibinfo{year}{2011}), \eprint{1111.5059}.

\bibitem{Hasenbusch:2001ne}
\bibinfo{author}{\bibfnamefont{M.}~\bibnamefont{Hasenbusch}},
  \bibinfo{journal}{Phys.Lett.} \textbf{\bibinfo{volume}{B519}},
  \bibinfo{pages}{177} (\bibinfo{year}{2001}), \eprint{hep-lat/0107019}.

\bibitem{Kennedy:2009fe}
\bibinfo{author}{\bibfnamefont{A.}~\bibnamefont{Kennedy}},
  \bibinfo{author}{\bibfnamefont{M.}~\bibnamefont{Clark}}, \bibnamefont{and}
  \bibinfo{author}{\bibfnamefont{P.}~\bibnamefont{Silva}},
  \bibinfo{journal}{PoS} \textbf{\bibinfo{volume}{LAT2009}},
  \bibinfo{pages}{021} (\bibinfo{year}{2009}), \eprint{0910.2950}.

\bibitem{deForcrand:1997sq}
\bibinfo{author}{\bibfnamefont{P.}~\bibnamefont{de~Forcrand}},
  \bibinfo{author}{\bibfnamefont{M.}~\bibnamefont{Garcia~Perez}},
  \bibnamefont{and} \bibinfo{author}{\bibfnamefont{I.-O.}
  \bibnamefont{Stamatescu}}, \bibinfo{journal}{Nucl. Phys.}
  \textbf{\bibinfo{volume}{B499}}, \bibinfo{pages}{409} (\bibinfo{year}{1997}),
  \eprint{hep-lat/9701012}.

\bibitem{Sharpe:2007yd}
\bibinfo{author}{\bibfnamefont{S.~R.} \bibnamefont{Sharpe}}
  (\bibinfo{year}{2007}), \eprint{0706.0218}.

\bibitem{Blum:2001xb}
\bibinfo{author}{\bibfnamefont{T.}~\bibnamefont{Blum}} \emph{et~al.}
  (\bibinfo{collaboration}{RBC}), \bibinfo{journal}{Phys. Rev.}
  \textbf{\bibinfo{volume}{D68}}, \bibinfo{pages}{114506}
  (\bibinfo{year}{2003}), \eprint{hep-lat/0110075}.

\bibitem{32cubedpaper}
\bibinfo{author}{\bibfnamefont{Y.}~\bibnamefont{Aoki}} \emph{et~al.}
  (\bibinfo{collaboration}{RBC and UKQCD Collaborations}),
  \bibinfo{journal}{Phys.Rev.} \textbf{\bibinfo{volume}{D83}},
  \bibinfo{pages}{074508} (\bibinfo{year}{2011}), \eprint{1011.0892}.

\bibitem{Eichten:1978tg}
\bibinfo{author}{\bibfnamefont{E.}~\bibnamefont{Eichten}},
  \bibinfo{author}{\bibfnamefont{K.}~\bibnamefont{Gottfried}},
  \bibinfo{author}{\bibfnamefont{T.}~\bibnamefont{Kinoshita}},
  \bibinfo{author}{\bibfnamefont{K.~D.} \bibnamefont{Lane}}, \bibnamefont{and}
  \bibinfo{author}{\bibfnamefont{T.-M.} \bibnamefont{Yan}},
  \bibinfo{journal}{Phys. Rev.} \textbf{\bibinfo{volume}{D17}},
  \bibinfo{pages}{3090} (\bibinfo{year}{1978}).

\bibitem{Colangelo:2005gd}
\bibinfo{author}{\bibfnamefont{G.}~\bibnamefont{Colangelo}},
  \bibinfo{author}{\bibfnamefont{S.}~\bibnamefont{Durr}}, \bibnamefont{and}
  \bibinfo{author}{\bibfnamefont{C.}~\bibnamefont{Haefeli}},
  \bibinfo{journal}{Nucl. Phys.} \textbf{\bibinfo{volume}{B721}},
  \bibinfo{pages}{136} (\bibinfo{year}{2005}), \eprint{hep-lat/0503014}.

\bibitem{Colangelo:2010et}
\bibinfo{author}{\bibfnamefont{G.}~\bibnamefont{Colangelo}},
  \bibinfo{author}{\bibfnamefont{S.}~\bibnamefont{Durr}},
  \bibinfo{author}{\bibfnamefont{A.}~\bibnamefont{Juttner}},
  \bibinfo{author}{\bibfnamefont{L.}~\bibnamefont{Lellouch}},
  \bibinfo{author}{\bibfnamefont{H.}~\bibnamefont{Leutwyler}}, \emph{et~al.},
  \bibinfo{journal}{Eur.Phys.J.} \textbf{\bibinfo{volume}{C71}},
  \bibinfo{pages}{1695} (\bibinfo{year}{2011}), \eprint{1011.4408}.

\bibitem{Bazavov:2010hj}
\bibinfo{author}{\bibfnamefont{A.}~\bibnamefont{Bazavov}} \emph{et~al.}
  (\bibinfo{collaboration}{MILC Collaboration}), \bibinfo{journal}{PoS}
  \textbf{\bibinfo{volume}{LATTICE2010}}, \bibinfo{pages}{074}
  (\bibinfo{year}{2010}), \eprint{1012.0868}.

\bibitem{Bazavov:2010yq}
\bibinfo{author}{\bibfnamefont{A.}~\bibnamefont{Bazavov}},
  \bibinfo{author}{\bibfnamefont{C.}~\bibnamefont{Bernard}},
  \bibinfo{author}{\bibfnamefont{C.}~\bibnamefont{DeTar}},
  \bibinfo{author}{\bibfnamefont{X.}~\bibnamefont{Du}},
  \bibinfo{author}{\bibfnamefont{W.}~\bibnamefont{Freeman}}, \emph{et~al.},
  \bibinfo{journal}{PoS} \textbf{\bibinfo{volume}{LATTICE2010}},
  \bibinfo{pages}{083} (\bibinfo{year}{2010}), \eprint{1011.1792}.

\bibitem{Baron:2010bv}
\bibinfo{author}{\bibfnamefont{R.}~\bibnamefont{Baron}},
  \bibinfo{author}{\bibfnamefont{P.}~\bibnamefont{Boucaud}},
  \bibinfo{author}{\bibfnamefont{J.}~\bibnamefont{Carbonell}},
  \bibinfo{author}{\bibfnamefont{A.}~\bibnamefont{Deuzeman}},
  \bibinfo{author}{\bibfnamefont{V.}~\bibnamefont{Drach}}, \emph{et~al.},
  \bibinfo{journal}{JHEP} \textbf{\bibinfo{volume}{1006}}, \bibinfo{pages}{111}
  (\bibinfo{year}{2010}), \eprint{1004.5284}.

\bibitem{Arthur:2010ht}
\bibinfo{author}{\bibfnamefont{R.}~\bibnamefont{Arthur}} \bibnamefont{and}
  \bibinfo{author}{\bibfnamefont{P.}~\bibnamefont{Boyle}}
  (\bibinfo{collaboration}{RBC and UKQCD Collaborations}),
  \bibinfo{journal}{Phys.Rev.} \textbf{\bibinfo{volume}{D83}},
  \bibinfo{pages}{114511} (\bibinfo{year}{2011}), \eprint{1006.0422}.

\bibitem{Sturm:2009kb}
\bibinfo{author}{\bibfnamefont{C.}~\bibnamefont{Sturm}} \emph{et~al.},
  \bibinfo{journal}{Phys. Rev.} \textbf{\bibinfo{volume}{D80}},
  \bibinfo{pages}{014501} (\bibinfo{year}{2009}), \eprint{0901.2599}.

\bibitem{Gorbahn:2010bf}
\bibinfo{author}{\bibfnamefont{M.}~\bibnamefont{Gorbahn}} \bibnamefont{and}
  \bibinfo{author}{\bibfnamefont{S.}~\bibnamefont{Jager}},
  \bibinfo{journal}{Phys.Rev.} \textbf{\bibinfo{volume}{D82}},
  \bibinfo{pages}{114001} (\bibinfo{year}{2010}), \eprint{1004.3997}.

\bibitem{Almeida:2010ns}
\bibinfo{author}{\bibfnamefont{L.~G.} \bibnamefont{Almeida}} \bibnamefont{and}
  \bibinfo{author}{\bibfnamefont{C.}~\bibnamefont{Sturm}},
  \bibinfo{journal}{Phys.Rev.} \textbf{\bibinfo{volume}{D82}},
  \bibinfo{pages}{054017} (\bibinfo{year}{2010}), \eprint{arXiv:1004.4613}.

\bibitem{Martinelli:1994ty}
\bibinfo{author}{\bibfnamefont{G.}~\bibnamefont{Martinelli}},
  \bibinfo{author}{\bibfnamefont{C.}~\bibnamefont{Pittori}},
  \bibinfo{author}{\bibfnamefont{C.~T.} \bibnamefont{Sachrajda}},
  \bibinfo{author}{\bibfnamefont{M.}~\bibnamefont{Testa}}, \bibnamefont{and}
  \bibinfo{author}{\bibfnamefont{A.}~\bibnamefont{Vladikas}},
  \bibinfo{journal}{Nucl. Phys.} \textbf{\bibinfo{volume}{B445}},
  \bibinfo{pages}{81} (\bibinfo{year}{1995}), \eprint{hep-lat/9411010}.

\bibitem{Arthur:2011cn}
\bibinfo{author}{\bibfnamefont{R.}~\bibnamefont{Arthur}},
  \bibinfo{author}{\bibfnamefont{P.}~\bibnamefont{Boyle}},
  \bibinfo{author}{\bibfnamefont{N.}~\bibnamefont{Garron}},
  \bibinfo{author}{\bibfnamefont{C.}~\bibnamefont{Kelly}}, \bibnamefont{and}
  \bibinfo{author}{\bibfnamefont{A.}~\bibnamefont{Lytle}}
  (\bibinfo{collaboration}{RBC and UKQCD Collaborations}),
  \bibinfo{journal}{Phys.Rev.} \textbf{\bibinfo{volume}{D85}},
  \bibinfo{pages}{014501} (\bibinfo{year}{2012}), \eprint{1109.1223}.

\bibitem{Lehner:2011fz}
\bibinfo{author}{\bibfnamefont{C.}~\bibnamefont{Lehner}} \bibnamefont{and}
  \bibinfo{author}{\bibfnamefont{C.}~\bibnamefont{Sturm}},
  \bibinfo{journal}{Phys.Rev.} \textbf{\bibinfo{volume}{D84}},
  \bibinfo{pages}{014001} (\bibinfo{year}{2011}), \eprint{1104.4948}.

\bibitem{Boyle:2011cc}
\bibinfo{author}{\bibfnamefont{P.}~\bibnamefont{Boyle}},
  \bibinfo{author}{\bibfnamefont{N.}~\bibnamefont{Garron}}, \bibnamefont{and}
  \bibinfo{author}{\bibfnamefont{A.}~\bibnamefont{Lytle}}
  (\bibinfo{collaboration}{RBC and UKQCD collaborations}),
  \bibinfo{journal}{PoS} \textbf{\bibinfo{volume}{LATTICE2011}},
  \bibinfo{pages}{227} (\bibinfo{year}{2011}), \eprint{1112.0537}.

\bibitem{Aoki:1997xg}
\bibinfo{author}{\bibfnamefont{S.}~\bibnamefont{Aoki}} \bibnamefont{and}
  \bibinfo{author}{\bibfnamefont{Y.}~\bibnamefont{Taniguchi}}
  (\bibinfo{collaboration}{RBC and UKQCD Collaborations}),
  \bibinfo{journal}{Phys. Rev.} \textbf{\bibinfo{volume}{D59}},
  \bibinfo{pages}{054510} (\bibinfo{year}{1999}), \eprint{hep-lat/9711004}.

\bibitem{Aoki:2007xm}
\bibinfo{author}{\bibfnamefont{Y.}~\bibnamefont{Aoki}} \emph{et~al.}
  (\bibinfo{collaboration}{RBC and UKQCD Collaborations}),
  \bibinfo{journal}{Phys. Rev.} \textbf{\bibinfo{volume}{D78}},
  \bibinfo{pages}{054510} (\bibinfo{year}{2008}), \eprint{0712.1061}.

\bibitem{Chetyrkin:1999pq}
\bibinfo{author}{\bibfnamefont{K.~G.} \bibnamefont{Chetyrkin}}
  \bibnamefont{and} \bibinfo{author}{\bibfnamefont{A.}~\bibnamefont{Retey}},
  \bibinfo{journal}{Nucl. Phys.} \textbf{\bibinfo{volume}{B583}},
  \bibinfo{pages}{3} (\bibinfo{year}{2000}), \eprint{hep-ph/9910332}.

\bibitem{vanRitbergen:1997va}
\bibinfo{author}{\bibfnamefont{T.}~\bibnamefont{van Ritbergen}},
  \bibinfo{author}{\bibfnamefont{J.~A.~M.} \bibnamefont{Vermaseren}},
  \bibnamefont{and} \bibinfo{author}{\bibfnamefont{S.~A.} \bibnamefont{Larin}},
  \bibinfo{journal}{Phys. Lett.} \textbf{\bibinfo{volume}{B400}},
  \bibinfo{pages}{379} (\bibinfo{year}{1997}), \eprint{hep-ph/9701390}.

\bibitem{Chetyrkin:1997sg}
\bibinfo{author}{\bibfnamefont{K.}~\bibnamefont{Chetyrkin}},
  \bibinfo{author}{\bibfnamefont{B.~A.} \bibnamefont{Kniehl}},
  \bibnamefont{and}
  \bibinfo{author}{\bibfnamefont{M.}~\bibnamefont{Steinhauser}},
  \bibinfo{journal}{Phys.Rev.Lett.} \textbf{\bibinfo{volume}{79}},
  \bibinfo{pages}{2184} (\bibinfo{year}{1997}), \eprint{hep-ph/9706430}.

\bibitem{Beringer:1900zz}
\bibinfo{author}{\bibfnamefont{J.}~\bibnamefont{Beringer}} \emph{et~al.}
  (\bibinfo{collaboration}{Particle Data Group}), \bibinfo{journal}{Phys.Rev.}
  \textbf{\bibinfo{volume}{D86}}, \bibinfo{pages}{010001}
  (\bibinfo{year}{2012}).

\bibitem{Bazavov:2009fk}
\bibinfo{author}{\bibfnamefont{A.}~\bibnamefont{Bazavov}} \emph{et~al.}
  (\bibinfo{collaboration}{MILC Collaboration}), \bibinfo{journal}{PoS}
  \textbf{\bibinfo{volume}{CD09}}, \bibinfo{pages}{007} (\bibinfo{year}{2009}),
  \eprint{0910.2966}.

\bibitem{McNeile:2010ji}
\bibinfo{author}{\bibfnamefont{C.}~\bibnamefont{McNeile}},
  \bibinfo{author}{\bibfnamefont{C.}~\bibnamefont{Davies}},
  \bibinfo{author}{\bibfnamefont{E.}~\bibnamefont{Follana}},
  \bibinfo{author}{\bibfnamefont{K.}~\bibnamefont{Hornbostel}},
  \bibnamefont{and} \bibinfo{author}{\bibfnamefont{G.}~\bibnamefont{Lepage}},
  \bibinfo{journal}{Phys.Rev.} \textbf{\bibinfo{volume}{D82}},
  \bibinfo{pages}{034512} (\bibinfo{year}{2010}), \eprint{1004.4285}.

\bibitem{Aubin:2009jh}
\bibinfo{author}{\bibfnamefont{C.}~\bibnamefont{Aubin}},
  \bibinfo{author}{\bibfnamefont{J.}~\bibnamefont{Laiho}}, \bibnamefont{and}
  \bibinfo{author}{\bibfnamefont{R.~S.} \bibnamefont{Van~de Water}},
  \bibinfo{journal}{Phys.Rev.} \textbf{\bibinfo{volume}{D81}},
  \bibinfo{pages}{014507} (\bibinfo{year}{2010}), \eprint{0905.3947}.

\bibitem{Durr:2011ap}
\bibinfo{author}{\bibfnamefont{S.}~\bibnamefont{Durr}},
  \bibinfo{author}{\bibfnamefont{Z.}~\bibnamefont{Fodor}},
  \bibinfo{author}{\bibfnamefont{C.}~\bibnamefont{Hoelbling}},
  \bibinfo{author}{\bibfnamefont{S.}~\bibnamefont{Katz}},
  \bibinfo{author}{\bibfnamefont{S.}~\bibnamefont{Krieg}}, \emph{et~al.},
  \bibinfo{journal}{Phys.Lett.} \textbf{\bibinfo{volume}{B705}},
  \bibinfo{pages}{477} (\bibinfo{year}{2011}), \eprint{1106.3230}.

\bibitem{Bae:2011ff}
\bibinfo{author}{\bibfnamefont{T.}~\bibnamefont{Bae}},
  \bibinfo{author}{\bibfnamefont{Y.-C.} \bibnamefont{Jang}},
  \bibinfo{author}{\bibfnamefont{C.}~\bibnamefont{Jung}},
  \bibinfo{author}{\bibfnamefont{H.-J.} \bibnamefont{Kim}},
  \bibinfo{author}{\bibfnamefont{J.}~\bibnamefont{Kim}}, \emph{et~al.},
  \bibinfo{journal}{Phys.Rev.Lett.} \textbf{\bibinfo{volume}{109}},
  \bibinfo{pages}{041601} (\bibinfo{year}{2012}), \eprint{1111.5698}.

\bibitem{Laiho:2011dy}
\bibinfo{author}{\bibfnamefont{J.}~\bibnamefont{Laiho}} \bibnamefont{and}
  \bibinfo{author}{\bibfnamefont{R.~S.} \bibnamefont{Van~de Water}},
  \bibinfo{journal}{PoS} \textbf{\bibinfo{volume}{LATTICE2011}},
  \bibinfo{pages}{293} (\bibinfo{year}{2011}), \eprint{1112.4861}.

\bibitem{Aubin:2004wf}
\bibinfo{author}{\bibfnamefont{C.}~\bibnamefont{Aubin}} \emph{et~al.},
  \bibinfo{journal}{Phys. Rev.} \textbf{\bibinfo{volume}{D70}},
  \bibinfo{pages}{094505} (\bibinfo{year}{2004}), \eprint{hep-lat/0402030}.

\bibitem{Nakamura:2010zzi}
\bibinfo{author}{\bibfnamefont{K.}~\bibnamefont{Nakamura}} \emph{et~al.}
  (\bibinfo{collaboration}{Particle Data Group}), \bibinfo{journal}{J.Phys.G}
  \textbf{\bibinfo{volume}{G37}}, \bibinfo{pages}{075021}
  (\bibinfo{year}{2010}).

\bibitem{Boyle:2005qc}
\bibinfo{author}{\bibfnamefont{P.}~\bibnamefont{Boyle}} \emph{et~al.},
  \bibinfo{journal}{IBM Journal of Research and Development}
  \textbf{\bibinfo{volume}{49, number 2/3}}, \bibinfo{pages}{351}
  (\bibinfo{year}{2005}).

\bibitem{Boyle:2003mj}
\bibinfo{author}{\bibfnamefont{P.~A.} \bibnamefont{Boyle}},
  \bibinfo{author}{\bibfnamefont{C.}~\bibnamefont{Jung}}, \bibnamefont{and}
  \bibinfo{author}{\bibfnamefont{T.}~\bibnamefont{Wettig}}
  (\bibinfo{collaboration}{QCDOC Collaboration}), \bibinfo{journal}{in
  \textit{Computing in High Energy and Nuclear Physics 2003 Conference
  Proceedings, eConf}} \textbf{\bibinfo{volume}{C0303241}},
  \bibinfo{pages}{THIT003} (\bibinfo{year}{2003}), \eprint{hep-lat/0306023}.

\bibitem{Boyle:2005fb}
\bibinfo{author}{\bibfnamefont{P.~A.} \bibnamefont{Boyle}} \emph{et~al.},
  \bibinfo{journal}{J. Phys. Conf. Ser.} \textbf{\bibinfo{volume}{16}},
  \bibinfo{pages}{129} (\bibinfo{year}{2005}).

\bibitem{Boyle:2009bagel}
\bibinfo{author}{\bibfnamefont{P.~A.} \bibnamefont{Boyle}},
  \bibinfo{journal}{Computer Physics Communications}
  \textbf{\bibinfo{volume}{180/12}}, \bibinfo{pages}{2739}
  (\bibinfo{year}{2009}).

\bibitem{Balog:2009yj}
\bibinfo{author}{\bibfnamefont{J.}~\bibnamefont{Balog}},
  \bibinfo{author}{\bibfnamefont{F.}~\bibnamefont{Niedermayer}},
  \bibnamefont{and} \bibinfo{author}{\bibfnamefont{P.}~\bibnamefont{Weisz}},
  \bibinfo{journal}{Phys.Lett.} \textbf{\bibinfo{volume}{B676}},
  \bibinfo{pages}{188} (\bibinfo{year}{2009}), \eprint{0901.4033}.

\end{thebibliography}

\end{document}